\documentclass[12pt]{article}
\pdfoutput=1

\usepackage{amssymb}
\usepackage{amsmath}
\usepackage{amsthm}
\usepackage{mathtools}
\usepackage[usenames,dvipsnames]{xcolor}
\usepackage{epsfig}
\usepackage{dcolumn}
\usepackage{upgreek}
\usepackage{setspace}
\usepackage{enumitem}
\usepackage{array,multirow,bigdelim,arydshln}
\usepackage{appendix}
\usepackage[export]{adjustbox}
\usepackage{xparse}
\usepackage[utf8]{inputenc}
\usepackage{microtype}
\usepackage{bm}
\usepackage{braket}
\usepackage{dsfont}
\usepackage{nccmath}
\usepackage{datetime}
\usepackage{bm}
\usepackage{multirow}

\usepackage{jheppub}

\usepackage[usenames,dvipsnames]{xcolor}
\usepackage{hyperref}
\hypersetup{
    pdfencoding=unicode,
	colorlinks=true,
	urlcolor=Maroon,
	linkcolor=RoyalBlue,
	citecolor=Maroon,
	pdftitle={What is the iε for the S-matrix?},
	pdfauthor={Holmfridur S. Hannesdottir, Sebastian Mizera},
	pdfdisplaydoctitle=true,
	pdfstartview=FitH,
	linktocpage=true
}
\DeclareMathAlphabet{\mathbbold}{U}{bbold}{m}{n}
\newcommand*{\boldone}{\mathbbold{1}}
\usepackage{physics}
\usepackage{lastpage}

\allowdisplaybreaks
\graphicspath{{figures/}}
\numberwithin{figure}{section}
\DeclareMathOperator{\sgn}{sgn}
\DeclareMathOperator{\Disc}{Disc}
\DeclareMathOperator{\disc}{disc}

\def \be {\begin{equation}}
\def \ee {\end{equation}}
\def \nn {\nonumber}
\def \C {\mathbb{C}}
\def \CC {C}
\def \CP {\mathbb{CP}}
\def \R {\mathbb{R}}
\def \D {\mathrm{D}}
\def \E {\mathrm{E}}
\def \L {\mathrm{L}}
\def \U {\mathcal{U}}
\def \F {\mathcal{F}}
\def \d {\mathrm{d}}
\def \eps {\varepsilon}
\def \V {\mathcal{V}}

\def \leq {\leqslant}
\def \geq {\geqslant}

\def \CC {\mathrm{C}}

\def \JJ {\mathcal{J}}

\def \I {\mathcal{I}}
\def \Z {\mathbb{Z}}
\def \Im {\mathrm{Im}}
\def \GL {\mathrm{GL}}
\def \N {\mathcal{N}}
\def \bubble {\mathrm{bub}}
\def \triangle {\mathrm{tri}}
\def \boxx {\mathrm{box}}
\def \normal {\mathrm{norm}}
\def \Cut {\mathrm{Cut}}

\def \J {\mathrm{J}}
\def \inn {\mathrm{in}}
\def \out {\mathrm{out}}
\def \T {\mathbf{T}}
\def \m {\mathfrak{m}}
\def \Li {\mathrm{Li}}
\def \NN {\mathrm{N}}

\def \VV {\mathrm{V}}
\def \B {\mathrm{B}}
\def \LL {\mathbf{L}}
\def \RR {\mathcal{I}^\ast}
\def \upgamma {\rho}
\def \Y {\mathbf{Y}}
\def \UHP {\mathrm{UHP}}
\def \LHP {\mathrm{LHP}}

\newcommand*\Bell{\ensuremath{\boldsymbol\ell}}
\newcommand{\mc}[1]{\mathcal{#1}}
\newcommand{\cut}{\text{Cut}}
\newcommand{\bdelta}[2]{\bm{\delta}_{#1 , \, #2}}

\DeclareMathOperator*{\sumint}{%
\mathchoice%
  {\ooalign{$\displaystyle\sum$\cr\hidewidth$\displaystyle\int$\hidewidth\cr}}
  {\ooalign{\raisebox{.14\height}{\scalebox{.7}{$\textstyle\sum$}}\cr\hidewidth$\textstyle\int$\hidewidth\cr}}
  {\ooalign{\raisebox{.2\height}{\scalebox{.6}{$\scriptstyle\sum$}}\cr$\scriptstyle\int$\cr}}
  {\ooalign{\raisebox{.2\height}{\scalebox{.6}{$\scriptstyle\sum$}}\cr$\scriptstyle\int$\cr}}
}

\newcommand{\RN}[1]{\textup{\uppercase\expandafter{\romannumeral#1}}}

\title{What is the $i\varepsilon$ for the S-matrix?}
\author{Holmfridur Sigridar Hannesdottir,}\emailAdd{hofie@ias.edu}
\author{Sebastian Mizera}\emailAdd{smizera@ias.edu}
\affiliation{Institute for Advanced Study, Einstein Drive, Princeton, NJ 08540, USA}

\abstract{%
    Can the S-matrix be complexified in a way consistent with causality?
    Since the 1960's, the affirmative answer to this question has been well-understood for $2 \to 2$ scattering of the lightest particle in theories with a mass gap at low momentum transfer, where the S-matrix is analytic everywhere except at normal-threshold branch cuts.
    We ask whether an analogous picture extends to realistic theories, such as the Standard Model, that include massless fields, UV/IR divergences, and unstable particles.
    Especially in the presence of light states running in the loops, the traditional $i\varepsilon$ prescription for approaching physical regions might break down, because causality requirements for the individual Feynman diagrams can be mutually incompatible. We demonstrate that such analyticity problems are not in contradiction with unitarity.
	Instead, they should be thought of as finite-width effects that disappear in the idealized $2\to 2$ scattering amplitudes with no unstable particles, but might persist at higher multiplicity. 
    To fix these issues, we propose an $i\varepsilon$-like prescription for deforming branch cuts in the space of Mandelstam invariants without modifying the analytic properties.
    This procedure results in a complex strip around the real part of the kinematic space, where the S-matrix remains causal.
    In addition to giving a pedagogical introduction to the analytic properties of the perturbative S-matrix from a modern point of view,
    we illustrate all the points on explicit examples, both symbolically and numerically.
	To help with the investigation of related questions, we introduce a number of tools, including holomorphic cutting rules, new approaches to dispersion relations, as well as formulae for local behavior of Feynman integrals near branch points.
}

\arxivnumber{2204.02988}

\begin{document}
	
	\setcounter{tocdepth}{3}
	\maketitle
	\setcounter{page}{3}
	
	\newpage
	\section{\label{sec:introduction}Introduction}
	
	Imprints of causality on the S-matrix remain largely mysterious. In fact, there is not even an agreed-upon definition of what \emph{causality} is supposed to entail in the first place, with different notions including microcausality (vanishing of commutators at space-like separations), macrocausality (only stable particles carrying energy-momentum across long distances), Bogoliubov causality (local variations of coupling constants not affecting causally-disconnected regions), or the absence of Shapiro time advances.
	At the mechanical level, there is presently no check that can be made on S-matrix elements that would \emph{guarantee} that it came from a causal scattering process in space-time.
	Motivated by the intuition from $(0{+}1)$-dimensional toy models, where causality implies certain analyticity properties of \emph{complexified} observables \cite{Toll:1956cya,Nussenzveig:1972tcd}, 
	it is generally believed that its extrapolation to relativistic $(3{+}1)$-dimensional S-matrices will involve similar criteria \cite{Martin:1969ina,Sommer:1970mr,Itzykson:1980rh,bogolubov1989general,Iagolnitzer:1994xv}. Converting this insight into precise results has proven enormously difficult, leaving us with a real need for making analyticity statements sharper, especially since it is expected that they impose stringent conditions on the space of allowed S-matrices. Progress in such directions includes \cite{PhysRev.83.249,PhysRev.91.1267,GellMann:1954db,Wanders:1959rxn,doi:10.1063/1.1704697,Eden:1965aww,rohrlich1965microcausality, PERES1966179,PhysRev.135.B1255,Wanders1965,PhysRev.146.1123,PhysRev.174.1749,AIHPA_1967__6_2_89_0,Chandler:1969bd,Iagolnitzer1969,Peres:1970tr,Adams:2006sv,Grinstein:2008bg,Giddings:2009gj,Camanho:2014apa,Tomboulis:2017rvd,Capatti:2020xjc,Chandorkar:2021viw,deJesusAguilera-Verdugo:2021mvg,Haring:2022cyf}, often under optimistic assumptions on analyticity. This work takes a step towards answering an even more basic question: how do we consistently uplift the S-matrix to a complex-analytic function in the first place?
	
	\paragraph*{\bf Complexification.} Let us decompose the S-matrix operator in the conventional way, $S = \boldone + iT$, into its non-interacting and interacting parts and call the corresponding matrix elements $\T$.
	We want to ask how to extend $\T$ to a function of complex Mandelstam invariants $\T_\C$.
	For example, for $2\to 2$ scattering $\T_\C(s,t)$ would be a function in the two-dimensional complex space $\C^2$ parametrized by the center-of-mass energy squared $s$ and the momentum transfer squared $t$.
	
	Among many motivations we can mention exploiting complex analysis to derive physical constraints via the theory of dispersion relations, complex angular momenta, or on-shell recursion relations; see, e.g., \cite{Martin:1969ina,Sommer:1970mr,Nussenzveig:1972tcd,Itzykson:1980rh,bogolubov1989general,Iagolnitzer:1994xv}. Another incentive stems from the conjectural property of the S-matrix called \emph{crossing symmetry}, which can be summarized by the following practical problem. Let us say that we performed a difficult computation for the positron-electron annihilation process $e^+ e^- \to \gamma \gamma$ at a given number of loops. The question is whether we can recycle this result to obtain the answer for the crossed process, Compton scattering $\gamma e^- \to \gamma e^-$, ``for free'', i.e., by analytic continuation. Unfortunately, the two $S$-matrix elements are defined in disjoint regions of the kinematic space: for $s>0$ and $s<0$ respectively, so in order to even ponder such a connection, one is forced to uplift $s$ to a complex variable.
	
	\begin{figure}
	\centering
	\includegraphics[scale=1.1]{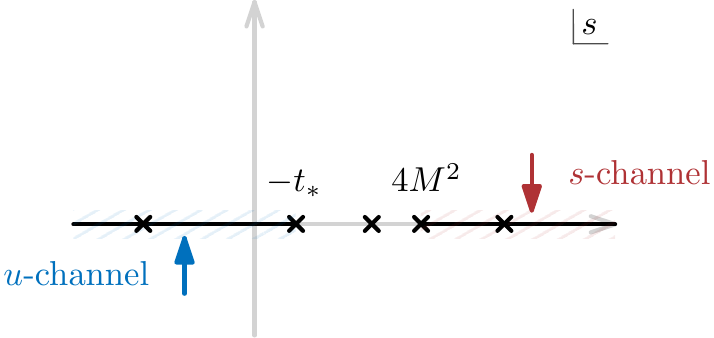}
	\caption{Analytic structure of the matrix element $\mathbf{T}_\C(s,t_\ast)$ for $2\to 2$ scattering of the lightest state of mass $M$ in theories with a mass gap in the complex $s$-plane at sufficiently small fixed $t = t_\ast < 0$. There are two sets of branch cuts (thick lines) corresponding to normal thresholds in the $s$-channel ($s > 4M^2$) and $u$-channel ($u > 4M^2$ or $s < - t_\ast$). The amplitude is real in the Euclidean region between them, which can also feature single-particle poles. The causal way of approaching the physical channels is indicated with arrows. The purpose of this work is to investigate how this picture generalizes to more realistic theories.}
	\label{fig:lightest}
	\end{figure}
	
    Alas, complexifying the S-matrix opens a whole can of worms because it now becomes a multi-valued function with an enormously-complicated branch cut structure. Notwithstanding this obstruction, a lot of progress in understanding the analytic structure has been made for $2\to2$ scattering of the lightest state in theories with a mass gap at low momentum transfer, see, e.g., \cite{Eden:1966dnq,Sommer:1970mr}. An often-invoked application is the pion scattering process $\pi \pi \to \pi \pi$ \cite{Martin:1965jj,Roy:1971tc,Colangelo:2001df,Caprini:2003ta,Pelaez:2004vs,Paulos:2017fhb,Martin:2017ndt,Guerrieri:2018uew,Guerrieri:2020bto,Albert:2022oes}. This setup gives rise to the classic picture of the complex $s$-plane for sufficiently small physical $t = t_\ast < 0$ illustrated in Fig.~\ref{fig:lightest}. In this toy model, there are branch cuts extending along the real axis with $s > 4M^2$ responsible for $s$-channel resonances and similarly for $s < -t_\ast$ for the $u$-channel ones (by momentum conservation $s+t+u = 4M^2$, so $u > 4M^2$, where $M$ is the mass of the lightest particle), with possible poles responsible for single-particle exchanges. In principle, this structure can be argued for non-perturbatively, see, e.g., \cite{Bogolyubov:104088}.
    
    It turns out that, in this case, the causal matrix element $\T$ in the $s$-channel is obtained by approaching $\mathbf{T}_\C$ from the upper-half plane:
    \be\label{eq:S-C}
    \mathbf{T}(s,t_\ast) = \lim_{\eps \to 0^+} \mathbf{T}_\C(s + i\eps, t_\ast)
    \ee
    for $s>4M^2$. Similarly, the $u$-channel needs to be approached from the $s-i\eps$ direction. Because of the branch cut, it is important to access the physical region from the correct side: the opposite choice would result in the $\T$-matrix with anti-causal propagation. Establishing such analyticity properties hinges on the existence of the ``Euclidean region'', which is the interval $-t_\ast < s < 4M^2$ where the amplitude is real and meromorphic; see, e.g., \cite{Eden:1966dnq}.
    
    A closely related question is whether the imaginary part of the amplitude,
	\be\label{eq:ImT-C}
	\Im\, \T(s,t_\ast) = \tfrac{1}{2i}\Big(\T(s, t_\ast) - \overline{\T(s, t_\ast)}\Big)
	\ee
	for physical $s$ is always equal to its discontinuity across the real axis
	\be\label{eq:DiscT-C}
	\Disc_s \T_\C(s,t_\ast) = \lim_{\eps \to 0^+} \tfrac{1}{2i}\Big(\T_\C(s+i\eps, t_\ast) - \T_\C(s-i\eps, t_\ast)\Big).
	\ee
	Recall that the former is the absorptive part of the amplitude related to unitarity, while the latter enters dispersion relations. So far, the only way for arguing why \eqref{eq:ImT-C} equals \eqref{eq:DiscT-C} relies on the application of the Schwarz reflection principle when the Euclidean region is present, but whether this equality persists in more general cases is far from obvious.
    
    It might be tempting to draw a parallel between \eqref{eq:S-C} and the Feynman $i\eps$ prescription, though at this stage it is not entirely clear why the two should be related: one gives a small imaginary part to the external energy, while the other one to the propagators. So what is the connection between \eqref{eq:S-C} and causality? One of the objectives of this work is studying this relationship and delineating when \eqref{eq:S-C} is valid and when it is not.
    
    More broadly, the goal of this paper is to investigate the extension of Fig.~\ref{fig:lightest} to more realistic scattering processes, say those in the Standard Model (possibly including gravity or other extensions), that might involve massless states, UV/IR divergences, unstable particles, etc. Little is known about general analyticity properties of such S-matrix elements. The most naive problem one might expect is that the branch cuts in Fig.~\ref{fig:lightest} start sliding onto each other and overlapping, at which moment the Euclidean region no longer exists and many of the previous arguments break down. But at this stage, why would we not expect other singularities that used to live outside of the $s$-plane to start contributing too? What then happens to the $i\eps$ prescription in \eqref{eq:S-C}? Clearly, before starting to answer such questions we need to understand the meaning of singularities of the S-matrix in the first place. This question is tightly connected to unitarity.
	
\begin{figure}
	    \centering
	    \includegraphics[scale=1]{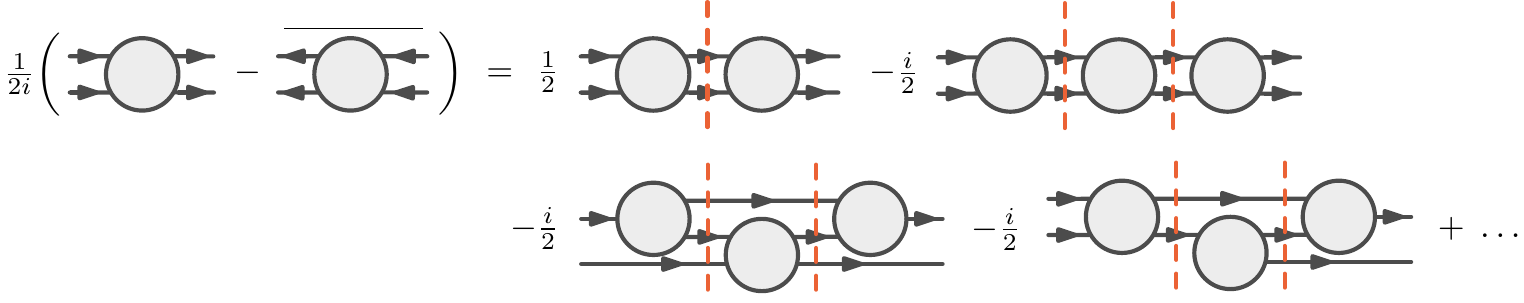}
	    \caption{Schematic illustration of the first few terms of \eqref{eq:ImT2}. The orange dashed lines represent holomorphic cuts. The first two terms on the right-hand side are normal thresholds, while the latter two are examples of anomalous thresholds. Note that the constituent $\T$ elements can have disconnected components, as long as they are not $\boldone$. In Sec.~\ref{sec:cutting} we show that these cutting rules carry over to individual Feynman integrals.}
	    \label{fig:unitarity}
	\end{figure}

	\paragraph{Unitarity and analyticity.}
	Unitarity of the $S$-matrix, $S S^\dagger = \boldone$, encodes the physical principle of probability conservation. Expanded in terms of $T$ and $T^\dagger$, it implies the constraint
	\be\label{eq:ImT1}
	\tfrac{1}{2i}(T - T^\dagger) = \tfrac{1}{2} T T^\dagger.
	\ee
	This statement is useful because it allows us to relate the right-hand side to the total cross-section, in a result known as the optical theorem; see, e.g., \cite{Itzykson:1980rh}. However, in order to be able to probe \emph{complex}-analytic properties of $T$ and manifest all its singularities, it is much more convenient to express the right-hand side as a holomorphic function. To this end, using \eqref{eq:ImT1} we can eliminate $T^\dagger = (\boldone+iT)^{-1}T$, which plugging back into the same equation gives
	\be\label{eq:ImT2}
	\tfrac{1}{2i}(T - T^\dagger) = -\tfrac{1}{2}\sum_{c=1}^{\infty} (-i T)^{c+1},
	\ee
	where we expanded the right-hand side perturbatively in $T$. This equation is central to understanding the analytic structure of the matrix elements $\mathbf{T}$, because it gives an on-shell relation expressing the left-hand side as an infinite sum over multiple glued copies of $\mathbf{T}$ by $c$ unitarity cuts. Here, a \emph{unitarity cut} means a sum over all intermediate on-shell states propagating with positive energy together with an integration over their phase space. When the initial and final states are the same, the left-hand side equals $\Im\, \mathbf{T}$.  The first few terms are illustrated graphically in Fig.~\ref{fig:unitarity}.
	
	One can easily show that \eqref{eq:ImT2} 
	carries over to individual Feynman diagrams in perturbation theory, where the sum on the right-hand side truncates at a finite number of terms, see Sec.~\ref{sec:cutting}. We call them \emph{holomorphic} unitarity cuts to stress that, unlike those stemming from \eqref{eq:ImT1}, there is no complex-conjugate involved; see also \cite{Coster:1970jy,Bourjaily:2020wvq,Blazek:2021olf,Blazek:2021zoj}. They manifest analyticity and the singularity structure of the S-matrix, at a cost of obscuring positivity.
	
	The place in the kinematic space where each individual term starts contributing is called a \emph{threshold}. At the threshold, the phase space for the corresponding process just opens up and allows for only classical propagation \cite{Coleman:1965xm}. It can be a violent event giving rise to singularities, though it does not have to. Moreover, since threshold configurations correspond to classical scattering, determining their positions in the kinematic space is a concrete algebraic problem \cite{Bjorken:1959fd,Landau:1959fi,10.1143/PTP.22.128}. It turns out there is a loophole in the above logic: there might also exist terms in \eqref{eq:ImT2} that always contribute and hence do not have a corresponding threshold where they switch on. This feature will be important in what follows.
	
	\paragraph*{\bf Normal and ``anomalous'' thresholds.}
	We can classify the terms appearing on the right-hand side of \eqref{eq:ImT2} as follows.
	Those giving rise to a one-dimensional chain of $\mathbf{T}$'s are called \emph{normal} thresholds, while all the remaining contributions are known as the \emph{anomalous} thresholds, cf. Fig.~\ref{fig:unitarity} \cite{PhysRev.111.1187,PhysRev.114.376,Nambu1958}.\footnote{Historically, at the time when consequences of unitarity were not fully understood, the term \emph{anomalous} was employed to mean \emph{unexpected} and does not have anything to do with anomalies in the modern physics parlance.} They are interchangeably called \emph{Landau singularities} \cite{Landau:1959fi}. Whenever allowed kinematically, each term possibly contributes a divergence and makes the imaginary part of the amplitude jump discontinuously.
	Existence of both types of thresholds is a direct consequence of unitarity and is an exact non-perturbative statement, see Sec.~\ref{sec:unitarity} for more details.
	
	Unlike normal thresholds, which are resonances localized in space that can be easily detected in particle colliders, anomalous thresholds are spatially spread out. Together with the fact that their contributions are often suppressed, their presence was never \emph{directly} confirmed experimentally, though their observable effects have been recently studied, e.g., in hadron spectroscopy \cite{Liu:2015taa,Liu:2017vsf,Guo:2017wzr,Abreu:2020jsl}, tetraquark physics \cite{Braaten:2022elw,Achasov:2022onn}, $\mathrm{b}\bar{\mathrm{b}}\mathrm{H}$ production or $\mathrm{ZZ} \to \mathrm{ZZ}$ scattering in the Standard Model \cite{Denner:1996ug,Boudjema:2008zn,Ninh:2008cz}, and LHC phenomenology more broadly \cite{NLOMultilegWorkingGroup:2008bxd,Passarino:2018wix}, among others. 
	A surprising result is that the complexified S-matrix elements $\mathbf{T}_\C$ do not have any conceptually new types of singularities, at least not in perturbation theory. One encounters only a generalization of normal and anomalous thresholds with complexified momenta \cite{Bjorken:1959fd,Landau:1959fi,10.1143/PTP.22.128}.

	How is this structure consistent with the simple picture in Fig.~\ref{fig:lightest}, which shows only normal thresholds? It is straightforward to show that for $2\to 2$ scattering of \emph{stable} particles, anomalous thresholds do not have support in the physical regions, see Sec.~\ref{sec:analyticity-stable}. They can, however, contribute anywhere else in the $(s,t)$-planes. Additionally, if one assumes that the external masses are strictly positive and considers sufficiently small $t_\ast$ (such that the Euclidean region exists), the $s$-plane does not have any singularities beyond the normal-threshold branch cuts at all, leading to Fig.~\ref{fig:lightest}. Even so, one is often interested in computing the discontinuity $\Disc_s \T_{\C}(s,t_\ast)$ or double-spectral density $\Disc_t \Disc_s \T_\C (s,t)$, both of which correspond to an analytic continuation onto unphysical sheets, where anomalous thresholds abound. Naturally, they also enter algebraic expressions for scattering amplitudes as branch points and poles, regardless of whether they contribute to physical scattering or not.
	
	Next to nothing is known about physical-sheet analyticity of the S-matrix once the above assumptions are relaxed. The most general result is that for any $t_\ast < 0$, the complexified $\mathbf{T}_\C$ is analytic for sufficiently large complex $s$ \cite{Bros:1964iho,Bros:1965kbd}. However, the sufficiently-large condition was never quantified in terms of the mass gap or other scales in the problem, nor was a physical explanation for this result understood. The condition \eqref{eq:S-C} was taken as an assumption. The results of \cite{Bros:1972jh} already hint at difficulties with defining $\T$ as a boundary value of a single analytic function $\T_\C$ at higher multiplicity.

	To recapitulate, there are several directions in which the intuition of Fig.~\ref{fig:lightest} might break down, all connected with the appearance of singularities implied by unitarity beyond the normal thresholds. The arena for future exploration involves higher-multiplicity processes, scattering of massless and heavy states, or unstable particles.
	For instance, one might worry that consistent implementation of unitarity equations, \eqref{eq:ImT1} or \eqref{eq:ImT2}, means that once unstable particles propagate as internal particles, one also needs to consider matrix elements $\mathbf{T}$ with unstable external particles as well. Nonetheless, one can show that these cuts resum to zero, see Sec.~\ref{sec:shifts-widths} and \cite{Veltman:1963th}.
	The discussion of external unstable particles was often dismissed in the S-matrix literature on the grounds of not providing well-defined asymptotic states (similar logic would lead us to exclude massless particles). At any rate, the topic of analytic properties of such processes was not pursued seriously in the 1960's, because no unstable elementary particles were discovered at that stage \cite{Mandelstam_1962}.

	But in practice we do not prepare scattering states at past infinity and we certainly do scatter unstable particles in collider experiments, so it is important to understand analytic properties of such scattering amplitudes for phenomenological purposes.
	Examples include scattering of charged pions or kaons, and electroweak processes involving external W/Z bosons.

	\paragraph{Causality and analyticity.}
	In this work we are primarily interested in the question of how to extend $\T$ to $\T_\C$ in a way consistent with causality. In particular: is the former always a boundary value of the latter according to \eqref{eq:S-C}? If so, is the imaginary part $\Im\, \T$ equal to the discontinuity $\Disc_s \T_\C$?
	To have a concrete handle on such questions, we work in perturbation theory, where one can isolate such issues from UV/IR divergences, since the latter can be easily regularized.
	
	At the level of the individual Feynman diagrams, one can in principle employ the textbook $i\eps$ prescription, guaranteeing causal propagation along each internal line.
	One would naively expect that this puts an end to the discussion. However---as we shall see---not all the $i$'s have been dotted and not all $\eps$'s have been curled in the understanding of this prescription, even for simple examples at low-loop orders.
	
\begin{figure}
		\centering
		\includegraphics[scale=1.04]{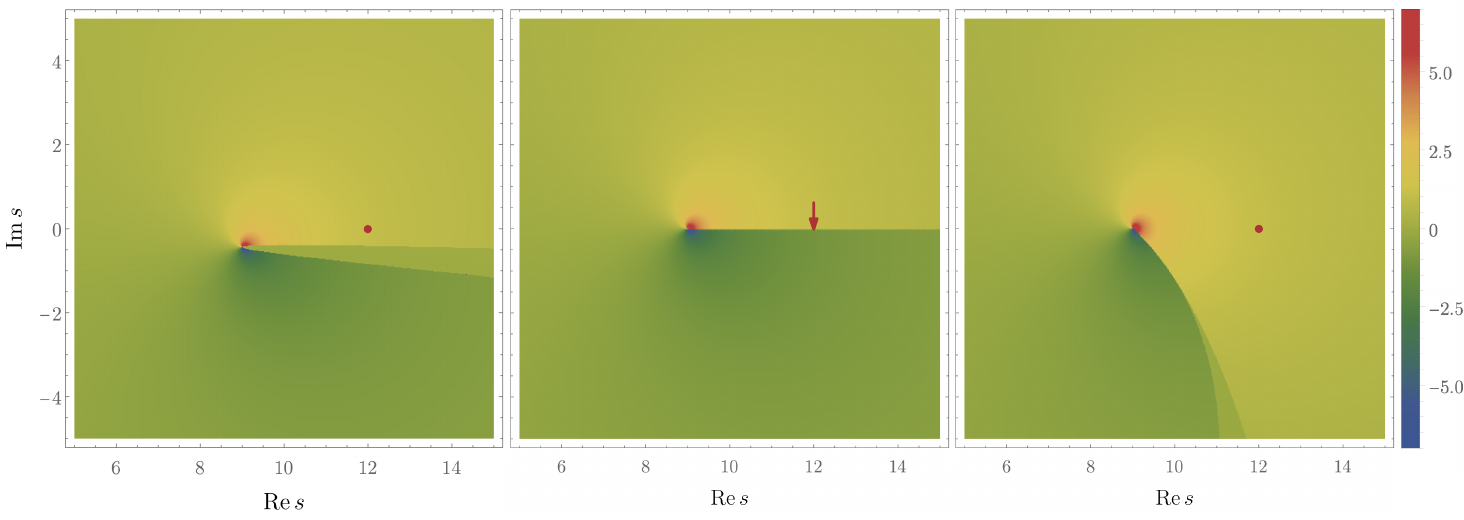}
		\caption{\label{fig:Fig1}Three distinct ways of implementing an $i\eps$ prescription, illustrated on the imaginary part of the bubble diagram with masses $m_1=1$ and $m_2=2$. The physical value, say at $s=12$, is indicated in red. Left: Feynman $i\eps$ displaces the original branch point at $s = (m_1+m_2)^2 = 9$ to $s \approx 9.003 - 0.450i$. Middle: The $s+i\eps$ prescription for approaching the branch cut from the upper-half plane. Right: Branch cut deformation revealing the causal amplitude without modifying analytic properties. 
		We set $\eps = \tfrac{1}{10}$ to make the effects more visible. See Sec.~\ref{sec:primer} for more details.}
	\end{figure}

	Before jumping to the general story, let us illustrate three options of how to implement an $i\eps$ prescription on the simplest example of the bubble integral with a single normal-threshold resonance, see Fig.~\ref{fig:Fig1}. Sec.~\ref{sec:primer} goes into much more detail of this example, which also serves as a primer to the ideas of the analytic S-matrix.

	\paragraph{Feynman $i\eps$ prescription.}
	Recall that the aforementioned Feynman $i\eps$ prescription deforms each propagator to be proportional to $1/(q^2 - m^2 + i\eps)$,
	where $q^\mu$ and $m$ are the momenta and masses of the internal states.
	Equivalently, one can think of it as giving the internal particle a tiny width, so that $m^2 \to m^2 - i\eps$. It is important to remember that the $i\eps$ does not have any physical meaning, and is merely a trick that lets us choose the correct integration contour for the Feynman propagator. The limit $\eps \to 0^+$, if it exists, has to be taken at the end of the computation. This prescription leads to two significant problems.
	
	On the conceptual side, it inevitably \emph{modifies} the analytic structure of the amplitude by an unphysical dimensionful parameter $\eps$, e.g., by deforming the positions of resonances, see Fig.~\ref{fig:Fig1} (left). In this example, the branch cut starts at $s \approx 9.003 - 0.450i$ instead of $s=9$. As exemplified in Fig.~\ref{fig:Fig1}, it also has other undesirable properties, such as introducing multiple branch cuts sprouting from a single branch point, which ``condense'' onto each other in the $\eps \to 0^+$ limit. Therefore, it cannot be taken as a fundamental concept if we were to think of the full S-matrix without a reference to perturbation theory. Moreover, the Feynman $i\eps$ only works for kinematics in the physical region (when all the external energies are real), and might require modifications once the kinematics is complexified.
	
	On a more practical side, while Feynman integrals are guaranteed to converge for fixed $\eps > 0$, we are ultimately interested in the limit $\eps \to 0^+$, where the numerical errors become worse and worse close to branch cuts. We can do better.
	
	\paragraph{Kinematic $s+i\eps$ prescription.} As an alternative, one may attempt to perform an $i\eps$ prescription purely in terms of the external kinematics (masses, of course, cannot be deformed because that would correspond to an off-shell continuation of the S-matrix). For instance, the prescription mentioned earlier for $2\to 2$ scattering is to approach the $s$-channel physical region from the direction of $s+i\eps$ according to \eqref{eq:S-C}, see Fig.~\ref{fig:Fig1} (middle).
	We are not aware of any convincing physical explanation why such a prescription should be equivalent to the Feynman $i\eps$ or be related to causality in some other way. Its previous derivations were tied to the existence of the Euclidean region \cite{Wu:1961zz,Eden:1966dnq}, which happens for example, for scattering of the lightest state in a theory with a mass gap, with no more than $7$ external particles \cite{Mizera:2021ujs}.

	As a matter of fact, the purpose of this work is to point out that the $s + i\eps$ prescription is not always the choice consistent with causality. What is more, whether to choose $s \pm i\eps$ for a given $s$ might differ from diagram to diagram (or even not be well-defined in the first place), which means that there are situations in which there is no global way of assigning it uniformly for the full $\T_\C$.

	So what is the $i\eps$ for the S-matrix? Given the above discussion, one is essentially forced to find a consistent way of \emph{deforming} branch cuts in the kinematic space in order to access the scattering amplitude on the sheet consistent with causality.
	
	\paragraph{Branch cut deformations.}
	The final approach is based on contour deformations for Feynman integrals, which is indeed what the Feynman $i\eps$ was supposed to imitate in the first place. This prescription does not modify the analytic structure, but instead has the effect of ``bending away'' branch cuts in the kinematic space to reveal the physical value of $\T_\C$ coinciding with that of $\T$, see Fig.~\ref{fig:Fig1} (right).
	
	Recall that scattering amplitudes are multi-sheeted functions of complexified kinematic invariants. The choice of branch cuts has absolutely no physical significance, as long as the correct, physical, sheet has been selected. However, a specific \emph{representation} of the scattering amplitude---such as that using Feynman integrals---does select a particular form of branch cuts and the associated way they need to be $i\eps$-deformed. We review it in Sec.~\ref{sec:branch-cuts}. As we will see, this approach also has the advantage of fixing the numerical convergence issues in practical computations. It seems to be the most invariant way of understanding the $i\eps$, since it does not modify the physical content of the S-matrix.
	
	However, beyond the simplest of examples, one encounters situations with multiple---possibly overlapping---branch cuts contributing to the same process. At present there is no systematic way of understanding which way the cuts are supposed to be deformed (or, in fact, whether they can be deformed at all). The goal of this paper is to investigate this issue and rephrase it as an algebraic problem.
	
	\paragraph{Algebraic formulation.}
	A convenient language for phrasing the above questions is the worldline formalism, where each internal edge $e$ is endowed with a Schwinger proper time $\alpha_e$. Recall that scattering amplitudes in this language can be understood as integrals over all worldline topologies and geometries: in other words, summing over all Feynman diagrams and integrating over all Schwinger parameters. The analytic structure of each individual diagram is governed by the on-shell action $\V$, which is a function of Mandelstam invariants, masses, and the Schwinger parameters.
	
	Threshold singularities can then be understood as classical saddle points obtained by extremizing $\V$. Moreover, a branch cut appears whenever $\V=0$ for some values of Schwinger parameters. One is then left with a choice of how to deform away from it, and $\Im\, \V > 0$ turns out to be the only choice consistent with causality, or more precisely, the Feynman $i\eps$ prescription. They give an extension of Landau equations \cite{Bjorken:1959fd,Landau:1959fi,10.1143/PTP.22.128} that not only talk about branch points, but also branch cuts and causal branches. We can summarize them as follows:
    \begin{alignat}{2}
    \V = 0 \quad&\text{ for any $\alpha$'s } \quad&&\Leftrightarrow\quad \text{branch cut}\\
    \partial_{\alpha_e}\V = 0 \quad&\text{ for any $\alpha$'s } \quad&&\Leftrightarrow\quad \text{branch point}\\
    \Im\,\V > 0 \quad&\text{ for all $\alpha$'s} \quad&&\Leftrightarrow\quad \text{causal branch}
    \end{alignat}
    Indeed, the three options presented in Fig.~\ref{fig:Fig1} can be seen as different ways of implementing the final constraint, respectively: putting the $i\eps$ by hand, deforming the external kinematics, and deforming the Schwinger parameters.
    There are multiple other applications in which properties of $\V$ can be exploited to study analyticity. For example, the sign of $\partial_s \V$ on the cut $\V=0$ can be used to determine the difference between the imaginary part of the amplitude \eqref{eq:ImT-C} and its discontinuity \eqref{eq:DiscT-C}, see Sec.~\ref{sec:im-disc}.
	The main takeaway is that the sheet structure, at least close to the physical regions, can be largely determined without explicit computations! 
	
	To complete the story, we give a prescription for deforming Schwinger parameters with small phases so that $\Im\, \V > 0$ is always satisfied away from branch points \cite{Mizera:2021fap}, which is tantamount to deforming branch cuts in the kinematic space revealing the causal branch of the S-matrix.
	
	\begin{figure}
	    \centering
	    \includegraphics[scale=1.1]{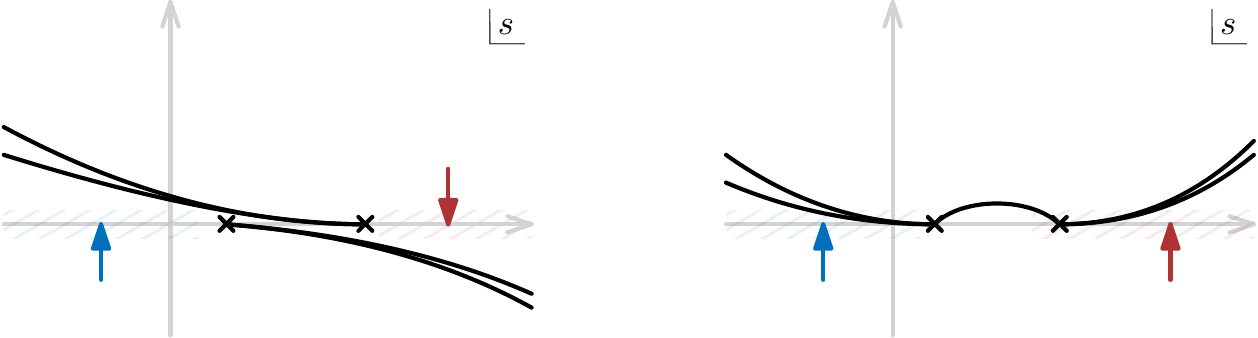}
	    \caption{Examples of branch cut deformations is the complex $s$-plane at fixed $t=t_\ast < 0$. Causal directions for approaching the physical regions are indicated with red arrows. Left: The simplest case in which cuts are deformed so that analytic continuation between the upper- and lower-half planes is possible, see Sec.~\ref{sec:ExampleI}. Right: The opposite situation in which the two half-planes are disconnected by branch cuts, see Sec.~\ref{sec:ExampleII}. Note that, in general, $i\eps$ procedures introduce multiple branch cuts sprouting from every branch point.}
	    \label{fig:deformations}
	\end{figure}
	
	This picture clarifies how the conventional use of the term ``physical sheet''---which is normally defined to mean the region of the complexified kinematic space accessible without crossing branch cuts---is tied to the Schwinger parameters $\alpha_e$. Since contour deformations in the Schwinger-parameter space move around branch cuts in the kinematic space, physical-sheet singularities are those and only those corresponding to saddles with $\alpha_e \geq 0$.
	
	\paragraph{How to approach the physical regions?}
	
	In order to illustrate the above points, we consider deformations of branch cuts that otherwise run across the whole real axis in the complex $s$-plane. We can distinguish between two possibilities: either branch cut deformations allow us to reconnect the upper- and lower-half planes, or they do not, see Fig.~\ref{fig:deformations}.
	
	Using these techniques we can settle the question of how to approach the physical regions in a way consistent with causality, see Fig.~\ref{fig:deformations} (left). First of all, we close a gap in the previous literature by proving that \eqref{eq:S-C} is always valid for $2 \to 2$ scattering of \emph{stable} external particles regardless of the masses and momentum transfer $t_\ast$ at every order in perturbation theory, see Sec.~\ref{sec:analyticity-stable} (previously this was either taken as an assumption \cite{Bros:1964iho,Bros:1965kbd}, or hinged on the existence of the Euclidean region \cite{Wu:1961zz,Eden:1966dnq}).

	However, when \emph{unstable} particles are included on the external states, we no longer expect that branch cuts can be deformed to connect the upper- and lower-half planes, see Fig.~\ref{fig:deformations} (right). This issue can be traced back to the fact that certain unitarity cuts are supported everywhere along the real $s$-axis. The additional discontinuity across the real $s$-axis grows with the decay width of the external particles, which is also why it is not present in the stable-particle case. Moreover, \eqref{eq:S-C} might no longer be true and discontinuities no longer have to equal the imaginary parts. A simple example is one for which both the $s$- and $u$-channels ought to be approached from the $s-i\eps$ direction, cf.~Fig.~\ref{fig:deformations} (right), which is why it does not constitute a counterexample to crossing symmetry. We can equivalently think of such a process as being embedded in a higher-point scattering amplitude, in which case we are still probing analyticity of the stable S-matrix elements, except at higher multiplicity.
	
\begin{figure}
	    \centering
	    \includegraphics[scale=1.1]{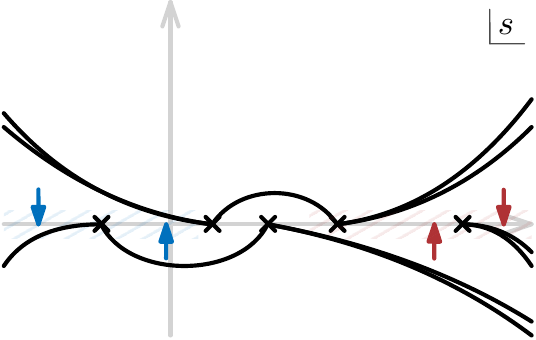}
	    \caption{Illustration of the analytic structure in the $s$-plane after summing over multiple diagrams involving external unstable particles. The physical sheet is confined to a region near the real axis between the branch cuts, see Sec.~\ref{sec:ExampleIII}. Once decay widths are included, the branch points move into the complex plane, but the qualitative picture remains the same.}
	    \label{fig:generic}
	\end{figure}

	Unfortunately, causality requirements for different diagrams might be mutually exclusive: certain diagrams might call for approaching the $s$-channel from the $s-i\eps$ direction, while others from the $s+i\eps$. At the level of the full amplitude $\T_\C$, after summing over individual diagrams, branch cut deformations are necessary to reveal the physical sheet in the strip around the real axis. 
	Hence, we expect the analytic structure of such S-matrix elements to be closer to that depicted in Fig.~\ref{fig:generic}. We anticipate that such problems disappear in the idealized zero-width limit of $2\to2$ scattering amplitudes, but might in general persist at higher multiplicity.
	
	\paragraph{Case studies.}
	
	To demonstrate these phenomena, we consider a couple of one-loop case studies in Sec.~\ref{sec:branch-cut-deformations}.
	These specific examples are chosen because they are complicated enough to exemplify the features we want to emphasize, while at the same time being tractable computationally.
	
	Before starting, one may reasonably ask why such difficulties have not shown up in practical computations of Feynman integrals over the last decades. The main reason is that for phenomenological purposes, one is only interested in the S-matrix elements $\T$ in the physical region, which do not have to be analytic. But more importantly, computing Feynman integrals symbolically is extremely difficult and the way they generally come out of the computation is with branch cuts placed ``randomly'' in the complex $(s,t)$-planes, not necessarily following the neat pictures in Fig.~\ref{fig:deformations} or even Fig.~\ref{fig:lightest}.
	In particular, we stress that none of the problems described in this paper are in contradiction with any physical property of $\T$. They only affect how we extend $\T$ to a complex-analytic function $\T_\C$.
	
	The strongest cross-check on the above assertions comes from unitarity. To this end, we verify that the holomorphic cutting rules would not work unless we approach the real axis from the directions indicated above.
	We also check our conclusions from other points of view, including numerical evaluation and a direct computation of analytic expressions, together with a set of other tools for Feynman integrals introduced in Sec.~\ref{sec:general}.
	
	In particular, we would like to highlight two complementary tools we introduce to understand the global and local analytic aspects of Feynman integrals.
	
	\paragraph{Generalized dispersion relations.} 
	One of the hallmarks of the analytic S-matrix program are the dispersion relations \cite{GellMann:1954db,Mandelstam_1962}. They relate $\T_\C(s_\ast,t_\ast)$ for some point $s_\ast$ to an $s$-integral over the discontinuity $\Disc_s \T_\C(s,t_\ast)$, and can be realized when the Euclidean region exists. While in such cases the discontinuity equals the imaginary part $\Im\, \T_\C(s,t_\ast)$, this might no longer be true in general, leaving us with two possibilities for how to write down generalized dispersion relations: either integrating the discontinuity or the imaginary part. We consider both options and write down the corresponding relations in  Sec.~\ref{sec:dispersion}.
	
	Our derivation exploits a trick, in which contour deformations in the $s$-plane are performed at the level of the \emph{integrand}, as opposed to the full Feynman \emph{integral}. This allows us to write down dispersion relations in any linear combination of Mandelstam invariants (for example, fixed-impact-parameter or fixed-angle) without restrictions on the masses of particles or momenta. Moreover, they give a new handle on studying analyticity of Feynman integrals.
	
	\paragraph{Nature of singularities.}
	Once singularities of Feynman integrals are understood as coming from the worldline saddle points, it is natural to ask whether one can study fluctuations around such classical configurations. Note that in our applications, saddles are already tied with specific points in the kinematic space. This means that studying fluctuations around saddles also captures the local behavior of Feynman integrals near the singularities. 
	
	We consider a special class of singularities coming from isolated and non-degenerate saddle points, i.e., those that do not lie on continuous manifolds in the Schwinger-parameter space, along the lines of \cite{Landau:1959fi}. In Sec.~\ref{sec:bound} we show that such singularities of multi-loop Feynman integrals cannot be worse than at tree level. To be more precise, calling the position of the singularity $\Delta = 0$ (for example, $\Delta = s - 4m^2$ for a normal threshold) we can, in local quantum field theories, get a behavior proportional to
	\be
	\frac{1}{\Delta}, \qquad \frac{1}{\sqrt{\Delta}}, \qquad \log \Delta,
	\ee
	but never worse, for every one-vertex irreducible component of a Feynman diagram. We show that this statement is a direct consequence of analyticity of the S-matrix. Moreover, in Sec.~\ref{sec:threshold-expansion} we provide an explicit expression that determines the coefficient of such a leading divergence and specifies the direction of the branch cut extending from it.
	
	\paragraph{Conventions and notation.} We use mostly-minus signature in $\D$ space-time dimensions. The incoming momenta are $p_i^\mu$, while the outgoing ones $-p_i^\mu$, such that the overall momentum conservation reads $\sum_{i=1}^{n} p_i^\mu = 0$ for an $n$-point process. Masses of the external particles are denoted by $M_i$ with $M_i^2 = p_i^2$ and internal ones $m_e$. For $n=4$ we use the notation
	\be
	s = (p_1 + p_2)^2, \qquad t = (p_2 + p_3)^2, \qquad u = (p_1 + p_3)^2,
	\ee
	and the momentum conservation $s+t+u=\sum_{i=1}^{4} M_i^2$ is always solved for $u$, leaving $s$ and $t$ as the only independent Mandelstam invariants. For a given Feynman diagram, $\L$ is the number of loops, while $n$ and $\E$ denote the number of external legs and internal edges (propagators) respectively.
	
	\newpage
	\section{\label{sec:unitarity}Unitarity implies anomalous thresholds}
	
	Unitarity of the S-matrix, $S S^\dag = \boldone$, embodies the physical principle of probability conservation. In this section, we show how unitarity underpins the singularity structure of scattering amplitudes, leading to both normal- and anomalous-threshold singularities. Using a repeated expansion of the unitarity equation in terms of the transfer matrix $T$, we formulate a holomorphic version of the traditional cutting rules, that have the advantage of manifesting the analytic structure of the cuts. We discuss the implications of the cutting rules when unstable particles are involved in the scattering process: we derive how to cut through dressed propagators, and show that cuts through undressed propagators for unstable particles resum to zero.
	
	\subsection{Holomorphic unitarity equation}
	
	It is traditional to separate the non-scattering contribution from $S$ and call the interacting term $T$:
	\be
	S = \boldone + i T.
	\ee
	The normalization is a convention.
	Recall that the individual matrix elements of $T$ contain momentum-conserving delta function $\bdelta{\inn}{\out} = \left(2\pi\right)^\D \delta^\D\! \left(P_{\out}+P_{\inn}\right)$ and can be normalized to
	\be
	\bra{\out}T\ket{\inn} = \bdelta{\inn}{\out}\, \T_{\inn \to \out},
	\ee
	where $P_\inn$ and $-P_\out$ denote the total incoming and outgoing momenta respectively and $\D$ is the space-time dimension. Similarly, matrix contractions can be resolved using the identity
	\be
	\boldone = \sumint_I \ket{I} \bra{I},
	\ee
	where the sum-integral goes over all the intermediate states in a given theory as well as their phase space (we will return to the normalization in due time).
	
	As a consequence of unitarity, $T$ has to satisfy
	\be\label{eq:unitarityT}
	\tfrac{1}{2i}(T - T^\dagger) = \tfrac{1}{2} T T^\dagger.
	\ee
	For example, when the incoming and outgoing states are identical, $\ket{\inn} = \ket{\out}$, the left-hand side is the imaginary part of $T$. It corresponds to forward scattering, explicitly:
	\be
	\Im\, \T_{\inn\to\inn} = \tfrac{1}{2} \sumint_I \bdelta{\inn}{I} |\T_{\inn \to I}|^2 ,
	\label{eq:Impositive}
	\ee
	where $\bdelta{\inn}{I}$ imposes that the momenta of the intermediate states sum to the total momentum.
	Since the argument of the sum-integral is a modulus squared and the measure is positive, the right-hand side is manifestly positive too. In fact, a textbook result (see, e.g., \cite{Itzykson:1980rh}) is that the right-hand side is proportional to the total cross-section for the decay of $\ket{\inn}$. This is known as the optical theorem, guaranteeing that $\Im\, \T_{\inn\to\inn} \geq 0$.
	
\begin{figure}
		\centering
		\includegraphics[scale=1]{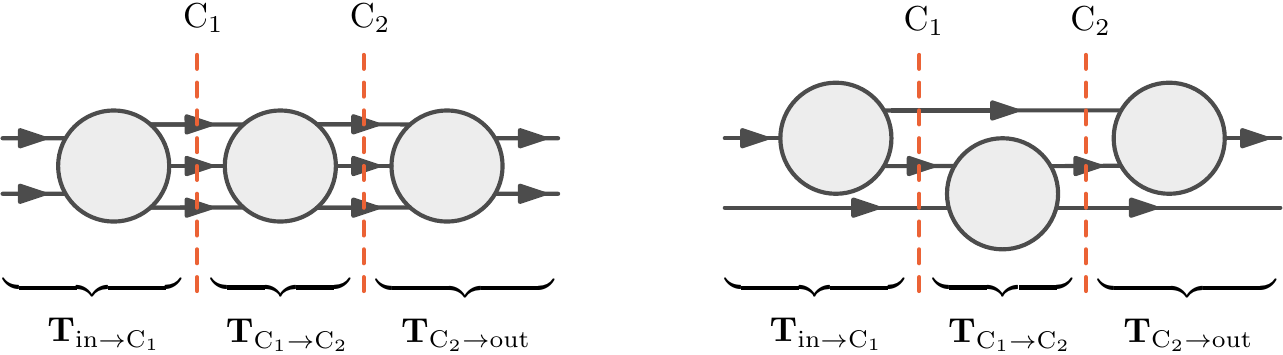}
		\caption{\label{fig:cuts}Examples of two $c=2$ terms contributing to \eqref{eq:ImT3}: a normal threshold (left) and an anomalous one (right).}
	\end{figure}

	For the purpose of studying the analytic properties, however, we can extract more information from \eqref{eq:unitarityT}. Notice that its right-hand side contains holomorphic and anti-holomorphic contributions. We can fix this by plugging in $T^\dagger$ from the left-hand side into the right-hand side as $T^\dagger = T(\boldone - iT^\dagger)$: 
	\begin{align}\label{eq:ImT2a}
	\tfrac{1}{2i}(T - T^\dagger) &= \tfrac{1}{2} (T^2 - iT^2 T^\dagger)\nn\\
	&= \tfrac{1}{2} (T^2 - iT^3 - T^3 T^\dagger)\\
	&= \tfrac{1}{2} (T^2 - iT^3 - T^4 + i T^4 T^\dagger)\nn\\
	&= \cdots,\nn
	\end{align}
	where starting from the second line we kept plugging in the same expression over and over. Since the trend is clear, formally we can write
	\be\label{eq:ImT3}
	\boxed{
	\tfrac{1}{2i}(T - T^\dagger) =   -\tfrac{1}{2}\sum_{c=1}^{\infty} (-i T)^{c+1}.}
	\ee
	Of course, the same result could have been obtained by eliminating
	$T^\dagger = (\boldone + iT)^{-1}T$ from \eqref{eq:unitarityT}, plugging it back in, and expanding the result perturbatively in $T$ (note that each line in \eqref{eq:ImT2a} gives a valid non-perturbative constraint with a finite number of terms). More explicitly, at the level of matrix elements $\T$, this sum becomes
	\begin{align}\label{eq:ImT4}
	\tfrac{1}{2i}(\T_{\inn\to\out} - \overline{\T_{\out\to\inn}}) = \tfrac{i}{2} \!\!\!\! \sum_{\substack{\text{holomorphic}\\ \text{cuts C}}}\!\!\! \left(-i\right)^c \, \cut_{\mathrm{C}} \, \T_{\inn \to \out},
	\end{align}
	where each $\mathrm{C} = \mathrm{C}_1 \cup \mathrm{C}_2 \cup \cdots \cup \mathrm{C}_c$ is a collection of $c$ unitarity cuts, see Fig.~\ref{fig:cuts}. Each individual term is simply given by
	\be\label{eq:holomorphic-cuts}
	\cut_{\mathrm{C}} \T_{\inn \to \out} = \sumint_{\mathrm{C}_1, \mathrm{C}_2,\ldots,\mathrm{C}_c} \bdelta{\inn}{\mathrm{C}_1} \bdelta{\mathrm{C}_1}{\mathrm{C}_2} \cdots \bdelta{\mathrm{C}_{c-1}}{\mathrm{C}_c}\,
	\T_{\inn\to \mathrm{C}_1} \T_{\mathrm{C}_1\to \mathrm{C}_2} \cdots \T_{\mathrm{C}_c\to \out} \,.
	\ee 
	Note that we label cuts by the set of edges that are on shell, and there might be different ways of gluing the diagrams that result in the same $\CC$. In such cases, \eqref{eq:ImT4} might have cancellations. To distinguish between the cuts obtained with \eqref{eq:Impositive}, we call \eqref{eq:cuts} \emph{holomorphic cuts}, since they do not have an additional complex conjugate.
	
	Notice that \eqref{eq:ImT3} is much less economical than \eqref{eq:unitarityT}, so in practical computations of the left-hand side, one would prefer the latter. The advantage of writing out \eqref{eq:ImT3} is that it makes the answer holomorphic, and the singularity structure completely manifest. See also \cite{Coster:1970jy} for related work.
	
	Each term in the sum corresponds to $c$ unitarity cuts \cite{Cutkosky:1960sp}, through which only on-shell states with positive energies are allowed to flow.\footnote{We remind the reader that the cutting rules of Cutkosky are not directly related to generalized unitarity cuts in the sense of \cite{Bern:2011qt}, which do not impose positivity of energies.} Of course, a given propagator can be put on-shell only once, so neither cutting the external legs nor cutting the internal edges twice has any effect. Two examples of terms with $c=2$ are shown in Fig.~\ref{fig:cuts}. In both cases, gluing of the matrix elements of $T$ is done by $3$-particle intermediate states, but they are wired differently. Of course, each diagram contributes to the sum only if it is allowed kinematically. The critical value of kinematics, where a given term is on the verge of being allowed, is called a \emph{threshold}. At such a kinematic point, the phase space for propagation of internal particles is vanishingly small and hence allows only for configurations where all the scattered particles propagate as if the process was 
	classical~\cite{Coleman:1965xm}. 
	This configuration can lead to a violent event, causing a singularity of the S-matrix, although whether it actually does depends on the details of the interaction vertices and the space-time dimension. 
	Unitarity, therefore, predicts an infinite number of potential singularities.
	
	\subsection{Normal and anomalous thresholds}
	
	We will distinguish between two basic classes of thresholds. Whenever the constituents $\T_{\mathrm{C}_{i-1}\to \mathrm{C}_i}$ and $\T_{\mathrm{C}_{i} \to \mathrm{C}_{i+1}}$ are connected by all the intermediate particles in each cut $\mathrm{C}_i$---such as on the left panel of Fig.~\ref{fig:cuts}---the threshold happens when the center-of-mass energy squared $s$ equals
	\be\label{eq:s-normal}
	s = (m_1 + m_2 + \ldots)^2,
	\ee
	where $m_1, m_2, \ldots$ are the masses of intermediate states. These are known as the \emph{normal} thresholds and are completely understood in the S-matrix theory. One of their characteristics is that they are purely localized in space: momenta of all the intermediate on-shell states are proportional to the total incoming momentum $P_{\inn}^\mu$ and hence we can choose the center-of-mass frame, in which they do not have any spatial components.
	
	All the remaining thresholds---such as the one on the right panel of Fig.~\ref{fig:cuts}---are called \emph{``anomalous''} \cite{PhysRev.111.1187,PhysRev.114.376,Nambu1958}. This term is an unfortunate quirk of history and has nothing to do with anomalies in quantum field theories in the modern sense. As implied by \eqref{eq:ImT3}, anomalous thresholds are a central part of understanding the S-matrix as a consequence of unitarity. In contrast with normal thresholds, they are no longer localized in space. Unfortunately, much less is known about them, and in particular, their kinematic dependence does not admit a simple solution as \eqref{eq:s-normal}. One of the points of this work is to better understand their properties. In Sec.~\ref{sec:analyticity-stable} we will show that for $2\to2$ scattering of external stable particles specifically, anomalous thresholds do not contribute in physical processes. 
	
	\begin{figure}
		\centering
		\includegraphics[scale=1]{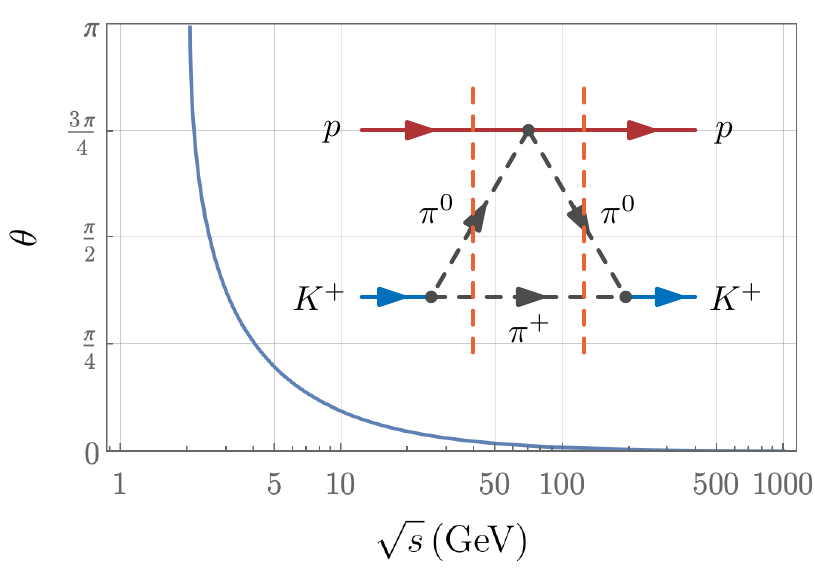}
		\caption{\label{fig:kaon}Example of an anomalous threshold in the $K^+ p \to K^+ p$ scattering by pions (inset). The plot represents the position of the corresponding resonance in the phase space parametrized by the center-of-mass energy $\sqrt{s}$ and the scattering angle $\theta$. The masses are $(m_{K^+}, m_p, m_{\pi^0}, m_{\pi^+}) \approx (494, 938, 135, 140) \, \mathrm{MeV}$ \cite{ParticleDataGroup:2020ssz}.}
	\end{figure}
	
	Let us illustrate the simplest appearance of anomalous thresholds on an example of kaon-proton scattering $K^+ p \to K^+ p$ mediated by pions $\pi^{0}$ and $\pi^+$ \cite{RevModPhys.33.448}, illustrated in Fig.~\ref{fig:kaon}.
	Here the kaon decays into two pions (at the branching ratio $\sim 21\%$ \cite{ParticleDataGroup:2020ssz}), one of which scatters elastically off a proton and then together with the remaining pion decays to a kaon.
	It is straightforward to work out that the triangle scattering is only viable as a classical configuration when the center-of-mass energy $\sqrt{s}$ is related to the scattering angle $\theta$ through
	\be
	\cos \theta = 1 - \frac{2s \left(m_{K^+}^2 - (m_{\pi^0} + m_{\pi^+})^2\right)\left(m_{K^+}^2 - (m_{\pi^0} - m_{\pi^+})^2\right)}{m_{\pi^+}^{2}\left(s- (m_{K^+}+m_p)^2\right)\left(s -(m_{K^+}-m_p)^2\right)}.
	\ee
	The position of the anomalous threshold in terms of these variables is plotted in Fig.~\ref{fig:kaon}. If such an experiment was feasible, quantum field theory would predict a resonance peak at those values of energies and angles. The threshold starts being kinematically allowed at $\sqrt{s}\approx 2\, \mathrm{GeV}$ giving backward scattering, $\theta = \pi$, while in the high-energy limit $s \to \infty$, the scattering angle  tends to zero. Note that in this example the kaons and pions are unstable (with mean lifetimes of $c(\tau_{K^+}, \tau_{\pi^0}, \tau_{\pi^+}) = (3.7, 2.5 \times 10^{-8}, 7.8)\mathrm{m}$ \cite{ParticleDataGroup:2020ssz}). Including their widths would shift the peak to complex values of $s$ and $\theta$, though it might still have observable effects within physical kinematics as a Breit--Wigner-like resonance. Measuring it experimentally does not seem presently accessible, see, e.g., \cite{Morrison:1969zzb,ALICE:2021szj}, and even if it were, it would be numerically suppressed compared to tree-level diagrams for $K^+p$ scattering: since this threshold corresponds to a one-loop diagram, and in addition involves weak interactions that do not conserve strangeness at the $K\pi\pi$ vertices, its contribution will be suppressed compared to tree-level diagrams, including strangeness-conserving weak-interaction diagrams along with ones involving only strong interactions.
	
	Despite the fact that the above example involves unstable particles on external states, one might easily come up with other instances (even by embedding the $K^+ p \to K^+ p$ diagram in a larger process) with long-lived external states. 
	
	Finally, let us stress that all the above manipulations are valid only for physical scattering amplitudes, where all the energies and momenta of the external particles are real. The image of these kinematic points in the Mandelstam invariants $s_{ij\ldots} = (p_i + p_j + \ldots)^2$ is called the \emph{physical region}. However, majority of the interest in the S-matrix program comes from exploiting tools in complex analysis to study constraints on scattering amplitudes. For this purpose, it is necessary to complexify the Mandelstam invariants and study the behavior of the S-matrix in an enlarged space. As we will see, this opens up a whole can of worms since the S-matrix becomes multi-valued with a complicated branch-cut structure. Alas, the unitarity equation $SS^\dagger = \boldone$ no longer makes sense. Under certain circumstances one can also use the notion of \emph{extended unitarity} a little bit outside of the physical regions, see, e.g.,
	\cite{Olive1962b,Olive1963,Boyling1964,Homrich:2019cbt,Guerrieri:2020kcs,Correia:2021etg} (see also \cite{PhysRevLett.4.84,Goddard1969,doi:10.1063/1.1665233,doi:10.1063/1.1703897} for related work).
	Nevertheless, one of the surprises of the complexified S-matrix is that no qualitatively-new singularities can appear beyond the normal and anomalous thresholds introduced above. 
	
	\subsection{\label{sec:shifts-widths}Mass shifts and decay widths}
    
    In scattering experiments involving unstable particles as internal states, unitarity forces us to understand the matrix elements $\T$ with unstable external particles.
    Therefore, let us briefly pause to discuss unstable particles, which will play a prominent role in the rest of the paper. We will verify directly that cutting rules can be straightforwardly applied to unstable particles (or just stable particles with mass shifts) without further subtleties arising from resummation. Previous literature includes \cite{Veltman:1963th,Denner:2014zga,Donoghue:2019fcb}.
    
    Recall that particles can get a mass shift and a decay width coming from a resummation of an infinite sum over one-particle irreducible (1PI) diagrams.
    Here we focus on the case of scalar particles, since one can easily derive analogous formulae for particles with spin by using the appropriate forms of the propagators.
    Denoting each 1PI self-energy diagram with $\Sigma(p^2)$ and the original mass $m$, this sum converts the original propagator $1/(p^2 - m^2 + i\eps)$ with infinitesimal $\eps$ to its shifted version $\mathbf{T}_{1 \to 1}$ given by the geometric series
    \be\label{eq:geometric}
        i \mathbf{T}_{1 \to 1} = \lim_{\eps \to 0^+}\sum_{n=0}^\infty \frac{i}{p^2-m^2 + i\eps} \left[i \Sigma \frac{i}{p^2-m^2 + i\eps} \right]^n ,
    \ee
    where the factors of $i$ are included to account for the $i$ in the definition of $\T_{1\to 1}$ as $S = \boldone + i \bdelta{1}{1} \T_{1 \to 1}$.
    Including the string of self-energy diagrams will therefore result in a modified propagator of the form
    \be\label{eq:dressed}
    \mathbf{T}_{1 \to 1} = \lim_{\eps \to 0^+} \frac{1}{p^2-\m^2 +i(m \Gamma + \eps)} \,,
    \ee
    where the shifted mass is given by
    \be 
    \m^2 = m^2 - \Re  \Sigma \,,
    \label{eq:polemass}
    \ee
    and the decay width is defined to be
    \be
    \Gamma = \frac{\Im\, \Sigma}{m}
    \label{eq:width}
    \ee
    at the corresponding scale.
    For stable particles, i.e., those that cannot decay to any other particles in the theory, we see from~\eqref{eq:width} that $\Im\, \Sigma = 0$ and hence their width vanishes. The self-energy diagrams of unstable particles, on the other hand, generally have both real and imaginary parts. 
    Note that according to~\eqref{eq:Impositive}, the decay width is always non-negative, $\Gamma \geq 0$. Note that the $i\eps$ was necessary for the convergence of the geometric series \eqref{eq:geometric}. We also see that endowing a particle with tiny width $\Gamma = \mathcal{O}(\eps)$ is equivalent to imposing the Feynman $i\eps$ prescription, and hence for unstable particles we may drop the $i\eps$.
    
    \begin{figure}
        \centering
        \includegraphics{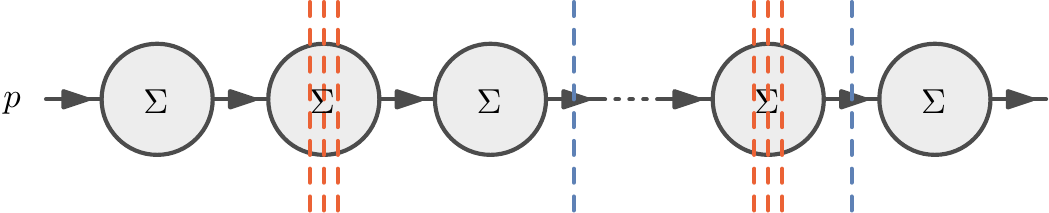}
        \caption{A single contribution to the cutting rules for the dressed propagator $\T_{1\to 1}$. For every chain of $n$ 1PI diagrams $\Sigma$, we cut through $k$ instances of $\Sigma$ (those with orange lines), as well as $l$ propagators (those with blue lines). The sum over all cuts \eqref{eq:cut-sum} includes $\binom{n}{k}$ ways of choosing which $k$ $\Sigma$'s and $\binom{n+1}{l}$ propagators to cut.}
        \label{fig:unstable}
    \end{figure}
    
    At this stage, let us return back to the cutting formula \eqref{eq:ImT4}, which,
    for the shifted propagator $\T_{1\to 1}$, gives the constraint
    \be
	\Im\, \T_{1\to 1} =
	\;
	\tfrac{i}{2} 
	\sum_{\CC}
	\left(-i\right)^c \, \cut_{\mathrm{C}} \, \T_{1 \to 1}.
	\label{eq:Imfromcuts}
	\ee
    The right-hand side is the sum over all possible unitarity cuts through chains of $n$ 1PI self-energy diagrams $\Sigma$, summed over all positive $n$. We can reorganize it as follows. Let us say that exactly $k$ out of $n$ instances of $\Sigma$ have been cut, while the remaining $n{-}k$ were not. This can happen in $\binom{n}{k}$ ways and accounts for all ways of cutting $\T_{1\to 1}$. Likewise, let us take $l$ out of the $n+1$ propagators to be cut, which can happen in $\binom{n+1}{l}$ different ways, see Fig.~\ref{fig:unstable}. Recall that cutting an individual $\Sigma$ gives, by definition,
    \be
    \sum_{C} (-i)^c\, \Cut_{\mathrm{C}} \Sigma = -2i\, \Im\, \Sigma  =  -2i m \Gamma.
    \ee
    Cutting a propagator is much easier and simply gives
    \be
    -2i\, \Im \left[ \frac{1}{p^2 - m^2 + i\eps} \right] = \frac{2i\eps}{(p^2 - m^2)^2 + \eps^2}.
    \ee
    The counting of $k$ and $l$ is almost completely independent, except we have to make sure to exclude the case $k=l=0$, since at least one cut has to happen on the right-hand side of \eqref{eq:Imfromcuts}.
    It can therefore be rewritten as
    \begin{align}\label{eq:cut-sum}
    \Im\, \T_{1\to 1} = \tfrac{i}{2} \lim_{\eps \to 0^+} \sum_{n=0}^{\infty} (-1)^n &\sum_{l=0}^{n+1} \binom{n{+}1}{l}  \left[\frac{2i \eps}{(p^2 - m^2)^2 + \eps^2}\right]^{l} \left[ \frac{1}{p^2 - m^2 + i\eps} \right]^{n+1-l} \nn\\
    \times &\!\!\sum_{\substack{k=0\\ k+l > 0}}^{n} \binom{n}{k} \left[ -2i m\Gamma \right]^{k} \Sigma^{n-k} .
    \end{align}
    Finally, performing the sums is straightforward and gives
    \be 
    \Im\, \T_{1\to 1} = - \lim_{\eps \to 0^+}
        \frac{m \Gamma + \eps}{(p^2-\m^2)^2 + (m\Gamma + \eps)^2},
    \ee 
    which is indeed the imaginary part of the dressed propagator \eqref{eq:dressed}, verifying the cutting rules. As a special case, unitarity cut for stable particles with $\Gamma = 0$ amounts to the on-shell delta function on the shifted mass. In summary, we have
    \be\label{eq:ImT11}
    \Im \T_{1 \to 1}  = \begin{dcases}
    \Im \left[ \frac{1}{p^2 - \m^2 - im\Gamma} \right] \quad &\mathrm{for }\; \Gamma>0,\\
    -\pi \delta(p^2 - \m^2) &\mathrm{for }\; \Gamma=0,
    \end{dcases}
    \ee
    which is the expected result. The positive-energy step function $\Theta(p^0)$ is implicit because the unitarity equations we used are valid only for physical energy $p^0 > 0$.
    
    At this stage, it is important to draw a distinction between cutting dressed and undressed propagators. The former is governed by \eqref{eq:ImT11} and must be included in the cutting rules regardless of the decay width. The latter is more subtle.
    
    Indeed, one might notice a curious feature of the cutting formula \eqref{eq:cut-sum}: every term that cuts through an unstable particle at least once, i.e., $l \geq 1$, features a delta function raised to the $l$-th power. However, one can check that  summing over the whole series \eqref{eq:cut-sum} with the second sum $\sum_{l=0}^{n+1}$ replaced with $\sum_{l=1}^{n+1}$ gives a result proportional to $\eps$ whenever $\Gamma>0$, namely
    \be
    -\lim_{\eps \to 0^+}\frac{\eps}{[p^2 - \m^2 - i(m\Gamma + \eps)][p^2 - \m^2 - i(m\Gamma -\eps)]} = 0.
    \ee
    In other words, unitarity cuts through the internal \emph{undressed} unstable particles resum to zero!
    We can easily apply the above arguments to a theory with multiple unstable species, say with masses $m_1 \leq m_2 \leq \cdots \leq m_N$: once the arguments are applied to the propagators with the lightest mass $m_1$, they also carry over to the next-to-lightest $m_2$ (even if it may decay into $m_1$), since all possible cuts through $m_1$ resum to zero and can thus be ignored. Repeating this logic for $m_3, m_4, \ldots, m_N$, and
    for every on-shell propagator of a given threshold, we learn that none of the thresholds involving internal unstable particles actually contribute to the unitarity equation \eqref{eq:ImT3}.
    In particular, this resummation means that once we start with an on-shell S-matrix with external stable particles, unitarity does not force us to talk about S-matrix elements with external unstable states, at least in perturbation theory, see also \cite{Veltman:1963th}.
    
    However, to make contact with collider physics, it is often important to consider scattering of unstable particles, which comprise a substantial part of the Standard Model. Their inclusion in perturbative computations is a rather subtle issue (and not fully understood theoretically), but roughly speaking it amounts to replacing every appearance of $m^2$ with $\m^2 - im\Gamma$ in the Feynman rules. This procedure turns out to be self-consistent, in that, e.g., it does not spoil gauge invariance in the electroweak sector; see, e.g., \cite{Veltman:1992tm,Denner:1997kq,Denner:2006ic,Denner:2014zga} and \cite[Sec.~6]{Denner:2019vbn} for a review. Aspects of wavefunction renormalization of external unstable particles were discussed in \cite{Kniehl:2001ch,Espriu:2002xv,Bharucha:2012nx,Collins:2019ozc}. In the context of the S-matrix theory, resonances associated to unstable particles can be identified with poles on unphysical Riemann sheets, see, e.g., \cite{Levy1959,PhysRev.119.1121,PhysRev.123.692,Landshoff1963,Olive1963b,PhysRev.131.888,Stapp1964,doi:10.1063/1.1704342,doi:10.1063/1.1704343}. We will consider processes with unstable particles in Sec.~\ref{sec:ExampleII}, which can be thought of as either a contribution to a $2 \to 2$ scattering on its own, or as being embedded in a higher-multiplicity process with external stable particles only.
	
	\subsection{\label{sec:cutting}Holomorphic cutting rules}
	
	At this stage, we would like to convert \eqref{eq:ImT3} into a prescription that can be used to check unitarity in practice for individual Feynman integrals. First, note that
    we can always design a Lagrangian reproducing any given Feynman diagram. For instance, for an $n$-point scalar process with $\E$ internal propagators and $\mathrm{V}$ vertices, we can write
	\be
	{\cal L} = \frac{1}{2} \sum_{i=1}^{n} \left(\partial_\mu \phi_i \partial^\mu \phi_i - M_i^2 \phi_i^2 \right) + \frac{1}{2} \sum_{e=1}^{\E} \left(\partial_\mu \psi_e \partial^\mu \psi_e - m_e^2 \psi_e^2 \right) + \sum_{v=1}^{\mathrm{V}} g_v\, V_v[\phi_i, \psi_e].
	\label{eq:lagrangian}
	\ee
	Here we introduced a field $\phi_i$ for every external line with mass $M_i$, and likewise $\psi_e$ for every internal propagator with mass $m_e$. The potentials $V_v$ are designed to reproduce interaction vertices of the Feynman diagram, and $g_v$ are auxiliary coupling constants. Clearly, one can design analogous Lagrangians for theories with spin. With such a quantum field theory, the contribution at order $\prod_{v=1}^{\mathrm{V}}g_v$ to the S-matrix is an individual Feynman diagram.
	When factoring out the overall coupling constant on both sides of \eqref{eq:ImT4}, the left-hand side
	equals $\tfrac{1}{2i} (\I - \I^\dagger)$ with the Feynman integral $\I$ given by,
	\be
    \I = \lim_{\eps \to 0^+} \int \prod_{a=1}^{\L} \frac{\d^\D \ell_a}{ i\pi^{\D/2}} \frac{(-1)^\E\, \mathrm{N}}{ \prod_{e=1}^{\E}(q_e^2 - m_e^2 + i\eps)},
    \label{eq:FeynInt}
    \ee
    where $\ell_a$ are the $\L$ loop momenta and we separated the numerator $\mathrm{N}$ from the propagators $-1/(q_e^2 - m_e^2 + i\eps)$ for each of the $\E$ internal edges. Above, $\I^\dagger$ amounts to using $\mathrm{N}|_{\inn \leftrightarrow \out}$ (obtained by exchanging the in/out states),
    and complex conjugating the resulting expression.
    
    \begin{figure}
        \centering
        \includegraphics{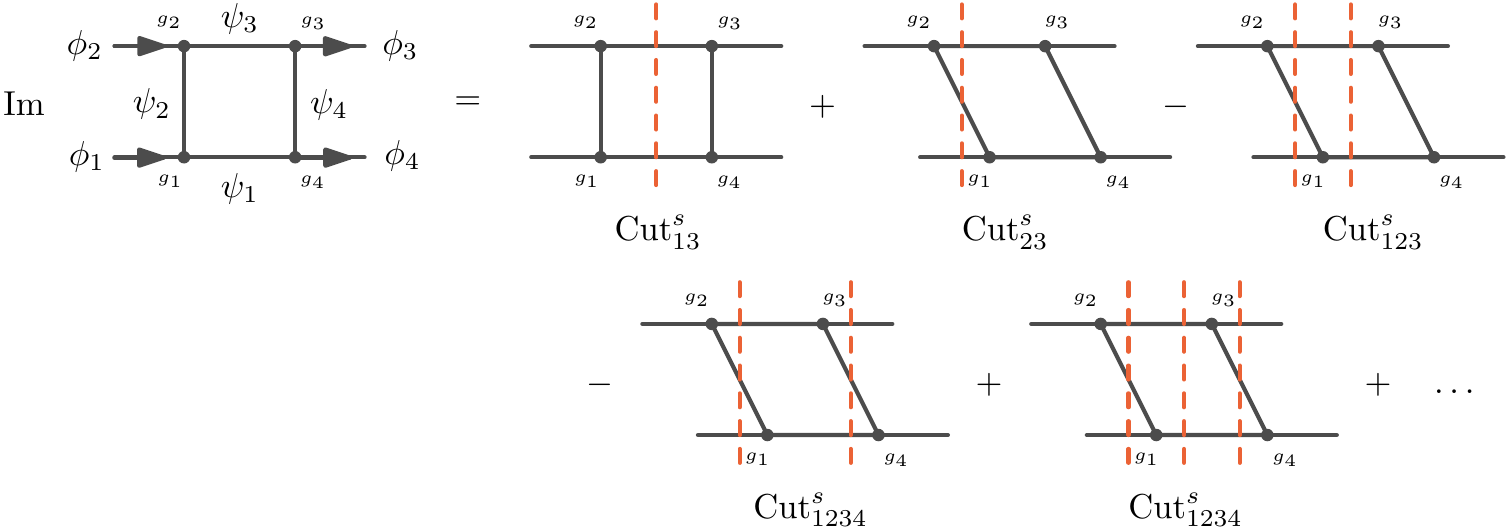}
        \caption{Example application of holomorphic unitarity cuts to a single Feynman integral generated by the Lagrangian \eqref{eq:lagrangian}. The external legs and internal edges have fields $\phi_i$ and $\psi_e$ attached to them, respectively. Each vertex comes with a factor of $g_i$, Extracting the coefficient of $g_1 g_2 g_3 g_3$ gives a relation between the imaginary part of a single diagram on the left- and a sum of its cuts according to \eqref{eq:cuttingrules} on the right-hand side, where the ellipsis denotes permutations. Note that the sum might involve cancellations and individual terms might also vanish.}
        \label{fig:holomorphic-cuts}
    \end{figure}
    
	While multiple terms can appear on the right-hand side of \eqref{eq:ImT4}, at the order $\prod_{v=1}^{\mathrm{V}}g_v$ we have only a finite number of terms corresponding matrix elements with $\psi_e$ fields on external states. These are of course nothing else but unitarity cuts through the diagram $\I$ that put the corresponding edges on shell with positive energies. See Fig.~\ref{fig:holomorphic-cuts} for an example.
	The right-hand side of \eqref{eq:ImT4} gives in our case
	\begin{align}\label{eq:ImTL}
	\tfrac{1}{2i}( \I - \I^\dagger ) = & \tfrac{1 }{2}\sumint_{\CC_1} \bdelta{\inn}{\CC_1}\, \I_{\inn\to \CC_1} \I_{\CC_1\to \out}
	\\
	& -\tfrac{i}{2}\sumint_{\CC_1,\CC_2} \bdelta{\inn}{\CC_1} \bdelta{\CC_1}{\CC_2}\, \I_{\inn\to \CC_1} \I_{\CC_1 \to \CC_2} \I_{\CC_2\to \out} + \ldots.\nn
	\end{align}
	Recall that $\bdelta{I}{J}$ is the $\D$-dimensional momentum-conserving delta function, and the integration in $\sumint$ runs over the
	Lorentz-invariant phase space, such that
	\be
	    \int_I \bdelta{I}{J} = \int \left[ \prod_{i \in I} \frac{\d^{\D} q_i}{(2 \pi)^{\D-1}} \delta^+(q_i^2-m_i^2) 
	    \right] \left(2\pi\right)^{\D} \delta^\D(P_I-P_J),
	\ee
	where $q_i$ are the momenta of each particle going through the cut, summing up to $P_I = \sum_{i \in I} q_i$. The positive-energy delta functions
	\be
	\delta^+(q_i^2-m_i^2) = \delta(q_i^2-m_i^2) \Theta(q_i^0)
	\ee
	select only the on-shell states propagating in the causal direction.
	We can use the momentum-conserving delta function to perform the integral over a single arbitrary $r\in I$, resulting in
	\be
	    \int_I \bdelta{I}{J} = \int \Bigg[\prod_{j \in I \backslash \{r\}} \frac{\d^{\D} q_j}{(2 \pi)^{\D}} \Bigg] \left[ \prod_{i \in I} 2 \pi \, \delta^+(q_i^2-m_i^2) \right],
	    \label{eq:delta}
	\ee
    which points to the correct normalization of the cut propagators.

	In other words, the imaginary part combination $\tfrac{1}{2i}( \I - \I^\dagger)$ of each Feynman diagram is given by the sum over all possible ways of cutting through the diagram by putting intermediate particles on shell. To account for the difference in normalization between the Feynman integral $\I$, and the combination that appears in \eqref{eq:ImTL}, we add an overall factor of $(-i)^{c+\vert \CC \vert}$.%
	\footnote{To see how this factor comes about, first note that the normalization of \eqref{eq:ImTL} contains a factor of $i^{\L_{\text{cut}}}(-1)^{\vert \CC \vert}$ compared to $\I$, where $\L_{\text{cut}}$ is the number of cut loops. Then, using Euler's identity $\L_i=\E_i{-}\VV_i{+}1$ for each of the $c+1$ intermediate diagrams corresponding to $\I_{\CC_i \to \CC_{i+1}}$, we can write $\L_{\text{cut}}=\vert \CC \vert-V+1+V-(c+1)$. When changing between \eqref{eq:ImTL} and $\I$, we therefore need to multiply by a factor of $i^{\L_{\text{cut}}}(-1)^{\vert \CC \vert}=(-i)^{c+\vert \CC \vert}$ to compensate for the difference in normalization.}
	We therefore get,
    \be\boxed{
        \tfrac{1}{2i}( \I - \I^\dagger) =
        \sum_{\substack{\mathrm{holomorphic}\\ \mathrm{cuts}\,\CC}} (-1)^{c+1}\, \cut_\CC \, \I
        \,,}
        \label{eq:cuttingrules}
    \ee
	where the individual holomorphic cuts are given by
    \be
    \Cut_{\CC} \, \I = \frac{1}{2 i} (-2\pi i)^{|\CC|} \left(-1\right)^\E  \lim_{\eps \to 0^+} \int \prod_{a=1}^{\L} \frac{\d^\D \ell_a}{ i\pi^{\D/2}} \frac{\mathrm{N}\prod_{e \in \CC} \delta^+(q_e^2 - m_e^2)}{\prod_{e\notin \CC}(q_e^2 - m_e^2 + i\eps)}.
    \label{eq:cuts}
    \ee
    This is a holomorphic version of Cutkosky rules \cite{Cutkosky:1960sp}, see also \cite{Coster:1970jy,Bourjaily:2020wvq,Blazek:2021olf,Blazek:2021zoj}. Whenever, the numerator satisfies $\mathrm{N} = \mathrm{N}|_{\inn \leftrightarrow \out}$, the left-hand side of \eqref{eq:cuttingrules} is simply the imaginary part $\Im\, \I$, as will be the case in all the examples considered in this paper.

    By contrast, we could have started with \eqref{eq:unitarityT}, which instead puts only edges along a single cut on shell, but complex-conjugates the propagators and vertices to the right of the cut. A derivation analogous to the one above would have given the traditional Cutkosky rules (see \cite{Veltman:1963th,tHooft:1973wag,Sterman:1993hfp,Bloch:2015efx,Pius:2018crk} for previous proofs). Note, however, that sign of the imaginary part of a given Feynman diagram cannot be fixed by such arguments, because positivity of the $\Im\, \T_{\inn\to\inn}$ does not in general imply positivity of its $\prod_{v=1}^{\mathrm{V}}g_v$ coefficient.

    \newpage
	\section{\label{sec:primer}Primer on the analytic S-matrix}
	
	The purpose of this section is to give a gentle introduction to the analytic aspects of Feynman integrals on the simple example of the bubble diagram, many of which have never appeared in the previous literature. In particular, one of the goals is to understand the ingredients leading to different ways of implementing $i \varepsilon$ prescriptions for scattering amplitudes in perturbation theory. While rather elementary, the bubble example will allow us to develop intuition and illustrate the key points relevant for general multi-loop diagrams. Throughout this section, we point out which facets of the analysis generalize straightforwardly and which are more intricate.

	\subsection{From loop momenta to Schwinger parameters}
	
\begin{figure}
	    \centering
	    \raisebox{-0.5\height}{\includegraphics[scale=1.1]{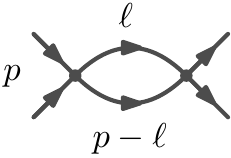}}
	    \hspace{4em}
	    \raisebox{-0.5\height}{\includegraphics[scale=1.1]{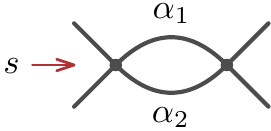}}
	    \caption{Bubble diagram labelled by the loop momentum $\ell$ (left) and the Schwinger parameters $\alpha_1$ and $\alpha_2$ (right). The internal particles have masses $m_1$ and $m_2$.}
	    \label{fig:bubble-diagram}
	\end{figure}

	Arguably the simplest Feynman integral we can write down corresponds to the scalar bubble diagram. In the loop-momentum space, it is given by
	\be\label{eq:I-bubble-mom}
	\I_\bubble(s) = \lim_{\eps \to 0+} \int_{\R^{1,\D-1}} \frac{\d^\D \ell}{i\pi^{\D/2}} \frac{1}{[\ell^2 - m_1^2 + i\eps] [(p-\ell)^2 - m_2^2 + i\eps]},
	\ee
	where $m_e$ are the masses of the two internal edges and $p^\mu$ is the incoming momentum, see Fig.~\ref{fig:bubble-diagram}. We ignore coupling constants: the overall normalization is introduced only for later convenience. The propagators feature the Feynman $i\eps$ prescription ensuring causality. Strictly speaking, the above integral is only defined in the physical region, requiring for example, that the center-of-mass energy $\sqrt{s} = \sqrt{p^2}$ is positive.

	The representation \eqref{eq:I-bubble-mom} is rarely used in practical computations. For one thing, it does not manifest the fact that the result is a function of the Mandelstam invariant $s$. But more importantly, the number of integrations can be drastically reduced. To this end, we use the Schwinger trick:%
	\footnote{For future reference, let us spell out the identity:
		\be
		\int_0^{\infty} \frac{\d \lambda}{\lambda^{1-B}}\, e^{\frac{i}{\hbar}\lambda A} = (-i\hbar)^B \frac{\Gamma(B)}{(-A)^B} \qquad\mathrm{for}\qquad \mathrm{Re}\, B >0 \quad\mathrm{and}\quad \Im\, A > 0,
		\ee
		where for non-integer $B$'s we chose the branch cut to run just below the positive real axis in the $A$-plane.
	}
	\be
	\frac{-1}{q_e^2 - m_e^2 + i\eps} =
	\frac{i}{\hbar}
	\int_0^\infty \d \alpha_e \exp\left[ \frac{i}{\hbar} (q_e^2 - m_e^2 + i\eps) \alpha_e \right]
	\ee
	for each propagator, $e=1,2$. Note that from this perspective, the $i\eps$ is needed for convergence of the integral as $\alpha_e \to \infty$. This gives us
	\begin{align}\label{eq:I-bubble3}
		\I_\bubble =  \frac{i}{ \pi^{\D/2} \hbar^2} \lim_{\eps \to 0+} \int_{\R^{1,\D-1}} \d^\D \ell\, &
		\int_0^{\infty}
		\d\alpha_1 \int_0^{\infty} \d \alpha_2 \exp\bigg[ \frac{i}{\hbar} \Big( \ell^2 (\alpha_1 {+} \alpha_2)  \\ & + (p^2- 2\ell\cdot p) \alpha_2
		- (m_1^2 - i\eps) \alpha_1 - (m_2^2 - i\eps) \alpha_2  \Big) \bigg].\nn
	\end{align}
	Since the exponent is quadratic in the loop momentum $\ell^\mu$, we can attempt to perform the Gaussian integral
	which has its peak at
	\be\label{eq:ell}
	\ell^\mu = p^\mu \frac{\alpha_2}{\alpha_1 {+} \alpha_2}.
	\ee
	However, since the integration contour in the loop-momentum space runs over $\ell^\mu \in \mathbb{R}^{1,\D-1}$ with $\ell^2 = (\ell^0)^2-\vec{\ell}\,^2$, the argument of the exponent in~\eqref{eq:I-bubble3} varies between being positive and negative along this contour. But in order to perform the Gaussian integral, we must make sure that the integrand decays sufficiently fast as $\ell^\mu \to \pm\infty$, which is ambiguous when $\ell^2$ does not have a definite sign. Hence, before evaluating the Gaussian integral, we use Wick rotation to deform the integration contour from Lorentzian $\mathbb{R}^{1,\D-1}$ to Euclidean $\mathbb{R}^\D$. One needs to be careful, because this deformation must be done without encountering any singularities.
	
\begin{figure}
	    \centering
	    \includegraphics[scale=1.1]{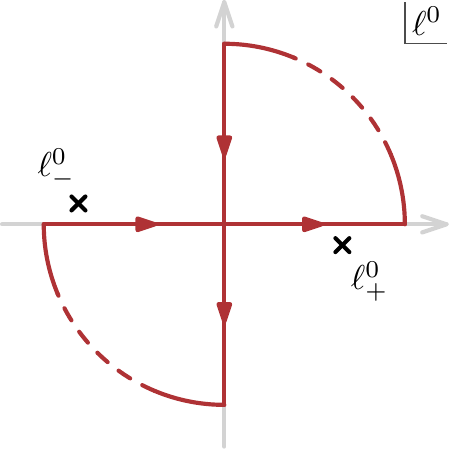}
	    \caption{Figure-eight contour in the $\ell^0$-plane showing equivalence of the Lorentzian contour $\ell^0 \in \R$ and its Wick rotation $\ell^0 = i\ell_E^0 \in i\R$, defining the Euclidean loop momentum $\ell_E^\mu = (\ell_E^0, \vec{\ell})$.}
	    \label{fig:ell-contour}
	\end{figure}

	Let us explain how to perform the Wick rotation. We start with changing variables with $\ell^\mu \to \ell^\mu + p^\mu \frac{\alpha_2}{\alpha_1+\alpha_2}$, resulting in
	\begin{align}\label{eq:I-bubble3p5}
		\I_\bubble = \frac{i}{ \pi^{\D/2} \hbar^2} \lim_{\eps \to 0+}  \int_{\R^{1,\D-1}} \d^\D \ell \, &
		\int_0^{\infty} \d\alpha_1
		\int_0^{\infty} \d \alpha_2 \exp\bigg[ \frac{i}{\hbar} \Big( \ell^2 (\alpha_1 {+} \alpha_2) \\ & + s \frac{\alpha_1 \alpha_2}{\alpha_1+\alpha_2}
		- (m_1^2 - i\eps) \alpha_1 - (m_2^2 - i\eps) \alpha_2  \Big) \bigg].\nn
	\end{align}
	The roots of the polynomial in $\ell^0$ in the numerator are at
	\begin{equation}\label{eq:ell-0}
	    \ell^0_\pm = \pm  \sqrt{\vec{\ell}\,^2 - s \frac{ \alpha_1 \alpha_2}{\left( \alpha_1+\alpha_2\right)^2}  + \frac{m_1^2 \alpha_1 + m_2^2 \alpha_2}{\alpha_1+\alpha_2}  - i \varepsilon }.
	\end{equation} 
	By expanding in $\varepsilon$, we see that for any real value of $\vec{\ell}$, the root $\ell^0_+$ lies in the fourth quadrant of the complex $\ell^0$-plane: it is either right below the positive real axis or to the right of the negative imaginary axis. Similarly, the root $\ell^0_-$ always lies in the second quadrant. We therefore see by Cauchy's residue theorem that the contour integral of the figure-eight contour in Fig.~\ref{fig:ell-contour} is zero. Since the integrand goes like $\sim \d \ell^0 (\ell^0)^{\D-5}$ at large $\vert \ell^0\vert$, the arcs at infinity vanish in $\D < 4$, and hence the integral over the real  $\ell^0$-axis is the same as the one along the imaginary $\ell^0$-axis. The failure of this procedure in $\D \geq 4$ is precisely a sign of a UV divergence.
	
	Finally, we change variables to $\ell^0 = i \ell^0_E$, so that in terms of the Euclidean loop momentum $\ell^\mu_E = (\ell^0_E, \vec{\ell})$, the contour in~\eqref{eq:I-bubble3p5} can be taken over $\mathbb{R}^{\D}$ with all-minus signature, i.e., $\ell_E^2 = -\vert \vec{\ell}_E\vert^2$. Accounting for the Jacobian factor of $i$, the integral becomes:
	\begin{align}\label{eq:I-bubble3p6}
		\I_\bubble = - \frac{1}{\pi^{\D/2} \hbar^2} \lim_{\eps \to 0+} \int_{\R^{\D}} \d^\D \ell_E \,
		& \int_0^\infty
		\d\alpha_1 \int_0^\infty \d \alpha_2 \exp\bigg[ \frac{i}{\hbar} \Big( \ell_E^2 (\alpha_1 {+} \alpha_2) \\
		& + s \frac{\alpha_1 \alpha_2}{\alpha_1+\alpha_2}- (m_1^2 - i\eps) \alpha_1 - (m_2^2 - i\eps) \alpha_2  \Big) \bigg]\,.\nn
	\end{align}
	Now, we can simply perform the Gaussian integral in $\ell_E^\mu$, which gives
	\be\label{eq:I-bubble2}
	\I_\bubble = \left(-i \hbar \right)^{\D/2-2} \lim_{\eps \to 0^+} \int_0^\infty \frac{\d\alpha_1\, \d\alpha_2}{(\alpha_1 {+} \alpha_2)^{\D/2}} \exp \left[ \frac{i}{\hbar}\Big(\V + i\eps (\alpha_1 {+} \alpha_2)\Big)\right] \,,
	\ee
	with
	\be
	\V = s\frac{\alpha_1 \alpha_2}{\alpha_1 {+} \alpha_2} - m_1^2 \alpha_1 - m_2^2 \alpha_2.
	\ee
	It is now manifest that the result depends only on the Mandelstam invariant $s = p^2$ and the internal masses, $m_1$ and $m_2$. This is precisely the form of the integral we would have obtained from the worldline formalism, where $\alpha_e$ should be thought of as measuring the lengths of the internal edges, and $\V$ as the action.
	
	Notice that the above integrand transforms covariantly under dilations of all the edges, $\alpha_e \to \lambda \alpha_e$ with $\lambda > 0$. More concretely, we can make a change of variables
	\be\label{eq:alpha-tilde}
	(\alpha_1, \alpha_2) = (\lambda \tilde{\alpha}_1, \lambda \tilde{\alpha}_2).
	\ee
	Since this procedure gives one too many variables on the right-hand side, we can fix one of the tilded ones. While any choice will lead to an identical result for $\I_{\bubble}$, different options might be more or less convenient, depending on the problem at hand.
	
	Let us briefly discuss two choices, both of which will be used later to illustrate different points. 
	The first one is $(\tilde{\alpha}_1, \tilde{\alpha}_2) = (\alpha, 1 {-} \alpha)$.
	The Jacobian gives $\d\alpha_1 \d\alpha_2 = \lambda\, \d\alpha\, \d\lambda$ and the integral proceeds over $\alpha \in [0,1]$ as well as $\lambda > 0$. Therefore, we obtain
	\be
	\label{eq:I-bub-lambda}
	\I_\bubble = \left(-i \hbar \right)^{\D/2-2} \lim_{\eps \to 0^+} \int_0^1 \d\alpha \int_0^{\infty} \frac{\d\lambda}{\lambda^{\D/2-1}}\exp \bigg[ \frac{i}{\hbar} \lambda \left( \tilde{\V} + i\eps\right) \bigg],
	\ee
	where
	\be\label{eq:V-tilde}
	\tilde{\V} = \V\big|_{(\alpha_1,\alpha_2) = (\alpha, 1-\alpha)} = s \alpha (1-\alpha) - m_1^2 \alpha - m_2^2 (1-\alpha).
	\ee
	The second one sets $(\tilde{\alpha}_1, \tilde{\alpha}_2) = (\alpha, 1)$, with a Jacobian factor of $\d\alpha_1 \d\alpha_2 = \lambda\, \d\alpha\, \d\lambda$. In this case, the integration contour is $\alpha \in [0,\infty]$ and $\lambda>0$, so
	\be
	\label{eq:I-bub-lambda2}
	\I_\bubble = \left(-i \hbar \right)^{\D/2-2} \lim_{\eps \to 0^+} \int_0^\infty \frac{\d\alpha}{(1+\alpha)^{\D/2}} \int_0^{\infty} \frac{\d\lambda}{\lambda^{\D/2-1}}\exp \bigg[ \frac{i}{\hbar} \lambda \left( \tilde{\V} + i\eps\right) \bigg],
	\ee
	where
	\be\label{eq:V-tilde2}
	\tilde{\V} = \V\big|_{(\alpha_1,\alpha_2) = (\alpha, 1)} = s \frac{\alpha}{\alpha+1} - m_1^2 \alpha - m_2^2.
	\ee
	The freedom in fixing different Schwinger parameters is sometime called the Cheng--Wu theorem. Physically, it tells us about the reparametrization invariance of the worldline parameters. We can thus think of $(\alpha_1:\alpha_2) = (\lambda \alpha_1 : \lambda \alpha_2) \in \CP^1$ as projective coordinates and the overall $\lambda$ is the $\GL(1)$ degree of freedom acting by dilating the whole diagram by an overall factor.
	
	At any rate, the point of the above exercise was to separate the projective content of the Feynman integral from the overall scaling $\lambda$. At this stage, we can simply integrate $\lambda$ out using
	\be
	\int_0^{\infty} \frac{\d\lambda}{\lambda^{\D/2-1}}\exp \bigg[ \frac{i}{\hbar} \lambda \left( \tilde{\V} + i\eps\right) \bigg] = \left( - i \hbar \right)^{2-\D/2} \frac{\Gamma(2 {-} \D/2)}{(-\tilde{\V} - i\eps)^{2 - \D/2}} \,,
	\ee
	which holds for $\eps>0$.
	Note that the limit $\eps \to 0^+$ of the above expression does not exist when $\tilde{\V} = 0$ for some value of $\alpha$. This fact will play an important role later. In summary, we find using the $\GL(1)$ fixing in~\eqref{eq:I-bub-lambda},
	\be\label{eq:I-bubble}
	\I_\bubble =  \,  \Gamma(2{-}\D/2) \lim_{\eps \to 0^+} \int_{0}^{1} \frac{\d\alpha}{( -\tilde{\V} - i\eps)^{2-\D/2}} \,.
	\ee
	This form is called the Schwinger (or Feynman) parametrization of the bubble integral. Following essentially identical steps, it can be generalized to an arbitrary Feynman diagram with more bookkeeping, see Sec.~\ref{sec:general} and App.~\ref{app:parametric} for more details.
	
	The above integral can easily be evaluated. For example, in $\D=2$ it gives
	\begin{align}
	\label{eq:bub2d}
	\I_{\bubble}\big|_{\D=2} =\, &\frac{2i}{\sqrt{-[s - (m_1{+}m_2)^2][s-(m_1{-}m_2)^2]}}\nn
	\\
	&\times \log \left( \frac{\sqrt{(m_1{+}m_2)^2 -s } - i \sqrt{s - (m_1{-}m_2)^2}}{\sqrt{(m_1{+}m_2)^2 -s } + i \sqrt{s - (m_1{-}m_2)^2}} \right),
	\end{align}
	while in $\D=3$
	\be\label{eq:bub3d}
	\I_{\bubble}\big|_{\D=3} = \frac{\sqrt{\pi}}{\sqrt{s}} \log \left(\frac{m_1+m_2+\sqrt{s}}{m_1+m_2 - \sqrt{s}}\right).
	\ee
	Note that in these forms, the square roots and the logarithms are evaluated on their principal branches whenever their arguments are negative. When $\D \geq 4$, the bubble diagram has an overall UV divergence, which is made obvious in \eqref{eq:I-bubble} by the $\Gamma(2-\D/2)$ function up front. This is not a coincidence: the number $d = 2-\D/2$ is the superficial degree of divergence of the bubble diagram. For $d\leq 0$ (or $\D\geq 4$) the integral over the overall scale $\lambda$ is singular near $\lambda \to 0$, which is a reflection of the UV singularity we encountered below~\eqref{eq:ell-0} in the loop momentum space.
	This divergence occurs at any value of $s$, and can be regularized, say by dimensional regularization $\D \to \D - 2\epsilon$, where $\epsilon$ is a generic complex parameter that is sent to zero at the end of the computation. Even though a systematic study of UV/IR divergences in the Schwinger-parameter space can be achieved \cite{smirnov1991renormalization,zavialov2012renormalized,Arkani-Hamed:2022cqe}, it will not concern us here since it is rather orthogonal to our discussion.
	
	While the bubble integral is simple enough to be evaluated directly, many other Feynman diagrams lead to integrals that become too involved to compute directly using their Schwinger-parametrized form.
	Indeed, Feynman integrals can feature seemingly arbitrarily-complicated functions, with most complex examples known presently involving periods of Calabi--Yau manifolds, see, e.g., \cite{brown2010periods,BrownSchnetz2012,Bloch:2014qca,Bourjaily:2019hmc,Bonisch:2021yfw}. The purpose of this section is
	learning how to understand the analytic structure of Feynman integrals without needing to perform any explicit computations. In the case at hand, this is a rather elementary task, but it will teach us about the complications we might expect in Feynman integrals contributing to more realistic scattering processes. We will study \textit{threshold singularities} that, contrary to UV divergences mentioned above, occur at specific values $s$ and cannot be regularized.

	\subsection{\label{sec:branch-cuts}Where are the branch cuts?}

	Analytic properties of \eqref{eq:I-bubble} are entirely determined by the roots of the denominator $\tilde{\V} + i\eps$.
	In our case, this gives a quadratic equation and hence yields two roots:
	\be\label{eq:alpha-1}
	\alpha_\pm = \frac{s - m_1^2 + m_2^2 \pm i\sqrt{-\Delta}}{2s},
	\ee
	where $\Delta = \Delta_+\Delta_-$ is the discriminant with
	\be
	\Delta_{\pm} = s - \left(\sqrt{m_1^2 - i\eps} \pm \sqrt{m_2^2 - i\eps}\right)^2.
	\ee
	For the time being let us consider $\eps>0$ to be fixed and study what happens as it is tuned to be smaller and smaller. This perspective will be quite crucial to learn about the branch cuts of Feynman integrals.

\begin{figure}
		\centering
		\includegraphics[scale=1]{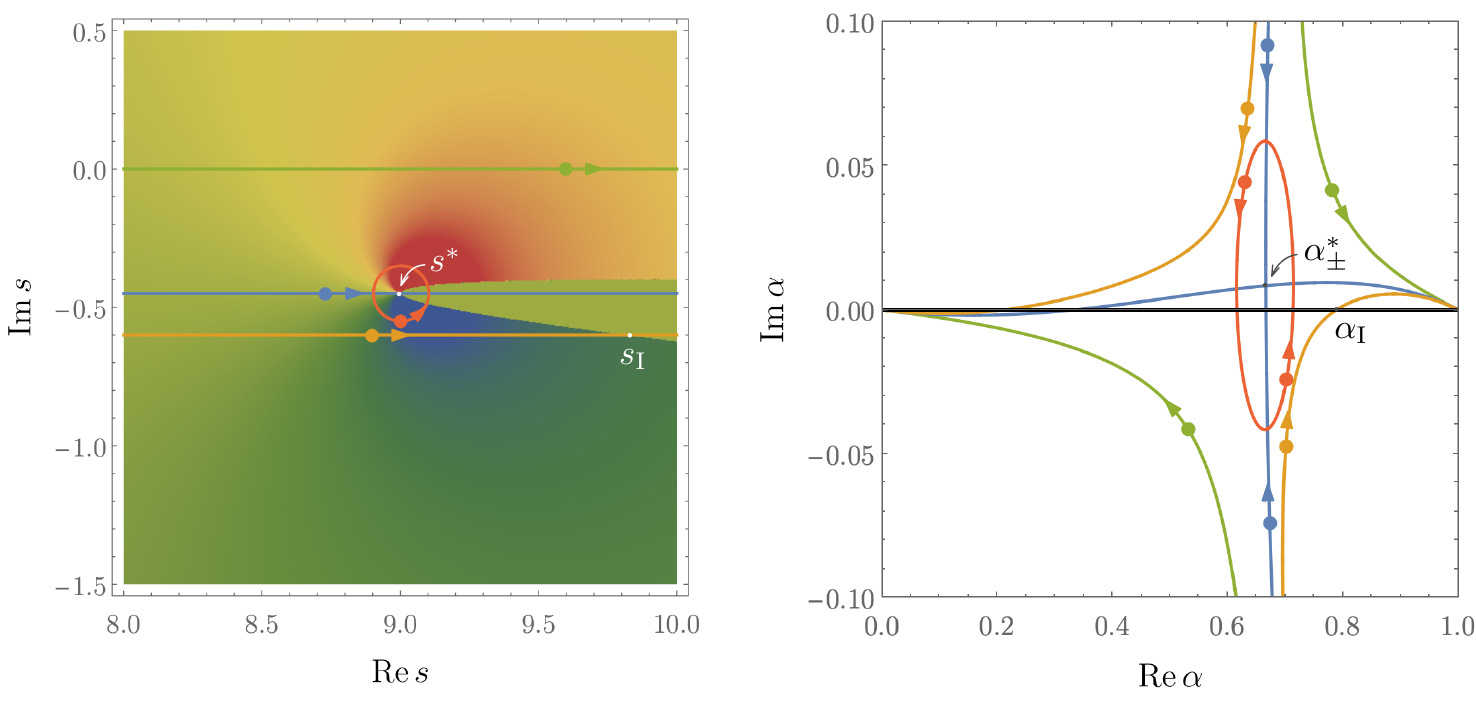}
		\caption{\label{fig:bubble}Positions of the roots $\alpha_\pm$ (right) along four families of curves in the $s$-plane (left): $\Im\, s=0$ (green), $\Im\, s=-0.45$ (blue), $\Im\, s=-0.60$ (orange), and a small circle centered around $s = s^\ast \approx 9.00 - 0.45i$ (red). The integration contour (black) is the interval $[0,1]$. Two branch cuts extend from the branch point at $s = s^\ast$ to the right, where the value of $\Im\, \I_{\bubble}$ jumps discontinuously. The point where the trajectory of the orange root intersects the real axis at $\alpha_{\RN{1}}$ corresponds to crossing on of the branch cuts at $s_{\RN{1}}$. For the purpose of the numerical plot we used $m_1 =1$, $m_2=2$, and $\eps = \tfrac{1}{10}$.}
	\end{figure}

    The roots $\alpha_\pm$ can cause singularities when they are on the integration contour. It is thus instructive to consider how they move around in the complex $\alpha$-plane as we scan over different values of the Mandelstam invariant $s$. Let us treat $s$ as a complex variable. This is illustrated in Fig.~\ref{fig:bubble} (right), where we show the trajectories of $\alpha_\pm$, as $s$ is varied. In particular, the integrand in~\eqref{eq:I-bubble} is singular whenever $\alpha_{\pm}$ is on the integration contour (thick black in Fig.~\ref{fig:bubble} (right)). For the numerical plots we set $m_1 = 1$, $m_2 = 2$, and $\D=2$, together with $\eps = \tfrac{1}{10}$. Note that Fig.~\ref{fig:bubble} is a zoomed-in version of Fig.~\ref{fig:Fig1} (left). We have chosen $\D$ to be even such that the Feynman integrand~\eqref{eq:I-bubble} has poles, as opposed to odd $\D$, where the integrand has branch cuts. The situation in odd $\D$ is slightly more complicated, and we will return to it in Sec.~\ref{sec:discontinuities}.
	
	First, let us look at what happens as we move along the real $s$-axis (green in Fig.~\ref{fig:bubble}). In the $\alpha$-plane, the two roots $\alpha_\pm$ travel from the upper- and lower-half planes respectively towards the integration contour $[0,1]$, but do not come near it until very large values of $s \gg 10$ are reached. In particular, the integration contour does not pass through the roots for any finite values of $s$, and hence the integral converges everywhere within the range of the $s$-plane plot and can be computed numerically. Note, however, that the roots get closer and closer to the integration contour as we tune $\eps \to 0^+$, so the convergence becomes poorer and poorer due to the increasingly larger integrand in~\eqref{eq:I-bubble}.
	
	Second, we consider a different straight path along the real part of $s$ axis: one with $\Im\, s = -0.60$ (orange in Fig.~\ref{fig:bubble}). While the trajectories of $\alpha_{\pm}$ look qualitatively similar to the previous $\Im\, s=0$ (green) path, we note that they now pass through the integration contour in two places. With our choice of numerical values, the first intersection is at $s = s_{\RN{1}}^\ast \approx 9.80 - 0.60\,i$ and the second is at $s = s_{\RN{2}}^\ast \approx 20.20 - 0.60\,i$, for $\alpha_+$ and $\alpha_-$ respectively.
	Fortunately, we can deform the integration contour to avoid these singularities, while preserving the endpoints. Since the integrand is holomorphic, the value of the integral in~\eqref{eq:I-bubble} remains unchanged for any deformation that avoids singularities.
	
	However, notice that in deforming the contour we have to make a \emph{choice}: in Fig.~\ref{fig:bubble} (right) we can either integrate over a contour between $0$ and $1$ that goes above or below the root $\alpha_+$. For us, it is natural to deform the integration contour to always go below $\alpha_+$, such that the value of the integral remains continuous as we travel along the orange line to the right. In contrast, if we did not know about the history of how the root $\alpha_+$ had crossed the integration contour, we could instead have evaluated the Feynman integral directly along the interval $[0,1]$, so that for $s>s_+^\ast$ the contour would go above the root $\alpha_+$. Since the two discrete choices differ by a residue around $\alpha_+$, this signals the presence of a \textit{branch cut} in the $s$-plane. A similar situation occurs when the other root $\alpha_-$ crosses the integration path, giving rise to another branch cut.
	
	Third, to drive this point home further, we look at a small circle around $s \approx 9.00 - 0.45\,i$ corresponding to the solution of $\Delta_+ = 0$ (red in Fig.~\ref{fig:bubble}). As we make one turn in the $s$-plane, the roots $\alpha_\pm$ make half a turn and exchange their positions. In doing so, both of them pass through the integration contour, one by one. Thus, each turn around $\Delta_+ = 0$ changes the value of the integral by adding an additional residue in the $\alpha$-space. We will return to this way of computing discontinuities in Sec.~\ref{sec:discontinuities}.
	
	The \emph{physical sheet} in the $s$-plane is defined as precisely the one selected by the real integration contour $[0,1]$ together with the $i\eps$ prescription that displaced $\alpha_\pm$ in the appropriate way. If we instead decide to start with a deformed contour in the $\alpha$-space, the branch cuts in the $s$-space will move accordingly.
	
	This simple exercise already teaches us a couple of general lessons. First, Feynman integrals are multi-valued functions in terms of the kinematic variables.
	Second, positions of branch cuts (but not branch points) are physically meaningless: they are simply a consequence of the choice of the integration contour.
	We might attempt to exploit this connection to systematically avoid poles hitting the integration contour and thus replace the Feynman $i\eps$ prescription altogether. Before we do so, let us learn a bit more about singularities.
	
	\subsection{\label{sec:singularities}Where are the singularities?}
	
	Going back to Fig.~\ref{fig:bubble}, we look at the last contour, along $\Im\, s = -0.45$ (blue). This contour is special, since the two roots \emph{collide} at
	\be\label{eq:alpha-normal}
	\Delta_+ = 0, \qquad \alpha_\pm^\ast = \frac{\sqrt{m_2^2 - i\eps}}{\sqrt{m_2^2 - i\eps} + \sqrt{m_1^2 - i\eps}},
	\ee
	which corresponds to $\alpha_\pm^\ast \approx 0.666 + 0.008 \, i$ and $s=s^\ast\approx 9.003 - 0.450\, i$ in our numerical setup. Hence, we cannot escape the pole of the integrand by contour deformations and the integral in~\eqref{eq:I-bubble} develops a divergence at $s=s^\ast$ in $\D<4$. This is the normal threshold.
	
	Let us explore other possibilities for the two roots colliding. Looking at the explicit form \eqref{eq:alpha-1}, we might also have had a pinch singularity at 
	\be\label{eq:alpha-pseudo}
	\Delta_- = 0, \qquad \alpha_\pm^\ast = \frac{\sqrt{m_2^2 - i\eps}}{\sqrt{m_2^2 - i\eps} - \sqrt{m_1^2 - i\eps}},
	\ee
	which after plugging in the numbers gives $s \approx 0.997 + 0.050i$ and $s \approx 2.000 - 0.075i$.
	Following the positions of $\alpha_\pm$ along a given path in the $s$-plane, one can determine whether they actually pinch the integration contour. For example, if we consider walking from $s=9$ to $s  \approx 0.997 + 0.050i$ in a straight line, the two roots do not pinch the contour. If we instead make an excursion around $\Delta_+ = 0$ before approaching $\Delta_- = 0$, we encounter a singularity. This singularity in the complex $s$ plane is called the pseudo-normal threshold.
	
	A similar singularity happens at $s=0$. Comparing with \eqref{eq:alpha-1}, we see that it corresponds to both $\alpha_\pm$ shooting off to infinity. Like in the case of the pseudo-normal threshold, this singularity is not on the physical sheet connected to $s=9$ by a straight line, but it is present on other sheets. We still need $\D<4$ for $\alpha_\pm$ to be poles in the first place, but now also $\D>2$ for the integrand of \eqref{eq:I-bubble} to have a pole at infinity, hence, this singularity only exists in $\D=3$. It corresponds to the collinear divergence for external particles. Note that the reason why this singularity was placed at infinity was just an artifact of the choice we made in~\eqref{eq:I-bub-lambda} for the $\GL(1)$ fixing. Had we made an alternative choice, say $\tilde{\alpha}_2 = 1$ as in~\eqref{eq:I-bub-lambda2}, we would have gotten the integral representation
	\be
	\I_\bubble = \Gamma(2{-}\D/2) \lim_{\eps \to 0^+} \int_{0}^{\infty} \frac{\d\alpha}{(\alpha + 1)^{\D/2}( -\tilde{\V} - i\eps)^{2-\D/2}},
	\ee
	where now $\tilde{\V} = \V|_{(\alpha_1,\alpha_2) = (\alpha,1)}$. In these variables, the collinear singularity is at a finite value: $\alpha = -1$. In fact, this is another sign that singularities are really best thought of in the projective sense, i.e., $(\alpha_1 : \alpha_2) = (\lambda \alpha_1 : \lambda \alpha_2)$ are treated as equivalent for any $\lambda > 0$. Written in those terms, the singularity at $s=0$ happens at
	\be\label{eq:bubble-2nd}
	(\alpha_1 : \alpha_2) = (\infty : 1{-}\infty) = (1 : -1),
	\ee
	where the former choice corresponds to the fixing $\tilde{\alpha}_2 = 1-\alpha$ and the latter to $\tilde{\alpha}_2 = 1$. Consulting~\eqref{eq:ell}, we also see that in the loop-momentum space, this singularity corresponds to a pinch as $\ell^\mu \to \infty$.
	
	Next, there are singularities associated to the roots $\alpha_\pm$ colliding with the endpoints of the integration domain. The endpoints, at $0$ and $1$ in the gauge fixing in which $\alpha_1=\alpha$ and $\alpha_2=1-\alpha_1$, cannot be displaced. Consider, for example, what happens near $\alpha = 0$. We can simply look at the function \eqref{eq:V-tilde}, which gives
	\be\label{eq:bubble-subleading}
	\tilde{\V}|_{\alpha = 0} = -m_2^2.
	\ee
	Hence such a singularity occurs only if $m_2^2 - i\eps = 0$. Similarly, there is a singularity at $m_1^2 - i\eps = 0$, corresponding to $\tilde{\alpha}_2 = 1$.
	
	Last, we can explore the singularity in the Regge limit, as $s \to \infty$ (potentially, on different sheets of the branch cuts we understood previously). At this stage we can explore yet another projective invariance of the problem, this time in the kinematic variables $(s : m_1^2{-}i\eps: m_2^2{-}i\eps)$. In those terms, we want to explore
	\be
	(\infty : m_1^2{-}i\eps : m_2^2{-}i\eps) = (s : 0 : 0) \,.
	\ee
	That is, at the level of determining singularities, the asymptotic limit $s \to \infty$ is equivalent to the simultaneous massless limit, $m_1^2 - i\eps = m_2^2 - i\eps = 0$, which we already explored above.
	
	To summarize, we have found three singularities associated to pinching the contour: the normal and pseudo-normal thresholds, as well as the collinear singularity. In the limit as $\eps\to0^+$, they can happen at
	\be
	\lim_{\eps \to 0^+} s = (m_1 {\pm} m_2)^2\quad\mathrm{or}\quad \lim_{\eps \to 0^+} s = 0 \,,
	\ee
	and are located at
	\be
	\lim_{\eps \to 0^+} (\alpha_1 : \alpha_2) = (\tfrac{1}{m_1} : \pm\tfrac{1}{m_2}) \quad\mathrm{or}\quad \lim_{\eps \to 0^+} (\alpha_1 : \alpha_2) = (1,-1) \,,
	\ee
	respectively.
	We also found singularities in the degenerate limits, as $m_1 = 0$, $m_2 = 0$, or both at the same time. Whether each of these singularities appears on a given sheet has to be determined using a more detailed analysis. For example, the pseudo-normal thresholds and the collinear singularity do not contribute on the physical sheet. Moreover, in the strict $\eps \to 0^+$ limit, the two branch cuts present at finite $\eps$ merge into a single one, and we recover Fig.~\ref{fig:Fig1} (middle).
	
	Note that the above discussion allows one to determine possible positions of singularities in a way that was independent of the space-time dimension $\D$. However, as we will find out in Sec.~\ref{sec:thimbles}, the type of those singularities (for example, if it gives a logarithmic or square-root branch point, or whether it is present at all), does depend on the specific $\D$.
	
	The problem of identifying branch cuts and branch points becomes substantially more involved for more complicated Feynman diagrams. This is because in general the ``roots'' $\tilde{\V}=0$ create a complex codimension-$1$ submanifold of the Schwinger-parameter space (in our case, a set of two points). A singularity of the corresponding Feynman integral may develop only if this submanifold becomes singular (in our case, two points colliding). Mathematically, the condition for this to happen can be formalized as the \emph{discriminant} of the $\{\tilde{\V}=0\}$ surface \cite{Mizera:2021icv}. Examples will be given in Sec.~\ref{sec:general-thresholds}.
	
	While being very precise, the discussion has been rather mathematical up to this stage. We should therefore pause and ask ourselves: what is the physical meaning of the above singularities? There are at least two natural answers, to which we turn now.
	
	\subsection{\label{sec:interpretation}Physical interpretations}
	
	Let us first consider singularities coming from pinching the contour with the two roots $\alpha_+ = \alpha_-$. Equivalently, we can think of it as
	\be
	\tilde{\V}_\eps =
	\tilde{\V}+i\eps = -s(\alpha - \alpha_+)(\alpha - \alpha_-)
	\ee
	developing a double zero. The condition for the double zero is that $\tilde{\V}_\eps$ and its first derivative vanish, i.e.,
	\be
	\tilde{\V}_\eps = \partial_{\alpha} \tilde{\V}_\eps = 0.
	\ee
	To understand the meaning of this statement more invariantly, let us return back to the form \eqref{eq:I-bubble2} before gauge fixing the $\GL(1)$ redundancy. In terms of the original variables $(
	\alpha_1, \alpha_2) = \lambda(\alpha, 1{-}\alpha)$, we have 
	\begin{align}
		\partial_{\alpha_1}\big[\V {+} i\eps(\alpha_1 {+} \alpha_2)\big] &= (1{-}\alpha) \partial_{\alpha} \tilde{\V}_\eps + \tilde{\V}_\eps = 0,\\
		\partial_{\alpha_2}\big[\V {+} i\eps(\alpha_1 {+} \alpha_2)\big] &= -\alpha\, \partial_{\alpha} \tilde{\V}_\eps + \tilde{\V}_\eps = 0.
	\end{align}
	So, the pinch singularity is nothing else than a condition for the saddle point of $\V$.
	
	Recall from~\eqref{eq:I-bubble2} that the factor $\V + i\eps(\alpha_1 {+} \alpha_2)$ appears in the exponent of the integrand of the Feynman integral in~\eqref{eq:I-bubble2}:
	\be
	\I_\bubble = \left(- i \hbar \right)^{\D/2 - 2}  \lim_{\eps \to 0^+} \int_0^\infty \frac{\d\alpha_1\, \d\alpha_2}{(\alpha_1 {+} \alpha_2)^{\D/2}} \exp \bigg[ \frac{i}{\hbar} \bigg( \V + i\eps (\alpha_1 {+} \alpha_2)\bigg) \bigg].
	\ee
	Just as in any quantum-mechanical system, saddle points dominate the integral in the classical limit, $\hbar \to 0$. Therefore, the corresponding singularities of the Feynman integral are to be understood as contributions to the scattering process where the path integral becomes sharply peaked around configurations in which the virtual particles become physical on-shell states. The saddles passing through the integration contour are guaranteed to contribute, while the question becomes more complicated on complex saddles: we will return back to this point in Sec.~\ref{sec:thimbles} in order to phrase the problem more precisely in the language of Lefschetz thimbles.
	
	Actually, there is a small caveat in the above discussion. While saddles indeed dominate the integral when $\hbar \to 0$, as a consequence of the $\GL(1)$ redundancy, the final expression turns out to be $\hbar$-independent. We can still retain the physical interpretation as the classical limit, but it is important to remember that singularities happen in the \emph{kinematic} limit (for example, $\Delta\to0$), and its value is independent of $\hbar$. This is a special property of saddle points in the worldline formalism, which distinguishes it from other kinds of path integrals.
	
	Let us keep going to see if we can trace back the interpretation of singularities in term of the loop momenta, which would give us yet another point of view. Indeed, following the treatment of singularities as saddle points, we can easily come back to the form \eqref{eq:I-bubble3}. In fact, performing the Gaussian integral in $\ell^\mu$ was noting else than an (exact) saddle-point expansion localized around the specific value of the loop momentum determined by
	\be\label{eq:bubble-loop}
	\partial_{\ell^\mu} \!\left[ (\ell^2 - m_1^2 + i\eps)\alpha_1 + ((p-\ell)^2 - m_2^2 + i\eps)\alpha_2 \right] = 0.
	\ee
	This condition indeed yields the solution \eqref{eq:ell}. After plugging in the values of $\alpha$'s for the normal \eqref{eq:alpha-normal} and pseudo-normal \eqref{eq:alpha-pseudo} thresholds (for the collinear singularity we already established around~\eqref{eq:bubble-2nd} that $\ell^\mu \to \infty$) we find
	\be\label{eq:ell-bubble}
	\ell^\mu = p^\mu \frac{\sqrt{m_1^2 - i\eps}}{\sqrt{m_1^2 - i\eps} \pm \sqrt{m_2^2 - i\eps}}.
	\ee
	Using $p^2 = s$, the inverse propagators read on the support of this saddle
	\begin{align}
		\ell^2 - m_1^2 + i\eps = \frac{(m_1^2 - i\eps)\Delta_\pm  }{s - \Delta_\pm},\qquad (p-\ell)^2 - m_2^2 + i\eps = \frac{(m_2^2 - i\eps)\Delta_\pm}{s - \Delta_\pm}.
	\end{align}
	That is, both of them are put on shell when either $\Delta_+ = 0$ or $\Delta_-=0$. Note that the converse is not true: solely putting propagators on shell does not necessarily lead to a singularity. In fact, we have already identified the physical origin of singularities as classical configurations. But on-shell particles are not enough to create a physical scattering process: we also have to impose that the momenta enforce the particles to meet at the interaction points, which is a version of locality.
	
	The space-time distance $\Delta x_e^\mu$ that a classical particle travels is proportional to its momentum times proper time. Note that orientation of $\Delta x^\mu$ is inherited from the (auxiliary) orientation of the corresponding propagator. For the bubble integral, we have
	\be
	\Delta x_1^\mu = \ell^\mu \alpha_1, \qquad \Delta x_2^\mu = (p^\mu-\ell^\mu) \alpha_2.
	\ee
	Therefore, to impose that the particles interact locally, we need
	\be
	\Delta x_1^\mu = \Delta x_2^\mu.
	\ee
	One can easily check that this constraint is equivalent to \eqref{eq:bubble-loop}.
	This perspective consolidates the interpretation of singularities as classical configurations of on-shell particles interacting at isolated space-time points. It is known as the Coleman--Norton theorem \cite{Coleman:1965xm}.
	
	Finally, with the above intuition, the meaning of subleading singularities becomes clear. For example, consider the case $\alpha_1=0$. The problem becomes identical to that of the bubble diagram, except the first edge gets contracted to a point. We end up with a tadpole diagram with a single loop. We already established around \eqref{eq:bubble-subleading} that the on-shell constraint can be only solved when $m_2^2 - i\eps = 0$. Moreover, the explicit solution \eqref{eq:ell-bubble} indicates that $\ell^\mu = p^\mu$. Note that $\alpha_2$ remains unconstrained. Hence this singularity comes from the massless propagator of the second edge becoming soft.

	\subsection{\label{sec:thimbles}Lefschetz thimbles}
	
	Now that we know about the interpretation of singularities as coming from saddle points of $\V$, let us present another perspective on the $i\eps$, which will serve as a segue to the later discussion of contour deformations. It will also teach as something new about local behavior of Feynman integrals near their thresholds.
	
	The idea is to reverse the logic and first classify the possible integration contours giving rise to singularities. Then, the remaining question is whether the original integration contour is equivalent to one of them. This is a standard procedure in studying analytic continuation of path integrals with complex Morse theory, previously applied to similar problems, including Chern--Simons theory \cite{Witten:2010cx}, Liouville theory \cite{Harlow:2011ny}, lattice QCD \cite{Cristoforetti:2012su}, wavefunction of the Universe \cite{Feldbrugge:2017kzv}, and tree-level scattering in string theory \cite{Mizera:2019gea}. 
	
	In the present application, we are interested in the behavior of the exponent in \eqref{eq:I-bubble2}. More specifically, let us define
	\be
	\V_\eps = \V + i\eps(\alpha_1 + \alpha_2),
	\ee
	whose saddles are determined by solving $\partial_{\alpha_e} \V_\eps = 0$.
	For each saddle point, we associate the contour of steepest descent $\cal J$, which is the manifold along which $\Re (i\V_\eps) = - \Im \V_\eps$ decreases the most rapidly away from the saddle, while keeping the value of $\Im (i\V_\eps) = \Re \V_\eps$ fixed. In other words, the exponent $\exp(\tfrac{i}{\hbar}\V_\eps)$ damps the integrand as much as possible along $\cal J$, while its phase remains stationary. Similarly, one can associate the contour of steepest ascent $\cal K$ with the same property, except $\Re(i\V_\eps)$ increases the most rapidly away from the saddle. The contours $\cal J$ and $\cal K$ are referred to as Lefschetz thimbles and anti-thimbles respectively (Cauchy--Riemann equations guarantee that they have the same dimension, i.e., the critical point is a saddle).
	
\begin{figure}
		\centering
		\includegraphics[scale=0.37]{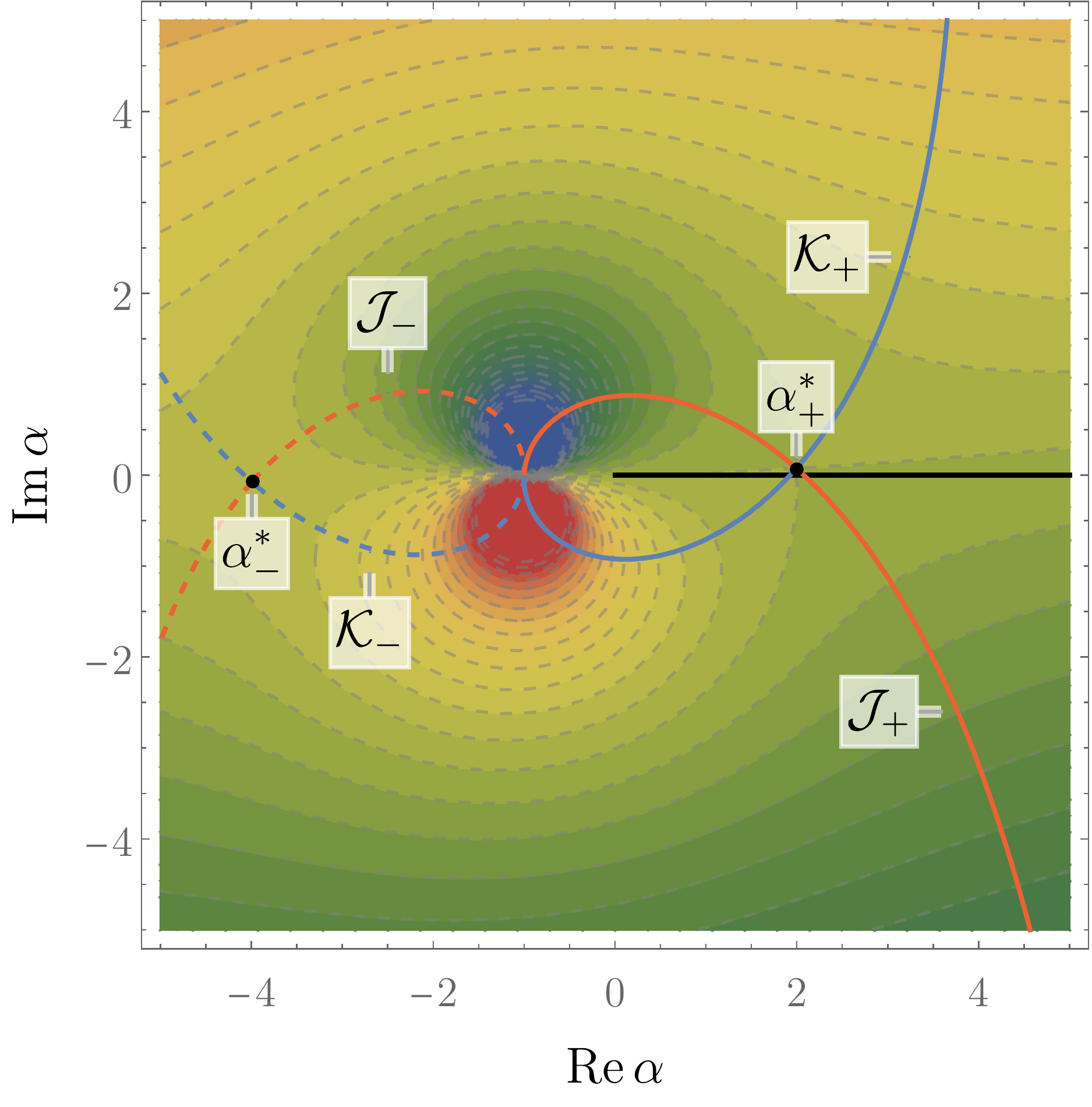}\quad
		\includegraphics[scale=0.37]{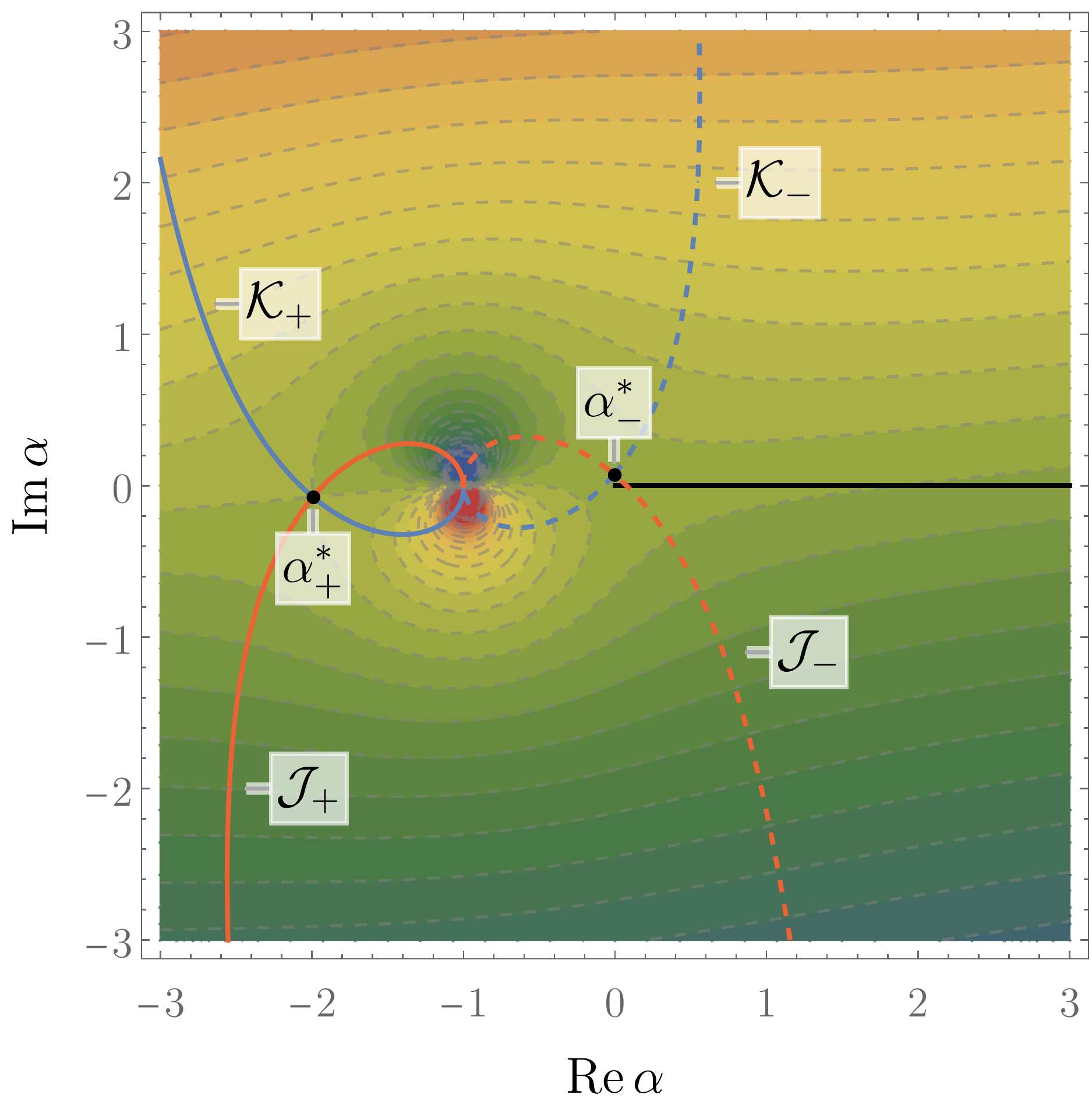}
		\caption{\label{fig:thimbles}Lefschetz thimbles for the normal (left) and pseudo-normal (right) thresholds of the bubble diagram with the $\GL(1)$ fixing $(\alpha_1,\alpha_2) = (\alpha,1)$. The value of $\Re(i\V_\eps)$ is plotted with gradient lower values in blue and higher in red. In each case, the integration contour $[0,\infty]$ is shown in black, while the thimbles ${\cal J}_\pm$ and ${\cal K}_\pm$ is in red and blue respectively. Pseudo-thimbles are dashed. The corresponding (pseudo-)saddle points $\alpha_\pm^\ast$ are indicated as black dots. For numerics we used $m_1 =1$, $m_2=2$, and $\eps = \tfrac{1}{10}$.}
	\end{figure}

	\paragraph{Classifying contours.}
	The situation is complicated slightly by the presence of the $\GL(1)$ redundancy. We can fix it, e.g., by $(\alpha_1,\alpha_2) = (\alpha,1)$ and analyze the thimbles in the complex $\alpha$-plane, see Fig.~\ref{fig:thimbles}. Crucially, the new saddle-point condition $\partial_{\alpha}\V_\eps = 0$ is only a single equation and hence constraints only $\alpha$:
	\be
	\alpha^\ast_\pm = \pm \frac{\sqrt{s}}{\sqrt{m_1^2 - i\eps}} - 1,
	\ee
	on which $\V^\ast_\eps = (\sqrt{s} \mp \sqrt{m_1^2 - i\eps})^2 - m_2^2 + i\eps$.
	The Mandelstam invariant $s$ is left free. On the normal threshold $\Delta_+ = 0$, i.e., $ \sqrt{s} = \sqrt{m_1^2 - i\eps} + \sqrt{m_2^2 - i\eps}$, we find that only the solution $\alpha^\ast$ with the plus sign is the singular saddle point that leads to $\V^\ast_\eps = 0$. It can be simplified to
	\be
	\label{eq:apstar}
	\alpha^\ast_+ = \frac{\sqrt{m_2^2 - i\eps}}{\sqrt{m_1^2 - i\eps}},
	\ee
	which is equivalent to \eqref{eq:alpha-normal} in the current $\GL(1)$ fixing. Since the solution $\alpha_-$ does not correspond to one for which $\V_{\eps} = 0$ at the normal threshold, we call it a \emph{pseudo-saddle}. In other words, the true saddle is the place where the pinch singularity happens, while the pseudo one does not give any divergence. Similarly, on the pseudo-normal threshold, we have one of each types of saddles.
	
	The situation is summarized in Fig.~\ref{fig:thimbles}, which presents the thimbles on the value of $s$ corresponding to the normal (left) and pseudo-normal (right) thresholds. In both cases, the Lefschetz thimbles ${\cal J}_\pm$ and anti-thimbles ${\cal K}_{\pm}$ are drawn in red and blue respectively. Those associated to the true saddle are in solid, while those corresponding to the pseudo-saddle are dashed. In both cases, they join together regions where $\Re(i\V_\eps) \to \pm \infty$, which happen either when $\alpha$ reaches $-1$ or goes to infinity in various complex directions. The solid black line represents the original integration contour, which is $[0,\infty]$ in the current $\GL (1)$ fixing.
	
	Let us first consider the normal threshold. Because of the $i\eps$ factor in $\V_\eps$, the integrand decays along the real axis as $\alpha \to \infty$. Therefore, the original integration contour can be deformed downwards, such that it coincides with the Lefschetz thimble ${\cal J}_+$ (red, solid) without spoiling convergence of the integral. In particular, it can be deformed to pass through the true saddle point at $\alpha = \alpha_\pm$ given in \eqref{eq:alpha-normal}. Of course, the endpoint at $\alpha=0$ has to remain fixed, but within the vicinity of the saddle, ${\cal J}_+$ provides an excellent approximation to the original integration contour. This is what gives the normal threshold singularity. As an aside, let us mention that ${\cal J}_{+}$ asymptotically tending to $\alpha \to -i\infty$ instead of $\alpha \to +\infty$ has the physical interpretation of Wick-rotating the associated worldline from Lorentzian to Euclidean signature, cf. \cite{Witten:2013pra}.
	
	The situation changes qualitatively for the pseudo-normal threshold. While the integration contour can still be deformed to the pseudo-thimble ${\cal J}_-$ (red, dashed), 
	the corresponding pseudo-saddle does not lead to a singularity of the integral.
	However, when exploring singularities on other sheets in the $s$-plane, we will see 
	in Sec.~\ref{sec:discontinuities} how the integration contour is forced to be deformed, which can potentially lead to singularities. For example, on the second sheet of the normal threshold, the original contour needs to swirl around the origin, then pass through the true saddle of the pseudo-normal threshold $\alpha = \alpha_\pm$ (given in \eqref{eq:alpha-pseudo}), resulting in a singularity, before shooting off to infinity.
	
	An advantage of
	using Lefschetz thimbles to approximate the integration contour
	is that it allows us to compute the leading approximation to the Feynman integral around its singularity. Just like in any other path-integral approach to quantum-mechanical problems, this simply corresponds to
	studying fluctuations around saddle points. As we have learned in Sec.~\ref{sec:interpretation},
	saddle points of Feynman integrals do not localize in the Schwinger-parameter space only, but also put constraints on the kinematic invariants, such as $\Delta_+ = 0$. This means that the saddle-point expansion will not only analyze where the integral is singular, but also give a local expansion of Feynman integrals around thresholds.
	
	\paragraph{Threshold expansion.}
	Let us explain how this expansion works for the normal threshold. Since we already determined that ${\cal J}_+$ provides a good approximation near the saddle, we can write~\eqref{eq:I-bub-lambda2} close to $\Delta_+ = 0$:
	\be
	\I_{\bubble} \approx (-i\hbar)^{\D/2-2} \lim_{\eps \to 0^+} \int_{0}^\infty \frac{\d \lambda}{\lambda^{\D/2-1}} \int_{{\cal J}_+} \frac{\d \alpha}{(\alpha + 1)^{\D/2}} \exp\left[ \frac{i\lambda}{\hbar} \V_\eps \right].
	\ee
	with $\V_{\eps} = s \frac{\alpha}{\alpha+1} - m_1^2 \alpha - m_2^2 + i \varepsilon (\alpha+1)$.
	Around the solution $\alpha = \alpha_+$, given explicitly in \eqref{eq:apstar}, the action $\V_{\eps}$ admits the expansion
	\be
	\V_\eps = \frac{\Delta_+ \sqrt{m_2^2 - i\eps} - (\alpha - \alpha_+^\ast)^2 (m_1^2 - i \eps)^{3/2}}{\sqrt{m_1^2 - i\eps} + \sqrt{m_2^2 - i\eps}} + \ldots,
	\ee
	where, to stay consistent, we expanded in both $\Delta_+$ and $\alpha-\alpha_+^\ast$. Note that the latter starts at the quadratic order since both $\V_\eps^\ast$ and $\partial_{\alpha} \V_\eps^\ast$ vanish.
	Let us first localize the ${\cal J}_+$ integration using the standard saddle-point approximation (see, e.g, \cite{doi:10.1098/rspa.1991.0119}):
	\be
	\int_{{\cal J}_+}  \frac{\d \alpha}{(\alpha + 1)^{\D/2}} \exp\left[ \frac{i\lambda}{\hbar} \V_\eps \right] \approx  \left(-\frac{\pi i \hbar
	}{m_1^2 \lambda} \right)^{1/2} \left( \frac{m_1}{m_1 {+} m_2} \right)^{\!(\D-1)/2} \exp\left[ \frac{i\lambda}{\hbar}  \frac{m_2}{m_1{+}m_2} \Delta_+ \right].
	\ee
	We suppressed the $\eps$-dependence, since
	it is only needed inside of $\Delta_+$ (the integral over ${\cal J}_+$ converges by definition).
	In the next step we simply perform the leftover integral over $\lambda$.
	Note that the $\alpha$-localization results in the additional power of $\lambda^{-1/2}$. This means that the $\lambda$-integral is convergent at the lower limit $\lambda \to 0$ only when $\D < 3$, leaving us with
	\be
	\I_\bubble \approx \sqrt{\pi} \left( -i\hbar\, m_1 \right)^{\!(\D-3)/2} (m_1 {+} m_2)^{(1-\D)/2} \lim_{\eps \to 0^+} \int_{0}^{\infty} \frac{\d\lambda}{\lambda^{(\D-1)/2}} \exp\left[ \frac{i\lambda}{\hbar} \frac{m_2}{m_1 {+} m_2} \Delta_+ \right].
	\ee
    Performing the integral yields 
    \be
	\I_\bubble^{\D < 3} \approx \frac{\sqrt{\pi}\, \Gamma(\tfrac{3-\D}{2})}{(m_1 + m_2)^{\D-2}}  \left( - m_1 m_2 \Delta_+ \right)^{(\D-3)/2}
	\ee
	in $\D<3$. One can also make sense of this result as $\D \to 3^-$ by taking the limit and extracting the leading piece in $\Delta_+ \to 0$, giving
	\be
	\I_\bubble^{\D=3} \approx -\frac{\sqrt{\pi}}{m_1 + m_2}
    \log\left( -\Delta_+\right).
	\ee
	One can check that these two approximations agree with expanding the explicit answers \eqref{eq:bub2d} and \eqref{eq:bub3d} around the normal threshold. We will generalize this procedure to more general singularities of Feynman integrals in Sec.~\ref{sec:fluctuations}. 
	
	The above manipulations taught us an important lesson: deformations of Schwinger parameters can serve a similar role as the Feynman $i\eps$ prescription. This motivates approaching the subject from a more algebraic point of view, which is what we move on to now.
	
	\subsection{\label{sec:deformations}Contour deformations}
	
\begin{figure}
		\centering
		\includegraphics[scale=0.5]{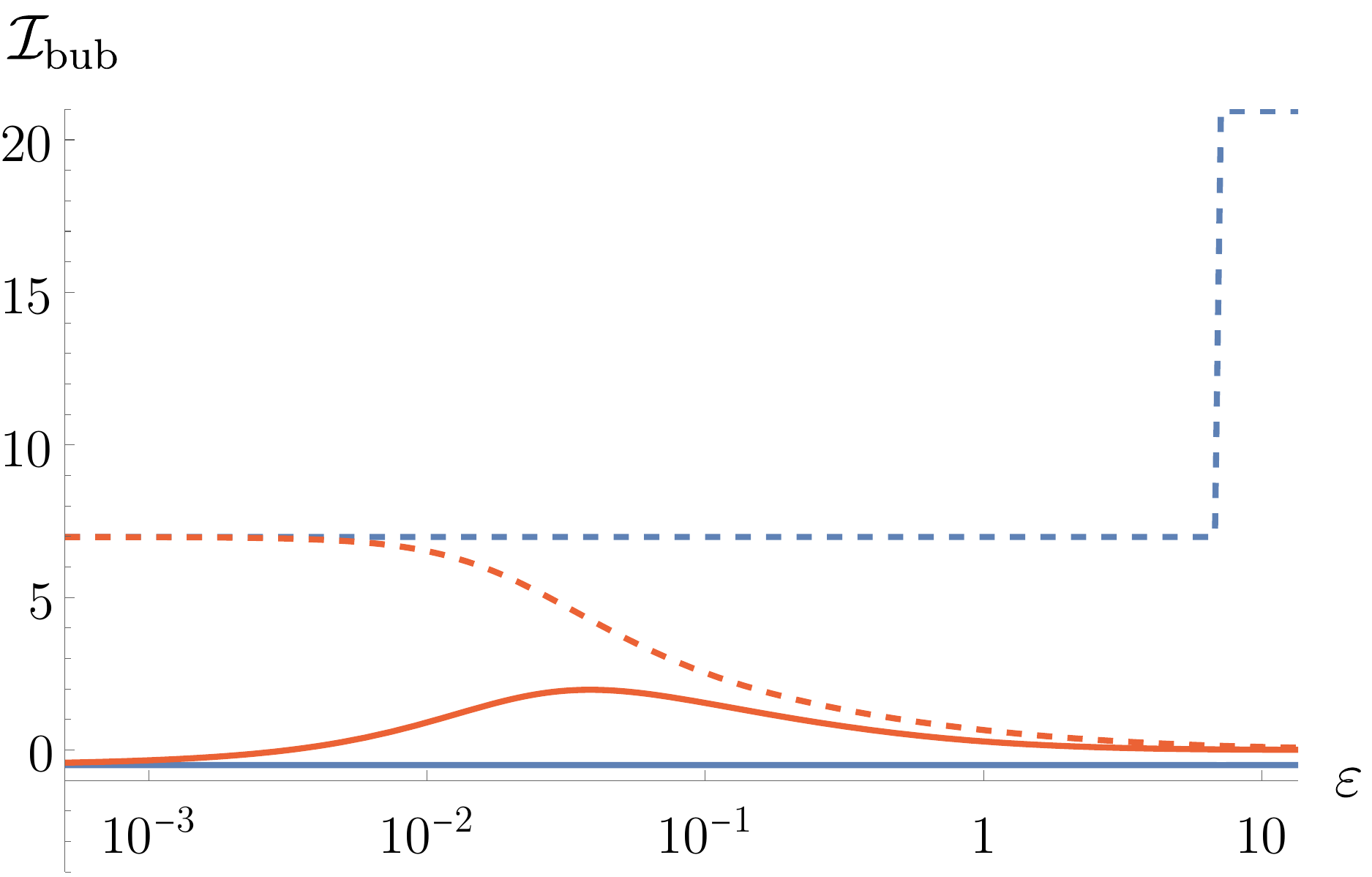}
		\caption{\label{fig:epsilon}Real (solid) and imaginary (dashed) parts of the bubble diagram $\I_\bubble(s)$ at $s=9.1$ obtained using the Feynman $i\eps$ (red) and contour deformations (blue) for different values of $\eps$. The latter gives an exact answer until $\eps$ reaches values of ${\cal O}(10)$.}
	\end{figure}

	Up to this point, we have kept $\eps$ as an infinitesimal parameter that had to be tuned to zero at the end of the computation. This prescription suffers from two serious problems. First, as exemplified in Sec.~\ref{sec:branch-cuts}, it modifies the analytic structure by displacing branch points and introducing spurious branch cuts. Second, it makes the numerical evaluation troublesome, as the poles $\alpha_\pm$ get arbitrarily close to the integration contour in the limit as $\eps \to 0^+$, resulting in large numerical errors. As an example, see Fig.~\ref{fig:epsilon} for $\I_\bubble$ as a function of $\eps$. However, we have learned important lessons about why the $i\eps$ was needed in the first place: it chooses the causal branch of the Feynman integral and ensures its convergence. To achieve these same goals, while also fixing the problems of modified analytic structure and poor numerical convergence, we can instead attempt to perform contour deformations. In the loop-momentum space, this method is the standard one, see, e.g., \cite{Itzykson:1980rh}, but unfortunately, it breaks Lorentz invariance. The goal of this subsection is to explain how to implement contour deformations in the Schwinger-parameter space, which has the advantage of being manifestly Lorentz-invariant \cite{Mizera:2021fap}. It leads to a more canonical way of understanding branch cuts and branch points for arbitrary Feynman integrals.
	
	\paragraph{Causality requirements.}
	Contour deformations in the space of Schwinger parameters can be implemented in many different ways, but
	in view of making
	the resulting expressions as practical as possible, we will start with the form \eqref{eq:I-bubble2}:
	\be
	\label{eq:I-bubble-contour}
	\I_\bubble = \left(-i \hbar \right)^{\D/2-2} \lim_{\eps \to 0^+} \int_0^\infty \frac{\d\alpha_1\, \d\alpha_2}{(\alpha_1 {+} \alpha_2)^{\D/2}} \exp \bigg[ \frac{i}{\hbar} \bigg( \V + i\eps (\alpha_1 {+} \alpha_2) \bigg) \bigg]
	\ee
	Recall that the $i\eps$ was needed only to ensure convergence of the integral as Schwinger parameters become large. We can achieve the same goal by deforming $\V$ such that
	\be\label{eq:ImV}
	\Im\, \V > 0
	\ee
	for all values of the $\alpha$'s (strictly speaking, this is only needed for asymptotically-large $\alpha$'s). There are
	two ways of implementing this constraint: deforming the external kinematic invariants or the internal integration variables.
	
	In a sense, deforming $m_e^2 \to m_e^2 - i\eps$ was an example of the first option. However, this deformation has a disadvantage of being unphysical: masses are often fixed in an on-shell process, e.g., by a Higgs mechanism or gauge invariance. In the case of the bubble integral, with $\V = s \frac{\alpha_1 \alpha_2}{\alpha_1+\alpha_2} -m_1^2 \alpha_1 - m_2^2 \alpha_2$, we can instead deform the kinematic invariant $s$ without breaking any symmetry. Since the coefficient of $s$ in $\V$ is always positive, we simply need
	\be
	\Im\, s > 0.
	\ee
	In the case at hand, this deformation is a perfectly good way of imposing causality, ensuring that the physical region is approached from the causal direction. However, as explained in Sec.~\ref{sec:introduction}, deforming the kinematic invariants by simply adding a small imaginary part does not extend to general S-matrix elements, and hence we need to seek alternatives.
	
	\paragraph{Contour deformations.} Let us instead pursue the second option: the avenue of contour deformations, simply implemented as a change of variables $\alpha_e \to \hat{\alpha}_e$. To make the deformations maximally simple, we work at the level of the formula \eqref{eq:I-bubble-contour} so that $\alpha_1$ and $\alpha_2$ can be deformed independently. Since we only need infinitesimal deformations, we can take
	\be
	\hat{\alpha}_e = \alpha_e + \eps\, \delta \alpha_e
	\ee
	with $\eps$ small, for $e=1,2$. Here both $\hat{\alpha}$'s are parametrized by $\alpha$'s running from $0$ to $\infty$.
	The goal is to determine the appropriate $\delta \alpha_e$ that implements the constraint~\eqref{eq:ImV}. Of course, we need $\delta\alpha_e \to 0$ as $\alpha_e \to 0$ so that the endpoint of integration at the origin is preserved. The behavior of $\delta\alpha_e$ at infinity needs be precisely such that~\eqref{eq:ImV} holds. Recall that the condition for a branch cut was $\V = 0$, so the contour deformations should be relevant only near such points in the $(\alpha_1, \alpha_2)$-space.
	
	After the deformation, we have a new $\hat{\V} = \V(\hat{\alpha}_e)$, which we can expand in $\eps$ to leading orders:
	\be
	\hat{\V} = \V + \eps \left( \delta\alpha_1\, \partial_{\alpha_1} \V + \delta\alpha_2\, \partial_{\alpha_2} \V \right) + \ldots.
	\ee
	If $s$ has an imaginary part, then $\Im\,\V = \Im s \frac{\alpha_1\alpha_2}{\alpha_1 + \alpha_2}$ could not be zero, so contour deformations are not necessary. Thus, we only need to understand what happens near the real $s$-axis, in which case $\V$ and all its derivatives are real. As a consequence, a natural candidate for the deformation is
	\be
	\delta \alpha_e = i \alpha_e \partial_{\alpha_e} \V,
	\ee
	since it leads to
	\be
	\Im \hat{\V} = \eps \left[ \alpha_1 (\partial_{\alpha_1}\V)^2 + \alpha_2 (\partial_{\alpha_2}\V)^2 \right] + \ldots.
	\ee
	Hence, for sufficiently small $\eps>0$, this choice leads to $\Im \hat{\V} > 0$, as required. This contour deformation is the Schwinger-parametric analogue of that for Feynman propagators, although the present form has the advantage of being Lorentz-invariant.
	
	In order to wrap up the discussion, let us spell out the full formula for the contour-deformed bubble integral:
	\be\label{eq:I-bubble4}
	\I_\bubble = \left(-i \hbar \right)^{\D/2-2} \int_{0}^{\infty} \frac{\JJ\, \d\alpha_1\, \d\alpha_2}{(\hat{\alpha}_1 {+} \hat{\alpha}_2)^{\D/2}} \exp \left[\tfrac{i}{\hbar} \hat{\V}\right],
	\ee
	where the Jacobian $\JJ = \det \mathbf{J}$ is the determinant of the $2\times 2$ matrix
	\be
	\mathbf{J}_{e e'} = \frac{\partial \hat{\alpha}_{e}}{\partial \alpha_{e'}} = \delta_{ee'} + i\eps \left(1 + \alpha_e \partial_{\alpha_{e'}}\right) \partial_{\alpha_e} \V.
	\ee
	Let us emphasize that even though it was sufficient to expand $\hat{\V}$ to leading orders to guarantee convergence of the integral, $\hat{\V} = \V(\hat{\alpha}_e)$ in the above expression has the full $\eps$-dependence. Explicitly, the deformed Schwinger parameters are given by
	\be
	\hat{\alpha}_1 = \alpha_1 \left[ 1 + i\eps \left( s \big(\tfrac{\alpha_2}{\alpha_1{+}\alpha_2}\big)^{2} - m_1^2 \right) \right], \quad \hat{\alpha}_2 = \alpha_2 \left[ 1 + i\eps \left( s \big(\tfrac{\alpha_1}{\alpha_1{+}\alpha_2}\big)^{2} - m_2^2 \right)\right].
	\ee

	As an additional bonus, the contour deformation preserved the  $\GL(1)$ redundancy, i.e., under dilations $\alpha_e \to \lambda \alpha_e$ we still have
	\be
	\hat{\alpha}_e \to \lambda \hat{\alpha}_e, \qquad \hat{\V} \to \lambda \hat{\V}.
	\ee
	This means we can safely repeat all the steps following \eqref{eq:I-bubble2} after contour deformations, effectively performing one integration exactly. For example, we can set $(\alpha_1,\alpha_2) = \lambda(\alpha, 1 {-} \alpha)$ and integrate out $\lambda$ (alternatively we could have chosen $(\hat{\alpha}_1,\hat{\alpha}_2) = \lambda(\alpha, 1 {-} \alpha)$, but this constraint would be difficult to express in terms of the resulting integration variable $\alpha$).
	After the dust settles, we are left with:
	\be\label{eq:I-bubble5}
	\I_\bubble = \Gamma(2{-}\D/2) \int_{0}^{1} \frac{\JJ\, \d\alpha}{(\hat{\alpha}_1 {+} \hat{\alpha}_2)^{\D/2} ( -\hat{\V})^{2-\D/2}} \bigg|_{(\alpha_1, \alpha_2) = (\alpha, 1-\alpha)}
	\ee
	This is the form most appropriate for numerical computations.
	
	\paragraph{Infinitesimal vs. sufficiently small $\eps$.}
	Note that in \eqref{eq:I-bubble4} and \eqref{eq:I-bubble5} we dropped the $\lim_{\eps \to 0^+}$ from the expression. This is because $\eps$ no longer needs to be an infinitesimal variable taken to vanish at the end of the computation. Instead, it only needs to be sufficiently small for the integral to converge to the correct value. Recall from Fig.~\ref{fig:Fig1} that the branch-cut deformation reveals an analytic region around the physical value of $\I_\bubble$, in contrast with the Feynman $i \eps$ prescription, which displaces the location of the branch point and thus modifies the physical value for any finite $\eps$. This crucial distinction is illustrated in Fig.~\ref{fig:epsilon}, where we evaluated $\I_\bubble$ at $s=9.1$ as a function of $\eps$ using the Feynman $i\eps$ prescription (orange) and contour deformations (blue). The former approaches the correct
	answer $\I_\bubble \approx -0.496 + 6.981\,i$ to three decimal places only for $\eps \lesssim 10^{-5}$, where numerical errors are so large that high numerical precision is required. The latter converges and gives the exact value even for $\eps = {\cal O}(1)$. 
	
	\paragraph{Summary.}
	Let us summarize the big picture. We have learned that the analytic structure of the bubble Feynman integral can be entirely fixed without having to perform the integration directly. Branch cuts are determined by asking for which values of the kinematic invariants $\V=0$ has a solution in terms of the $\alpha$'s. Freedom in deforming the $\alpha$-contour away from this situation is equivalent to studying different sheets in the kinematic space. Branch points further require that all derivatives of $\V$ vanish for some set of $\alpha$'s, in which case no deformation away from a singularity is possible. Finally, the causal branch is selected by requiring that $\Im\, \V > 0$, which is equivalent to the Feynman $i\eps$ prescription. Stated in this way, all three criteria generalize to an arbitrary Feynman integral, and in fact will be crucial in the following sections.
	Moreover, contour deformations of the type described above can also easily be generalized and are further discussed in Sec.~\ref{sec:deformation}.

	\subsection{\label{sec:discontinuities}Discontinuity, imaginary part, and unitarity cuts}
	
	Finally, let us return to the question of imaginary parts and discontinuities. The goal is to establish explicit formulae which compute these quantities without knowing the full answer, and show their equivalence.
	
	\paragraph{Discontinuity around the normal threshold.}
	Recall that we learned from Sec.~\ref{sec:branch-cuts} that the discontinuity around the normal threshold at $s = (m_1 + m_2)^2$ (which with our choice of numerical values becomes $s=9$), can be understood as integrating the original integrand over a different contour. In fact, to illustrate the general formalism, let us not restrict ourselves to $\D=2$ as in Sec.~\ref{sec:branch-cuts}, but rather take $\D<4$ (since we are deforming the external kinematics, we can safely forget about the $i\eps$ prescription for now). The integrand of \eqref{eq:I-bubble} contains $\tilde\V$ raised to the power $\D/2$, which can introduce either two poles $\alpha_\pm$, or a branch cut with branch points $\alpha_\pm$, when $\D$ is even and odd respectively.
	
\begin{figure}
		\centering
		\includegraphics[scale=0.85]{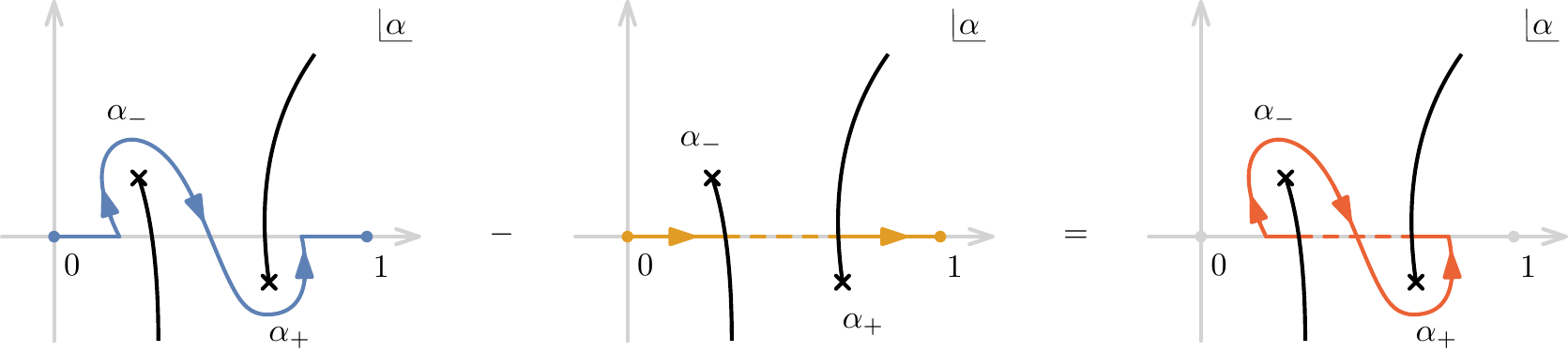}
		\caption{Contour prescription for computing discontinuity in the $\alpha$-space. The figure-eight contour $\gamma$ (right) is the difference between the integration contour after deformations (left) and the original one (middle).}
		\label{fig:discontinuity}
	\end{figure}

	As explained in Sec.~\ref{sec:branch-cuts}, the discontinuity across the normal threshold is given by a difference between the  integration contour deformed to avoid singularities and the original contour from $0$ to $1$, where we now use the $\GL (1)$ fixing $(\tilde{\alpha}_1,\tilde{\alpha}_2)=(\alpha,1-\alpha)$ as in~\eqref{eq:I-bub-lambda}. To stay general, we might as well pretend that branch cuts extending from $\alpha_\pm$ are always there, since the final prescription will only simplify more in their absence. This procedure leads to the figure-eight contour $\gamma$ illustrated in Fig.~\ref{fig:discontinuity}. Therefore, the resulting formula for discontinuity can be obtained from~\eqref{eq:I-bubble} by changing the integration contour:
	\be
	\disc_{s=(m_1 + m_2)^2} \I_{\bubble} = \frac{1}{2 i} \Gamma(2{-}\D/2) \int_{\gamma} \frac{\d\alpha}{(-\tilde{\V})^{2-\D/2}},
	\ee
	where the normalization of $1/2i$ is, for later convenience, added as a part of the definition of the discontinuity operator $\disc$. 
	Here, the symbol ``$\disc$'' on the left-hand side denotes the difference between the value of $\I_{\bubble}$ after analytically continuing anti-clockwise around the normal threshold once (without crossing any branch cuts) and the original value of $\I_{\bubble}$ divided by $2i$. In other words, we should think about a discontinuity as the difference between the value of the Feynman integral on the second sheet of the corresponding threshold cut and that on the physical sheet.
	
	One can easily see that the contour $\gamma$, and hence the value of the discontinuity, does not depend on the choice of the $\alpha$-branch cut.
	Note that when $\D$ is even, the cut disappears and the formula simply gives a difference between the residues around $\alpha_+$ and $\alpha_-$, as explained in Sec.~\ref{sec:branch-cuts}.
	In either case, the formula can be evaluated and gives
	\be\label{eq:disc-bub}
	\disc_{s=(m_1 + m_2)^2} \I_{\bubble} = \begin{dcases}
		0 \quad&\text{if}\quad \D=1,\\
		-\frac{2\pi}{i \sqrt{-\Delta}} \quad&\text{if}\quad \D=2,\\
		\frac{\pi^{3/2}}{\sqrt{s}} \quad&\text{if}\quad \D=3.\\
	\end{dcases}
	\ee
	Recall that $\Delta$ is the discriminant, which we now use in the $\eps \to 0^+$ limit:
	\be
	\Delta = \left(s - (m_1 + m_2)^2\right) \left(s - (m_1 - m_2)^2\right).
	\ee
	Note that we wrote the discontinuity such that the square root branch cut for the normal threshold in the $\D=2$ case is consistent with the one prescribed by the original integration contour, dictating a branch cut for $s>(m_1+m_2)^2$. We are, however, free to choose any directions for the branch cuts that are not on the physical sheet, i.e., for the one extending from the pseudo-normal threshold at $s=(m_1-m_2)^2$ in $\D=2$, as well as for the one that starts at the second-type singularity at $s=0$ in $\D=3$.
	
	\paragraph{Imaginary part.}
	As an alternative prescription, let us compute the imaginary part of the Feynman integral directly.
	For concreteness, we can specialize to $\D=2$. Starting from \eqref{eq:I-bubble}, we have
	\be
	\Im\, \I_{\bubble} = 
	-\lim_{\eps \to 0^+} \frac{1}{2i} \int_{0}^{1} \d \alpha \left( \frac{1}{\tilde{\V}+i\eps} - \frac{1}{\tilde{\V}-i\eps} \right)\\
	= \pi \int_{0}^{1} \d\alpha\, \delta(\tilde{\V}),
	\ee
	where we used the identity
	\be\label{eq:Im-identity}
	\lim_{\eps \to 0^+} \left( \frac{1}{x+i\eta\eps} - \frac{1}{x-i\eta\eps} \right) = -2\pi i \sgn(\eta)\delta(x)
	\ee
	with $\eta = 1$. Finally, we need to localize the integration variable $\alpha$ on the roots of $\tilde{V} = -s (\alpha - \alpha_+)(\alpha - \alpha_-) = 0$ lying within the interval $\alpha_{\pm} \in (0,1)$. First, notice that this can only happen for $s > (m_1 + m_2)^2$, i.e., above the threshold. We then find
	\be\label{eq:Im-bub2}
	\Im\, \I_{\bubble} = \pi \, \Theta[s - (m_1{+}m_2)^2] \underbrace{\int_{0}^{1} \d\alpha\, \frac{\delta(\alpha - \alpha_+) + \delta(\alpha - \alpha_-)}{|\partial_{\alpha}\tilde{\V}|}}_{2/\sqrt{\Delta}}.
	\ee

	\paragraph{Total discontinuity.}
	
	From the discussion in Sec.~\ref{sec:deformations}, we know that the imaginary part of the bubble integral is guaranteed to be equal to the discontinuity across the upper- and lower-half planes, i.e.,
	\be\label{eq:Disc-bub}
	\Disc_{s} \I_{\bubble} = \frac{1}{2i} \lim_{\eps \to 0^+}\left[\I_{\bubble}(s+i\eps) - \I_{\bubble}(s-i\eps) \right].
	\ee
	Notice that the evaluation of the discontinuity \eqref{eq:Disc-bub} would involve an almost identical computation: instead of $\tilde{\V} \pm i\eps$, the denominator would have involved $\tilde{\V} \pm i\eps \alpha(1-\alpha)$. But the combination $\alpha(1-\alpha)$ is always positive and the identity \eqref{eq:Im-identity} is only sensitive to its sign. Therefore, we the result is the same formula and hence
	\be
	\Disc_s \I_{\bubble} = \Im\, \I_{\bubble}.
	\ee

	\paragraph{Unitarity cuts.}
	Finally, we can use unitarity to relate the imaginary part of the bubble integral to its cut. To calculate the unitarity cut, we apply the cut formula from~\eqref{eq:cuttingrules} to the momentum-space expression in~\eqref{eq:I-bubble-mom}, giving
	\be 
	    \Cut_{12}^s\, \I_{\bubble}(s) = \frac{\left(-2\pi i \right)^2 }{2 i} 
	    \int \frac{\d^\D \ell}{i \pi^{\D/2}}\,
	    \delta^+ \!\left[\ell^2-m_1^2\right] 
	    \delta^+ \!\left[(p-\ell)^2-m_2^2\right] \,.
	\ee 
	Note that it is only valid in the physical region, denoted with superscript ${}^s$, which in particular means $s>0$. To evaluate this integral, we work in the center-of-mass frame of the incoming particles, in which $p^\mu = (\sqrt{s},\vec{0})$, and change to spherical coordinates for the spatial components of the loop momentum $\vec{\ell}$. This results in
	\begin{multline} 
	    \Cut_{12}^s\, \I_{\bubble} =
	    2 \pi^{2-\D/2}
	    \int_{-\infty}^{\infty} \d \ell^0 
	    \int_0^\infty \vert \vec{\ell} \vert^{\D-2} \d \vert \vec{\ell}  \vert \int \d \Omega_{\D-1}
	    \\ \times
	    \delta^+ \!\big[(\ell^0)^2-|\vec{\ell}|^2-m_1^2\big] 
	    \delta^+ \!\left[-2 \ell^0 \sqrt{s} + s + m_1^2-m_2^2\right] \,,
	\end{multline} 
	where in the second delta function we used $\ell^2 = m_1^2$ on the support of the first one. Here $\d \Omega_{\D-1}$ is the differential $(\D{-}1)$-dimensional solid angle, which integrates to the volume of a unit sphere, $\Omega_{\D-1} = 2 \pi^{\frac{\D-1}{2}}/\Gamma(\frac{\D-1}{2})$. We are thus left with two integrals: over $\ell^0$ and $|\vec{\ell}|$, as well as two delta functions.
	
	The solutions of the delta-function constraints are
	\be\label{eq:ell0-cut}
    \ell^0 = \frac{s + m_1^2 - m_2^2}{2\sqrt{s}}, \qquad |\vec{\ell}| = \pm \frac{\sqrt{\Delta}}{2\sqrt{s}} \,,
	\ee
	although the negative root of $|\vec{\ell}|$ does not have support on the integration contour.
	In addition, we also have the positive-energy conditions of $\delta^+$'s, which impose that
	\be
	\ell^0 > 0, \qquad \sqrt{s} - \ell^0 > 0.
	\ee
	Plugging in the values of $\ell^0$ from \eqref{eq:ell0-cut}, they only imply
	\be
	s > |m_1^2 - m_2^2|.
	\ee
	However, a stronger constraint comes from imposing that $|\vec{\ell}| \geq 0$ on the integration contour, which gives $\Delta \geq 0$ (since $s$ is already positive), which in fact implies
	\be
	s \geq (m_1 + m_2)^2.
	\ee
	The result of the cut is therefore proportional to $\Theta[s - (m_1 {+} m_2)^2]$, i.e., non-zero only above the production threshold.
	
	At this stage we can simply evaluate the first delta function in $|\vec{\ell}|$, at a cost of a Jacobian $1/(2 |\vec{\ell}|)$, followed by the second delta function in $\ell^0$, with a Jacobian of $1/(2 \sqrt{s})$. The result is
	\be 
	    \Cut_{12}^s\, \I_{\bubble} =
	    \frac{\pi^{3/2}}{\sqrt{s} \, \Gamma(\tfrac{\D-1}{2})}  \left\vert \frac{\sqrt{\Delta}}{2\sqrt{s}} \right\vert^{\D -3 } \! \Theta[s - (m_1 {+} m_2)^2] \,.
	\ee 
	Specializing to $\D=1,2,3$ dimensions gives
	\be \label{eq:Cut12-bub}
	    \Cut_{12}^s\, \I_{\bubble} =
	    \Theta[s - (m_1 {+} m_2)^2]\begin{dcases}
	    0 \quad&\text{if}\quad \D=1,\\
		\frac{2\pi}{\sqrt{\Delta}} \quad&\text{if}\quad \D=2,\\
		\frac{\pi^{3/2}}{\sqrt{s}} \quad&\text{if}\quad \D=3.\\
	\end{dcases}
	\ee 
	In conclusion, in this section we found the chain of equalities for real values of $s$:
	\be\label{eq:Im-Disc-Cut}
	\disc_{s = (m_1+m_2)^2} \I_{\bubble} = \Im\, \I_{\bubble} = \Disc_s \I_{\bubble} = \Cut_{12}^s\, \I_{\bubble} \,.
	\ee
	One can verify these equations using the explicit computations in \eqref{eq:disc-bub}, \eqref{eq:Im-bub2}, and \eqref{eq:Cut12-bub}.
	The first equality also holds above the threshold because there is only one branch cut starting at $s=(m_1+m_2)^2$. An understanding of relations such as \eqref{eq:Im-Disc-Cut} will be important later, in the discussion of dispersion relations in Sec.~\ref{sec:dispersion}.
	
	\paragraph{Mathematical perspective.}
	The equality between the imaginary part of $\I_\bubble$ and its discontinuity can be also seen from a more mathematical perspective. The Schwarz reflection principle states that if an analytic function $\I_\bubble(s)$ is defined in the upper-half $s$-plane and is real along a certain portion of the real $s$-axis, then its analytic extension to the rest of the complex plane is given by
	\be\label{eq:Schwarz}
	\I_\bubble(\overline{s}) = \overline{\I_\bubble(s)} .
	\ee
	The bubble integral $\I_\bubble(s)$ has this property because below the threshold at $s=(m_1+m_2)^2$, it does not have any imaginary part. Hence \eqref{eq:Schwarz} holds and we have
	\be
	\Im\, \I_\bubble = \frac{1}{2i}\Big( \I_\bubble(s) - \overline{\I_\bubble(s)} \Big) = \frac{1}{2i}\Big( \I_\bubble(s) - \I_\bubble(\overline{s}) \Big),
	\ee
	where the last expression is precisely the definition of $\Disc_s \I_\bubble$ given in \eqref{eq:Disc-bub}, when $s$ is close to the real axis.
	
	Finally, note that the discontinuity itself can have singularities and branch cuts. In fact, our previous discussion in Sec.~\ref{sec:singularities} for determining conditions for singularities, did not depend on the choice of the integration contour, and hence it applies to $\gamma$ as well. As before, one has to do more work to determine which of these singularities do actually contribute in practice. For example, in $\D=2$ the discontinuity contains only the normal and pseudo-normal thresholds coming from the roots $\alpha_\pm$ coinciding at $\Delta=0$, while in $\D=3$, only the collinear singularity at $\alpha_\pm = \infty$ ($s=0$) is present.

	\newpage
	\section{\label{sec:general}Singularities as classical saddle points}

	In this section, we explain how the conceptual points illustrated in Sec.~\ref{sec:primer} on the example of the bubble diagram generalize to arbitrary Feynman integrals. Using the Schwinger-parameterized form, we explain how to determine their analytic properties in terms of algebraic conditions on the worldline action $\mathcal{V}$. In particular, we find that branch cuts of the integral occur when $\mathcal{V}=0$, and the causal branch is obtained by taking $\Im \mathcal{V} > 0$. We analyze conditions for when the imaginary parts of $2\to2$ scattering amplitudes are equal to its discontinuities, which tell us under what circumstances imaginary parts (computed using unitarity cuts) can be used in dispersion relations.
	
	\subsection{\label{sec:parametric}Parametric representation}
	Let us denote the total number of external legs and internal edges with $n$ and $\E$ respectively, as well as the number of loops with $\L$. In terms of Schwinger parameters, the
	Feynman integral can be written as
	\be\label{eq:general-I}
		\I = (-i\hbar)^{-d} \lim_{\eps \to 0^+} \int_0^\infty \frac{\d^{\E} \alpha}{\U^{\D/2}}\, \N\, \exp \left[ \frac{i}{\hbar} \left(\V + i\eps \textstyle\sum_{e=1}^{\E}\alpha_e \right)\right].
	\ee
	The integration goes over all non-negative values of the Schwinger parameters $\alpha_e$, with one associated to every propagator. The infinitesimal $i\eps$ only enters the exponent to ensure suppression at infinity and is taken to zero at the end of the computation. The action
	\be\label{eq:V}
	\V = \frac{\F}{\U}
	\ee
	is a ratio of two \emph{Symanzik polynomials}, which are determined entirely in terms of combinatorics of the corresponding Feynman diagram. The first polynomial does not have any kinematic dependence and reads 
	\be\label{eq:U}
	\U = \sum_{\substack{\mathrm{spanning}\\ \mathrm{trees }\,T}} \prod_{e \notin T} \alpha_e,
	\ee
	where the sum runs over all possible spanning trees $T$ of the Feynman diagram obtained by removing exactly $\L$ propagators. For each $T$, the corresponding term is weighted with a product of the Schwinger parameters not present in that specific spanning tree. Note that \eqref{eq:U} is manifestly positive on the integration contour.
	The second Symanzik polynomial is defined similarly:
	\be\label{eq:F}
	\F = \sum_{\substack{\mathrm{spanning}\\ \text{two-trees}\\ T_L \sqcup T_R}} p_L^2 \prod_{e \notin T_L \sqcup T_R} \alpha_e - \U \sum_{e=1}^{\E} m_e^2 \alpha_e,
	\ee
	where the sum goes over all spanning two-trees $T_L \sqcup T_R$, that is, disjoint unions of spanning trees $T_L$ and $T_R$. Each of them separates the Feynman diagram into two sets of external particles $L$ and $R$. Therefore, the total momentum flowing across the two trees is $p_L^\mu = - p_R^\mu$. Every term in the first sum is weighted with the square of this momentum (a Mandelstam invariant $s_{ij}$ or an external mass squared $M_i^2$) times the product of Schwinger parameters of the removed propagators, while the second sum involves the internal masses $m_e$. The polynomials $\U$ and $\F$ are homogeneous in Schwinger parameters, with degrees $\L$ and $\L{+}1$ respectively, as is evident from their definitions. As a result, the action $\V$ has degree $1$, and it is also linear in the Mandelstam invariants and masses-squared of the internal and external particles.
	
	The most general Feynman integral \eqref{eq:general-I} also contains a numerator $\N$ encoding the interaction vertices, polarizations, colors, etc. 
	One can show that it is always a polynomial in the Schwinger parameters $\alpha_e$ and $\U^{-1}$. Because of this, the numerator $\N$ cannot introduce new kinematic singularities (though it can remove them), so its detailed discussion is not necessary for our purposes. We review the derivation of \eqref{eq:general-I} from the loop-momentum representation of Feynman integrals in App.~\ref{app:parametric}, and refer interested readers to \cite{smirnov1991renormalization,AHHM} for more details. For us, the only relevant fact is that $\N$ is a sum of homogeneous functions with degrees $d_\N, d_\N{+}1, \ldots, 0$ with $d_\N \leq 0$.
	
	There is an overall rescaling symmetry $\alpha_e \to \lambda \alpha_e$ associated to dilating the whole Feynman diagram by an overall scale $\lambda$. Integrating it out results in the expression
	\be\label{eq:I-GL}
		\I
		= \Gamma(d) \lim_{\eps \to 0^+} \int \frac{\d^{\E} \alpha}{\GL(1)} \frac{\widetilde\N}{\U^{\D/2} (-\V - i\eps)^{d}},
	\ee
	where $\widetilde{\N}$ is now homogeneous with degree $d_\N$ in Schwinger parameters (for scalar diagrams with no derivative interactions we have $\N = \widetilde{\N} = 1$).
	We have also used the degree of divergence
	\be
	d = \E - \L\D/2 + d_\N.
	\ee
	In the above formula it is important to not pull out the minus sign from $(-\V - i\eps)^d$, because we want to be careful about the branch cut of the integrand whenever $d$ is not an integer, for example when using dimensional regularization. 
	In particular, the above expression manifests the \emph{overall} UV divergence since $\Gamma(d)$ diverges whenever $d$ is a non-positive integer. As is conventional, we expressed the result as an integral ``modded out'' by the action of the $\GL(1)$. Two particular choices for implementing this in practice are the ones we already encountered in Sec.~\ref{sec:primer}: either setting a single Schwinger parameter to $1$ or their sum to $1$, i.e.,
	\be
	\int \frac{\d^{\E} \alpha}{\GL(1)} (\cdots) = \int_0^{\infty} \d^{\E-1} \alpha\, (\cdots) \Big|_{\alpha_{\E} = 1} = \int_{\Delta_{\E-1}} \d^{\E-1} \alpha\, (\cdots) \Big|_{\alpha_{\E} = 1-\sum_{e=1}^{\E-1} \alpha_{e}}
	\ee
	where $\Delta_{\E-1} = \{0 < \alpha_1 < \alpha_2 < \cdots < \alpha_{\E-1} < 1\}$ is a simplex.
	
	\paragraph{Physical regions.} Feynman integrals \eqref{eq:general-I} are defined in the physical regions, which are the ones corresponding to Mandelstam invariants that could
	have come from a physical scattering process
	with real energies and angles. The physical-region constraints are carved out by the inequality
	\be\label{eq:detG}
	{\det}' \mathbf{G} > 0,
	\ee
	where the entries of the Gram matrix are $\mathbf{G}_{ij} = p_i \cdot p_j$ and the prime means we only take the determinant after removing one column and row from $\mathbf{G}$ (the full determinant vanishes by momentum conservation). The constraint \eqref{eq:detG} decomposes kinematic space into disconnected crossing regions.
	
	For example, for elastic scattering of two particles with masses $M_1$ and $M_2$,
	\eqref{eq:detG} gives
	\be
	u \left(st - (M_1^2 - M_2^2)^2 \right) > 0,
	\ee
	which carves out four crossing regions in the real $st$-plane by the line $u=0$ and the hyperbolae asymptoting to $s=0$ and $t=0$, e.g., the $s$-channel region is given by $u<0$, $st<(M_1^2 - M_2^2)^2$, $s>(M_1^2 + M_2^2)/2$. These regions are denoted in Fig.~\ref{fig:envelope} in gray.
	
    One of the main challenges in the S-matrix program is to understand how to analytically continue the definition of Feynman integrals away from the physical regions.
	
	\paragraph{Examples.} Let us illustrate the formulae leading to the representation of Feynman integrals in~\eqref{eq:I-GL} on some explicit examples. For any $n=4$ diagram we can write $\V$ according to \eqref{eq:F},
	\be\label{eq:V4}
	\V = s \V_s + t \V_t + u \V_u + \sum_{i=1}^{4} M_i^2 \V_i - \sum_{e=1}^{\E} m_e^2 \alpha_e,
	\ee
	where each term is proportional to one of the Mandelstam invariants ($s$, $t$, or $u$), or a mass squared. For instance, computing $\V_s$ involves summing over all ways of clipping $\L{+}1$ propagators, such that the diagram decomposes into two spanning trees with the momentum $p_1{+}p_2$ flowing between them.
	Each individual term in \eqref{eq:V4} is homogeneous with degree $1$ and is manifestly positive for $\alpha_e > 0$,
	\be
	\V_s > 0, \quad \V_t > 0, \quad \V_u > 0, \quad \V_i > 0.
	\ee
	However, because of the momentum conservation, the Mandelstam invariants are not all independent, so we can use $u = \sum_{i=1}^{4} M_i^2 - s - t$ to get
	\be
	\V = s(\V_s - \V_u) + t(\V_t - \V_u) + \sum_{i=1}^{4} M_i^2 (\V_i + \V_u) - \sum_{e=1}^{\E} m_e^2 \alpha_e.
	\ee
	The fact that coefficients of the independent Mandelstam invariants $s$ and $t$ do not have a fixed sign will play a central role in understanding how to implement the $i\eps$ prescription consistently.
 
	\begin{figure}
	    \centering
	    \raisebox{-0.5\height}{\includegraphics{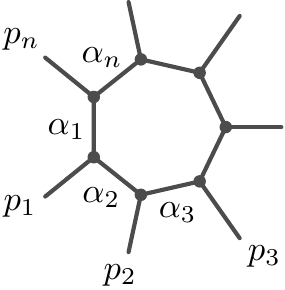}}\qquad\qquad
	    \raisebox{-0.5\height}{\includegraphics{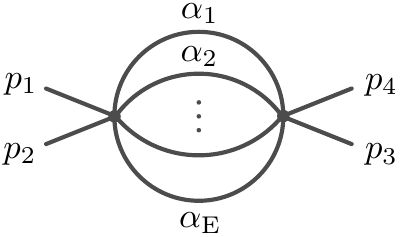}}
	    \caption{Left: One-loop $n$-gon diagrams with $n$ external legs and internal edges. Right: $\E$-banana diagrams with $\E$ internal edges and $\E{-}1$ loops.}
	    \label{fig:ngon-banana}
	\end{figure}
	
	As the simplest example at higher-multiplicity, let us consider the family of one-loop $n$-gon diagrams illustrated in Fig.~\ref{fig:ngon-banana} (left). Spanning trees and $2$-trees are obtained by clipping one and two edges respectively, which means that the corresponding Symanzik polynomials $\U_{n\text{-gon}}$ and $\F_{n\text{-gon}}$ are linear and quadratic in Schwinger parameters, respectively.
	Reading off the coefficient of each monomial, in the conventions of Fig.~\ref{fig:ngon-banana}, we can organize them as follows:
	\be\label{eq:ngon}
	\U_{n\text{-gon}} = \sum_{i=1}^{n} \alpha_i, \qquad
	\F_{n\text{-gon}} = \frac{1}{2} \sum_{i,j=1}^{n} \Y_{ij} \alpha_i \alpha_j,
	\ee
	where all the kinematic dependence can be put together into the symmetric matrix $\mathbf{Y}$, with entries defined as follows for $i\leq j$,
	\be\label{eq:Y-entries}
		\Y_{ij} = p_{i,i+1,\ldots,j-1}^2 - m_i^2 - m_j^2,
    \ee
	where $p_{i,i+1,\ldots,j-1} = {\textstyle\sum}_{k=i}^{j-1}p_{k}$.
	Let us illustrate these formulae even more explicitly on the two examples we will focus on later in Sec.~\ref{sec:branch-cut-deformations}: the $u$-triangle and $su$-box (see Fig.~\ref{fig:triangle-diagram} and \ref{fig:box}). They give
	\begin{align}
	\V_{\triangle} &= \frac{u \alpha_1 \alpha_2 + M_1^2 \alpha_3 \alpha_1 + M_3^2 \alpha_2 \alpha_3}{\alpha_1 + \alpha_2 + \alpha_3} - \sum_{e=1}^{3} m_e^2 \alpha_e,\label{eq:V-tri2}\\
	\V_{\boxx} &= \frac{s \alpha_1 \alpha_3 + u \alpha_2 \alpha_4 + \sum_{i=1}^{4} M_i^2 \alpha_i \alpha_{i+1}}{\alpha_1 + \alpha_2 + \alpha_3 + \alpha_4} - \sum_{e=1}^{4} m_e^2 \alpha_e,\label{eq:V-box2}
	\end{align}
	where in the last line $\alpha_5 = \alpha_1$.
	
	As another simple class of examples, consider the $\E$-banana diagrams from Fig.~\ref{fig:ngon-banana} (right), where $\E$ denotes the number of intermediate edges. Here, $\U_{\E\text{-ban}}$ and $\F_{\E\text{-ban}}$ have degrees $\E{-}1$ and $\E$ respectively: the former is obtained by  cutting through all but one edge in all possible ways, while the latter expects us to cut through every intermediate edge, resulting in
	\be\label{eq:E-ban}
	\U_{\E\text{-ban}} = \prod_{e=1}^{\E}\alpha_e \sum_{e'=1}^{\E}\frac{1}{\alpha_{e'}},\qquad
	\F_{\E\text{-ban}} = s\prod_{e=1}^{\E}\alpha_e - \U_{\E\text{-ban}}\sum_{e=1}^{\E} m_e^2 \alpha_e.
	\ee
	For instance, the case $\E=2$ gives the bubble diagram studied in Sec.~\ref{sec:primer}.
	
	\subsection{\label{sec:general-thresholds}Thresholds and Landau equations}

    Just as in Sec.~\ref{sec:interpretation}, one can interpret kinematic singularities as configurations for which propagators of a Feynman diagram become trajectories of classical particles interacting at vertices. There are two complementary perspectives on how to find these singularities: we can either solve kinematic constraints of momentum conservation, locality, and mass-shell conditions directly, or we can look for saddle points of the action $\V$. We start with the latter.
    
    \paragraph{Saddle points.}
    We can distinguish between saddle points which are in the bulk of the integration, i.e., $\alpha_e \neq 0$ for all $e$, or those that lie on the boundaries with a subset of $\alpha_e$'s vanishing.
    They give rise to \emph{leading} and \emph{subleading} singularities of Feynman integrals respectively.
    
    The bulk saddles are simply determined by solving the variation problem
    \be\label{eq:partial-V}
    \partial_{\alpha_e} \V = 0
    \ee
    for all $e=1,2,\ldots,\E$. Here we work in the $\eps \to 0^+$ limit directly. Since $\V$ is homogeneous with degree $1$, the action of the dilation operator $\sum_{e=1}^{\E} \alpha_e \partial_{\alpha_e}$ on $\V$ leaves it invariant, so on the saddle we have
    \be\label{eq:V0}
    \V = \sum_{e=1}^{\E} \alpha_e \partial_{\alpha_e} \V = 0.
    \ee
    In other words, $\V=0$ is a necessary but not a sufficient condition for a saddle point.

    \begin{figure}
    \centering
    \includegraphics[scale=1]{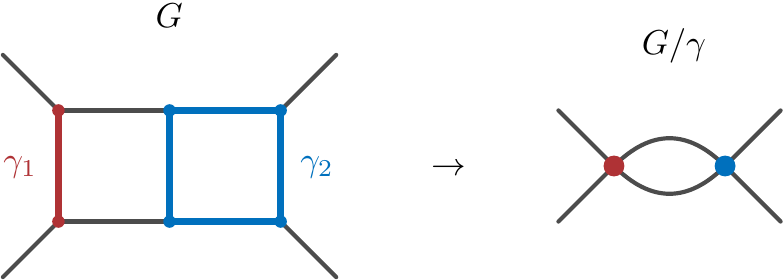}
    \caption{\label{fig:reduced}Left: Example subdiagram $\gamma=\gamma_1 \sqcup \gamma_2$ of $G$ consists of a single edge $\gamma_1$ and a loop $\gamma_2$. Right: Shrinking all the edges in $\gamma$ results in the reduced diagram $G/\gamma$. The corresponding subleading singularity of $G$ is the leading singularity of $G/\gamma$ and is responsible for one of the normal thresholds of $G$.}
    \end{figure}
    
    Boundary saddle points can be labelled by the Schwinger parameters that vanish.  Let us call the subdiagram created by these edges $\gamma$. Such $\gamma$ can itself contain loops and does not need be connected. The contracted (or reduced) diagram $G/\gamma$ is obtained by shrinking edges of each connected component of $\gamma$ to a point, see Fig.~\ref{fig:reduced}. At this stage we are left with finding bulk saddle points of $G/\gamma$. In other words, every subleading singularity of $G$ is labelled by $\gamma$ and is identical to a leading singularity of $G/\gamma$. In addition, it might also occur that both $\alpha_e = \partial_{\alpha_e} \V = 0$ for a subset of edges, i.e., a given point looks like a saddle from the bulk and boundary direction simultaneously.
    
    Recall that $\V = \V(\alpha_e; s_{ij}, M_i, m_e)$ and the solutions in terms of Schwinger parameters $\alpha_e^\ast$ are projective, i.e., they are only defined up to an overall scale
    \be
    (\alpha_1^\ast : \alpha_2^\ast : \cdots : \alpha_\E^\ast) = (\lambda\alpha_1^\ast : \lambda\alpha_2^\ast : \cdots : \lambda\alpha_\E^\ast)
    \ee
    with $\lambda > 0$. It means that the conditions \eqref{eq:partial-V} put $\E$ constraints on $\E{-}1$ independent Schwinger parameters, generically resulting in \emph{at least} one constraint on the external kinematics. This can be always written as vanishing of a polynomial
    \be\label{eq:Delta}
    \Delta(s_{ij}, M_i, m_e) = 0,
    \ee
    defined up an overall constant.
    However, it turns out that solutions of higher codimension exist, starting at two loops. Higher-codimension singularities are closely related to continuous solutions in terms of Schwinger parameters (where there are manifolds of solutions $\alpha_e^\ast$): additional constraints are put on the external kinematics instead of the internal variables. The simplest example is the triangle-box diagram \cite{Mizera:2021ujs}. We review this example in Sec.~\ref{sec:codimension}, where we also conjecture that such higher-codimension singularities always lie on the intersection of codimension-$1$ subleading singularities. This proposal implies that the full set of singularities of any Feynman integral can be written as a union of varieties of the form \eqref{eq:Delta}.

    \paragraph{Condition for branch cuts.}
    We can also discuss what happens when $\V=0$ for some values of $\alpha$'s but \eqref{eq:partial-V} is not satisfied.
    From the perspective of \eqref{eq:general-I}, $\V=0$ corresponds to the integral no longer being exponentially suppressed at infinity. Moreover, after integrating out the overall scale, setting $\V=0$ in \eqref{eq:I-GL} makes the denominator vanish (at least for positive $d$). From either point of view, there is a potential ambiguity in defining the Feynman integral.
    The freedom of being able to deform away from it (by varying kinematic invariants or Schwinger parameters) is precisely a sign of a potential \emph{branch cut} in the kinematic space. Indeed, in Sec.~\ref{sec:im-disc} we will derive explicit formulae for discontinuities across branch cuts that localize on the submanifold $\V=0$ in the Schwinger-parameter space.
    
    If the additional condition \eqref{eq:partial-V} is satisfied, there are no means of deforming away from a branch cut. In other words, the surface $\{\V=0\}$ becomes singular. This signifies a potential singularity, such as a branch point, if such a singularity happens to pinch the integration contour. This means the specific type of singularity, or whether it is present at all, depends on the parameters of the individual Feynman integral such as its number of loops and edges as well as the space-time dimension. We will return back to this question in Sec.~\ref{sec:fluctuations}.
    
    \paragraph{Causality.}
    Once we are at a branch cut determined by $\V=0$, we can ask: what is a good contour deformation consistent with causality? From \eqref{eq:general-I} it is already apparent that for all points close to $\V = 0$ we need
    \be\label{eq:ImV-pos}
    \Im \V > 0,
    \ee
    which has the same effect as adding the Feynman $i\eps$ by hand. Another point of view is that deformation of masses $m_e^2 \to m_e^2 - i\eps$, according to \eqref{eq:F}, has the same effect of adding a small positive part to $\V$. In Sec.~\ref{sec:deformation} we will explain how to consistently deform Schwinger parameters such that \eqref{eq:ImV-pos} is satisfied.
    
    Let us point out that solutions of Landau equations that are directly relevant for Feynman integrals in the physical region are those with $\alpha_e \geq 0$, the so-called \emph{$\alpha$-positive} singularities. There exist some graphical techniques for finding these classes of solutions, because they can be mapped onto Euclidean-geometry problems, see, e.g., \cite[Ch.~18]{Bjorken:100770} and \cite{Bjorken:1959fd,Eden:1966dnq,todorov2014analytic,Correia:2021etg}.
    However, most solutions to Landau equations are not $\alpha$-positive, and once we analytically continue away from physical regions, singularities with $\alpha_e \in \C$ may contribute. In particular, analytic expressions for Feynman integrals will feature branch points coming from generic, complex solutions. Determining which singularities contribute on which sheet in the kinematic space is an extremely difficult problem.
    
    \paragraph{Landau equations.}
    As an alternative interpretation of kinematic singularities, we can re-introduce the loop momenta $\ell_a^\mu$ for $a=1,2,\ldots,\L$. After Schwinger parametrization, we obtain the function
    \be\label{eq:V-alpha-ell}
    \V(\alpha_e, \ell_a^\mu; p_i, m_e) = \sum_{e=1}^{\E} \alpha_e (q_e^2 - m_e^2),
    \ee
    which appears in the exponent of the Feynman integral, cf.~\eqref{eq:I-Schwinger-2}.
    The internal momenta $q_e^\mu = q_e^\mu(\ell_a,p_i)$ are themselves expressed in terms of the loop momenta as well as the external kinematics in the standard way by using momentum conservation at every vertex. From this perspective, varying with respect to the Schwinger parameters $\alpha_e$ results in
    \be\label{eq:q2m2}
    q_e^2 = m_e^2,
    \ee
    which is putting all the internal propagators on their mass shell. Varying with respect to the loop momenta we have
    \be\label{eq:Landau-locality}
    \sum_{e \in L} \pm\underbrace{\alpha_e q_e^\mu}_{\Delta x_e^\mu} = 0,
    \ee
    for every loop $L$. Here, the sign depends on whether orientation of the edge matches the orientation of the loop $L$. This amounts to the requirement that the sum of space-time displacements $\Delta x_e^\mu$ adds up to zero, for every edge along the loop. Hence this constraint encodes locality: the vertices of the Feynman diagram at the singularity are uniquely determined up to an overall scale. Together, \eqref{eq:q2m2} and \eqref{eq:Landau-locality} are known as \emph{Landau equations} \cite{Bjorken:1959fd,Landau:1959fi,10.1143/PTP.22.128}.

    To translate between \eqref{eq:V-alpha-ell} and \eqref{eq:V}, we need to eliminate the loop momenta $\ell_a^\mu$ using \eqref{eq:Landau-locality}. The solution is given by \cite{Mizera:2021fap}:
    \be\label{eq:qe}
    q_e^\mu = \frac{1}{\U} \sum_T p_{T,e}^{\mu} \prod_{e' \notin T} \alpha_{e'},
    \ee
    where $p_{T,e}^\mu$ is the total momentum flowing through the edge $e$ through the spanning tree $T$ in the direction of this edge. Feynman diagrams also have so-called \emph{second-type} singularities, which are limiting cases when $\U=0$ on the solution. The Eq.~\eqref{eq:qe} immediately tells us that they have to correspond to configurations with infinite loop momenta (the literature often makes a distinction between \emph{pure} and \emph{mixed} second-type singularities, depending on whether all or only a subset of the independent loop momenta blow up \cite{Cutkosky:1960sp,doi:10.1063/1.1724262,Drummond1963}).
    They also have a simple on-shell interpretation, because \eqref{eq:qe} implies that the external momenta satisfy an extra linear identity beyond the momentum conservation. Thus the necessary condition for a second-type singularity can be stated as
    \be\label{eq:detG0}
    {\det}^\prime \mathbf{G} = 0.
    \ee
    Hence, they lie on the boundaries of the physical regions when a subset of external momenta become collinear.
    Note that second-type singularities are never $\alpha$-positive because $\U > 0$ on the original integration contour. On top of that, the condition  \eqref{eq:detG0} is always satisfied in dimensions $\D \leq n{-}2$.
    
    Using \eqref{eq:V}, the saddle-point conditions can be also restated as the polynomial set of equations $\alpha_e \partial_{\alpha_e} \F = 0$ as long as we impose $\U \neq 0$. The second-type singularities correspond to the closure obtained by adding the solutions with $\U = 0$ back in, or, in other words, those where $\F \propto \U$. Geometrically, it means that the surface $\{ \F = 0 \}$ not only degenerates itself, but also pinches $\{ \U = 0 \}$ at the same time. By adding an auxiliary parameter $\alpha_0$ (with mass dimension $2$), everything can be neatly packaged into the variation problem:
    \be\label{eq:FU-vary}
    \alpha_e \partial_{\alpha_e} (\F + \alpha_0\, \U) = 0
    \ee
    for all $e = 0,1,2,\ldots,\E$. The branch of solutions with $\alpha_0 = 0$ is called first-type, while those with $\alpha_0 \neq 0$ are second-type singularities. However, in contrast with saddle-point equations, more care is required with subleading singularities in the above polynomial formulation for which $\gamma$ contains at least one loop, because in this limit $\F$ and $\U$ both vanish with the ratio $\F/\U$ remaining finite, cf. \cite{Mizera:2021icv} and Sec.~\ref{sec:boundary-saddles}.
    
    Note that solutions of Landau equations will in general not be continuous in special mass configurations, e.g., when they become equal or vanish. In particular, in the massless limit, some branches of solutions might disappear, degenerate, change type or codimension.
    
    \paragraph{Practical computations.}
    Unfortunately, solving Landau equations in practice turns out to be an insurmountable task beyond low-loop examples. In fact, extensive literature on this subject from the 1960's (see, e.g., \cite{Eden:1966dnq,todorov2014analytic,doi:10.1063/1.1703752,kolkunov1961b,kolkunov1961a,kolkunov1961positions,petrina,doi:10.1063/1.1705071,doi:10.1063/1.1704978}) contains many incomplete solutions, as can be verified nowadays with a broader availability of specialized algebraic-geometry software. For recent progress see \cite{Brown:2009ta,Muhlbauer:2020kut,Mizera:2021icv,Klausen:2021yrt,Correia:2021etg} (see also  \cite{Dennen:2015bet,Dennen:2016mdk,Prlina:2017azl,Prlina:2017tvx,Prlina:2018ukf,Gurdogan:2020tip} for twistor methods for massless scattering).
    
    Two families of Feynman integrals easily admit closed-form solutions: the aforementioned $n$-gon and $\E$-banana diagrams illustrated in Fig.~\ref{fig:ngon-banana}. Let us consider them in turn.
    
    Starting with the $\E$-banana diagrams from \eqref{eq:E-ban},  we find $2^{\E-1}$ first-type solutions with
    \be
    s = \left( \textstyle{\sum}_{e=1}^{\E} \pm\! m_e \right)^2
    \ee
    and each Schwinger parameter is fixed to $\alpha_e^\ast = \pm\frac{1}{m_e}$. The case with all plus signs is called the normal threshold, while the remaining ones are the pseudo-normal thresholds and do not appear on the physical sheet. A more detailed analysis is postponed until Sec.~\ref{sec:threshold-examples}.
    
    \begin{table}[!t]
    \centering
    \def\arraystretch{1.18}
\begin{tabular}{c|c|c|c}
$G/\gamma$                                          &  type                        & $\Delta_{G/\gamma}$ & $(\alpha_0^\ast : \alpha_1^\ast : \alpha_2^\ast : \alpha_3^\ast : \alpha_4^\ast)$ \\ \cline{1-4}
\multicolumn{1}{c|}{\multirow{4}{*}{\rotatebox{90}{$su$-box\;}}} & $1^{\text{st}}$ & $su+4m^2 t - 4M^4$         & $(\circ : 4m^2 {-} 2M^2 : s {-} 4m^2 : 4m^2 {-} 2M^2 : s {-} 4m^2)$       \\ \cline{2-4}
\multicolumn{1}{c|}{}                     & \multicolumn{1}{c|}{\multirow{3}{*}{$2^{\text{nd}}$}} & $s$         & $(0 : 1 : 0 : -1 : 0)$              \\
\multicolumn{1}{c|}{}                     & \multicolumn{1}{c|}{} & $t$         & $(s{-}2M^2 : -1 : 1 : -1 : 1)$              \\
\multicolumn{1}{c|}{}                     & \multicolumn{1}{c|}{}                     &  $u$        & $(0 : 0 : 1 : 0 : -1)$               \\ \cline{1-4}
\multicolumn{1}{c|}{\multirow{3}{*}{\rotatebox{90}{\parbox{3em}{\centering$s$-tri\; \scriptsize$(\alpha_4=0)$}}}} & $1^{\text{st}}$ & $m^2 s + M^2(M^2 - 4m^2)$         & $(\circ : m^2 : M^2 {-} 2m^2 : m^2 : \circ)$            \\ \cline{2-4}
\multicolumn{1}{c|}{}                     & \multicolumn{1}{c|}{\multirow{2}{*}{$2^{\text{nd}}$}} & $s$         & $(0 : 1 : 0 : -1 : \circ)$              \\ 
\multicolumn{1}{c|}{}                     & \multicolumn{1}{c|}{}                     &  $s-4M^2$        & $(2M^2 : -1 : 2 : -1 : \circ)$            \\ \cline{1-4}
\multicolumn{1}{c|}{\multirow{3}{*}{\rotatebox{90}{\parbox{3em}{\centering$u$-tri\; \scriptsize$(\alpha_1=0)$}}}} & $1^{\text{st}}$ & $m^2 u + M^2(M^2 - 4m^2)$         & $(\circ : \circ : m^2 : M^2 {-} 2m^2 : m^2 )$            \\ \cline{2-4}
\multicolumn{1}{c|}{}                     & \multicolumn{1}{c|}{\multirow{2}{*}{$2^{\text{nd}}$}} & $u$         & $(0 : \circ : 1 : 0 : -1)$              \\ 
\multicolumn{1}{c|}{}                     & \multicolumn{1}{c|}{}                     &  $u-4M^2$        & $(2M^2 : \circ : -1 : 2 : -1 )$            \\ \cline{1-4}
\multicolumn{1}{c|}{\multirow{2}{*}{\rotatebox{90}{\parbox{3em}{\centering$s$-bub\; \scriptsize$(\alpha_2 =0$, $\alpha_4 =0)$}}}} & $1^{\text{st}}$ & $s - 4m^2$         & $(\circ : 1 : \circ : 1 : \circ)$            \\ \cline{2-4}
\multicolumn{1}{c|}{}                     & \multicolumn{1}{c|}{$2^{\text{nd}}$}                    &  $s$        & $(0 : 1 : \circ : -1 : \circ)$            \\ \cline{1-4}
\multicolumn{1}{c|}{\multirow{2}{*}{\rotatebox{90}{\parbox{3em}{\centering$u$-bub\; \scriptsize$(\alpha_1 =0$, $\alpha_3 =0)$}}}} & $1^{\text{st}}$ & $u - 4m^2$         & $(\circ : \circ : 1 : \circ : 1 )$            \\ \cline{2-4}
\multicolumn{1}{c|}{}                     & \multicolumn{1}{c|}{$2^{\text{nd}}$}                    &  $u$        & $(0 : \circ : 1 : \circ : -1 )$            \\
\end{tabular}
\caption{Summary of the codimension-$1$ Landau singularities of the $su$-box diagram with generic $M$ and $m$. Subleading singularities for reduced diagrams $G/\gamma$ corresponds to the $s$- and $u$-triangles ($\alpha_4 = 0$ and $\alpha_1 = 0$ respectively), as well as the $s$- and $u$-bubbles ($\alpha_2 = \alpha_4 = 0$ and $\alpha_1 = \alpha_3 = 0$ respectively), together with their cyclic permutations. In each case, $\circ$ denotes the $\alpha$'s that are set to zero and not varied according to \eqref{eq:FU-vary}. For example, the first-type singularity of the $s$-triangle is obtained by setting $\alpha_0 = \alpha_4 = 0$ and then solving $\partial_{\alpha_e} \F = 0$ for $e=1,2,3$.}
\label{tab:su-box}
\end{table}
    
    In the $n$-gon case, leading Landau equations are obtained by varying \eqref{eq:ngon} with respect to all Schwinger parameters, giving $\sum_{i=1}^{n} \Y_{ij} \alpha_j = 0$ for every $j$. Reading it as a matrix equation, it is the requirement that the $n\times n$ matrix $\Y$ has a null vector, or in other words,
    \be
    \Delta_{n\text{-gon}} = \det \Y = 0,
    \ee
    where its entries were given in \eqref{eq:Y-entries}.
    This condition determines the position of the leading singularity in the kinematic space. The null vector of this matrix gives the values of Schwinger parameters on this solution. Subleading singularities are obtained in the same way and correspond to vanishing of every minor of the matrix $\mathbf{Y}$. Second-type singularities are a special case when the entries of the null vector add up to zero, which is equivalent to setting $\U_{n\text{-gon}} = 0$. We can isolate them more easily by varying $\F_{n\text{-gon}}$ for every $\alpha_i$ and using $\U_{n\text{-gon}} = 0$ there, which gives
    \be\label{eq:ngon-2nd}
    \sum_{j=1}^{n} \mathbf{y}_{ij} \alpha_j = \sum_{j=1}^{n} m_j^2 \alpha_j
    \ee
    for every $i$. Here $\mathbf{y}$ is the symmetric $n\times n$ matrix with the entries $\mathbf{y}_{ij} = ({\textstyle\sum}_{k=i}^{j-1}p_{k})^2$ for $i\leq j$ that only depends on the external kinematics. Since the right-hand side of \eqref{eq:ngon-2nd} is a constant independent of the choice of $i$, we can once again rewrite this constraint as a matrix equation
    \be
    \det \begin{bmatrix}
    0 & \vec{\mathbf{1}}^\intercal \\
    \vec{\mathbf{1}} & \mathbf{y}
    \end{bmatrix} = 0,
    \ee
    where $\vec{\mathbf{1}}$ is a length-$n$ vector of $1$'s. This is the so-called Cayley--Menger determinant and is equivalent to the Gram-determinant condition $\det^\prime \mathbf{G} = 0$ stated above. More details on the one-loop family of integrals will be given in Sec.~\ref{sec:one-loop}.
    
    As concrete examples of $n$-gon Landau singularities, we return back to the $u$-triangle and $su$-box diagrams considered in \eqref{eq:V-tri2} and \eqref{eq:V-box2}. When setting internal masses to $m$ and external masses to $M$ for simplicity, we find leading singularities at 
    \begin{align}
    \Delta_{u\text{-}\triangle} &= u[m^2 u + M^2 (M^2 - 4m^2)],\label{eq:Delta-tri}\\
    \Delta_{su\text{-}\boxx} &= su[su + 4m^2 t - 4M^4].\label{eq:Delta-box}
    \end{align}
    In both cases, the branches contained in the square brackets are the first-type singularities, while $u=0$ and $su=0$, respectively, are the second-type singularities. The full set of subleading singularities of the box diagram with generic $M$ and $m$ (containing also those of the triangle) is listed in Tab.~\ref{tab:su-box}.
    
    \begin{figure}
		\centering
		\raisebox{-0.5\height}{\includegraphics[scale=1]{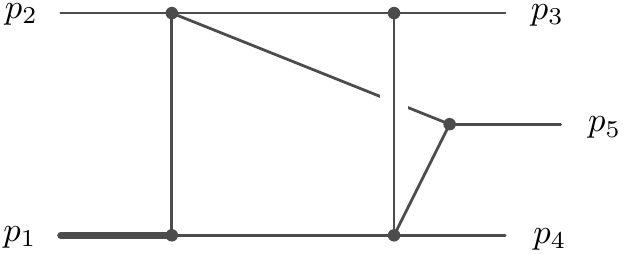}}\qquad\qquad
		\raisebox{-0.5\height}{\includegraphics[scale=1]{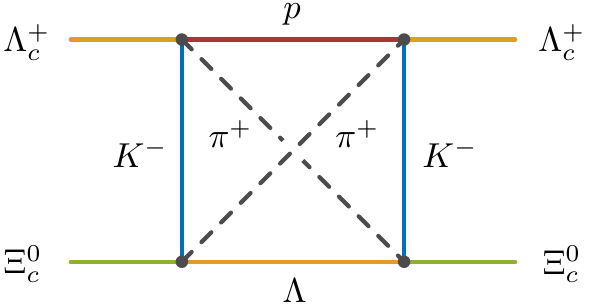}}
 		\caption{\label{fig:xi-lambda}Left: The two-box master integral with all the internal edges and external legs massless except for the corner mass $p_1^2 = M_1^2 \geq 0$, whose leading singularity is given in~\eqref{eq:tbox}. Right: A contribution to the $\Xi_c^0 \Lambda^+_c \to \Xi_c^0 \Lambda^+_c$ elastic scattering mediated by pions $\pi^+$, kaons $K^-$, a lambda baryon $\Lambda$, and a proton $p$. Recall that their masses are $(M_{\Xi_c^0}, M_{\Lambda_c^+}, m_{\pi^+}, m_{K^-}, m_p, m_{\Lambda}) \approx (2467, 2286, 140, 494, 938, 1116) \mathrm{MeV}$ \cite{ParticleDataGroup:2020ssz}. Its leading singularities are illustrated in Fig.~\ref{fig:envelope}.}
	\end{figure}
    
    As a more advanced example, consider the diagram shown in Fig.~\ref{fig:xi-lambda} (left), which is a master integral contributing to the  two-loop virtual QCD corrections for two-jet-associated W, Z or Higgs-boson production \cite{Abreu:2021smk}. Using the algorithm from \cite[Sec.~3.1]{Mizera:2021icv} one finds the leading singularity in terms of $s_{ij} = (p_i + p_j)^2$:
    \be\label{eq:tbox}
    \Delta_{\mathrm{tbox}} = \left(s_{12} s_{15}-s_{12} s_{23}-s_{15} s_{45}+s_{34} s_{45}+s_{23} s_{34}\right)^{2}-4 s_{23} s_{34} s_{45}(s_{34}{-}s_{12}{-}s_{15}),
    \ee
    together with $M_1 = 0$, which reproduces the branch point identified in \cite[Eq.~(2.7)]{Abreu:2021smk}.

\begin{figure}
		\centering
		\includegraphics[scale=0.95]{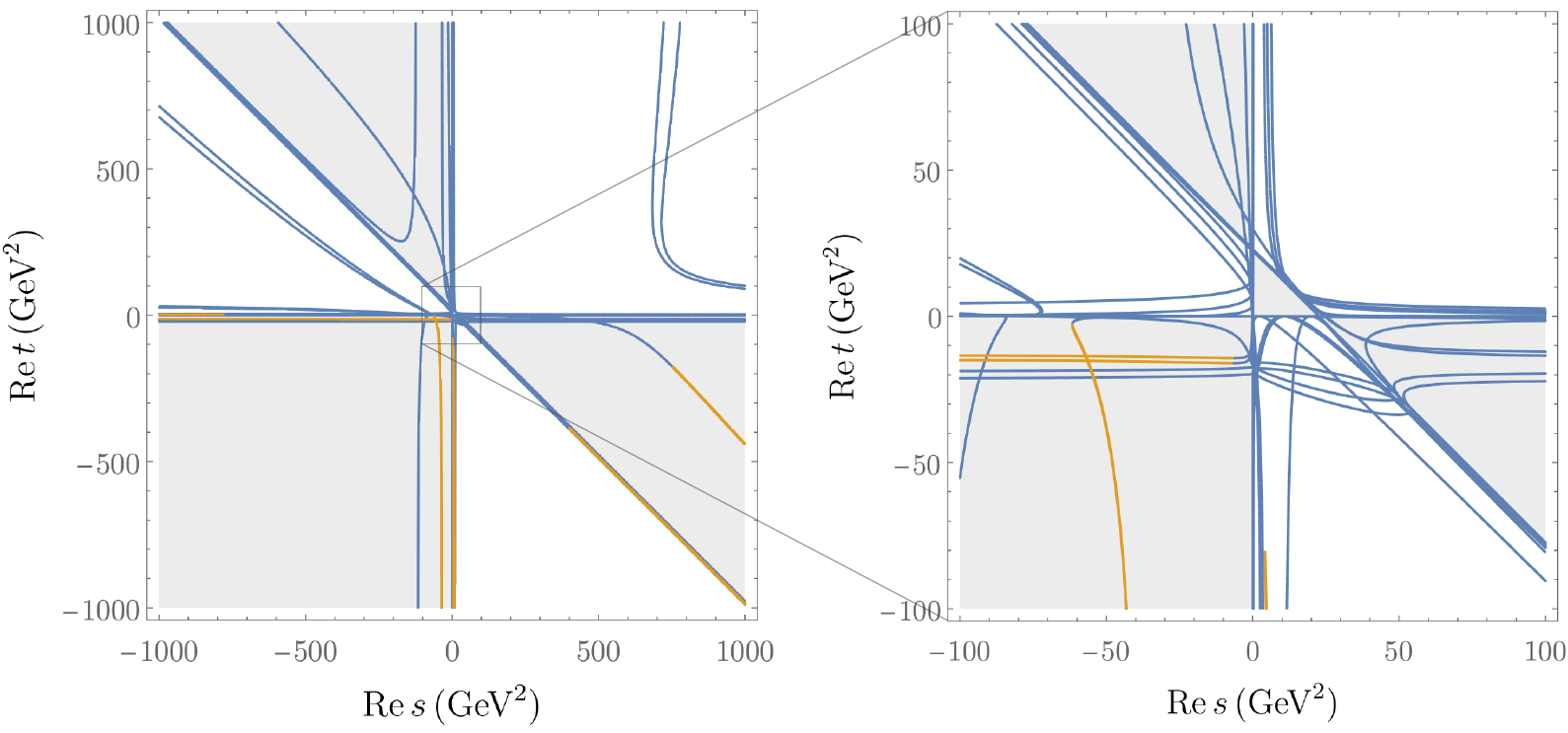}
		\caption{\label{fig:envelope}Example Landau curves: Leading singularity for the $\Xi_c^0 \Lambda^+_c \to \Xi_c^0 \Lambda^+_c$ process illustrated in Fig.~\ref{fig:xi-lambda} (right).}
	\end{figure}

    In order to give a hint at how quickly the Landau singularities get out of hand, let us consider the Standard Model process illustrated in Fig.~\ref{fig:xi-lambda} (right). Its analytic solution is presently intractable analytically, but we can numerically plot the singularities in the real $st$-plane using the tools from \cite{Mizera:2021icv}. The results for the leading singularities are shown in Fig.~\ref{fig:envelope} at two different magnifications. The $\alpha$-positive singularities are shown in orange and the remaining ones in blue. One can observe that some components of the $\alpha$-positive singularities lie within the physical regions denoted in gray, which means they contribute on the physical sheet. Of course this specific process is highly suppressed compared to lower-loop and lower-point processes for $\Xi_c^0 \Lambda^+_c$ scattering, because of the three-loop phase-space factor, and since each of the vertex factors indicated in Fig.~\ref{fig:xi-lambda} (right) corresponds to multiple interactions.

    \paragraph{Summary.}

    We can summarize the condition on $\V$ that determine different aspects of analyticity as follows:
    \begin{alignat}{2}
    \V = 0 \quad&\text{ for any $\alpha$'s } \quad&&\Leftrightarrow\quad \text{branch cut}\\
    \partial_{\alpha_e}\V = 0 \quad&\text{ for any $\alpha$'s } \quad&&\Leftrightarrow\quad \text{branch point}\\
    \Im\,\V > 0 \quad&\text{ for all $\alpha$'s } \quad&&\Leftrightarrow\quad \text{causal branch}
    \end{alignat}
    Together, they allow us to determine properties of Feynman integrals without explicit computations. As we will see in later sections, multiple other properties of Feynman integrals can be determined in terms of $\V$. For example, the sign of $\partial_s \V$ determines the direction of the branch cut, while the Hessian $\partial_{\alpha_e} \partial_{\alpha_{e'}}\V$ describes local behavior near thresholds.

	\subsection{\label{sec:deformation}Complexifying worldlines}
	
	It is now straightforward to generalize the derivation of Sec.~\ref{sec:deformations} to perform causal branch-cut deformations for an arbitrary Feynman integral. The consistent way of doing so is introducing small phases to each Schwinger parameter:
	\be\label{eq:alpha-hat}
	\hat{\alpha}_e = \alpha_e \exp\left( i\eps\, \partial_{\alpha_e} \V \right).
	\ee
	These are designed precisely to have the same effect as the Feynman $i\eps$, i.e., select the causal branch of the Feynman integral. To see this, let us consider the effect of \eqref{eq:alpha-hat} on the action, $\hat{\V} = \V(\hat{\alpha}_e)$: 
	\be
	\hat{\V} = \V + i\eps \sum_{e=1}^{\E} \alpha_e (\partial_{\alpha_e} \V)^2 + {\cal O}(\eps^2),
	\ee
	where one power of $\partial_{\alpha_e}\V$ came from a Taylor series around $\hat{\alpha}_e = \alpha_e$ and the other from the shift $\hat{\alpha}_e = \alpha_e (1+ i\eps \partial_{\alpha_e} \V + \ldots)$ itself. Since $\alpha_e \geq 0$ and $\V$ is real for real kinematics, for sufficiently small $\eps$ we must have
	\be
	\Im \hat{\V} \geq 0.
	\ee
	Moreover, the equality only happens when all $\partial_{\alpha_e}\V$ vanish identically, i.e., on the singularity. We thus conclude that the contour deformation \eqref{eq:alpha-hat} deforms branch cuts in the causal way, without affecting positions of branch points.\footnote{
	As a matter of fact, there is an alternative deformation $\hat{\alpha}_e(\eps)$, which works even for finite $\eps$. It is prescribed by the system of differential equations
	\be\label{eq:alpha-hat-cplx}
	\d_{\eps} \hat{\alpha}_e(\eps) = i \alpha_e\, \overline{\partial_{\hat{\alpha}_e} \hat{\V}},
	\ee
	together with the boundary conditions $\hat{\alpha}_e(0) = \alpha_e$. Here, $\hat{\V} = \V(\hat{\alpha}_e(\eps))$ depends on $\eps$ through the deformed Schwinger parameters. Let us consider the change in $\hat{\V}$ as $\eps$ is varied:
	\be
	\d_\eps \V(\hat{\alpha}_e(\eps)) = \sum_{e=1}^{\E} (\d_{\eps} \hat{\alpha}_e) (\partial_{\hat{\alpha}_e} \hat{\V}) = i \sum_{e=1}^{\E} \alpha_e |\partial_{\hat{\alpha}_e} \hat{\V}|^2,
	\ee
	where in the second equality we used \eqref{eq:alpha-hat-cplx}. This tells us that as $\eps$ grows, $\Im\, \hat{\V}$ cannot decrease and $\Re \hat{\V}$ remains constant. Moreover, $\Im\, \hat{\V}$ ceases to increase if and only if a Landau singularity is encountered, thus guaranteeing that the causality condition \eqref{eq:ImV-pos} remains satisfied for real kinematics. In practice, the non-linear system \eqref{eq:alpha-hat-cplx} can be solved either numerically or perturbatively in $\eps$. For example, at the leading order we can treat $\partial_{\hat{\alpha}_e} \hat{\V} = \partial_{\alpha_e} \V$ as constants, resulting in the solution
	\be
	\hat{\alpha}_e = \alpha_e \exp \left( i \eps\, \overline{\partial_{\alpha_e} \V} \right),
	\ee
	which is a complex version of \eqref{eq:alpha-hat}.
	In this work we prefer to use~\eqref{eq:alpha-hat} because holomorphic dependence on the external kinematics makes it manifest that the resulting integral is analytic.}
	
	The above contour deformations also give practical means of performing the integration numerically.
	The Jacobian for the change of variables~\eqref{eq:alpha-hat} is $\J = \det \mathbf{J}$ with
	\be
	\mathbf{J}_{e e'} = \frac{\partial \hat{\alpha}_{e}}{\partial \alpha_{e'}} = \left( \frac{\delta_{ee'}}{\alpha_e} + i\eps\, \partial_{\alpha_e} \partial_{\alpha_{e'}}\V \right) \hat{\alpha}_e.
	\ee
	Hence the Feynman integral~\eqref{eq:general-I} can be written as
	\be
	\I = (-i\hbar)^{-d} \int_0^\infty \frac{\d^{\E} \alpha}{ \hat{\U}^{\D/2}}\,\J\, \hat{\N}\, \exp\left[\frac{i}{\hbar}\hat{\V} \right],
	\ee
	where all the hatted variables are evaluated at $\hat{\alpha}_e$, keep the dependence on $\eps$ exact. Compared to~\eqref{eq:general-I}, no additional $i\eps$ factor is necessary in the exponent. We can simplify this further by noticing that the phase of the rotated Schwinger parameter~\eqref{eq:alpha-hat} does not carry any $\GL(1)$ weight, so the homogeneity properties are preserved:
	\be
	\hat{\alpha}_e \to \lambda \hat{\alpha}_e,\qquad \hat{\U}_e \to \lambda^{\L} \hat{\U}_e,\qquad \hat{\V}_e \to \lambda \hat{\V}_e \,.
	\ee
	This means we can still use the $\GL(1)$ freedom in the same way as before, for example by mapping the integration cycle to the unit simplex $\hat{\Delta}_{\E-1}$ with
	\be
	\hat{\alpha}_\E = 1 - \sum_{e=1}^{\E-1} \hat{\alpha}_e.
	\ee
	As a result, integrating out the overall scale leaves us with
	\be\label{eq:I-deformed}
	\I =\Gamma(d) \int_{\hat{\Delta}_{\E-1}}  \frac{\d^{\E-1} \alpha\,\J\, \hat{\widetilde\N}}{\hat{\U}^{\D/2} (-\hat{\V})^{d}} \bigg|_{\hat{\alpha}_{\E} = 1-\sum_{e=1}^{\E-1} \hat{\alpha}_{e}},
	\ee
	which is the form of integral suitable for practical evaluation. See also \cite{Nagy:2006xy,Anastasiou:2007qb} for previous approaches to contour deformations.

	\subsection{\label{sec:im-disc}When is the imaginary part a discontinuity?}
	
	We are now well-equipped to investigate how the Feynman $i\eps$ is related to complex deformations of the Mandelstam invariants. Recall that former is equivalent to deforming the action such that $\Im \V > 0$, which will serve as a benchmark for a causal kinematic deformation.
	
	For instance, if we consider four-particle scattering and complexify the Mandelstam invariant $s$ with $t$ remaining real, we have
	\be\label{eq:ImV2}
	\Im \V = \Im s\, \partial_s \V \qquad\text{with}\qquad \partial_s \V = \V_s - \V_u
	\ee
	since by momentum conservation $\Im\, u = - \Im\, s$. Since both $\V_s$ and $\V_u$ are positive, we cannot make a blanket statement about which sign of $\Im\, s$ is needed to extract the causal amplitude: the answer to this question might differ along the integration contour in the Schwinger-parameter space. So how do we decide? As will be exemplified by explicit computations in Sec.~\ref{sec:branch-cut-deformations}, unfortunately, different Feynman diagrams might require mutually exclusive directions of approaching the branch cuts. This motivates the need for branch cut deformations as an alternative procedure for defining the complexified S-matrix.
	
	The problem is already visible when we specialize to the planar sector. If a given Feynman diagram is $st$-planar, i.e., planar with respect to the variables $s$ and $t$, we have $\V_u=0$ because there are no spanning $2$-trees that would separate the legs $1$ and $3$ from $2$ and $4$. According to~\eqref{eq:ImV}, this unambiguously specifies that $\Im\, s > 0$ is the causal direction. In fact, this logic guarantees that $st$-planar amplitudes do not have singularities or branch cuts in the $s$ upper-half plane at all \cite{Mizera:2021fap}. However, using the same logic, the $tu$-planar diagrams require $\Im\, s <0$ to be consistent with causality. This is not yet a problem, because the causal directions need to be determined only close to branch cuts. Hence, as long as $st$- and $tu$-planar diagrams do not have overlapping branch cuts, there is no problem with extending the amplitudes to the complex $s$-plane. As shown in Sec.~\ref{sec:analyticity-stable}, this is precisely what happens in the case of $2\to2$ scattering of stable particles near the physical regions. However, once we go beyond the neighborhood of the physical regions, include unstable particles, or consider higher-multiplicity processes, the problem will in general persist.
	
	Let us make these statements more quantitative. As a concrete question, we can ask when the imaginary part of the amplitude coincides with its discontinuity in the $s$-plane. Recall that the imaginary part is the one that is more naturally related to unitarity, see Sec.~\ref{sec:cutting}, while the discontinuity is related to dispersion relations, see Sec.~\ref{sec:dispersion} for more details. It is therefore important to understand the conditions under which they are equal. In fact, it will be a simple application of the distributional identity
	\begin{align}\label{eq:Im-identity2}
	\lim_{\eps \to 0^+} \Im \frac{1}{(x+i\eta\eps)^d}
	&=
	\lim_{\eps \to 0^+} \frac{1}{2i}\left( \frac{1}{(x+i\eta\eps)^d} - \frac{1}{(x-i\eta\eps)^d} \right)\\
	&= \frac{(-1)^d \pi}{\Gamma(d)}  \sgn(\eta) \underbrace{\partial_{x}^{d-1} \delta(x)}_{\delta^{(d-1)}(x)}\nn
	\end{align}
	in different ways. Here $x$ and $\eta$ must be real, but the right-hand side depends only on the sign of $\eta$. For simplicity, we will only consider integer $d$ from now on (covering the case of finite integrals in $\D=4$) and assume that the numerator $\widetilde{\N}$ is real. Using the above identity with $x=-\V$ and $\eta=-1$ on~\eqref{eq:I-GL}, the imaginary part of a Feynman integral can be written as
	\be\label{eq:ImI}
	\boxed{
	\Im\, \I = \pi \int \frac{\d^{\E} \alpha}{\GL(1)} \frac{\widetilde\N}{\U^{\D/2}} \delta^{(d-1)}(\V).}
	\ee
	This is precisely according to the expectations: the imaginary part of the integral localizes on the surface $\V=0$ responsible for branch cuts. Of course, we could have equivalently written the above formula as a residue integral around $\V=0$. Recall that here we demand that the degree of divergence $d$ is a positive integer.
	
	One the other hand, we could have computed the discontinuity across the $s$-plane, defined as
	\be
	\Disc_s \I = \frac{1}{2i} \lim_{\eps \to 0^+} \Big( \I(s+i\eps, t) - \I(s-i\eps, t) \Big).
	\ee
	The denominator of~\eqref{eq:I-GL} now deformed to $-\V \mp i\eps \partial_s\V$ instead of $-\V \mp i\eps$, so the sign of $\partial_s \V$ now plays an important role. Applying the identity~\eqref{eq:Im-identity2} with $x = -\V$ and $\eta = -\partial_s \V$ gives
	\be\label{eq:DiscI}
	\boxed{
	\Disc_s \I = \pi \int \frac{\d^{\E} \alpha}{\GL(1)} \frac{\widetilde\N}{\U^{\D/2}} \delta^{(d-1)}(\V) \sgn(\partial_s\V).}
	\ee
    At this stage we can state the difference between~\eqref{eq:ImI} and~\eqref{eq:DiscI} in a sharper way. We define
    \be\label{eq:I-pm}
    \I_D^\pm(s,t) = \pi \int \frac{\d^{\E} \alpha}{\GL(1)} \frac{\widetilde\N}{\U^{\D/2}} \delta^{(d-1)}(\V)\, \Theta(\pm \partial_s \V),
    \ee
    depending on which of the $\V_s$ and $\V_u$ is bigger. This gives
    \be\label{eq:Im-Disc}
    \Im\, \I = \I_D^+ + \I_D^-, \qquad \Disc_s \I = \I_D^+ - \I_D^-.
    \ee
    This give a clear criterion for when the two are equal: when $\I_D^- = 0$. A sufficient condition is that on the support of $\V = 0$, the $\V_s > \V_u$ for every value of the Schwinger parameters. Analogous formulae can be easily derived for higher-multiplicity processes.
	
	The formula~\eqref{eq:ImI} can certainly be used in practical computations. To further simplify it, we can first lower the exponent $d$ appearing in~\eqref{eq:I-GL} to $1$ by acting with kinematic derivatives:
	\be
	\frac{1}{(-\V - i\eps)^d} = \frac{(-1)^{d-1}}{\Gamma(d) \alpha_e^{d-1}} \partial_{m_e^2}^{d-1} \left( \frac{1}{-\V - i\eps}\right)
	\ee
	for any $e$.
	Of course, we could have instead used any other invariant, e.g., $\partial_{t}^{d-1}$ to the same effect. Applying~\eqref{eq:Im-identity2} gives
	\be
	\Im\, \I = (-1)^{d-1} \pi \partial_{m_e^2}^{d-1} \int \frac{\d^{\E} \alpha}{\GL(1)} \frac{\widetilde\N}{\U^{\D/2} \alpha_e^{d-1}} \delta(\V).
	\ee
	Example application of this formula will be given later in Sec.~\ref{sec:triangle-s-unitarity}. After multiplying the integrand with $\sgn(\partial_s \V)$, one obtains a similar representation for $\Disc_s \I$. Similar expressions appeared previously in \cite{10.1143/PTPS.18.1,10.1143/PTP.25.277}, \cite[Sec.~7.6]{Schultka:2019tfi}, and \cite[App.~B]{Arkani-Hamed:2020blm}. Let us stress that $\Disc_s \I$ computes the \emph{total} discontinuity along the real axis, but one can also attempt to compute the \emph{individual} discontinuities $\disc_{\Delta}$ around each branch point $\Delta=0$ separately, as in Sec.~\ref{sec:discontinuities}. Rich literature on this topic includes \cite{Cutkosky:1960sp,FOTIADI1965159,Hwa:102287,AIHPA_1967__6_2_89_0,pham1967introduction,Pham:1968wxy,10.1007/BFb0062917,doi:10.1063/1.1704822,Boyling1966,Boyling1968,pham2011singularities,Bourjaily:2020wvq,Hannesdottir:2021kpd}. Alternatively, we could have derived a formula using the contour-deformation prescription from Fig.~\ref{fig:discontinuity}, which would be valid even if $d$ is not an integer and thus applicable in dimensional regularization.
	
	As a matter of fact, the above derivation makes it natural to introduce two modified Feynman integrals
	\be\label{eq:I-UHP-LHP}
	\I^{\pm}
	(s,t) = \Gamma(d) \lim_{\eps \to 0^+} \int \frac{\d^{\E} \alpha}{\GL(1)} \frac{\widetilde\N\, \Theta(\pm\partial_s \V)}{\U^{\D/2} (-\V - i\eps)^{d}},
	\ee
	which differ by inserting the step function selecting parts of the Schwinger integration with $\partial_s \V \gtrless 0$. Because of this $\I^+$ is completely analytic in the upper-half plane and is consistent with the Feynman $i\eps$ (since $\Im\, \V \geq 0$ for all Schwinger parameters) and likewise for $\I^-$.
    This allows us two write original Feynman integral as a sum over \emph{two} boundary values
	\be\label{eq:two-boundary}
	\I(s,t_\ast) = \lim_{\eps \to 0^+} \left[ \I^{+}(s+i\eps, t_\ast) + \I^{-}(s-i\eps, t_\ast) \right]
	\ee
	for any real $t_\ast$. Of course, we have
	\be
	\lim_{\eps \to 0^+} \Im\, \I^{\pm}(s\pm i\eps, t_\ast) = \I_D^\pm(s,t_\ast).
	\ee
	Further, if at least one of them, say $\I^{-}$, does not have cuts extending throughout the whole $s$-axis, it can be continued to the other half-plane, in which case the whole expression~\eqref{eq:two-boundary} becomes a boundary value of a \emph{single} function.
	
    In Sec.~\ref{sec:dispersion} we will return to the question of imaginary parts and discontinuities in the context of dispersion relations, where we also discuss the subtleties with the treatment of the edge case $\partial_s \V = 0$.
	
	Let us remark on the fact that the above manipulations can be also repeated in the case when $d$ is not a positive integer. In general, one uses the distributional identity
	\be\label{eq:ImGammaV}
	\Im \frac{\Gamma(d)}{(-\V - i\eps)^d} = \begin{dcases}
    \phantom{\frac{1}{1}} \pi \delta^{(d-1)}(\V) \quad&\mathrm{for}\; d \in \Z_{> 0},\\
	\frac{(-1)^d\pi}{(-d)! (-\V)^d} \Theta(\V) \quad&\mathrm{for}\; d \in \Z_{\leq 0},\\
	\;\;\frac{i \Gamma(d)}{(-\V)^d} \Theta(\V) \quad&\mathrm{for}\; d \in \Z/2,
	\end{dcases}
	\ee
	for the imaginary part and the analogous formulae for the discontinuity.
	Note that in the second case the corresponding Feynman integral has an overall UV divergence coming from the $\Gamma$-function, but its imaginary part is free of this divergence (though it might have other divergences), as can be checked by expanding the left-hand side in dimensional regularization around the non-positive integer $d$. Using \eqref{eq:ImGammaV}, one can derive generalizations of the formulae in this section with the essential conclusions unchanged.
	
	\paragraph{Euclidean region.} Let us review the standard result stating that the existence of the Euclidean region guarantees that the imaginary part equals the discontinuity in the $s$-channel. The Euclidean region for a given Feynman integral is defined as the region of the kinematic space $(s_\ast, t_\ast)$, such that the action is negative for any positive value of Schwinger parameters,
	\be\label{eq:V-ast}
	\V_\ast < 0.
	\ee
	For instance, for the bubble diagram from Sec.~\ref{sec:primer}, one can check that the requirement translates to $s_\ast < (m_1+m_2)^2$. More generally, the Euclidean region is guaranteed to exist when the internal masses $m_e$ are not too light \cite{Mizera:2021ujs}:
	\be
	m_e > \frac{1}{\sqrt{2}} \sqrt{\max_{i=1,2,3,4} \left( M_i^2, \tfrac{\sum_{j=1}^{4} M_j^2 - 2M_i^2}{2} \right)}.
	\ee
	For example, when all the external masses are equal to $M$, we need $m_e > M/\sqrt{2}$ and the Euclidean region is given by the triangle $s_\ast,t_\ast, 4M^2 {-} s_\ast {-} t_\ast < 4m_e^2$ \cite{10.1143/PTP.20.690,Wu:1961zz}. The term ``Euclidean'' is a bit of a misnomer used in the literature \cite{Eden:1965aww} and it does not mean to imply that the momenta can be realized in Euclidean space.
	
	At this stage we can take $(s_\ast, t_\ast)$, fix the physical momentum transfer $t_\ast<0$, and start varying $s$ away from the Euclidean region. For sufficiently larger (smaller) values of $s$, we encounter the physical $s$-channel ($u$-channel) region. Since the action is linear in $s$, it takes the form
	\be
	\V = \V_\ast + (s-s_\ast) \partial_s \V.
	\ee
	The functions $\I_\pm$ localize on $\V=0$, which by~\eqref{eq:V-ast} means we need $(s-s_\ast)\partial_s \V > 0$. Since $s>s_\ast$ in the $s$-channel, the step function in $\I_-$ does not have support, which means $\Im\, \I = \Disc_s \I$. By the same arguments, $s<s_\ast$ in the $u$-channel, meaning $\I_+$ does not have support and $\Im\, \I = - \Disc_s \I$.
	
	In Sec.~\ref{sec:analyticity-stable} we will extend these standard arguments to show that the imaginary part and discontinuity still agree even without the existence of the Euclidean region, as long as all the external particles are stable. We will then proceed to studying examples in which $\Im\, \I \neq \Disc_s \I$ in the $s$-channel.

	\newpage
	\section{\label{sec:branch-cut-deformations}Branch cut deformations}
	
    In this section we demonstrate how to exploit the branch cut deformations explained in Sec.~\ref{sec:general} in practice. We start by presenting general results about analyticity near the physical regions for $2\to2$ scattering of stable particles, and then move on to the analysis of concrete examples in later subsections. These examples illustrate cases in which branch cuts run along the whole real $s$-axis, and they reveal two different situations: one when the branch cuts can be deformed away to restore the connection between the upper- and lower-half planes, and one when they cannot. Finally, we explain how summing over different Feynman diagrams might result in incompatible $s \pm i\varepsilon$ prescriptions and one is forced to consider branch cut deformations to define the complexified amplitude in the first place.
	
	\subsection{\label{sec:analyticity-stable}Analyticity from branch cut deformations}
	
	We start by using branch cut deformations to prove analyticity in a certain neighborhood of the physical regions. In the case of $2\to2$ scattering with no unstable or massless external particles, we prove that the $s + i\eps$ deformation is always consistent with causality near the $s$-channel. This is an improvement over previous results, which had to assume the existence of the Euclidean region and thus were not applicable, e.g., to quantum field theories with massless particles or large momentum transfers \cite{Eden:1966dnq}.
	
\begin{figure}
	    \centering
	    \includegraphics[scale=1.1]{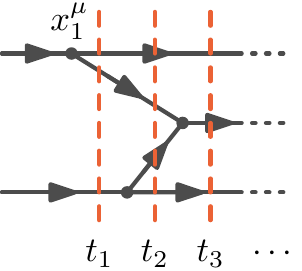}
	    \caption{Unitarity cuts with time ordering $t_1 < t_2 < t_3 < \cdots$. The first vertex occurring at $x_1^0 < t_1$ is classically forbidden if the incoming particle is stable. For $2\to 2$ scattering of stable particles, this argument shows that both incoming and both outgoing particles have to meet at a single vertex, which forbids anomalous thresholds from contributing in the physical regions.}
	    \label{fig:time-ordered}
	\end{figure}

	\paragraph{$2 \to 2$ scattering of stable particles.}
	Our starting point is determining which singularities appear in the physical regions. Without loss of generality, we will consider the $s$-channel region, i.e., $12\to34$ scattering.
	Since a singularity corresponds to a classical scattering process, the positions of the interaction vertices $x_v^\mu$ have some specific values and hence can be time-ordered. The earliest vertex, say $x_1^\mu$, has to have at least one incoming external momentum, since otherwise energy conservation would be violated, see Fig.~\ref{fig:time-ordered}. However, we assumed that all external particles are stable (meaning that $1 \to$ many processes do not exist), so if a vertex $x_1^\mu$ existed, it would have to have at least \emph{two} incoming external momenta. In particular, for $2\to 2$ scattering, both incoming momenta have to enter at the same vertex and by time-reflection symmetry the same holds for the outgoing ones. This implies that the only external momentum is $P_{\inn}^\mu = p_1^\mu + p_2^\mu$ and we can go to a frame in which it is purely timelike, $P_{\inn}^\mu = (\sqrt{s},0,0,\ldots)$. Since Landau equations are linear in the momenta, we must have $q_e^\mu \propto P_\inn^\mu$ for every internal momentum (the explicit solution was given in~\eqref{eq:qe}). This, by definition, is a normal threshold. Recall that its position in the kinematic space is simply determined by
	\be\label{eq:smm}
	s = (m_1 + m_2 + \ldots + m_R)^2,
	\ee
	where $m_1$, $m_2$, $\ldots$, $m_R$ are the masses of the intermediate particles with $R\geq 1$. Singularities might be branch points or poles. Recall that accounting for mass shifts and decay widths, we should replace
	\be
	m_e^2 \to \m_e^2 - i m_e \Gamma_e,
	\ee
	where $\mathfrak{m}_e \geq 0$ is the shifted mass and $\Gamma_e \geq 0$ is the particle width. Hence, $\alpha$-positive singularities in the $s$-channel are located either directly on the real $s$-axis when all the intermediate particles are stable ($\Gamma_e = 0$), or below it in the presence of unstable particles ($\Gamma_e > 0$). In addition, one can show \cite{doi:10.1063/1.1705398} that for a quantum field theory with a finite spectrum of particles, the number of singular points~\eqref{eq:smm} remains finite even after summing over all the Feynman diagrams.
	
	The goal is to prove that under the above assumptions, there exists an infinitesimal strip of analyticity right above the $s$-axis in the physical regions, which in addition allows the causal S-matrix to be defined is a way consistent with the Feynman $i\eps$ prescription. Note that naively one would simply want to deform the internal masses $m_e^2 \to m_e^2 - i\eps$ and conclude that all the normal-threshold branch cuts are deformed to the lower-half planes and hence $s+i\eps$ is the direction consistent with causality. This argument, however, is not complete because it does not take into account that adding the $i\eps$ by hand could have moved some complex anomalous thresholds to the real axis, and moreover it does not constrain branch cuts that do not have a branch point in the physical region.
	Therefore, more work is necessary for the proof.
	We split this problem into two stages: first considering the singular points~\eqref{eq:smm} lying on the real axis and then the remaining non-singular points. We will work diagram-by-diagram and show that all their $i\eps$ rules are consistent with each other, as a refinement of the arguments presented in \cite[Sec.~III]{Mizera:2021fap}.

\begin{figure}
		\centering
		\includegraphics[scale=1.1]{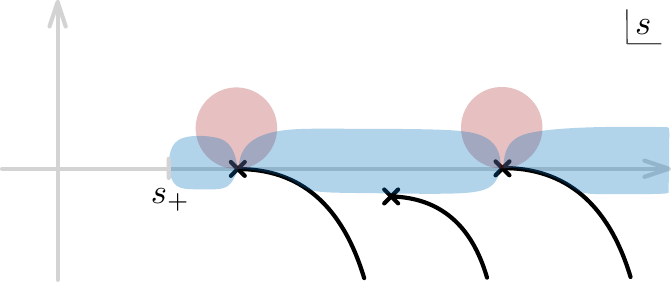}
		\caption{\label{fig:stable} Region of analyticity in the $s$-plane proven in Sec.~\ref{sec:analyticity-stable} (shaded). Here $s_+(t)$ is the boundary of the $s$-channel physical region for a given $t$. Above it, every real singular point has an $i\eps$-compatible neighborhood in the upper-half plane (shaded, red), and so does every non-singular point (shaded, blue). Branch cuts are deformed downwards (black) and likewise including internal unstable particles moves the singularities downwards.}
	\end{figure}

	Let us consider any Feynman diagram with a singularity at the point $s= s_\ast \in \R$.  In the neighborhood of the corresponding normal threshold, the action $\V$ takes the form
	\be
	\V^\ast = s \V_s^\ast - \sum_{e=1}^{\E} m_e^2 \alpha_e^\ast,
	\ee
	where explicitly $\alpha_e^\ast = 1/m_e$ is the solution of Landau equations and all the starred quantities are evaluated on its support. Recall that the causal prescription corresponds to $\Im \V > 0$, which applied here gives
	\be
	\Im \V^\ast = (\Im\, s) \V_s^\ast > 0.
	\ee
	As reviewed in Sec.~\ref{sec:parametric}, $\V_s$ is always positive on the integration contour, and in particular for $\alpha_e = \alpha_e^\ast$, which implies that $\Im\, s > 0$ is the causal direction. This argument shows analyticity in an infinitesimal neighborhood of $s=s_\ast$ in the direction of the upper-half plane, as indicated in Fig.~\ref{fig:stable}.
	
	Let us now consider non-singular points, i.e., those $s=s_\ast$ at which Landau equations are not satisfied. Here we can simply apply the branch cut deformation, as described in Sec.~\ref{sec:deformation} to obtain
	\be\label{eq:hatV}
	\hat{\V} = \V + i \eps \sum_{e=1}^{\E} \alpha_e (\partial_{\alpha_e} \V)^2 + {\cal O}(\eps^2).
	\ee
	Since $\partial_{\alpha_e} \V = 0$ cannot be all satisfied simultaneously, the action is deformed with a small positive imaginary part. As a result, the branch cuts of the normal thresholds move downwards, as illustrated in Fig.~\ref{fig:stable}. To complete the argument, it suffices to notice that simultaneously we can make a subleading (for example, order ${\cal O}(\eps^2)$) deformation to the $s$ variable without spoiling the $i\eps$ property~\eqref{eq:hatV}. Hence, there is an infinitesimal region close to any non-singular point $s=s_\ast$ within which the Feynman integral agrees with the $i\eps$ prescription, see Fig.~\ref{fig:stable}.
	
	To wrap up the proof, we only need to notice that all the above regions overlap and, moreover, are consistent for every Feynman integral. Analogous derivation can be given in the $t$- and $u$-channels. Branch cut deformation therefore lead to infinitesimal regions near the physical kinematics, where the complexified S-matrix $\T_\C$ can be defined perturbatively. In other words,
	\be\label{eq:T1}
	\T(s,t_\ast) = \lim_{\eps \to 0^+} \T_\C(s+i\eps, t_\ast)\,,
	\ee
    in the $s$-channel physical region, $s>s_+$.
	We could have repeated the same analysis with the anti-causal $-i\eps$ prescription, which shows that the complex-conjugate satisfies
	\be\label{eq:T2}
	\overline{\T(s,t_\ast)} = \lim_{\eps \to 0^+} \T_\C(s-i\eps, t_\ast) \,.
	\ee
	Importantly, note that it is not required that $\T_\C$ is analytic in the whole $s$-plane, or even that the upper- and lower-half planes are connected. Combining~\eqref{eq:T1} and~\eqref{eq:T2}, we find
	\be
	\Im\, \T(s,t_\ast) = \tfrac{1}{2i} \lim_{\eps \to 0^+} \left[ \T_\C(s+i\eps, t_\ast) - \T_\C(s-i\eps, t_\ast) \right] = \Disc_s \T_{\C}(s,t_\ast) \,,
	\ee
	and hence the imaginary part and discontinuity agree in the $s$-channel. Recall that we assume stability of the external particles and take $s$ to be in the $s$-channel physical region. The same analysis gives $\Im\, \T(s,t_\ast) = -\Disc_s \T_{\C}(s,t_\ast)$ in the $u$-channel region.

	\paragraph{Higher-point scattering and unstable particles.}
	At this stage we can ask how much of the above analysis holds more generally, i.e., beyond the case of $2\to 2$ scattering for stable particles.  Let us stress that the first part of the discussion still works: branch cut deformations can be use reveal the physical sheet of the S-matrix. Nevertheless, anomalous thresholds might contribute in the physical regions and it is not a priori known which $s \pm i\eps$ direction---if any---is consistent with the Feynman $i\eps$. In the following sections we consider some case studied that illustrate these points explicitly.

	\subsection{\label{sec:ExampleI}Example I: Necessity of deforming branch cuts}
	
	The simplest example of branch cuts extending throughout the whole $s$-axis is one with overlapping normal thresholds. Recall that the $s$-channel normal threshold gives a cut at $s \geq (m_1+ m_2)^2$, while the $u$-channel might have $u \geq (m_3 + m_4)^2$ for some values of the threshold masses. Note that the $u$-channel cut never enters the $s$-channel physical region and vice versa. But by momentum conservation we have $s+t+u = \sum_{i=1}^{4} M_i^2$, so the two cuts do overlap whenever
	\be
	\sum_{i=1}^{4} M_i^2 - t > (m_1+m_2)^2 + (m_3+m_4)^2.
	\ee
	For example, in a single-mass theory $M_i = m_e = m$, the cuts overlap when $t < -4m^2$, while setting external masses to zero gives $t < -8m^2$. It is then important how to understand how to resolve such overlapping branch cuts. This section studies the simplest instance of this procedure.
	
	\subsubsection{Box diagram}
	
	\begin{figure}
	    \centering
	    \includegraphics[scale=1.1]{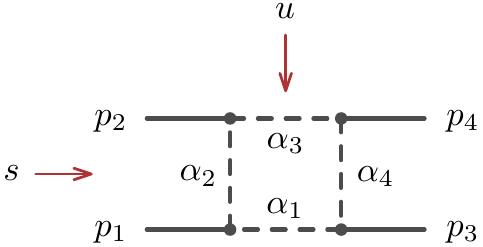}
	    \caption{The $su$-box diagram considered in Sec.~\ref{sec:ExampleI}. All the internal and external masses are set to $m$ and $M$ respectively. Throughout this section we set $M=0$.}
	    \label{fig:box}
	\end{figure}
	
	Let us consider the $su$-planar box with massless external particles, $M=0$, and internal particles of non-zero mass $m$, illustrated in Fig.~\ref{fig:box}. The corresponding Feynman integral is given by
	\be\label{eq:I-box}
	\I_{\boxx}(s,t) = \Gamma(4-\D/2) \lim_{\eps \to 0^+}\int \frac{\d^4 \alpha}{\GL(1)} \frac{1}{\U_\boxx^{\D/2} (-\V_\boxx - i\eps)^{4-\D/2}},
	\ee
    with
    \begin{align}
    \U_\boxx = \alpha_1 {+} \alpha_2 {+} \alpha_3 {+} \alpha_4,\quad
    \V_\boxx = \frac{s\, \alpha_1 \alpha_3 + u\, \alpha_2 \alpha_4}{\alpha_1 {+} \alpha_2 {+} \alpha_3 {+} \alpha_4} - m^2 (\alpha_1 {+} \alpha_2 {+} \alpha_3 {+} \alpha_4).
    \end{align}
    By momentum conservation $s+t+u=0$.
	We will treat $s$ and $t$ as the independent Mandelstam invariants and fix $t= t_\ast < -8m^2$, so that the normal-threshold cuts overlap in the $s$-plane.
	
\begin{figure}
	    \centering
	    \includegraphics[scale=1.2]{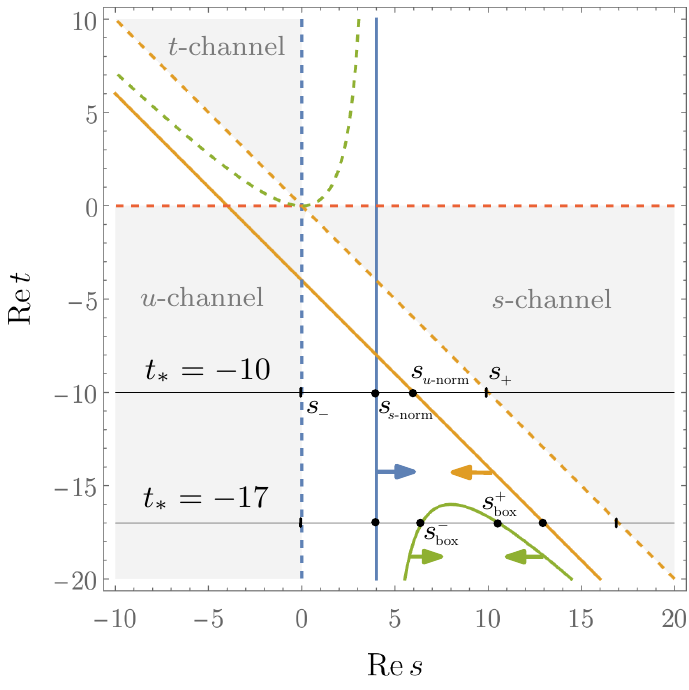}
	    \caption{Landau curves for the $su$-box diagram with external masses $M=0$ and internal $m=1$. The leading singularity $\Delta_{\boxx}=0$ (green) has branches: the physical (solid) when $s>4m^2$ and unphysical (dashed) for $s<4m^2$. Apart from it, there are normal thresholds in the $s$- (blue) and $u$-channel (orange), as well as their corresponding pseudo-normal thresholds and second-type singularities (dashed: blue, red, orange) at $s=0$, $t=0$, $u=0$. Physical regions are the three shaded areas. The complex $s$-planes for two fixed values of $t = t_\ast$ will be illustrated later in Fig.~\ref{fig:box10} and \ref{fig:box17}. The arrows indicate the direction of the corresponding branch cut in the $s$-variable.}
	    \label{fig:box-st}
	\end{figure}

	Recall from Sec.~\ref{sec:parametric} that the physical regions are carved out by the constraints $stu>0$ when the external particles are massless. The individual channels are characterized by which of the three Mandelstam invariants are positive, e.g., the $s$-channel is given by $s>0$ and $t,u<0$, which translates to $s > s_+ = -t_\ast$. Similarly, the $u$-channel is confined to $s < s_- = 0$, see Fig.~\ref{fig:box-st}. 
	
	Following Sec.~\ref{sec:general}, all the analytic features of this Feynman integral can be explained without performing explicit computations. First of all, let us understand why there is a cut along the whole $s$-axis. Recall that a branch cut happens when $\V=0$ for some value of the $\alpha$'s. Clearly, $\V$ can be negative, e.g., close to $\alpha_1 = \alpha_2 = \alpha_3 = 0$ we have $\V = -m^2 \alpha_4 < 0$. The question then becomes: for which values of $s$ does the maximum of $\V$ across the integration domain become positive? By continuity, this would imply that $\V=0$ somewhere between the minimum and maximum, indicating presence of a branch cut. To this end, let us consider the slice through the integration domain
	\be
	(\alpha_1 : \alpha_2 : \alpha_3 : \alpha_4) = (\rho : \tfrac{1}{2} {-}\rho : \rho : \tfrac{1}{2}{-}\rho),
	\ee
	parametrized by a variable $0 \leq \rho \leq \tfrac{1}{2}$. On this subspace, we have
	\be
	\V(\rho) = s \rho^2 + u \left(\rho - \tfrac{1}{2}\right)^2 - m^2.
	\ee
	The region where $\rho$ is close to $\tfrac{1}{2}$ is responsible for the $s$-channel branch cut because $\V(\tfrac{1}{2}) > 0$ when $s>4m^2$, and similarly $\rho$ close to $0$ gives the $u$-cut when $u>4m^2$ to ensure $\V(0) > 0$. Hence, regardless of the value of $s \in \R$, when $t = t_\ast < -8m^2$, $\V=0$ somewhere in the integration domain, guaranteeing a branch cut.

	The normal thresholds are thus located at
	\be
	s_{s\text{-norm}} = 4m^2, \qquad (\alpha_1^\ast : \alpha_2^\ast : \alpha_3^\ast : \alpha_4^\ast ) = (\tfrac{1}{m}: 0 : \tfrac{1}{m} : 0),
	\ee
	as well as
	\be
	\qquad s_{u\text{-norm}} = - t_\ast - 4m^2, \qquad (\alpha_1^\ast : \alpha_2^\ast : \alpha_3^\ast : \alpha_4^\ast ) = ( 0 : \tfrac{1}{m}: 0 : \tfrac{1}{m}).
	\ee
	Next, let us locate the anomalous thresholds. Positions of all Landau singularities of this diagram are plotted in Fig.~\ref{fig:box-st} in the real $st$-plane. The leading singularity is
	\be
    \Delta_{\boxx} = su + 4m^2 t = 0, \quad (\alpha_1^\ast : \alpha_2^\ast : \alpha_3^\ast : \alpha_4^\ast ) = (4m^2 : s{-}4m^2 : 4m^2 : s{-}4m^2),
	\ee
	where $\Delta_{\boxx}$ has already appeared in~\eqref{eq:Delta-box}.
	This solution is only $\alpha$-positive if $s > 4m^2 > 0$, which corresponds to a parabola asymptoting to $t \to 4m^2$ and $u \to 4m^2$ between the $s$- and $u$-channel physical regions, see Fig.~\ref{fig:box-st} (green). It has a tip at the point $(s,t) = (8m^2, -16m^2)$. Apart from that, setting $M=0$ gives a considerable simplification to Landau singularities: the remaining box, triangle and pseudo-normal thresholds are located at $s=0$, $t=0$, and $u=0$, cf. Tab.~\ref{tab:su-box}. All of them are second-type singularities since they only exist when $\U_{\boxx}^\ast = 0$ and hence are not on the physical sheet.
	
	\begin{figure}
	    \centering
	    \includegraphics[scale=1.05]{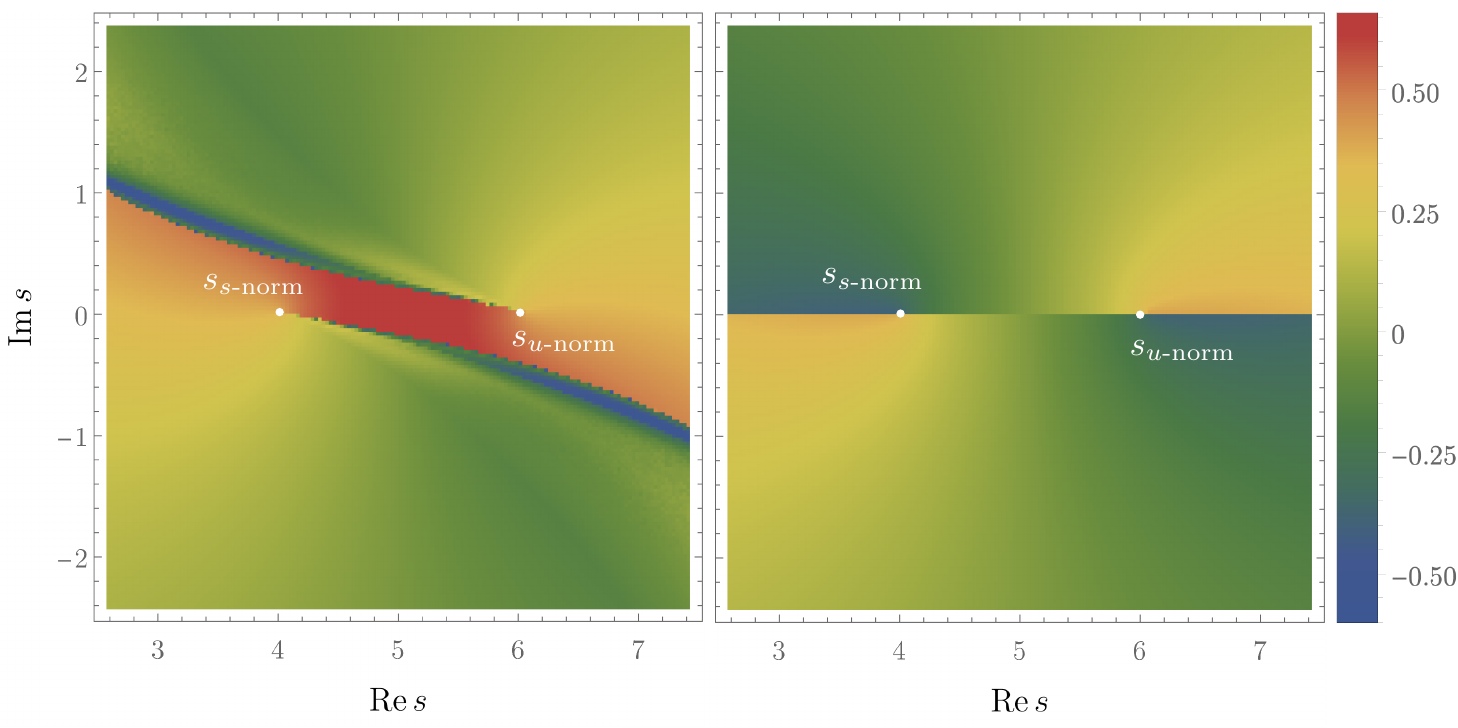}
	    \caption{Comparison between the numerical (left) and analytic (right) results for the imaginary part of the $su$-box Feynman integral, $\Im\, \I_{\boxx}(s,t_\ast)$, in the $s$-plane with $m=1$, $t_\ast = -10$, and $\eps = \tfrac{1}{10}$. Branch cuts originate at the $s$- and $u$-channel normal thresholds, $s_{s\text{-norm}}=4$ and $s_{u\text{-norm}} = 6$, respectively. Branch cut deformations allow us to reveal the value of $\I_\boxx(s,t_\ast)$ otherwise trapped between the two branch cuts.}
	    \label{fig:box10}
	\end{figure}
	
	We can verify these results directly using deformations of Schwinger parameters described in Sec.~\ref{sec:deformation}. To this end, we set $\eps = 0$ in~\eqref{eq:I-box} and perform the change of variables $\hat{\alpha}_e = \alpha_e \exp \left(i\eps \partial_{\alpha_e}\!\V_\boxx \right)$ together with the $\GL(1)$ fixing $\hat{\alpha}_4 = 1 - \hat{\alpha}_1 - \hat{\alpha}_2 - \hat{\alpha}_3$, according to the formula~\eqref{eq:I-deformed}. This procedure has the effect of deforming branch cuts in the kinematic space and allows us to perform the Feynman integral directly. The result for $t_\ast = -10 m^2$ is presented in Fig.~\ref{fig:box10} (left). One can immediately recognize two branch cuts responsible for the $s$- and $u$-channel normal thresholds. Decreasing the value of $\eps$ make the cuts lie closer and closer to the real $s$-axis. In the limit as $\eps \to 0^+$, they overlap on the strip $4m^2 < s < -t_\ast {-} 4m^2$. The situation changes once we consider high enough momentum transfer, $t_\ast < -16 m^2$,  such that the box threshold starts appearing. The corresponding situation is illustrated in Fig.~\ref{fig:box17}, where the branch cut structure becomes more involved.
	
	\subsubsection{\label{app:box}Analytic expression}
	
    We start with working out the analytic form of the box integral.
    Of course, this Feynman integral has known expressions in the physical regions or for low values of the momentum transfer $t$, see, e.g.,~\cite{karplus1951scattering,tHooft:1978jhc,Davydychev:1993ut}. Nevertheless, some of these results are no good in the region $t_\ast < -8m^2$ of our interest, while others are only valid close to the real axis, which means we have to compute $\I_{\boxx}(s,t_\ast)$ from scratch while being cautious about branch cuts. In principle, we could use the formula~\eqref{eq:two-boundary}, which naturally splits $\I_{\boxx}(s,t_\ast)$ into two contributions with the $s$- and $u$-channel each. However, in practice we found that we arrive at simpler expressions by direct integration of~\eqref{eq:I-box}.
    In the $\GL(1)$ fixing $\sum_{e=1}^4 \alpha_e = 1$, the box integral is given by
    \begin{multline}
        \I_\boxx = \Gamma\left(4-\D/2\right) \lim_{\eps \to 0^+} \int_0^1 \d \alpha_1 \int_0^{1-\alpha_1} \d \alpha_2 \int_0^{1-\alpha_1-\alpha_2} \d\alpha_3 \\ 
        \times \frac{1}{\left[ -s \alpha_1 \alpha_3 - u \alpha_2 (1 - \alpha_1 - \alpha_2 - \alpha_3) + m^2-i \varepsilon \right]^{4-\D/2}}\,.
    \end{multline}
    Performing the integral in $\alpha_3$,
    in $\D=4$ space-time dimensions, gives
    \be
        \I_\boxx = \lim_{\eps \to 0^+} \int_0^1 \d \alpha_1 \int_0^{1-\alpha_1} \!\!\d\alpha_2 \frac{1-\alpha_1-\alpha_2}{\left[s \alpha_1 (1{-}\alpha_1{-}\alpha_2) - m^2 + i\eps \right]\left[u \alpha_2 (1{-}\alpha_1{-}\alpha_2) - m^2 + i\eps \right]}\ .
    \ee
    To make the integrand manifestly symmetric in $s$ and $u$, we make the following change of variables: 
    \be
    \beta_1 = 1-\alpha_1-\alpha_2 \,, \qquad \beta_2 = \alpha_1-\alpha_2 \,.
    \ee
    In these variables, the integration proceeds over $0<\beta_1<1$ and $-1{+}\beta_1<\beta_2<1{-}\beta_1$, with a Jacobian factor of $\vert \det \frac{\partial \alpha_i}{\partial \beta_j} \vert = \frac{1}{2}$. This gives
    \be
    \I_\boxx = \lim_{\eps \to 0^+} \int_0^1 \d \beta_1 \int_{-1+\beta_1}^{1-\beta_1} \!\! \d \beta_2
        \frac{2\,\beta_1}{\left[s \, \beta_1 (1{-}\beta_1{+}\beta_2) - 2 m^2 + i\eps \right] \left[u \, \beta_1 (1{-}\beta_1{-}\beta_2) - 2 m^2 + i\eps \right]} \,.
    \ee 
    Now, we simply perform the remaining integrals. Starting with the one in $\beta_2$, we get
    \be 
        \I_\boxx = -
        \lim_{\eps \to 0^+} \int_0^1 \d \beta_1
        \frac{\log\left[\frac{s}{m^2} \beta_1 \left(\beta_1-1\right)+ 1 - i \eps \right]+\log\left[\frac{u}{m^2} \beta_1 \left(\beta_1 -1\right)+ 1 - i \eps \right]}{s u \beta_1\left( \beta_1 - 1\right) + (s+u) m^2 -i \eps\, (s+u)} \,.
    \ee
    This integral will have a branch cut starting at $s= (t_\ast \pm \sqrt{t_\ast-16m^2} \sqrt{t_\ast})/2$, whenever $t_\ast<-16 m^2$. To ensure that this branch cut is not on the principal sheet, we take $-16 m^2 < t_\ast < -8 m^2$, i.e., we choose kinematics such that the box threshold does not contribute, cf. Fig.~\ref{fig:box-st}.
    
    To perform the last integral in $\beta_1$, we notice that the integral naturally splits as a sum of two terms,
    \be 
	    \I_\boxx^\C (s,t_\ast) = \I_\boxx^{\C,s} (s,t_\ast) + \I_\boxx^{\C,u} (s,t_\ast) \,,
	    \label{eq:box_su}
	\ee 
	where the term with the $s$-channel branch cut is given by
    \be 
    \I_\boxx^{\C,s} = - \lim_{\eps \to 0^+} \int_0^1 \d \beta_1
        \frac{\log\left[\frac{s}{m^2} \beta_1 \left(\beta_1-1\right)+ 1 - i \eps \right]}{s u \beta_1 \left( \beta_1 - 1 \right) + (s+u) m^2} \,,
        \label{eq:I-boxS}
    \ee 
    and $\I_\boxx^{\C,u}(s,t_\ast) = \I_\boxx^{\C,s}(-s{-}t_\ast,t_\ast)$. Each of the terms has a branch cut in $s>4m^2$ or $u>4m^2$ respectively.
    We will treat each of them as a function in the complex $s$-plane, as indicated by the superscript $\C$. 
    By doing this split, we introduced spurious branch cuts for (complex) values of $s$ for which the denominator can go to zero, starting at $s= (t_\ast \pm \sqrt{t_\ast-16m^2} \sqrt{t_\ast})/2$. Of course, the spurious branch cut cancels between the two terms $\I_\boxx^{\C,s}$ and $\I_\boxx^{\C,u}$ since we take $t_\ast>-16 m^2$.

    To perform the last integral, we note that we can split the logarithm in $\I_\boxx^{\C,s}$ without spoiling
    the branch-cut structure by writing
    \be 
        \I_\boxx^{\C,s} = - \lim_{\eps \to 0^+} \int_0^1 \d \beta_1
        \frac{\log\Big[\sqrt{\tfrac{-s}{m^2}} \beta_1 - \frac{-\sqrt{4-\tfrac{s}{m^2}}+\sqrt{\tfrac{-s}{m^2}}}{2} \Big]+\log\Big[-\sqrt{\tfrac{-s}{m^2}} \beta_1 + \frac{\sqrt{4-\tfrac{s}{m^2}}+\sqrt{\tfrac{-s}{m^2}}}{2} \Big]}{s u \beta_1\left( \beta_1 - 1 \right) + (s+u) m^2} \,,
    \ee 
    where the square roots and logarithm are on their principal branch. The $i\varepsilon$ has been dropped from this expression, since it has the same sign as $\Im\, s$, as can bee seen from~\eqref{eq:I-boxS}.
    
    We can now simply perform the integral in $\beta_1$. In doing so, we must be careful to write the resulting functions on the right branches. More precisely, when we integrate over $\beta_1$ for some values of $s$, we may need to take into account that the logarithms and dilogarithms go on the second sheet, resulting in additional lower-transcendentality terms if we insist to write the functions on the principal sheet. In order to take into account the possibility of landing on different sheets, we define the analytically continued logarithms and dilogarithms as
    \begin{align} 
        \log^{\pm} (x,\varphi) &= \log (x) \pm 2 \pi i \, \Theta \left[-\Re\, x \right] 
        \Theta \left[\mp \Im\, x \right] 
        \Theta \left(\varphi \right) \,, \\
        \Li_2^{\pm} (x,\varphi) &= \Li_2(x) 
        \pm 2 \pi i \log (x) \Theta \left[\Re\, x-1 \right] 
        \Theta \left[\mp \Im\, x \right] \Theta \left(\varphi \right) \,.
    \end{align} 
    The logarithm $\log(x)$ and the dilogarithm $\Li_2(x)$ are always taken to be on their principal branch, and $\Theta$ functions in $\Re\, x$ and $\Im\, x$ come about since the imaginary part of the logarithm $\log(x)$ and the dilogarithm $\Li_2(x)$ on the principal sheet changes discontinuously as the branch cuts at $x<0$ and $x>1$ are crossed. Thus, we must compensate for the artificial discontinuity coming from insisting on writing $\log(x)$ and $\Li_2(x)$ on their principal branch with the negative of the discontinuities across the branch cuts. The theta function in $\varphi$ gives us a handle for defining the analytic continuation in specific regions of the $s$-plane. We stress that the above expressions are analytic functions and the only reason for introducing them is to be careful about the sheet structure.

	The above definitions allow us to compactly write the result as a function that is analytic in both the upper- and lower-half $s$-planes (apart from the spurious box branch cuts), explicitly,
    \begin{align} \label{eq:box_UHP}
        \I_{\text{box}}^{\C,s} 
        & =
        -
        \frac{x y}{8 m^4 \beta_{xy}}
        \bigg\{
            \log \left(\frac{\beta_x-1}{\sqrt{x}} \right)
            \left[ \log\left( \frac{\beta_{xy}-1}{\beta_{xy}-\beta_x} \right)
            -
            \log \left( \frac{\beta_{xy}+1}{\beta_{xy}+\beta_x}\right)
            \right]
            \\
            &  \hspace{1.05cm} +
            \log \left(\frac{\beta_x+1}{\sqrt{x}}\right)
            \left[
            \log^{-} \left( \frac{\beta_{xy}-1}{\beta_{xy}+\beta_x}, \Im\, s \right)
            -
            \log^{+}\left( \frac{\beta_{xy}+1}{\beta_{xy}-\beta_x} , \, - \Im\, s  \right)
            \right]
            \nn\\ 
            & \hspace{1.05 cm}
            +
            \Li_2 \left( \frac{-\beta_x+1}{\beta_{xy}-\beta_x} \right)
            +
            \Li_2^{+} \left( \frac{\beta_x+1}{\beta_{xy}+\beta_x}, \, \Im\, s \right)
            \nn\\ 
            & \hspace{1.05 cm}
            -
            \Li_2 \left( \frac{\beta_x-1}{\beta_{xy}+\beta_x} \right)
            -
            \Li_2^{-} \left( \frac{-\beta_x-1}{\beta_{xy}-\beta_x} ,\,- \Im\, s \right)
        \bigg\}\nn
    \end{align}
    with
    \begin{gather}
        x = -\frac{4 m^2}{s}, \, \qquad
        y = -\frac{4 m^2}{u}, \\ 
        \beta_x = \sqrt{1+x}
        \qquad 
        \beta_y = \sqrt{1+y}
        \qquad 
        \beta_{xy} = - i \sqrt{-1-x-y} \,.
    \end{gather}
	Note that the expression in~\eqref{eq:box_UHP} is written in such a way that the logarithms and dilogarithms can be evaluated on their principal sheet as $s$ is taken to have small positive imaginary part, and consequently $u$ with a small negative imaginary part.

	The analogous expression for $\I_\boxx^{\C,u}$ can be obtained by simply relabelling $s \leftrightarrow u$ from~\eqref{eq:box_su}.
	Since there is no Euclidean region for our choice of kinematics, we have to take into account how the logarithms and dilogarithms are approached in the upper- and lower-half $s$-planes, and make sure to land on the right branches. 
	For example, while the expression in~\eqref{eq:box_su} is separately analytic in the upper-half $s$ plane and lower-half $s$ plane, the value of the box amplitude within the interval $4m^2 < s < -4m^2 {-} t_\ast$ cannot be expressed as a boundary value of a single function. Instead, the calculation shows that the $i \varepsilon$ prescription dictates the following approach of the physical amplitude,
	\be\label{eq:box-final} 
	    \I_\boxx(s,t) =  \lim_{\varepsilon\to 0^+} \left[ \I_\boxx^{\C,s} (s+i\varepsilon,t) + \I_\boxx^{\C,u} (s-i\varepsilon,t) \right].
	\ee 
	The comparison of this analytic result with numerical integration is shown in Fig.~\ref{fig:box10} (right).
	
	Before we go on to compute the discontinuities, imaginary parts and cuts of this expression, let us comment on the structure of the amplitude. The leading singularity, located at the box branch point corresponding to $\beta_{xy}=0$, is a square root singularity of $\I_\boxx^{\C,s}$. As discussed earlier, this is an artifact of how we decided to split of the full integral in intermediate steps, since the box branch point is not on the principal sheet for our chosen kinematics of $-16 m^2 < t_\ast < - 8 m^2$. Hence this spurious singularity vanishes from the principal sheet as we add the contribution from $\I_\boxx^{\C,u}$. The branch point at $s=4 m^2$, i.e., at $\beta_x = 0$, is a square root singularity of $\I_\boxx^{\C,s}$, while the branch point at $u=4 m^2$, i.e., $\beta_y=0$, does not appear at all in the expression for $\I_\boxx^{\C,s}$. Thus, as discussed earlier, we have indeed split the amplitude into contributions that have separate branch cuts: the one in $\I_\boxx^{\C,s}$ starts at $s=4 m^2$ while the one in $\I_\boxx^{\C,u}$ starts at $u=4 m^2$. If we solve for the location of potential logarithmic branch points on the principal sheet, which amounts to solving for when the arguments of the logarithms are 0 and when the arguments of the dilogarithm are 1, we find the triangle branch points.

	\subsubsection{Discontinuities and imaginary parts} 
	
	Using the expression in~\eqref{eq:box_UHP}, we can compute discontinuities and imaginary parts. We start with computing the discontinuity of $\I_\boxx^{\C,s}$ across the branch cut at $s>4m^2$:
	\begin{equation}
	    \Disc_s \I_\boxx^{\C,s}(s,t_\ast) = \frac{1}{2 i} \lim_{\varepsilon \to 0^+} \left[ \I^{\C,s}_\boxx(s+i \varepsilon,t_\ast) - \I^{\C,s}_\boxx(s-i \varepsilon,t_\ast) \right] \,.
	\end{equation}
	As we take the discontinuity across the $s$-axis, we can use that on the branch cut at $s>4m^2$, the following relations hold for the arguments of the logarithms when $t_\ast < - 8 m^2$,\footnote{Note that for $\Re u <0$ some infinitesimal imaginary parts of the arguments of the logarithms and dilogarithms may change signs across the branch cut. However, whenever this happens, the arguments of the logarithms have a positive real part, and the arguments of the dilogarithms have a real part that is less than $1$, so this sign change will not result in a discontinuity across the branch cut at $s>4m^2$.}
	\begin{align} 
	    \lim_{\eps \to 0^+} &\left( \frac{\beta_x\pm 1}{\sqrt{x}},\;
	    \frac{\beta_{xy}\pm 1}{\beta_{xy}\pm \beta_x},\;
	    \frac{\beta_{xy}\pm 1}{\beta_{xy}\mp \beta_x}
	    \right) \bigg|_{s\to s-i \eps} \nn\\
	    & = \lim_{\eps \to 0^+} \left( -\frac{\beta_x \pm 1}{\sqrt{x}},\;
	    \frac{\beta_{xy} \pm \zeta }{\beta_{xy} \pm  \zeta \beta_x},\;
	    \frac{\beta_{xy}\pm \zeta }{\beta_{xy}\mp \zeta \beta_x}
	    \right) \bigg|_{s\to s+i \eps},
	   \label{eq:discLHPrel}
	   \end{align}
	   while for the arguments of the dilogarithm, we get
	   \be
	    \lim_{\eps \to 0^+}\left(
	    \frac{\pm \beta_x+1}{\beta_{xy}\pm \beta_x},\;
	    \frac{\mp \beta_x+1}{\beta_{xy}\pm \beta_x}
	    \right) \bigg|_{s\to s-i \eps} 
	    =
	    \lim_{\eps \to 0^+}
	    \left(
	    \zeta
	    \frac{\pm \beta_x+1}{\beta_{xy}\pm \zeta \beta_x},\;
	    \zeta
	    \frac{\mp \beta_x+1}{\beta_{xy}\pm \zeta \beta_x}
	    \right) \bigg|_{s\to s+i \eps},
	   \label{eq:discLHPrelLi}
	   \ee
	where we have defined $\zeta = \sgn(\Re u)$.
	
	Let us first look at the discontinuity coming from the dilogarithm terms in~\eqref{eq:box_UHP}. Close to the real $s$-axis for $s>4m^2$, the dilogarithms are always on their principal branch, and the real part of their arguments are always less than 1. For each term, the difference between the arguments in the upper- and lower-half planes appears in another dilogarithmic term, and the net result is that these terms do not result in any discontinuity.
	
	Next, we find the discontinuity from the logarithmic terms. To calculate the contribution from each term, we subtract the value of the lower-half plane term, using the relations in~\eqref{eq:discLHPrel}. For the first term, we write the discontinuity as the following upper-half plane limit:
	\begin{multline} 
	    \Disc_s \left[ - 
        \frac{x y}{8 m^4 \beta_{xy}}
        \log \left(\frac{\beta_x-1}{\sqrt{x}} \right)
        \log\left( \frac{\beta_{xy}-1}{\beta_{xy}-\beta_x} \right) \right]
	    \\ =
	    i \frac{x y}{16 m^4 \beta_{xy}} \left[\log \left(\frac{\beta_x-1}{\sqrt{x}} \right) \log\left( \frac{\beta_{xy}-1}{\beta_{xy}-\beta_x} \right)
	    -
	    \zeta
	    \log \left(-\frac{\beta_x-1}{\sqrt{x}}\right) \log\left( \frac{\beta_{xy}-\zeta}{\beta_{xy}-\zeta \beta_x} \right)
	    \right] \,.
	\end{multline} 
	The discontinuity of each of the remaining three logarithmic terms can be computed in an analogous way. When adding the contributions from the four logarithmic terms of this form, we get the total discontinuity of $\I_\boxx^{\C,s}$ across the branch cut:
	\be\label{eq:box-Disc-s}
	    \Disc_s \I_\boxx^{\C,s} (s,t) =  \frac{\pi x y}{16 \beta_{xy}}  \left\{
	    \log\left[- ( \beta_{xy}-\beta_x)^2 \right]
	    -
	    \log\left[- ( \beta_{xy}+\beta_x)^2 \right]
	    \right\} \Theta(s-4m^2)  \,.
	\ee 
	The contribution to the discontinuity across the real $s$-axis from $\I_\boxx^{\C,u}$ is obtained by symmetry of the integral in $s \leftrightarrow u$. At fixed $t_\ast$, the discontinuity in $s$ is obtained by subtracting the expression for $s-i \varepsilon$ from the one for $s+i\varepsilon$:
	\be 
	    \Disc_s \I_\boxx^{\C,u}(s,t_\ast) = \frac{1}{2 i} \lim_{\varepsilon \to 0^+} \left[ \I^{\C,u}_\boxx(s+i \varepsilon,t_\ast) - \I^{\C,u}_\boxx(s-i \varepsilon,t_\ast) \right] \,.
	\ee 
	When using the symmetry in $s \leftrightarrow u$, implying that $\I_\boxx^{\C,u}$ is obtained from $\I_\boxx^{\C,s}$ by replacing $s \leftrightarrow u$, and recalling that $u=-s-t_\ast$, we can relate the above expression to the discontinuity we have already computed:
	\be 
	    \Disc_s \I_\boxx^{\C,u} (s,t_\ast) = \frac{1}{2 i} \lim_{\varepsilon \to 0} \left[ \I_\boxx^{\C,s}(-s{-}t_\ast-i \varepsilon,t_\ast) - \I_\boxx^{\C,s}(-s{-}t_\ast+i \varepsilon,t_\ast) \right] \,.
	\ee 
	We see that $\Disc_s \I_\boxx^{\C,u}$ is obtained from $\Disc_s \I_\boxx^{\C,s}$ by adding an extra minus sign, in addition to taking $ s\leftrightarrow u$, accounting for approaching the former term from the lower-half plane in $u$,
	\be\label{eq:box-Disc-u}
	    \Disc_s \I_\boxx^{\C,u} = - \frac{\pi x y}{16 \beta_{xy}}  \left\{
	    \log\left[- ( \beta_{xy}-\beta_y)^2 \right]
	    -
	    \log\left[- ( \beta_{xy}+\beta_y)^2 \right]
	    \right\} \Theta(-s{-}t_\ast{-}4m^2) \,.
	\ee 
	The total discontinuity across the $s$-axis is the sum of the two contributions~\eqref{eq:box-Disc-s} and~\eqref{eq:box-Disc-u},
	\be 
	\Disc_s \I_\boxx (s,t_\ast) = \Disc_s \I_\boxx^{\C,s}\, \Theta(s{-}4m^2) + \Disc_s \I_\boxx^{\C,u}\, \Theta(-s{-}t_\ast{-}4m^2) \,.
	\ee 
	In this expression, the theta functions are repeated to emphasize that the expression naturally splits into a sum of two contributions, one which has a branch cut for $s>4 m^2$ and another which has a branch cut for $s<-t_\ast{-}4 m^2$. These will overlap whenever $t_\ast<-8 m^2$, resulting in a discontinuity across the whole $s$-axis.
	
	The imaginary part of $\I_\boxx^{\C,s} (s+i\varepsilon,t_\ast)$ is computed by taking the imaginary part of the upper-half plane expression. Depending on the signs of $\Re u$ and $\Re (4m^2-s)$, we get different contributions from the various logarithmic and dilogarithmic terms. For example, when $\Re u <0$, which also implies that $\Re(4-s)>0$, all the dilogarithmic terms have real arguments that are less than one, and the prefactor $\frac{-x y}{8 m^2 \beta_{xy}}$ is real, which shows that these terms do not give a contribution. The imaginary part of the product of logarithms can be found by using $\Re \log(x) = \log \vert x \vert$ and $\Im \log(x) = \arg (x)$, and expanding out their products. The result is:
	\be 
	    \Im \, \I_\boxx^{\C,s} = \frac{\pi x y}{16 \beta_{xy}}  \left\{
	    \log\left[- ( \beta_{xy}-\beta_x)^2 \right]
	    -
	    \log\left[- ( \beta_{xy}+\beta_x)^2 \right]
	    \right\} \Theta(s-4m^2) \,,
	\ee 
	so, in particular, $\Im \,\I_\boxx^{\C,s} (s,t_\ast) = \Disc_s \I_\boxx^{\C,s} (s,t_\ast)$. Similarly, the imaginary part of $\I_\boxx^{\C,u} (s,t)$ can, by symmetry in $s \leftrightarrow u$, be obtained from the discontinuity expression. But recall that the physical amplitude is obtained by approaching $\I_\boxx^{\C,u} (s,t_\ast)$ from the lower-half plane in $s$, so
	\be 
	\Im \, \I_\boxx^{\C,u} (s,t_\ast) = - \Disc_s \I_\boxx^{\C,u} (s,t_\ast) \,.
	\ee 
	We therefore get that the imaginary part of the physical amplitude and the discontinuity across the branch cut that runs along the real $s$-axis are related as
	\be\boxed{\label{eq:Im-box}
	    \Im \, \I_\boxx(s,t_\ast) = \Disc_s \I_\boxx^{\C,s} (s,t_\ast) - \Disc_s \I_\boxx^{\C,u} (s,t_\ast) \,,}
	\ee 
	for $-16 m^2 < t_\ast < -8 m^2$.
	
	\subsubsection{Unitarity cuts in the \texorpdfstring{$s$}{s}-channel}
	\label{sec:box-cuts-s}
	
	\begin{figure}
        \centering
        \includegraphics[scale=1.1]{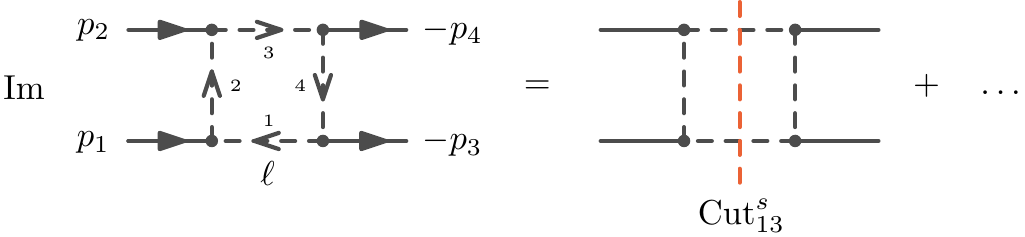}
        \caption{Unitarity of the $su$-box diagram in the $s$-channel. Out of all the unitarity cuts (cf. Fig.~\ref{fig:holomorphic-cuts}), only $\Cut_{13}^s\, \I_{\boxx}$ has support when $M=0$.}
        \label{fig:s-cut-box}
    \end{figure}
	
	In order to verify unitarity, we apply the holomorphic cutting rules from Sec.~\ref{sec:cutting}. Before presenting the calculations, we should note that an alternative way of calculating cuts of Feynman diagrams, which sometimes results in simplifications, is by changing variables from the loop momenta to the cut propagators, using the Baikov representation~\cite{Harley:2017qut,Papadopoulos:2018xgj}. For example at one-loop order, the positive-energy condition is trivially enforced when cutting at least two propagators in a physical kinematic region, and as a result, explicit formulae for the unitarity cuts can be worked out, see, e.g.,~\cite{Abreu:2017ptx,HMSV}. However, since our examples are easily tractable in momentum space, we illustrate the application of the momentum-space cutting rules by using \eqref{eq:cuttingrules} directly. 
	
	Without loss of generality, we can consider the $s$-channel, in which $s>0$ and $t,u<0$, or equivalently $s > -t_\ast$ for fixed $t_\ast$.
	Since the external particles are massless, the only allowed cut is the $s$-channel normal threshold, in which the particles labelled with $1$ and $3$ in Fig.~\ref{fig:s-cut-box} are put on shell.
	Hence unitarity amounts to
	\be\label{eq:box-unitarity}
	\Im\, \I_{\boxx} = \Cut_{13}^s\, \I_{\boxx}.
	\ee

	In the loop-momentum space, the $su$-box diagram is given by
	\begin{multline}
        \I_\boxx = \lim_{\eps \to 0^+} \int \frac{\d^\D \ell}{i \pi^{\D/2}}
        \frac{1}{[\ell^2 - m^2 + i\varepsilon][\left(\ell+p_{1}\right)^2 - m^2 + i\varepsilon]}
        \\ \times 
        \frac{1}{[\left(\ell+p_{12}\right)^2 - m^2 + i\varepsilon][\left(\ell+p_{124}\right)^2 - m^2 + i\varepsilon]}
        ,
        \label{eq:momboxintegral}
    \end{multline}
    where we have written $p_{24} = p_2 + p_4$ etc. The $s$-channel cut is given by
    \be
        \Cut_{13}^s \, \I_\boxx = \frac{(-2 \pi i )^2}{2 i} \int \frac{\d^\D \ell}{i \pi^{\D/2}}
        \frac{\delta^-\! \left[ \ell^2 - m^2 \right]
        \delta^+\!\left[ \left(\ell+p_{12}\right)^2 - m^2 \right]}{[\left(\ell+p_{1}\right)^2 - m^2][\left(\ell+p_{124}\right)^2 - m^2]}
        ,
        \label{eq:cutbox1}
    \ee
    where we have trivially taken the limit $\varepsilon \to 0^+$ since the remaining propagators cannot go on shell on the support of this cut. Note that causal propagation is guaranteed by the theta functions that are included in $\delta^+$ and $\delta^-$. We work in the center-of-mass frame of $p_{12}$:
    \begin{align} 
        p_{12}^\mu & = \left(\sqrt{s},0,\ldots,0\right) \,,
        \label{eq:box_coordinates_p12}
        \\
        p_1^\mu & = \tfrac{\sqrt{s}}{2} \left(1,0,\ldots,0,1 \right) \,,
        \label{eq:box_coordinates_p1}
        \\
        p_3^\mu & =  \tfrac{\sqrt{s}}{2} \left(-1,0,\ldots,\sin \theta ,\cos \theta \right) \,,
        \label{eq:box_coordinates}
    \end{align} 
    where the energy and the magnitude of three-momenta of the external particles are fixed to be $\sqrt{s}/2$ in this frame, by on-shell conditions and momentum conservation. We write out the cut integral from~\eqref{eq:cutbox1} as
    \begin{multline}
        \Cut_{13}^s \, \I_\boxx = \frac{2}{\pi^{\D/2-2}} \int_{-\infty}^{\infty} \!\d \ell^0\!\! \int_0^\infty |\vec{\ell}|^{\D-2} \d |\vec{\ell}| \int \d \Omega_{\D-1} \frac{1}{[\left(\ell+p_{1}\right)^2 - m^2][\left(\ell+p_{124}\right)^2 - m^2]}
        \\
        \times 
        \delta [(\ell^0)^2-|\vec{\ell}|^2-m^2 ]\, \Theta(-\ell^0)
        \,\delta \left[2 \ell^0 \sqrt{s} + s \right] \Theta(\ell^0+\sqrt{s})
        ,
    \end{multline}
    where use spherical coordinates for the loop momentum $\ell$, and $\Omega_{\D-1}$ is the  $\D{-}1$-dimensional solid angle.
    We perform the integrals in $\ell^0$ and $\vert \vec{\ell} \vert$ using the delta functions, which impose that
    \begin{equation}
    \label{eq:delta_fn_ell}
        \ell^0 = \frac{\sqrt{s}}{2} \,, \qquad \qquad
        \vert \vec{\ell} \vert = \frac{1}{2}\sqrt{s-4 m^2} \,,
    \end{equation}
    with the result
    \be
        \Cut_{13}^s \, \I_\boxx =
        \frac{\Theta(s-4m^2) (s-4 m^2)^{\frac{\D-3}{2}}}{2^{\D-2} \pi^{\D/2-2} \sqrt{s} }  
        \int \d \Omega_{\D-1} 
        \frac{1}{2 \, \ell \cdot p_1}
        \frac{1}{2 \, \ell \cdot p_{124}}
        \,.
    \ee
    Now, we combine the two denominators using an auxiliary Schwinger parameter $\alpha$, and let $\varphi$ be the angle between $\vec{\ell}$ and $\vec{p}_1+\alpha \, \vec{p}_{124}$. Then, we can write the cut as
    \begin{multline} 
        \Cut_{13}^s \, \I_\boxx  =
        \frac{\Theta(s-4m^2)(s-4 m^2)^{\frac{\D-3}{2}}}{2^\D \pi^{\D/2-2} \sqrt{s} }  
        \int_0^\infty \d\alpha
        \int_{-1}^1 \left(\sin \varphi\right)^{\D-4} \d \cos \varphi  
        \int \d \Omega_{\D-2} 
        \\ \times
        \frac{1}{\left[\ell^0 (p_1^0+\alpha \, p_{124}^0) - \vert \vec{\ell} \vert \, \vert \vec{p}_1 + \alpha \, \vec{p}_{124} \vert \cos \varphi \right]^2} \,.
    \end{multline}
    The angular integrals are easily evaluated; in $\D=4$ spacetime dimensions the result is
    \be
        \Cut_{13}^s \, \I_\boxx =
        \frac{\Theta(s-4m^2) 2 \pi \sqrt{s-4 m^2}}{s^{3/2} }
        \int_0^\infty 
        \frac{\d\alpha}{2 m^2 \alpha^2 - \alpha[s (1-\cos \theta)+4 \cos \theta \, m^2] +2 m^2} \,,
    \ee  
    where have written out the values of $\ell^0$, $\vert \vec{\ell} \vert$, $p_1^\mu$ and $p_{124}^\mu$ from~\eqref{eq:box_coordinates_p1},~\eqref{eq:box_coordinates} and~\eqref{eq:delta_fn_ell}, in addition to using that the solid angle in $\D{-}2$ dimensions is given by $\Omega_{\D-2} = 2 \pi^{\frac{\D-2}{2}} /\Gamma \left(\frac{\D-2}{2}\right)$. Performing the last integral in $\alpha$, and using that $u=(p_2+p_4)^2$ implies that $\cos \theta = 2 u/s+1$ gives
    \be 
	    \Cut_{13}^s \, \I_\boxx = \frac{\pi x y}{16 \beta_{xy}}  \left\{
	    \log\left[- ( \beta_{xy}-\beta_x)^2 \right]
	    -
	    \log\left[- ( \beta_{xy}+\beta_x)^2 \right]
	    \right\} \Theta(s-4m^2) \,,
	\ee  
    where we have used the variables $\beta_{xy} = - i \sqrt{-1+\frac{4m^2}{s}+\frac{4m^2}{u}}$ and $\beta_x=\sqrt{1-\frac{4m^2}{s}}$ as before.
    Together with~\eqref{eq:Im-box}, this verifies unitarity in the form of the holomorphic cut equation~\eqref{eq:box-unitarity}. The analogous check in the $u$-channel can be made by simply relabelling $s \leftrightarrow u$. 
    
    \subsubsection{Discussion}
	
	In this subsection, we have shown how normal-threshold branch cuts can overlap on the real $s$ axis at fixed $t=t_\ast$, resulting in a branch-cut along the whole real $s$-axis. As a concrete example where this overlap happens, we have studied the $su$-planar box diagram in the complex $s$-plane, for $2 \to 2$ scattering of massless particles, with internal particles of mass $m$, for which $t$ is fixed to $t_\ast<- 8 m^2$. For simplicity, we also chose $t_\ast$ to be large enough to prevent the box threshold (leading singularity) at $s u + 4m^2 t = 0$ from being on the physical sheet in the complex $s$-plane. Combining the two requirements amounts to taking $-16 m^2 < t_\ast < -8 m^2$.
	
	With an explicit example of overlapping normal-threshold branch cuts, we can ask whether it is still possible to analytically continue between the functions obtained in the upper- and lower-half planes. To show that this is indeed possible in this case, we split the complexified amplitude into two terms. Note that since $\partial_s \V$ has opposite signs for each of the branch cuts starting at $s=4 m^2$ and $s=-t_\ast - 4 m^2$,~\eqref{eq:I-UHP-LHP} guarantees that the box amplitude splits into a sum of two terms $\I^\pm (s,t_\ast)$, which have \textit{only} a single normal-threshold branch cut each. For practical purposes, however, we find a simpler expression when allowing for a sum of terms for which each has a spurious branch point at the location of the box Landau singularity, i.e., at $s u + 4m^2 t = 0$. Thus, we split the physical amplitude into a sum of terms of the form $\I_\boxx (s,t_\ast) = \lim_{\varepsilon\to 0^+} \!\big[ \I_\boxx^{\C,s} (s+i\varepsilon,t) + \I_\boxx^{\C,u} (s-i\varepsilon,t) \big]$, where $\I_\boxx^{\C,s}$ ($\I_\boxx^{\C,u}$) does not have a branch cut starting at $u=4 m^2$ ($s=4 m^2$), although both terms have branch cuts in the complex plane that do not intersect the real $s$ axis. When adding the two terms to get  $\I_\boxx (s,t_\ast)$, the spurious box branch cuts cancel out. This cancellation is expected, since the solutions to the Landau equations show that the box branch point is not on the principal sheet for $-16 m^2 < t_\ast < -8 m^2$. Since the spurious branch cuts do not cross the real $s$-axis, we can still use this split form to show that the upper- and lower-half expressions are given by the same complex function: Since there is no obstruction to analytically continuing each term $\I_\boxx^{\C,s}(s + i \varepsilon,t_\ast)$ and $\I_\boxx^{\C,u}(s - i \varepsilon,t_\ast)$ between the upper- and lower-half $s$ planes to get the term corresponding to the opposite $s \pm i \eps$ prescription, this form reveals an analytic-continuation path between the two branch cuts, consistent with Fig.~\ref{fig:box10}. 
	
	As expected from the analysis in Sec.~\ref{sec:im-disc}, we have verified explicitly using the analytic expression for $\I_\boxx$ that $\Im \, \I_\boxx = \Im \, \I_\boxx^{\C,s} + \Im \, \I_\boxx^{\C,u}$, and $\Disc_s \I_\boxx = \Disc_s \, \I_\boxx^{\C,s} - \Disc_s \, \I_\boxx^{\C,u}$. We computed the imaginary part using two different methods: from the analytic expression, and, in physical regions, using unitarity cuts. We can therefore ask whether we could have obtained the imaginary part from unitarity cuts anywhere on the real $s$-axis, without doing the computation of the full analytic form. Looking back at Fig.~\ref{fig:box-st}, we can easily argue that this is indeed possible when $t_\ast>-8 m^2$: we can analytically continue the value of the unitarity cuts without encountering other branch points, from the physical regions to the branch points at $s=4 m^2$ and $u=4 m^2$, and the imaginary part of $\I_\boxx$ vanishes between the two normal thresholds. Moreover, despite the obstruction of overlapping branch cuts whenever $-16 m^2 < t_\ast<-8 m^2$, we can still argue that the analytic continuation of cuts captures the imaginary part along the whole real $s$-axis, using the split of the amplitude introduced in Sec.~\ref{sec:im-disc}: since $\I_\boxx$ splits into two terms, each of which has a branch cut for either $s>4 m^2$ or $u>4 m^2$, we can analytically continue the values of the unitarity cuts into the region in which the branch cuts overlap. Thus, we have argued that whenever $t_\ast>-16 m^2$, the full analytic form is not needed for computing the imaginary part of $\I_\boxx$.
    
    \begin{figure}
	    \centering
	    \includegraphics[scale=1.1]{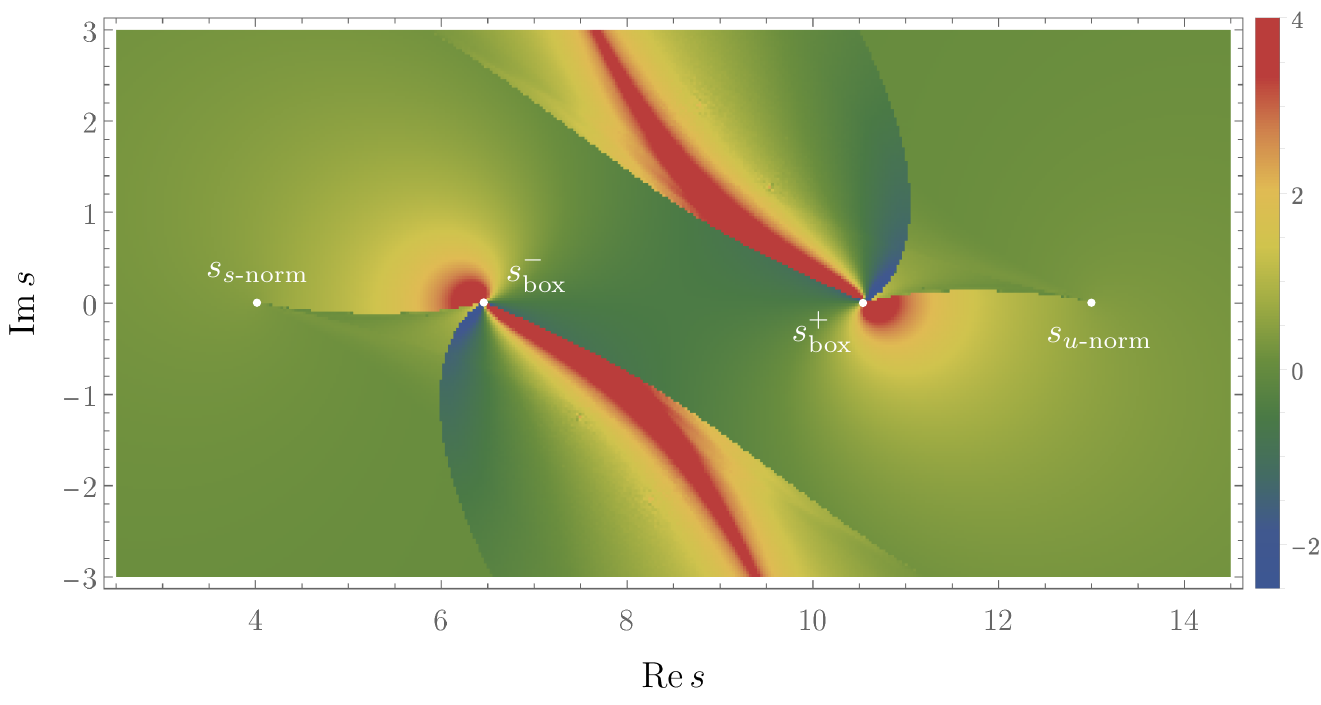}
	    \caption{Numerical plot of $\Im\, \I_{\boxx}(s,t_\ast)$ in the complex $s$-plane with $t_\ast = -17$, $m=1$, and $\eps = \tfrac{1}{10}$. The $i\eps$-deformed branch cuts extend from the normal thresholds in the $s$- and $u$-channels, $s_{s\text{-norm}} = 4$ and $s_{u\text{-norm}} = 13$, but also the box anomalous thresholds at $s_{\mathrm{box}}^\pm = (17 \pm \sqrt{17})/2$.}
	    \label{fig:box17}
	\end{figure}
	
    Finally, we can ask how the qualitative picture changes as we allow for values of $t_\ast$ for which $t_\ast<-16 m^2$, i.e., when the box branch point is on the principal sheet of $\I_\boxx$. A numerical plot is shown in Fig.~\ref{fig:box17}, where, in addition to the branch cuts extending from the normal thresholds, there are branch cuts extending from the box thresholds. We will not give a detailed analysis of this case.
	
	\subsection{\label{sec:ExampleII}Example II: Disconnecting the upper- and lower-half planes}
	
	Let us move on to another class of issues with the analytic S-matrix: branch cuts extending along the whole real $s$-axis. Before diving into details, let us explain more intuitively why such issues arise in the first place.
	
	\subsubsection{\label{sec:external-mass}External-mass singularities}

\begin{figure}
	    \centering
	    \includegraphics[scale=1.1]{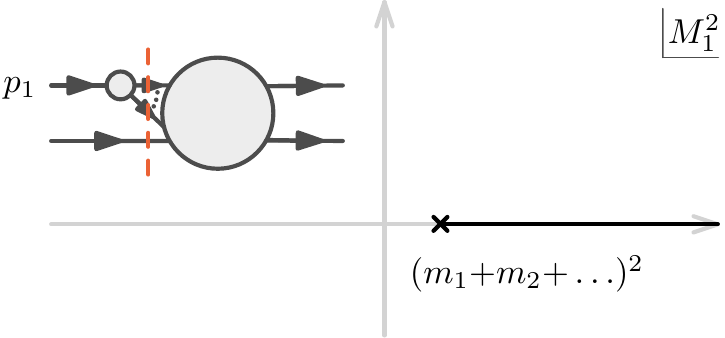}
	    \caption{External mass singularity occurs when an external particle with momentum $p_1$ can decay into a number of lighter particles (inset). For $p_1^2 = M_1^2$, the branch cut in the $M_1^2$-plane translates to a branch cut separating the upper- and lower-half $s$-planes.}
	    \label{fig:external-mass}
	\end{figure}

	Consider any Feynman diagram of the type given in Fig.~\ref{fig:external-mass}, where a number of edges with masses $m_1, m_2, \ldots$ are attached to an external leg with mass $M_1$ and
	\be\label{eq:M1}
	M_1 > m_1 + m_2 + \ldots.
	\ee
	The external mass $M_1$ is fixed and cannot be deformed for an on-shell observable, but let us briefly entertain the idea that it could. The reason to consider such a thought experiment is that there are simple statements we can make in the $M_1^2$ complex plane, see Fig.~\ref{fig:external-mass}. Apart from multiple anomalous thresholds that can be present, we know for certain that there will be a branch cut giving a discontinuity between $M_1^2 + i\eps$ from $M_1^2 - i\eps$ whenever~\eqref{eq:M1} is satisfied. But by momentum conservation
	$s+t+u = \sum_{i=1}^{4} M_i^2$,
	so deforming $M_1^2 \pm i\eps$ is indistinguishable from deforming $s \mp i\eps$ on shell. Therefore, we conclude that there has to be a discontinuity along the whole real $s$-axis, separating the upper- from lower-half planes. Its value is given by the cuts in Fig.~\ref{fig:external-mass}, which becomes larger and larger with higher width of the decaying particle. Study of such branch cuts dates back to \cite{doi:10.1063/1.1703936,10.1143/PTP.51.912}, where they were referred to as \emph{external-mass singularities}.
	
	We know that the $M_1^2 + i\eps$ deformation is the one consistent with the Feynman $i\eps$, since it always leads to $\Im \V > 0$, cf.~\ref{sec:parametric}. Alternatively, we see that it corresponds to approaching the branch cut in Fig.~\ref{fig:external-mass} from the upper-half plane. In the complex $s$-plane, this translates to approach the cut from the \emph{lower-half} plane.
	
	The above analysis does not yet say whether the cuts along the real axis can be deformed to connect the upper- and lower-half planes or not. For instance, in Sec.~\ref{sec:ExampleI} we have seen an example in which the two cuts could be deformed away to restore the connection. The rest of this subsection is dedicated to studying the simplest example in which this cannot be done. As a consequence, there are two separate analytic functions, which are \emph{not} continuations of each other on shell.

	\subsubsection{Triangle diagram}
	
\begin{figure}
	    \centering
	    \includegraphics[scale=1.1]{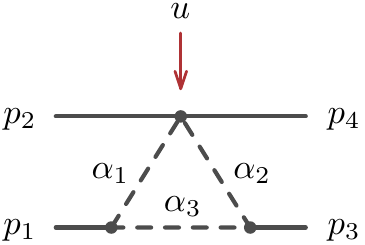}
	    \caption{The $u$-triangle diagram considered in Sec.~\ref{sec:ExampleII}. All the internal and external masses are set to $m$ and $M$ respectively. Throughout this section we take $M > 2m$.}
	    \label{fig:triangle-diagram}
	\end{figure}

	As the simplest concrete example, we consider the diagram from Fig.~\ref{fig:triangle-diagram} with all the internal and external masses equal to $m$ and $M$ respectively. To exhibit the analyticity features mentioned above, we set $M > 2m > 0$. The corresponding Schwinger-parametrized Feynman integral reads
	\be\label{eq:I-tri}
	\I_{\triangle}(s,t) = \Gamma(3-\D/2) \lim_{\eps \to 0^+}\int \frac{\d^3 \alpha}{\GL(1)} \frac{1}{\U_\triangle^{\D/2} (-\V_\triangle - i\eps)^{3-\D/2}},
	\ee
    where
    \be
    \U_\triangle = \alpha_1 + \alpha_2 + \alpha_3,\quad \V_\triangle = \frac{u \alpha_1 \alpha_2 + M^2 \alpha_3 (\alpha_1 + \alpha_2)}{\alpha_1 + \alpha_2 + \alpha_3} - m^2 (\alpha_1 + \alpha_2 + \alpha_3).
    \ee
    Recall that we treat $u = 4M^2 - s - t$ as fixed and study properties of $\I_\triangle(s,t)$ as a function of $s$ and $t$. The physical region in the $s$-channel is the one for which $s > 4M^2 - t$ and $t<0$, while in the $u$-channel we have $s < 0$ and $t<0$. For concreteness, let us work in $\D=4$, where the diagram is finite. All the analytic features of this Feynman integral can be understood without explicit computations.
    
    First of all, branch cuts exist when $\V_\triangle = 0$ for some positive values of Schwinger parameters. In this example, where we have chosen $M>2m>0$, such positive values always exist, regardless of the value of the kinematic invariants $s$ and $t$:
    Close to $\alpha_1 = \alpha_2 = 0$, the value of $\V_\triangle = -m^2 \alpha_3$ is always negative, while near $\alpha_1 = 0$ with $\alpha_2 = \alpha_3$ we have $\V_\triangle = \tfrac{1}{2}(M^2 - 4m^2) \alpha_3$, which is always positive. By continuity, there always exists a point in the Schwinger-parameter space for which $\V_{\triangle} = 0$, indicating a branch cut for any $s$ and $t$.
    
    Next, we can ask about which way of approaching the branch cuts is physical. Recall that the physical direction can be found as the one for which $\Im \V_\triangle > 0$, giving the constraint
    \be\label{eq:ImV-tri}
    \Im \V_\triangle = - \Im\, s \underbrace{\frac{\alpha_1 \alpha_2}{\alpha_1 + \alpha_2 + \alpha_3}}_{>0} > 0,
    \ee
    where $\Im\, u = - \Im\, s$ by momentum conservation, when we fix $t$ to be real. Note that~\eqref{eq:ImV-tri} does not depend on the physical region we work in, and in fact holds even away from the physical region. Hence, causality always requires that we approach the branch cuts along the real axis from the \emph{lower-half} $s$-plane. As a matter of fact, the above inequality guarantees that $\I_{\triangle}$ is analytic for any
    \be
    \Im\, s < 0,
    \ee
    i.e., throughout the whole lower-half plane.
    
    \begin{figure}
	    \centering
	    \includegraphics[scale=1.1]{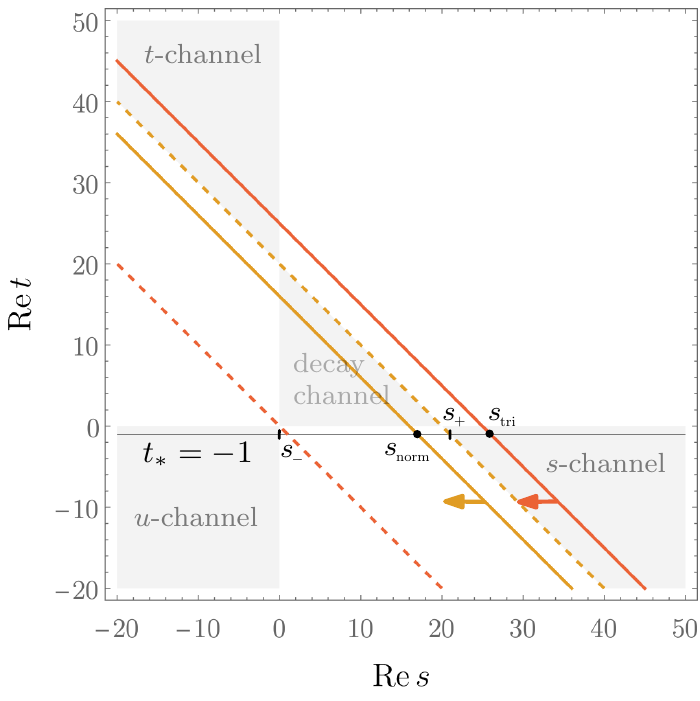}
	    \caption{Landau curves for the $u$-triangle diagram with external masses $M= \sqrt{5}$ and internal ones $m=1$. The leading singularity, triangle anomalous threshold, is located at $s = s_{\triangle}$ (red), while the $u$-channel normal threshold is at $s = s_{\normal}$ (orange). Second-type singularities at $s = -t$ and $s = 4M^2 - t$ on unphysical sheets are indicated with dashed lines. Physical regions are shaded. The complex $s$-plane for $t_\ast = -1$ will be illustrated in Fig.~\ref{fig:triangle}. The arrows indicate the direction of the branch cuts extending from their associated branch points. On top of them, the triangle diagram also has the external-mass singularities that contribute branch cuts for any value of $s$ and $t$.}
	    \label{fig:st-triangle}
	\end{figure}
    
    Furthermore, branch points can be obtained by extremizing the action $\V_{\triangle}$. Recall from Sec.~\ref{sec:general} that for this diagram we find two branch points on the physical sheet, see Fig.~\ref{fig:st-triangle}. Translated to the $s$ variable, they are at the normal threshold
    \be
    s_{\normal} = 4(M^2 - m^2) - t,\qquad (\alpha_1^\ast : \alpha_2^\ast : \alpha_3^\ast) = ( \tfrac{1}{m} : \tfrac{1}{m} : 0),
    \ee
    as well as the triangle threshold
    \be
    s_{\triangle} = \frac{M^4}{m^2} - t,\qquad (\alpha_1^\ast : \alpha_2^\ast : \alpha_3^\ast) = (m^2 : m^2 : M^2 {-} 2m^2).
    \ee
    Relative to the boundaries of the physical channels, we have
    \be
    s_- < s_{\normal} < s_+ < s_{\triangle}.
    \ee
    The normal threshold is always in the $u$-channel. The triangle threshold is only an $\alpha$-positive singularity if $M > \sqrt{2}m$. However, it only intersects the $s$- and $t$-channels when $M > 2m$, which is precisely the situation we are studying. This diagram
    also has the pseudo-normal threshold located at $u=0$, which is on unphysical sheets and hence need not concern us.
    
    \begin{figure}
        \centering
        \includegraphics[scale=1.05]{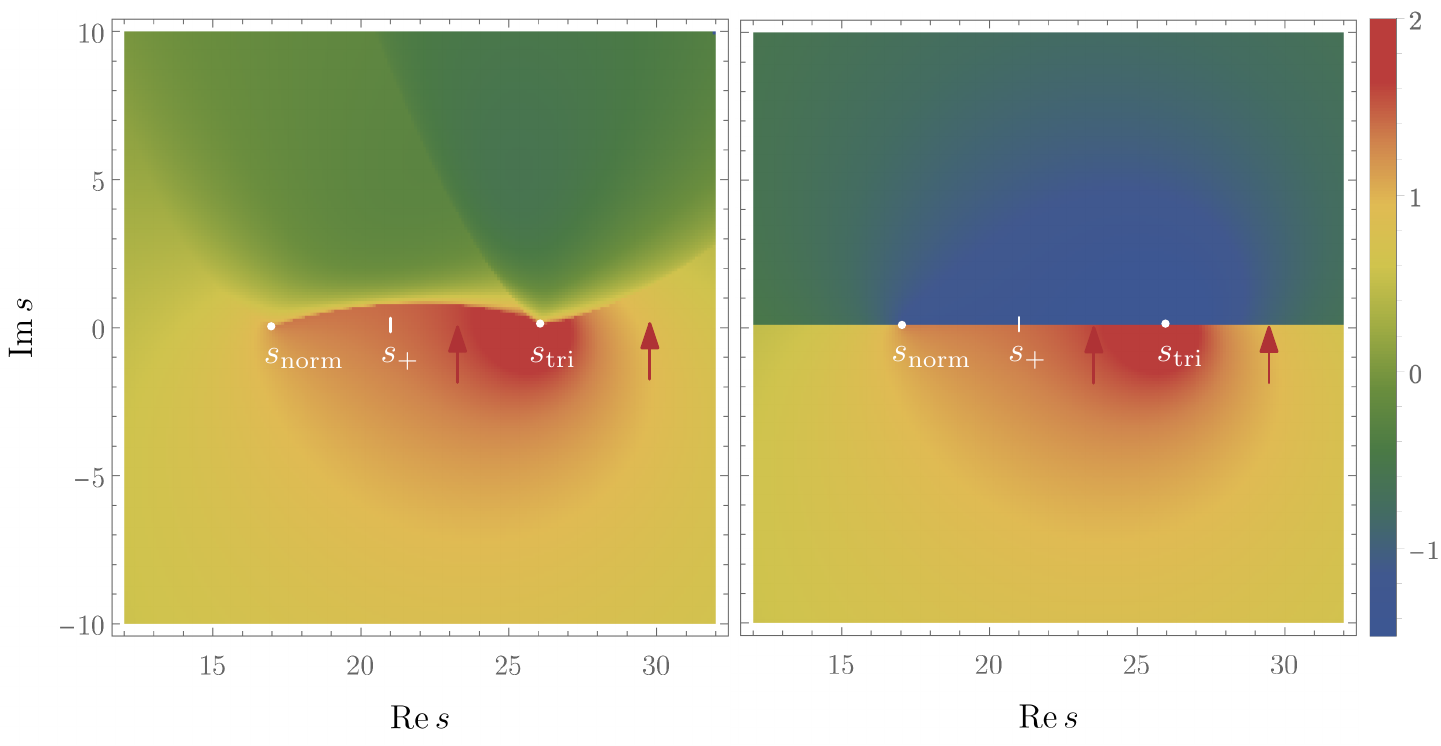}
        \caption{Left: Numerical plot of $\Im\, \I_{\triangle}(s,t_\ast)$ obtained using branch cut deformations in the complex $s$-plane with $t_\ast = -1$. We set $m=1$, $M=\sqrt{5}$, and $\eps = \tfrac{1}{10}$. In terms of these variables, the $s$-channel is in $s>s_+ = 21$ and the $u$-channel at $s< s_- = 0$ (not visible in the plot), while branch cuts extend from $s_{\normal}=17$ and $s_{\triangle}=26$. Right: Analytic plot with the same parameters. The lower- and upper-half planes are separated by branch cuts. The former agrees with the numerical evaluation, confirming that causal way of approaching the cuts is from the lower-half $s$-plane in both channels.}
        \label{fig:triangle}
    \end{figure}
    
    Finally, we note that one can easily verify the above assertions numerically using the branch cut deformation prescription from Sec.~\ref{sec:deformation}. This amounts to setting $\eps=0$ in~\eqref{eq:I-tri}, making the change of variables $\hat{\alpha}_e = \alpha_e \exp \left(i\eps \partial_{\alpha_e}\!\V_\triangle \right)$, followed by fixing $\hat{\alpha}_3 = 1 - \hat{\alpha}_1 - \hat{\alpha}_2$. The result of this integration, as a function of $s$, is shown in Fig.~\ref{fig:triangle}. We can recognize branch cuts extending from the normal and triangle thresholds and bending upwards in the upper-half plane. While the correct values for $\I_\triangle$ in the lower-half plane can be obtained by a finite but small $\eps$ (see Sec.~\ref{sec:deformations}), adjusting the value of $\eps$ to be smaller and smaller makes the branch cuts move closer to the real axis. The function does not have any non-analyticities in the lower-half plane.
    
    \subsubsection{Analytic expression}
    
    In order to get yet another perspective and an analytic handle on these problems, we can simply integrate~\eqref{eq:I-tri} directly. As in Sec.~\ref{sec:ExampleI}, we warn the reader that while this one-loop integral has been computed, see, e.g., \cite{tHooft:1978jhc,Patel:2015tea}, expressions available in the literature often place branch cuts arbitrarily away from physical regions. For our purposes of analyzing the complexified amplitude, it is therefore instructive to perform this computation from scratch.

    The Schwinger-parametrized form of $\mc{I}_\triangle$ is obtained from~\eqref{eq:I-GL} by taking $\tilde{\mathcal{N}}=1$, $\E=3$ and $\L=1$
    \begin{multline}
        \I_\triangle = \Gamma\left(3-\D/2\right) \lim_{\eps \to 0^+} \int_0^1 \d \alpha_1 \int_0^{1-\alpha_1} \d\alpha_2 \\ 
        \times \frac{1}{\left[ - \alpha_1 \alpha_2 (u-2 M^2) -\alpha_1 (1-\alpha_1) M^2 - \alpha_2 \left(1-\alpha_2\right) M^2 + m^2-i \varepsilon \right]^{3-\D/2}} \,,
        \label{eq:triintegral}
    \end{multline}
    where we have taken $u=(p_1+p_3)^2$, and fixed $p_2^2=p_3^2=M^2$. 
    To integrate this expression directly, we change variables from $(\alpha_1,\alpha_2)$ to $(\beta_1,\beta_2)$, such that the denominator is linear in each integration variable.
    We achieve this by setting
    \begin{align}
        \beta_1 & = a \, \alpha_1 + b \, \alpha_2 + c, \\
        \beta_2 & = b \, \alpha_1 + a \, \alpha_2 + c,
    \end{align}
    with the following values for $a$, $b$ and $c$,
    \begin{align}
        a & = \frac{\sqrt{-u}+\sqrt{-u+4 M^2}}{2} \,, \\
        b & = \frac{- u + 2 M^2 - \sqrt{-u \left(-u+4M^2\right)}}{2 M^2} a, \\
        c & = \frac{u - 4 M^2 + \sqrt{-u \left(-u+4M^2\right)}}{8 M^2-2 u} a,
    \end{align}
    where we have written the expressions such that the real parts of the square roots are manifestly positive for $\Re u <0$. When integrating the above expression for other values of $u$, we must be careful to choose the right branches of the square roots, that are consistent with the original $i \eps$ prescription. We will go through the computation of the analytic form of $\I_\triangle$ in detail for $s>s_\triangle$, assuming a slight positive or negative imaginary part of $s$, but the result for other values of $s$ can be achieved either by similar methods, or by analytic continuations in the upper half planes or lower half planes separately. With this change of variables, the integral becomes in $\D=4$ dimensions,
    \begin{multline}
        \I_\triangle \vert_{s>s_\triangle} = \frac{1}{\sqrt{-u \left(4 M^2{-}u\right)}}
        \Big[
        \big(
        \int_{c}^{b+c} \!\!\d \beta_1 \int_{\frac{a c- b c + b \beta_1}{a}}^{\frac{-ac + bc + a \beta_1}{b}}\! \d \beta_2
        +
        \int_{b+c}^{a+c} 
    \!\!\d \beta_1 \int_{\frac{a c- b c + b \beta_1}{a}}^{a+b+2c- \beta_1} \!\d \beta_2
        \big)
        \\
        \times
        \frac{1}{\beta_1  \beta_2 + m^2 - \frac{M^4}{4 M^2-u}- i \varepsilon} \Big]
        \,.
    \end{multline}

    Evaluating this integral in the region $s>s_\triangle$, noting that $ab + bc + ca = 0$, results in
    \begin{align}
        &\mathcal{I}_\triangle \vert_{s>s_\triangle} = \frac{1}{\sqrt{-u (4 M^2-u)}} \sum_{\pm} \bigg[ 
        \Li_2 \left( \frac{2(b+c)}{a+b+2c\pm \sqrt{\phi_1}} \right)
        \\
        &-
         \Li_2 \left( \frac{2(a+c)}{a+b+2c\pm \sqrt{\phi_1}} \right)
        + 2 \, \Li_2 \left( \frac{2 a c}{ac - bc \pm \sqrt{\phi_2}} \right)
        -
        2 \, \Li_2 \left( -\frac{2 b c}{ac - bc \pm \sqrt{\phi_2}} \right)
        \bigg] \,,\nn
    \end{align}
    where the arguments of the square roots are
    \begin{align}
        \phi_1 & = (a-b)^2+4c^2 + 4 m^2 - \frac{4 M^4}{4 M^2-u}- i \varepsilon \,, \\
        \phi_2 & = a^2 c^2 + b^2 c^2 - 2 a b \left(c^2 + 2 m^2 - \frac{2 M^4}{4 M^2-u} - i \varepsilon \right) \,.
    \end{align}
    Combining terms, and plugging in the expressions for $a$, $b$ and $c$, gives the following result,
    \begin{multline}
        \mathcal{I}_\triangle \vert_{s>s_\triangle} = \frac{z}{4 M^2 \beta_z} \Big\{ \sum_{\pm} \Big[  \Li_2 \left( \frac{1 +\frac{z}{2} -\beta_z}{1 + \frac{z}{2} \pm \beta_z \beta_y } \right) - \Li_2 \left( \frac{1 + \frac{z}{2}+\beta_z}{1 + \frac{z}{2} \pm \beta_z \beta_y} \right)
        \\
        +2 \, \Li_2 \left( \frac{1 + \beta_z }{1 \pm \beta_z \tilde \beta_{yz}} \right) - 2 \, \Li_2 \left( \frac{1 - \beta_z}{1 \pm \beta_z \tilde \beta_{yz} } \right) \Big] 
        - 4 \pi i \log \left( \frac{1+\beta_z}{1+\beta_z \tilde \beta_{yz}} \right)\Big\}\,,
    \end{multline}
    with
    \begin{gather}
        y  = - \frac{4 m^2}{u} \,, \qquad z = - \frac{4 M^2}{u} \,, \\
        \beta_y  = \sqrt{1+y}\,, \qquad \beta_z = \sqrt{1+z}\,, \qquad \tilde \beta_{yz}  = \sqrt{1-\frac{4y}{z}} \,.
        \label{eq:wpmA}
    \end{gather}
    In order to find which side of the branch cuts the dilogarithms and logarithm should be evaluated on to be consistent with the $i \eps$ prescription from~\eqref{eq:triintegral}, $s$ is taken to have a small negative imaginary part.
    We can now evaluate analogous expressions in other regions, which results in the following analytic expression for $\mathcal{I}_\triangle$ which gives the right form in all of the lower-half $s$-plane,
	\be
    \begin{aligned}
    \mathcal{I}_\triangle^{\LHP} (s,t) & = \frac{z}{4 M^2 \beta_z} 
        \sum_{\zeta \in \{-1,1\}} \Bigg\{
        \zeta \, \Li_2 \left( \frac{1+\frac{z}{2} - \zeta \beta_z}{1+ \frac{z}{2} + \zeta \beta_y \beta_z } \right)
        +
        \zeta \,
        \Li_2 \left(1- \frac{1+\frac{z}{2} + \zeta \beta_z}{1+ \frac{z}{2} + \zeta \beta_y \beta_z } \right) 
        \\ 
        & \hspace{0.5cm}
        +2 \, \Li_2 \left( \frac{1 + \beta_z}{1 + \zeta \beta_z \beta_{yz}} \right)  - 2 \, \Li_2 \left( \frac{1-\beta_z}{1 + \zeta \beta_z \beta_{yz}} \right) 
        -
        2 \pi i \log \left( \frac{1+\beta_z}{1+\beta_z \beta_{yz}} \right)
        \\ 
        & \hspace{0.5cm}
        + \zeta
        \log \left( \frac{1+\frac{z}{2}+\zeta \beta_z}{1+\frac{z}{2}+\zeta \beta_y \beta_z} \right) \left[\pi i + \log \left( -1 + \frac{1+ \frac{z}{2}+ \zeta \beta_z}{1+\frac{z}{2}+ \zeta \beta_y \beta_z} \right) \right]
        \Bigg\} \,,
    \label{eq:triLHP}
    \end{aligned}
    \ee
    where, in addition to the variables in~\eqref{eq:wpmA}, we have defined
    \begin{equation}
        \beta_{yz}  = - i \sqrt{-1+\frac{4y}{z}} \,.
        \label{eq:wpm}
    \end{equation}
    Note that in these expressions, all dilogarithms, logarithms, and square roots are on their principal sheet. Furthermore, in order to write a compact expression that is valid in the whole upper half plane, we have rotated the branch cut of two of the logarithms appearing in $\mathcal{I}_\triangle^{\LHP}$ by $\pi$, while leaving the lower-half plane values intact. As a consequence, the branch cuts of $\mathcal{I}_\triangle^{\LHP}$ are not the ones dictated by $\V_\triangle=0$ in the original integral in~\eqref{eq:triintegral}. The analytic continuation of this expression to the upper-half plane is therefore not the same as $\I_{\triangle}^{\UHP}$, which can be calculated with an analogous method as $\I_{\triangle}^{\LHP}$. The result for the analytic upper-half plane expression is
    \be
    \begin{aligned}
    \mathcal{I}_\triangle^{\UHP} (s,t) & = \frac{z}{4 M^2 \beta_z} 
        \sum_{\zeta \in \{-1,1\}} \Bigg\{
        \zeta \, \Li_2 \left( \frac{1+\frac{z}{2} - \zeta \beta_z}{1+ \frac{z}{2} + \zeta \beta_y \beta_z } \right)
        +
        \zeta \,
        \Li_2 \left(1- \frac{1+\frac{z}{2} + \zeta \beta_z}{1+ \frac{z}{2} + \zeta \beta_y \beta_z } \right) 
        \\ 
        & \hspace{0.5cm}
        +2 \, \Li_2 \left( \frac{1 + \beta_z}{1 + \zeta \beta_z \beta_{yz}} \right)  - 2 \, \Li_2 \left( \frac{1-\beta_z}{1 + \zeta \beta_z \beta_{yz}} \right) 
        +
        2 \pi i \log \left( \frac{1+\beta_z}{1+\beta_z \beta_{yz}} \right)
        \\ 
        & \hspace{0.5cm}
        + \zeta
        \log \left( \frac{1+\frac{z}{2}+\zeta \beta_z}{1+\frac{z}{2}+\zeta \beta_y \beta_z} \right) \left[-\pi i + \log \left( -1 + \frac{1+ \frac{z}{2}+ \zeta \beta_z}{1+\frac{z}{2}+ \zeta \beta_y \beta_z} \right) \right]
        \Bigg\} \, ,
    \label{eq:triUHP}
    \end{aligned}
    \ee
    where the arguments of the (di)logarithms feature the combinations in~\eqref{eq:wpmA} and~\eqref{eq:wpm}.
	These square-root arguments reveal the square-root branch point structure of $\I_\triangle$: the branch point at $\beta_y=0$ corresponds to the normal threshold at $s_{\normal} = 4(M^2 - m^2) - t$, while the variables $z$ and $\beta_{yz}$ would go to zero if $M^2=4 m^2$, so they do not correspond to branch points in $s$. The pseudo-normal threshold at $u=0$ corresponds to $\beta_x$ and $\beta_y$ going to infinity. Solving for potential logarithmic branch points, which could occur when the arguments of the dilogarithms go to 0, 1, or infinity, or the arguments of the logarithms go to 0 or infinity, reveals the triangle branch point at $s_{\triangle} = \frac{M^4}{m^2} - t$, in addition to the threshold branch points.
	
	Using these definitions, the physical amplitude along the real axis can be written as the boundary value of~\eqref{eq:triLHP}:
	\be\label{eq:tri-limit}
	\boxed{
	\I_{\triangle}(s,t) = \lim_{\eps \to 0^+} \I^{\LHP}_{\triangle} (s-i\eps, t).}
	\ee
	Let us stress that the above expressions~\eqref{eq:triLHP} and~\eqref{eq:triUHP} are valid in the lower- and upper-half planes respectively, but the directions of the cuts on along the real $s$-axis are arbitrary.
	
    The above expressions are distinct analytic functions, e.g., the last term in the second line of~\eqref{eq:triLHP} differs from that in~\eqref{eq:triUHP}. This means they cannot be analytically continued into each other in the $(s,t)$ variables.\footnote{Following the arguments in Sec.~\ref{sec:external-mass}, complexifying $M^2$ allows to connect~\eqref{eq:triLHP} and~\eqref{eq:triUHP} through the $M^2$-plane as a mathematical function, even though it corresponds to an off-shell continuation of the S-matrix. To be precise, if we allow to deform an external mass, say $M_j^2$, we can simply take
    \be
    (s,t) = \left(s_0 + x(s_1 {-} s_0),\; t_0 + x(t_1 {-} t_0)\right), \qquad M^2_j = M^2_j + i x (1-x).
    \ee
    The path $x \in [0,1]$ connects two arbitrary real points $(s_0,t_0)$ and $(s_1,t_1)$, while avoiding singularities and preserving the causality condition $\Im \V > 0$. Analogous analytic continuation can be applied to an arbitrary Feynman integral.
    } Let us now proceed with making further checks on $\I_{\triangle}$.
    
    \subsubsection{Discontinuities and imaginary parts}
    
    The discussion in Sec.~\ref{sec:im-disc} gives us an easy way of relating the imaginary part of $\I_\triangle$ to its discontinuity.
    In the notation of~\eqref{eq:Im-Disc}, the function $\I_+$ vanishes, because
    \be
    \partial_s \V = \V_s - \V_u = - \frac{\alpha_1\alpha_2}{\alpha_1+\alpha_2+\alpha_3} < 0
    \ee
    and hence $\Theta(\V_s - \V_u)$ does not have support. This immediately tells us that
    \be\label{eq:Im-minus-Disc}
    \Im\, \I_{\triangle}(s,t) = - \Disc_s \I_{\triangle}(s,t)
    \ee
    regardless of the physical region.
    
    Let us verify this mismatch explicitly using the analytic expressions. On the right-hand side we need 
    \begin{equation}
        \Disc_s \mathcal{I}_\triangle (s,t) = \frac{1}{2 i} \lim_{\varepsilon \to 0^+} \left[ \mathcal{I}_\triangle^{\UHP}(s+i \varepsilon,t) - \mathcal{I}_\triangle^{\LHP}(s-i \varepsilon,t) \right] \,,
    \end{equation}
    for any real $s$ and $t$.
    Since all logarithms and dilogarithms are written on their principal branches in~\eqref{eq:triLHP} and~\eqref{eq:triUHP}, we can evaluate the discontinuity directly using
    \be
        \Disc_{w} \log(w) = \pi \Theta(-w), \qquad 
        \Disc_{w} \Li_2(w) = \pi \log(w) \Theta(w-1).
    \ee
    Since the logarithms and dilogarithms, along with the square roots in $\beta_y$, $\beta_z$ and $\beta_{yz}$, can have branch points at the Landau branch points $s_{\normal}$ and $s_{\triangle}$, we split the computation into different regions on the real axis, depending on whether $s$ is above or below these branch points. For example, when $\Re s>s_{\triangle}$, the arguments of the square roots all have positive real parts, so none of the square roots in~\eqref{eq:wpm} change branches as $s$ crosses the real axis. The discontinuity comes from the dilogarithmic terms whose arguments close to the $s$ axis have a real part greater than 1 in addition to a sign change of the imaginary part across the branch cut. The discontinuities of the terms that contribute are 
    \begin{gather}
    \Disc_s \left[ -\frac{z}{4 M^2 \beta_z} \Li_2 \left(1- \tfrac{1+\frac{z}{2} - \beta_z}{1+ \frac{z}{2} - \beta_y \beta_z } \right) \right]
     =
    \frac{\pi z}{4 M^2 \beta_z} \log\left(1- \tfrac{1+\frac{z}{2} - \beta_z}{1+ \frac{z}{2} - \beta_y \beta_z } \right) \,, \\
    \Disc_s \left[\frac{z}{4 M^2 \beta_z} \sum_{\zeta \in \{-1,1\}} 2 \Li_2 \left(\tfrac{1+\beta_z}{1+\zeta \beta_z \beta_{yz}} \right) \right]
     =
    \frac{-2 \pi z}{4 M^2 \beta_z} \sum_{\zeta \in \{-1,1\}} \log\left(\tfrac{1+\beta_z}{1+\zeta \beta_z \beta_{yz}} \right) \,, \\
    \Disc_s \Bigg[\frac{z}{4 M^2 \beta_z} \sum_{\zeta \in \{-1,1\}} [
    2 \pi i \sgn(\Im s)
    ]  \log \Big( \tfrac{1+\beta_z}{1+\beta_z \beta_{yz}} \Big) \Bigg]
    = \frac{4 \pi z}{4 M^2 \beta_z} \log\left(\tfrac{1+\beta_z}{1+ \beta_z \beta_{yz}} \right) ,
    \end{gather}
    as well as
    \begin{gather}
    \Disc_s \Bigg[\frac{z}{4 M^2 \beta_z} \sum_{\zeta \in \{-1,1\}}  \zeta
        \log \left( \tfrac{1+\frac{z}{2}+\zeta \beta_z}{1+\frac{z}{2}+\zeta \beta_y \beta_z} \right) \bigg(-\sgn(\Im s) \pi i + \log \left( -1 + \tfrac{1+ \frac{z}{2}+ \zeta \beta_z}{1+\frac{z}{2}+ \zeta \beta_y \beta_z} \right) \bigg)\Bigg]\nn
    \\ =
    \frac{\pi z}{4 M^2 \beta_z} 
    \Bigg[
    -\log(1-\tfrac{1+\frac{z}{2}-\beta_z}{1+\frac{z}{2}-\beta_y \beta_z})
    \Bigg] \,. 
    \end{gather}
    Collecting and simplifying all terms results in
    \be 
        \Disc_{s > s_{\triangle}} \I_{\triangle} = 
        \frac{\pi z}{2 M^2 \beta_z} \log \left( 
        \frac{1-\beta_z \beta_{yz}}{1+\beta_z \beta_{yz}} \right) \,.
    \ee 
    Computation of discontinuities in the remaining regions is elementary but tedious. When the dust settles, the answer is
    
    \begin{equation}
         \Disc_s \mathcal{I}_\triangle = 
         \frac{\pi z}{4 M^2 \beta_z} 
         \begin{cases}
             2 \log \left( - \frac{1-\beta_z \beta_{yz}}{1+\beta_z \beta_{yz}}\right) + \log \left(- \frac{1+\frac{z}{2}+\beta_y \beta_z}{1+\frac{z}{2} -\beta_y \beta_z} \right) - \pi i  \hfill &\text{if }  s < s_{\normal},
             \\
             2 \log \left( - \frac{1-\beta_z \beta_{yz}}{1+\beta_z \beta_{yz}}\right) \hfill &\text{if }s_{\normal} < s < s_{\triangle},
             \\
             2 \log \left( \frac{1-\beta_z \beta_{yz}}{1+\beta_z \beta_{yz}}\right) \hfill &\text{if } s_{\triangle} < s,
         \end{cases}
         \label{eq:Disc_cases}
    \end{equation}
    where we take $s\to s- i\varepsilon$ to find the branches of the logarithms and square roots. Note that the analytic form of $\Disc_s \mathcal{I}_\triangle(s,t)$ changes at the branch points $s_{\normal}$ and $s_\triangle$.
    
    On the other hand, one can compute the imaginary part of $\I_{\triangle}$, which expressed in terms of~\eqref{eq:triLHP} reads
    \be
    \Im\, \I_{\triangle}(s,t) = \frac{1}{2 i} \lim_{\varepsilon \to 0^+}  \left[ \mathcal{I}_\triangle^{\LHP}(s-i \varepsilon,t) - \overline{\mathcal{I}_\triangle^{\LHP}(s-i \varepsilon,t)} \right].
    \ee
    Careful evaluation in each region along the $s$-axis indeed reveals agreement with~\eqref{eq:Im-minus-Disc}.

    \subsubsection{\label{sec:triangle-u-cuts}Unitarity cuts in the \texorpdfstring{$u$}{u}-channel}
    
    The final cross-check we make is unitarity, which provides a consistency condition that needs to be satisfied in the physical channels. We start with the $u$-channel: $s<0$ for fixed $t<0$.
    
    Following Sec.~\ref{sec:cutting}, unitarity gives the following relation in the $u$-channel:
    \be\label{eq:sum-u-cuts}
    \Im\, \I_{\triangle} = \Cut_{12}^u\, \I_{\triangle} + \Cut_{23}^u\, \I_{\triangle} + \Cut_{13}^u\, \I_{\triangle}  -2\,\Cut_{123}^u\, \I_{\triangle}.
    \ee
    The left-hand side equals the imaginary part because exchanging the incoming with outgoing states leaves scalar diagrams invariant. Recall that cuts are labelled by the propagators that are put on shell, as illustrated in Fig.~\ref{fig:s-cuts}, and we included the superscript to remember we are working in the $u$-channel. The final term features an additional minus sign, because it cuts through the diagram twice, i.e., $c=2$ in the notation of Sec.~\ref{sec:cutting}, and a factor of $2$ because there are two separate ways of slicing through the diagram that cut all three propagators.
    
    We will compute all four contributions in turn. First of all, notice that both contributions to $\Cut_{123}$ have a vertex where the lighter particle with mass $m$ decays into two new ones with masses $m$ and $M$. Such a configuration is not allowed kinematically, which immediately gives
    \be
    \Cut_{123}^u\, \I_{\triangle} = 0.
    \ee
    Moreover, by symmetry we also have
    \be
    \Cut_{23}^u\, \I_{\triangle} = \Cut_{13}^u\, \I_{\triangle}.
    \ee
    We are thus left with two non-trivial computations. To perform them, we go to the loop-momentum space, as in Sec.~\ref{sec:box-cuts-s}. In our conventions, we have
    \begin{equation}
        \I_\triangle(s,t) = \lim_{\eps \to 0^+} \int \frac{\d^\D \ell}{i \pi^{\D/2}}
        \frac{(-1)^3}{[\ell^2 - m^2 + i\varepsilon][\left(\ell+p_{24}\right)^2 - m^2 + i\varepsilon][\left(\ell+p_{234}\right)^2 - m^2 + i\varepsilon]},
        \label{eq:momtriintegral}
    \end{equation}
    where $p_{24} = p_2 + p_4$ etc.
    
    \begin{figure}
        \centering
        \includegraphics[scale=1]{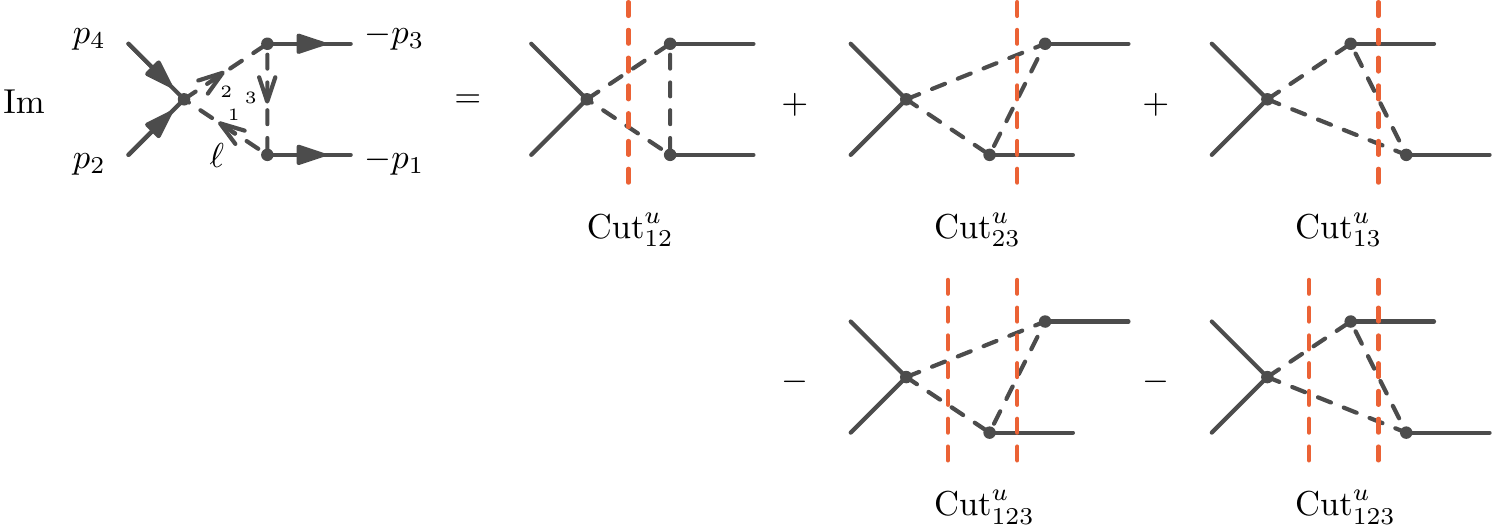}
        \caption{Unitarity for the triangle diagram in the $u$-channel, $s < 0$ and $t<0$. $\Cut_{12}^u$ is the normal threshold cut present in the whole of the $u$-channel since $s < 0 < s_{\normal}$. Both $\Cut_{13}^u$ and $\Cut_{23}^u$ correspond to the external-mass singularity and hence present everywhere. The final two cuts both equal to $\Cut_{123}^u$, but vanish as they are forbidden kinematically.}
        \label{fig:u-cuts}
    \end{figure}

    We start with the cut $\Cut_{12}^u$ computed by
    \begin{equation}\label{eq:Cut-12}
    \Cut_{12}^u\,\I_{\triangle}(s,t) = -\frac{\left(-2\pi i \right)^2}{2 i}  \int \frac{\d^\D \ell}{i \pi^{\D/2}}  \frac{\delta^-[\ell^2-m^2]\,  \delta^+[(\ell+p_{24})^2-m^2]}{(\ell+p_{234})^2-m^2} \,,
    \end{equation}
    where the negative-energy delta function $\delta^-$ is used in the conventions of Fig.~\ref{fig:u-cuts} to ensure causal propagation through the cut, i.e., $\ell^0 < 0$.
    The third propagator cannot go on shell (for the same reason why $\Cut_{123}^u$ is absent) and hence we can drop the $i\eps$ deformation.
    To evaluate the cuts, it will be convenient to work in the center-of-mass frame of the total incoming momentum $p_{24}^\mu$:
    \begin{align}
            p_{24}^\mu & = \left(\;\sqrt{u}\,,\;0,\ldots,0,0\right), \nn\\
            p_1^\mu & = \left(-\tfrac{\sqrt{u}}{2},0,\ldots,0,\tfrac{\sqrt{u-4M^2}}{2}\right), \label{eq:u-CM-frame}\\
            p_3^\mu & = \left(-\tfrac{\sqrt{u}}{2},0,\ldots,0,-\tfrac{\sqrt{u-4M^2}}{2}\right).\nn
    \end{align}
    In addition, we make use of spherical coordinates for $\ell^\mu$, setting
    \be
    \ell^\mu = \left(\ell^0,\,  |\vec{\ell}|\vec{n}  \sin \theta ,\, |\vec{\ell}| \cos \theta\right),
    \ee
    where $\vec{n}$ is a unit vector in $\D{-}2$ dimensions.
    The two delta functions in~\eqref{eq:Cut-12} together fix the energy of loop momentum $\ell^\mu$ to be $\ell^0 = -\frac{\sqrt{u}}{2}$,
    at which stage the positive-energy conditions have no further effect. We write out the loop momentum in spherical coordinates,
    \begin{multline}
    \Cut_{12}^u\,\I_{\triangle} = - \frac{2 \pi^2}{\pi^{\D/2}} \int_{-\infty}^{\infty} \d \ell^0 \int_0^\infty |\vec{\ell}|^{\D-2} \d |\vec{\ell}| \int_{-1}^1 \left(\sin \theta \right)^{\D-4} \d \cos \theta \int \d \Omega_{\D-2} 
        \\ 
        \times 
        \frac{
        \delta[\tfrac{u}{4}-|\vec{\ell}|^2-m^2]\,\delta[u + 2 \sqrt{u} \, \ell^0]}{(\ell+p_{234})^2-m^2} \,,
    \end{multline}
    where $\Omega_{\D-2}$ is the solid angle in $\D{-}2$ dimensions, explicitly, 
    $\Omega_{\D-2} = 2 \pi^{\frac{\D-2}{2}} /\Gamma \left(\frac{\D-2}{2}\right)$.
    Performing the delta functions gives
    \be
        \Cut_{12}^u\, \I_{\triangle} =
        - \pi^{2-\D/2} \Omega_{\D-2} \frac{\left(\frac{u}{4} - m^2 \right)^{\frac{\D-3}{2}}}{2\sqrt{u}} \int_{-1}^1 \frac{\left(\sin \theta\right)^{\D-4} \d \cos \theta\, \Theta(u-4m^2)}{M^2-\frac{u}{2} + 2 \sqrt{\frac{u}{4}-M^2}\sqrt{\frac{u}{4}-m^2} \cos \theta}\, ,
    \ee
    where the step function arises from the constraint $|\vec{\ell}|^2 \geq 0$, which means the cut has support in $s < s_{\normal} = 4(M^2 - m^2) - t$, which includes the whole $u$-channel physical region.
    The result of this integral in $\D=4$ in the $u$ channel when $s<0$ at fixed $t<0$ is,
    \begin{equation}
        \Cut_{12}^u\, \I_{\triangle} = - \frac{\pi z}{4 M^2 \beta_z} \Big[ \log \left(-\frac{1+\frac{z}{2}+\beta_y \beta_z}{1+\frac{z}{2}-\beta_y \beta_z} \right) + i \pi \Big] \Theta(s_\normal-s) ,
    \end{equation}
    where $z$, $\beta_y$ and $\beta_z$ were defined in~\eqref{eq:wpm}, and we take $s\to s- i \eps$ to find the branches of the square roots and logarithms. Note that despite the appearance of the $i \pi$ term, this expression is real as $\eps\to 0^+$, and written such that it can be trivially analytically continued up to $s=s_\normal$, despite having been calculated in the physical region.

    Next, we compute the cut $\Cut_{13}^u$ given by
    \begin{equation}
    \Cut_{13}^u\, \I_\triangle(s,t) = -\frac{\left(-2\pi i \right)^2}{2i}  \int \frac{\d^\D \ell}{i \pi^{\D/2}}   \frac{\delta^-[\ell^2-m^2]\, \delta^+[(\ell+p_{234})^2-m^2]}{(\ell + p_{24})^2-m^2},
    \end{equation}
    where, as before, we used $\delta^-$ to match the conventions of Fig.~\ref{fig:u-cuts} and the $i\eps$ can be dropped from the remaining propagator.
    Once again working in the center-of-mass frame~\eqref{eq:u-CM-frame}, we can rewrite it as
    \begin{multline}
    \Cut_{13}^u\, \I_\triangle = - \frac{2 \pi^2}{\pi^{\D/2}}
        \int_{-\infty}^{\infty} \d \ell^0 \int_0^\infty |\vec{\ell}|^{\D-2} \d |\vec{\ell}| \int_{-1}^1 \left(\sin \theta\right)^{\D-4} \d \cos \theta \; \d \Omega_{\D-2}
        \\
        \times \delta^- \left((\ell^0)^2-|\vec{\ell}|^2-m^2\right) 
        \delta^+ \left(M^2+\sqrt{u} \, \ell^0 + \sqrt{u-4 M^2} |\vec{\ell}| \cos \theta\right) \frac{1}{(\ell+p_{24})^2-m^2}.
    \end{multline}
    Performing the delta functions in $\ell^0$ and $\cos \theta$ results in
    \begin{multline}
    \Cut_{13}^u\, \I_\triangle = -  \frac{\pi^{2-\D/2}\Omega_{\D-2}}{\sqrt{u-4 M^2}} \int_{|\vec{\ell}|_-}^{|\vec{\ell}|_+} |\vec{\ell}|^{\D-3} \d |\vec{\ell}|
        \frac{1}{\sqrt{|\vec{\ell}|^2+m^2}}
        \frac{1}{u-2\sqrt{u}\sqrt{|\vec{\ell}|^2+m^2}}
        \\
        \times 
        \Bigg[1- \frac{\big(\sqrt{u}\sqrt{|\vec{\ell}|^2+m^2 }-M^2\big)^2}{|\vec{\ell}|^2(u-4M^2)}\Bigg]^{\frac{\D-4}{2}} ,
    \end{multline}
    where the boundaries on $|\vec{\ell}|$ arise from the condition  $-1<\cos \theta < 1$ placing restrictions on when the delta function can be satisfied, i.e.,
    \be
    |\vec{\ell}|_{\pm} = \frac{M \sqrt{u-4 M^2}\pm\sqrt{u (M^2-4 m^2)}}{4 M}.
    \ee
    In $\D=4$, the integral results in
    \begin{equation}
    \Cut_{13}^u\, \I_\triangle = \Cut_{23}^u\, \I_\triangle = - \frac{\pi z}{4 M^2 \beta_z}
        \log \left( 
        \frac{1-\beta_z \beta_{yz}}{1+\beta_z \beta_{yz}}
        \right),
    \end{equation}
    where $z$, $\beta_z$, and $\beta_{yz}$ were defined in~\eqref{eq:wpmA} and \eqref{eq:wpm}. The on-shell delta functions do not impose any extra constraints on the external kinematics, which is another sign these cuts contribute everywhere along the real $s$-axis.
    
    Summing over all the cuts in the $u$-channel according to~\eqref{eq:sum-u-cuts} results in
    \begin{equation}
        \Im\, \I_{\triangle}
        =
        - \frac{\pi z}{4 M^2 \beta_z} \left[ 2 \log \left( -
        \frac{1-\beta_z \beta_{yz}}{1+\beta_z \beta_{yz}} \right) + \log \left( -
        \frac{1+\frac{z}{2}+\beta_y \beta_{z}}{1+\frac{z}{2}-\beta_y \beta_{z}}
        \right) - i \pi \right] \;\; \text{if }  s < s_\normal\, ,
    \end{equation}
    which agrees with $\Im \, \I_{\triangle} = - \Disc_{s} \I_\triangle$ given in~\eqref{eq:Disc_cases}.
    
    \subsubsection{\label{sec:triangle-s-unitarity}Unitarity cuts in the \texorpdfstring{$s$}{s}-channel}
    
    Let us now check unitarity in the $s$-channel, $s>4M^2 -t$ with $t<0$. Following the steps in Sec.~\ref{sec:cutting}, we have
    \be\label{eq:Im-tri-s}
    \Im\, \I_{\triangle} = \Cut_{13}^s\, \I_{\triangle} + \Cut_{23}^s\, \I_{\triangle} - \Cut_{123}^s\, \I_{\triangle}.
    \ee
    Similarly to the $u$-channel, the left-hand side equals the imaginary part because exchanging the incoming with outgoing states leaves scalar diagrams invariant. The final term features an additional minus sign, because it cuts through the diagram twice, i.e., $c=2$ in the notation of Sec.~\ref{sec:cutting}. By symmetry we have
    \be
    \Cut_{13}^s\, \I_{\triangle} = \Cut_{23}^s\, \I_{\triangle},
    \ee
    hence we are left with evaluation of only two independent cuts.

    \begin{figure}
        \centering
        \includegraphics[scale=1]{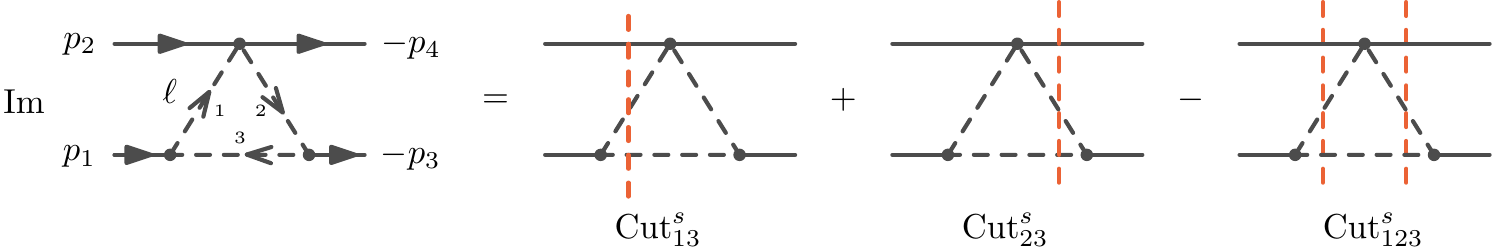}
        \caption{Unitarity for the triangle diagram in the $s$-channel, $s > 4M^2 - t$ and $t<0$. Both $\Cut_{13}^s$ and $\Cut_{23}^s$ are the external-mass singularity cuts that are present everywhere, while $\Cut_{123}^s$ is the triangle threshold cut contributing for $s > s_{\triangle}$.}
        \label{fig:s-cuts}
    \end{figure}
    
    Let us begin with $\Cut_{13}^s$ in the same conventions as in~\eqref{eq:momtriintegral}, which is given by
    \begin{equation}
        \Cut_{13}^s\, \I_{\triangle}(s,t) = -\frac{\left(-2\pi i \right)^2}{2 i}  \int \frac{\d^\D \ell}{i \pi^{\D/2}}   \frac{\delta^+(\ell^2-m^2) \delta^-((\ell+p_{234})^2-m^2)}{(\ell+p_{24})^2-m^2 + i\eps},
    \end{equation}
    where the $\delta^\pm$ are distributed according to the orientations of propagators from Fig.~\ref{fig:s-cuts}. 
    To evaluate this cut, we work in the lab frame of $p_1^\mu$ parametrized by
    \begin{align}
            p_1^\mu & = \left(M,0,\ldots,0,0\right),
            \nn\\
            p_{24}^\mu & = \left(E_{24},0,\ldots,0,k\right), \\
            p_3^\mu & = \left(-E_{3},0,\ldots,0,-k\right),\nn
    \end{align}
    where the on-shell conditions and momentum conservation fix
    \be
    E_{24} = -\frac{u}{2 M},\qquad E_3 =\frac{2 M^2-u}{2 M},\qquad k = \frac{\sqrt{-u(4M^2-u)}}{2 M}.
    \ee
    Using spherical coordinates for $\ell^\mu$, the cut integral reads
    \begin{align}
        &\Cut_{13}^s\, \I_{\triangle} = - \frac{2 \pi^2}{\pi^{\D/2}}
        \int_{-\infty}^{\infty} \d \ell^0 \int_0^\infty |\vec{\ell}|^{\D-2} \d |\vec{\ell}| \int_{-1}^1 \left(\sin \theta\right)^{\D-4} \d \cos \theta \; \d \Omega_{\D-2} 
        \\
        &\qquad\times 
        \delta^+ \left[(\ell^0)^2-|\vec{\ell}|^2-m^2\right]
        \delta^- \left[M^2 + 2(E_{24}{-}E_3) \ell^0 \right] \frac{1}{u+ 2 (E_{24}\ell^0 - |\vec{\ell}| k \cos \theta ) +i \varepsilon} \,.\nn
    \end{align}
    The delta functions set $|\vec{\ell}| = \sqrt{(\ell^0)^2-m^2}$ and $\ell^0=\frac{M}{2}$, so performing the integrals over $\ell^0$, $|\vec{\ell}|$ and $\cos \theta$,  analogously to Sec.~\ref{sec:triangle-u-cuts}, we get in $\D=4$
    \begin{equation}
        \Cut_{13}^s\, \I_{\triangle} = \Cut_{23}^s\, \I_{\triangle}
        =
        - \frac{\pi z}{4 M^2 \beta_z}
        \log \left(
        \frac{1-\beta_z \beta_{yz}}{1+\beta_z \beta_{yz}}
        \right)\,,
    \end{equation}
    where we take $s$ to have an infinitesimal imaginary part to find the branches of the square roots and logarithms.
    Note that in the region $4M^2 {-} t < s < s_\triangle$, this expression evaluates to have both real and imaginary parts. This is precisely the region in which $\Cut_{123}^s$ has support, and hence overlaps with the phase space for $\Cut_{13}^s$ and $\Cut_{23}^s$.
    
    To get a purely real expression from the sum of all cuts, we must add the triple cut contribution $\Cut_{123}^s$, which is given by
    \begin{equation}
        \Cut_{123}^s \I_{\triangle} = \frac{1}{2i} \left(-2\pi i \right)^3 (-1)^3 \!\!\int\! \frac{\d^\D \ell}{i \pi^{\D/2}}  \delta^+[\ell^2-m^2] \delta^+[(\ell+p_{24})^2-m^2] \delta^-[(\ell + p_{234})^2-m^2] \,.
    \end{equation}
    In addition to setting $|\vec{\ell}| = \sqrt{(\ell^0)^2-m^2}$ and $\ell^0=\frac{M}{2}$ as before, the delta function imposes that $4 M^2-t < s < s_{\triangle}$ from the integration region $-1<\cos \theta <1$. Physically, this corresponds to a constraint on $s$ for this cut diagram to have support with real values of energies and momenta.
    Thus, the result in $\D=4$ reads
    \begin{equation}
        \Cut_{123}^s\, \I_{\triangle} = 
        \frac{i \pi^2 z}{2 M^2 \beta_z} \Theta[4 M^2-t<s<s_{\triangle}]\,.
    \end{equation}
    The result of adding all cuts according to~\eqref{eq:Im-tri-s} therefore yields
    \begin{equation}\label{eq:Im-Itri}
        \Im\, \I_{\triangle}  = -\frac{\pi z}{2 M^2 \beta_z}
        \begin{cases}
        \log \left(-
        \frac{1-\beta_z \beta_{yz}}{1+\beta_z \beta_{yz}}
        \right) \hfill & \text{if } 4M^2{-}t < s < s_{\triangle},\\
        \log \left(
        \frac{1-\beta_z \beta_{yz}}{1+\beta_z \beta_{yz}}
        \right) \hfill & \text{if }  s_\triangle < s,
        \end{cases}
    \end{equation}
    where $z$, $\beta_z$ and $\beta_{yz}$ were spelled out in~\eqref{eq:wpmA} and \eqref{eq:wpm}.
    In particular, we see that subtracting the triple cut $\Cut_{123}^s$ in the region $4M^2 - t < s < s_{\triangle}$ renders the above combination purely real. 
    Hence in the $s$-channel we find agreement with~\eqref{eq:Disc_cases}.
    
    \paragraph{Check with the Im formula.}
    Let us take this opportunity to illustrate the use of the parametric formulae for the discontinuity and imaginary part from Sec.~\ref{sec:im-disc}:
    
    \be
    \Im\, \I_{\triangle} = - \Disc_s \I_{\triangle} = \pi \int \frac{\d^3 \alpha}{\GL(1)} \frac{\delta(\V_{\triangle})}{\U_{\triangle}^{\D/2}}.
    \ee
    Fixing $\alpha_3 = 1 - \alpha_1 - \alpha_2$ gives
    \be
    \Im\, \I_{\triangle} = \pi \int_{\Delta_2} \!\!\!\d \alpha_1 \d \alpha_2\, \delta\!\left(u \alpha_1 \alpha_2 + M^2 (1{-}\alpha_1 {-}\alpha_2)(\alpha_1 {+} \alpha_2) - m^2\right) \,,
    \ee
    where $\Delta_2 = \{ 0 < \alpha_1 < \alpha_2 < 1\}$. The delta function is satisfied at the points
    \be
        \alpha_2^{\pm} = \frac{1}{2} - \frac{(1+\beta_z^2)\, \alpha_1}{z} \pm \sqrt{\frac{\beta_{yz}^2}{4}-\frac{2 \, \alpha_1}{z} + \frac{4 \beta_z^2 \, \alpha_1^2}{z^2}} \,,
    \ee
    where we have used the variables $z$, $\beta_z$ and $\beta_{yz}$ from~\eqref{eq:wpm}.
    After performing the delta function in either $\alpha_2^+$ or $\alpha_2^-$, we are left with an integral of the form
    \be 
        \Im\, \I_{\triangle} = -
        \frac{\pi z}{4 M^2 \beta_z} \int_\Delta \frac{\d \alpha_1 }{\sqrt{\alpha_1^2 - \frac{z }{2 \beta_z^2} \alpha_1+\frac{z^2 \beta_{yz}^2}{16 \beta_z^2}}}\, ,
    \ee 
    where the integration region $\Delta$ is carved out by constraints on $\alpha_1$ arising from the delta function. More precisely, we integrate over all $\alpha_1 \in [0,1]$ with for which $0<\alpha_2^\pm<1-\alpha_1$ is satisfied. The region $\Delta$ takes a different form depending on the value of $s$ takes, so we choose the region
    \be 
        s > 4 M^2 - t + \frac{M^4 \beta_{yz}}{2 m^2}(\beta_{yz}+1) \,,
    \ee 
    for illustration.
    In this region, the integral can be written as
    \be
    \Im\, \I_{\triangle} = \frac{\pi z}{4 M^2 \beta_z}
    \left( 2 \int_{\alpha_{12}^-}^{\alpha_{12}^+} 
    \d \alpha_1
    +
    \int_{\alpha_{1}^-}^{\alpha_{1}^+} 
    \d \alpha_1
    \right)
    \frac{1}{\sqrt{\alpha_1^2 - \frac{z }{2 \beta_z^2} \alpha_1+\frac{z^2 \beta_{yz}^2}{16 \beta_z^2}}} \,,
    \ee
    with the integration bounds given by 
    \be
    \alpha_{12}^- = 0\,,
    \qquad
    \alpha_{12}^+ = \frac{z}{4 \beta_z^2} \big(1- \sqrt{1-\beta_{z}^2\beta_{yz}^2} \big) \,,
    \qquad 
    \alpha_{1}^\pm = \frac{1}{2} \pm \frac{1}{2} \beta_{yz} \,.
    \ee
    Performing the integrations gives
    \be 
        \Im\, \I_\triangle = 
        - \frac{\pi z}{2 M^2 \beta_z}
        \log \left( 
        \frac{1-\beta_z \beta_{yz}}{1+\beta_z \beta_{yz}}
        \right)
    \ee 
    in agreement with~\eqref{eq:Im-Itri} in this region.

    \subsubsection{Discussion}
    In this subsection, we illustrated that the amplitude for $2\to2$ scattering of unstable particles at fixed momentum transfer $t=t^\ast$ can have a branch cut along the whole real $s$-axis, with separate analytic expressions for the upper- and lower-half $s$-planes. In particular, this case is different from the one discussed in Sec.~\ref{sec:ExampleI}, since the branch cut \textit{cannot} be deformed to reveal a path of analytic continuation between the two half-planes. As a simplest example of this feature, we studied the amplitude $\I_\triangle (s,t)$, corresponding to the triangle diagram from Fig.~\ref{fig:triangle-diagram}, in $\D=4$ space-time dimensions. We showed why we expect a branch cut along the whole real $s$-axis whenever one of the external masses $M_1$ can decay into some internal masses $m_1$, $m_2$, $\ldots$ in a $2 \to 2$ scattering process (as is the case in the triangle example): if we were allowed to deform the external masses, the amplitude would have a discontinuity between $M_1^2+i\eps$ and $M_1^2-i\eps$, and a variation in the external mass $M_1^2$ can be captured as a deformation in $s$ in an on-shell process since $M_1^2 = s+t+u - \sum_{i=2}^4 M_i^2$. Of course, nothing prevents a cancellation between different terms of the expression, potentially leading to a vanishing discontinuity along parts of the $s$-axis. In fact, this cancellation happens for this triangle diagram in $\D=3$ space-time dimensions. In general, however, we expect from the arguments above that these processes should lead to a non-vanishing branch cut along the whole real $s$-axis, and---as a result---the amplitude splits into two separate analytic expressions: one in the upper-half plane and a different one in the lower-half plane.
    
    To verify explicitly that the upper- and lower-half planes are indeed entirely disconnected, we computed an analytic expression for the amplitude $\I_\triangle (s,t)$ in $\D=4$ dimensions in each of the half-planes, in which the external masses $M$ and internal masses $m$ satisfy $M>2m$. As expected, we found two different analytic functions, $\I^{\UHP}$ and $\I^{\LHP}$, each of which is valid in either the upper- or lower-half planes. With these kinematics, the physical amplitude $\I_\triangle(s,t)$, is obtained as the boundary value of the \textit{lower-half} plane expression in $s$ in both the $s$- and $u$-channels, i.e.,  $\I_\triangle(s,t)=\lim_{\eps\to 0^+}\I^{\LHP}_\triangle(s-i\varepsilon,t)$.
    
    Equipped with the analytic expressions, we computed the discontinuity across the $s$-axis, which is proportional to the difference between $\I^{\UHP}(s+i\eps, t)$ and $\I^{\LHP}(s-i\eps,t)$ as $\eps \to 0^+$. We compared the result with the imaginary part of the physical amplitude, which can be computed either using the full expression for $\I_\triangle(s,t)$, or, in physical regions, as a sum over all unitarity cuts through the triangle diagram. While the discontinuity and imaginary parts cannot always be easily related, we showed that this example is a special case in which $\V_s-\V_u<0$ along the whole integration region. Thus, we expect based on Sec.~\ref{sec:im-disc} that the imaginary part of the full expression is precisely \textit{negative} that of the discontinuity. Using the full expression we found that 
    \begin{equation}
         \Disc_s \mathcal{I}_\triangle = 
         \frac{\pi z}{4 M^2 \beta_z} 
         \begin{cases}
             2 \log \left( - \frac{1-\beta_z \beta_{yz}}{1+\beta_z \beta_{yz}}\right) + \log \left(- \frac{1+\frac{z}{2}+\beta_y \beta_z}{1+\frac{z}{2} -\beta_y \beta_z} \right) - \pi i  \hfill &\text{if }  s < s_{\normal},
             \\
             2 \log \left( - \frac{1-\beta_z \beta_{yz}}{1+\beta_z \beta_{yz}}\right) \hfill &\text{if }s_{\normal} < s < s_{\triangle},
             \\
             2 \log \left( \frac{1-\beta_z \beta_{yz}}{1+\beta_z \beta_{yz}}\right) \hfill &\text{if } s_{\triangle} < s,
         \end{cases}
         \label{eq:Disc_cases_summary}
    \end{equation}
    with $y = - 4 m^2/u$, $z = - 4 M^2/u$, $\beta_y = \sqrt{1+y}$, $\beta_z = \sqrt{1+z}$, and $\beta_{yz}  = - i \sqrt{-1+4y/z}$, and the branches of the logarithms and square roots are determined by taking $s\to s- i\varepsilon$. We also found that, indeed, $\Im \, \I_\triangle (s,t) = - \Disc_s \I_\triangle(s,t)$ for any real $s$.
    
    The full expression for the triangle diagram from~\eqref{eq:triLHP}, along with the one for its imaginary part which is the negative of~\eqref{eq:Disc_cases_summary}, exhibit how the Landau singularities manifest as square-root and logarithmic branch points. When solving the Landau equations, we find that the branch points on the physical sheet are at
    \begin{align}
        s_{\normal} = 4(M^2 - m^2) - t,\qquad s_{\triangle} = \frac{M^4}{m^2} - t \,.
    \end{align}
    Since the amplitude is polylogarithmic, we expect based on the arguments in~\cite{Landau:1959fi} that the normal threshold corresponds to a square-root singularity of $\I_\triangle(s,t)$, while the triangle branch point should be of logarithmic type.
    Looking at the arguments of the dilogarithms and logarithms, we see that $\beta_y=0$ corresponds to the normal threshold, $s=s_\normal$. It is easy to see that this singularity appears as a square-root singularity in the imaginary part of $\I_{\triangle}$, whenever $u$ above the threshold for the cut in the $u$ channel to be allowed, i.e., when $s<s_\normal$. We also see that $\beta_{yz}$ is a kinematic-independent constant, and hence not a branch point of $\I_{\triangle}$. It would be, however, if the diagram was instead embedded in a $3\to3$ process in which all vertices were quartic, so that the external masses squared $M^2$ were replaced with kinematic variables and hence could be deformed. Moreover, when solving for when the logarithms in~\eqref{eq:Disc_cases_summary} have branch points, we find logarithmic branch points precisely at the triangle singularity $s=s_\triangle$. 
    
    In the physical regions, where we can calculate the imaginary part of the amplitude $\mc{I}_{\triangle}$ along the real $s$-axis by evaluating the unitarity cuts through the triangle diagram, we can circumvent computing the full amplitude if we are only interested in its imaginary part. It is therefore natural to ask whether we can analytically the expression for the cuts to find the imaginary part for any real value of $s$. In this particular example, we can argue that this is indeed possible.
    Since the endpoints of the physical region, at $s_- = 0$ and $s_+ = 4 M^2- t$, do not coincide with any branch points on the physical sheet, we expect that we can analytically continue the imaginary part $\Im \, \I_{\triangle}$ from the region $s<s_-$ to the region $s \in [s_-, s_\normal]$. Similarly, there is no branch point at $s=s_+$, so we expect that we can analytically continue from $s \in [s_+, s_\triangle]$ to the region $s \in [s_\normal, s_+]$. In other words, since there is only one branch point between the boundaries of the physical region, we can argue that it is possible to analytically continue the cut expression to all real values of $s$, which is also in accordance with the imaginary part as computed from the full amplitude, see~\eqref{eq:Disc_cases} and~\eqref{eq:Im-minus-Disc}. This reasoning fails, however, whenever there are two or more physical-sheet branch points on the real $s$-axis between endpoints of the physical region. For a discussion on how to analytically continue cuts outside the physical region, see, e.g.,~\cite{PhysRevLett.4.84,Goddard1969,doi:10.1063/1.1665233,doi:10.1063/1.1703897}.

	\subsection{\label{sec:ExampleIII}Example III: Summing over multiple diagrams}
	
	Finally, we would like to illustrate features that can happen when multiple diagrams contributing to the same S-matrix element are taken into account, providing a more realistic picture of the analyticity region of the full S-matrix. This is already illustrated by adding the $s$-triangle contribution $\I_\triangle(u,t)$ to the $u$-triangle $\I_\triangle(s,t)$ considered in the previous subsection. The toy-model $\T$-matrix element is therefore given by
	\be
	\T = \I_{\triangle}(s,t) + \I_{\triangle}(u,t) + \ldots
	\ee
	Given that the first term has to be approached from the lower-half plane according to~\eqref{eq:tri-limit}, by momentum conservation $u = -s-t+4M^2$, the second term requires the opposite direction. In other words,
	\be
	\T(s,t_\ast) = \lim_{\eps \to 0^+} \left[ \I^{\LHP}_{\triangle}(s - i\eps,t_\ast) + \I^{\UHP}_{\triangle}({-}s{-}t_\ast{+}4M^2 + i\eps,t_\ast) + \ldots \right],
	\ee
	and there would be no consistent way of treating the causal S-matrix as a boundary value of an analytic function.
	
	\begin{figure}
	    \centering
	    \includegraphics[scale=0.8]{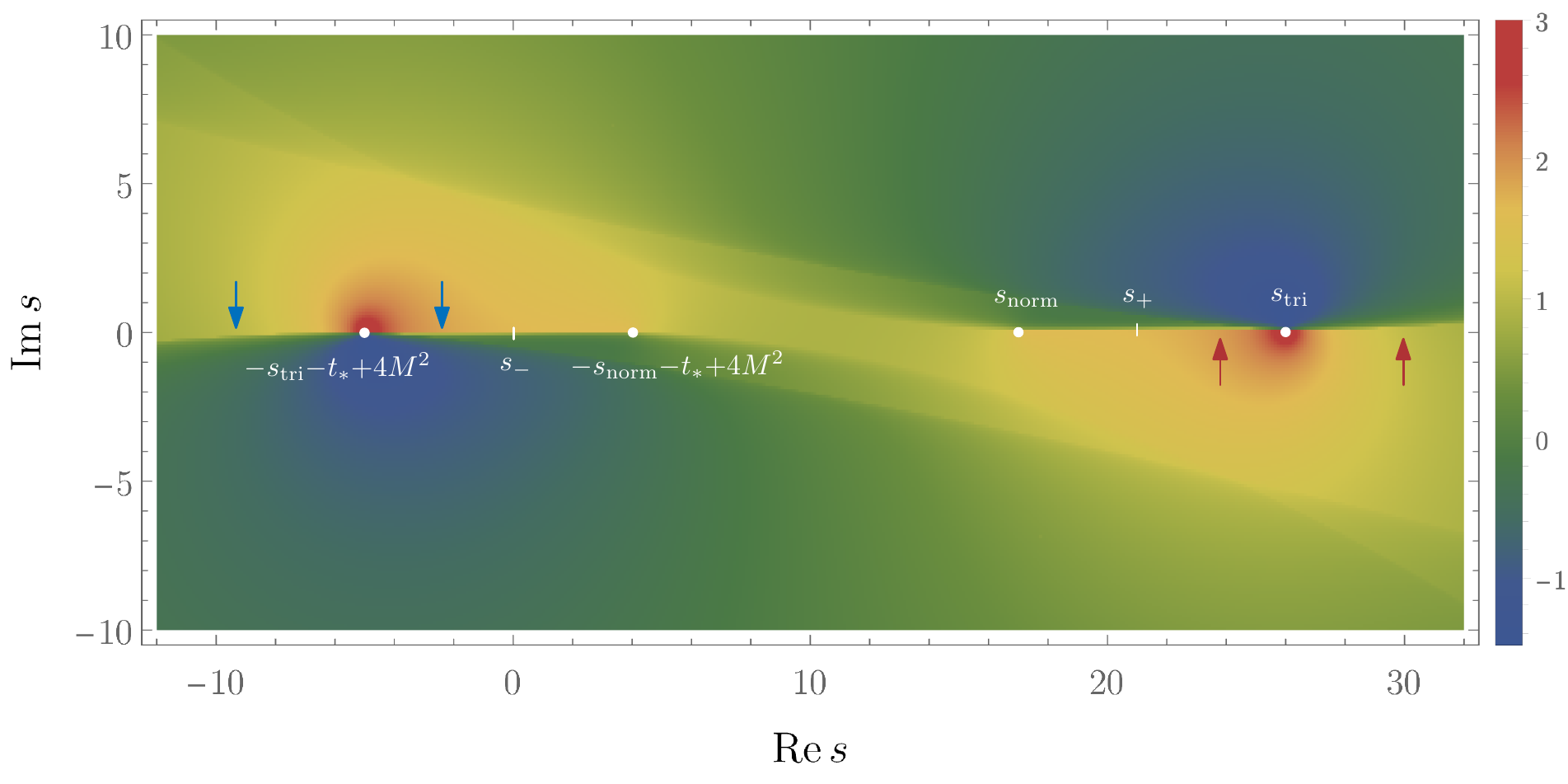}
	    \caption{Numerical plot of $\Im [ \I_{\triangle}(s,t_\ast) + \I_{\triangle}(u,t_\ast)]$ obtained using branch deformations in the complex $s$-plane with $t_\ast = -1$. We use $m=1$, $M=\sqrt{5}$, and $\eps = \tfrac{1}{100}$. The causal direction of approaching the $s$-channel physical region $s > s_+$ are indicated in red, while those for the $u$-channel $s< s_-$ are in blue. Discontinuities extend from the normal and triangles thresholds $s= s_{\normal}, s_\triangle$, as well as their $u$-channel counterparts.}
	    \label{fig:su-triangles}
	\end{figure}
	
	This gridlock is resolved by branch cut deformations, as illustrated with numerical data in Fig.~\ref{fig:su-triangles}. There is a strip in the neighborhood of the real axis where the complexified $\T_\C(s,t_\ast)$ can be defined. This strip defines the physical sheet.
	For example, the $s$-channel with $s>s_+$ is still approached from ``beneath'' the normal and $u$-triangle thresholds, while at the same time being ``over'' their $s \leftrightarrow u$ counterparts. Just as in Sec.~\ref{sec:ExampleII}, one can check that this is the only choice of approaching the physical regions consistent with unitarity.
	
	The physical sheet allows for an analytic continuation between the $s$- and $u$-channels, which shows crossing symmetry in this case. An apparent obstruction happens whenever two singularities lie on top of each other and have incompatible $i\eps$ directions. The simplest way for this to happen is when the $s$- and $u$-channel normal thresholds collide, i.e.,
	$s_\normal = - s_{\normal} - t_\ast + 4M^2$,
	which happens when
	\be
	t_\ast = 4M^2 - 8m^2.
	\ee
	However, in such situations one can wiggle the $t$-dependence slightly to establish the physical region in the complex $(s,t)$-planes. 
	
	Moreover, resummation of an infinite number of diagrams obtained by chains of 1PI insertions on each propagator will have the effect of adjusting the positions of branch points by mass shifts and decay widths. For example, in the simplest case when only the external particles acquire a width according to $M^2 \to M^2 - i M \Gamma$, we have
	\begin{align}
	s_{\normal} &\to s_{\normal} - 4 i M \Gamma,\\
	s_\triangle &\to s_{\triangle} - \frac{M^2 \Gamma}{m^2} ( \Gamma + 2 i M).
	\end{align}
	Since $\Gamma > 0$, both thresholds get shifted towards the lower-half plane.
	In this case we of course have to refine the above statements, because the physical sheet moves accordingly, but the essential features stay intact.
	It would be interesting to further study how the value of the corresponding discontinuity resums.

	\newpage
	\section{\label{sec:dispersion}Glimpse at generalized dispersion relations}
	
	In their traditional form, dispersion relations express scattering amplitudes in terms of integrals of their imaginary parts. This not only allows for practical computations of the S-matrix elements, but also, together with unitarity, provides an important consistency condition. While dispersion relations have mostly been implemented for scattering of the lightest particle, in this section, we take a step towards finding generalized relations for any $2\to 2$ scattering process. We find two types of identities: the first relates the scattering amplitude to an integral over its discontinuity, while the second relates parts of the scattering amplitude to an integral over its imaginary part.
	
	\subsection{Standard formulation}
	
\begin{figure}
    \centering
    \includegraphics[scale=1]{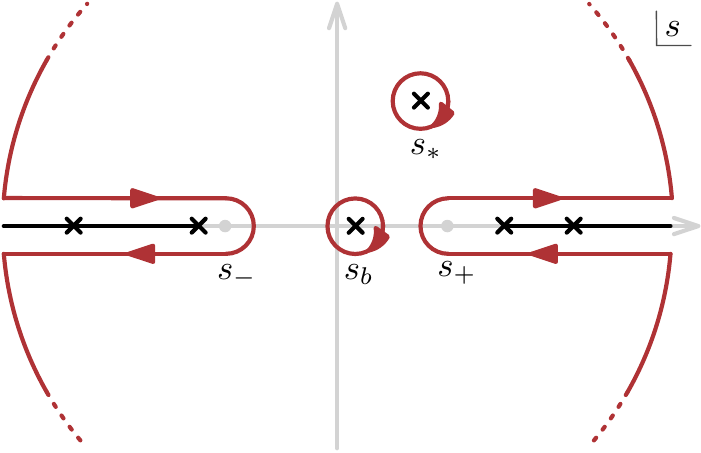}
    \caption{\label{fig:dispersion}Traditional form of dispersion relations for $2\to2$ scattering of the lightest particles in theories with a mass gap. At fixed $t=t_\ast < 0$, the amplitude evaluated at the point $s=s_\ast$ can be expressed in terms of the integrals over discontinuities in the $s$- and $u$-channels plus contributions from bound states and from infinity.}
    \end{figure}

	Validity of dispersion relations has so far been established only in special kinematic configurations, which for example exclude processes with large momentum transfer or involving massless external particles. To understand where these limitations come from, let us give a prototypical example at $n=4$:
	\be\label{eq:dispersion}
	\T(s_\ast, t_\ast) = -\frac{1}{\pi}\int_{-\infty}^{s_-} \frac{\d s\, \Im \T(s, t_\ast)}{s - s_\ast} + \frac{1}{\pi}\int_{s_+}^{\infty} \frac{\d s\, \Im \T(s, t_\ast)}{s - s_\ast} - \sum_{b} \frac{\Res_{s=s_b} \T(s,t_\ast)}{s_b - s_\ast}.
	\ee
	Fixing the momentum transfer $t = t_\ast < 0$, this relation comes from considering the function $\frac{1}{2\pi i} \tfrac{\T(s,t_\ast)}{s-s_\ast}$ in the complex $s$-plane, see Fig.~\ref{fig:dispersion}. Here, $s_\ast$ is some point away from singularities. The residue around $s=s_\ast$ gives the left-hand side of~\eqref{eq:dispersion}. The right-hand side comes from deforming this contour to enclose the remaining poles and branch cuts. If the $u$-channel and $s$-channel cuts are non-overlapping, we can separate their contributions. The contour around the $u$-channel cuts, running from $-\infty$ to some $s_-$, gives the first contribution on the right-hand side of~\eqref{eq:dispersion}, where we rewrote the discontinuity of $\T(s,t_\ast)$ as its imaginary part. Similar contributions come from the $s$-channel cuts, giving rise to the integral from $s_+$ to $\infty$. Note the difference in sign coming from the fact that the discontinuity in the $u$-channel equals \emph{minus} the imaginary part. Finally, the third term comes from possible one-particle exchanges contributing potential poles of $\T(s,t_\ast)$ at some $s = s_b$.
	
	In closing the contour we also made a tacit assumption that the contribution at infinity vanishes. In other words, we assumed that $\T(s,t_\ast)$ is bounded by $1/s$ at infinity. This is often not the case in practice, but can be easily accounted for by introducing $k$ \emph{subtractions}, which in practice amounts to starting with the function
	\be
	\frac{1}{2\pi i} \frac{\T(s,t_\ast)}{s-s_\ast} \frac{1}{\prod_{i=1}^{k} (s - s_i)},
	\ee
	for some $k$ points $s_i$ in the complex plane. For sufficiently large $k$, the contributions from infinity vanish and we are left with a generalization of~\eqref{eq:dispersion} with additional terms. Going through the same exercise as before gives
	\begin{align}
	\frac{\T(s_\ast, t_\ast)}{\prod_{i=1}^{k}(s_\ast - s_i)} = &-\frac{1}{\pi}\int_{-\infty}^{s_-} \frac{\d s\, \Im \T(s, t_\ast)}{(s - s_\ast)\prod_{i=1}^{k}(s-s_i)} + \frac{1}{\pi}\int_{s_+}^{\infty} \frac{\d s\, \Im \T(s, t_\ast)}{(s - s_\ast)\prod_{i=1}^{k}(s-s_i)}\\
	&- \sum_{b} \frac{\Res_{s=s_b} \T(s,t_\ast)}{(s_b - s_\ast) \prod_{i=1}^{k}(s_b - s_i)} - \sum_{i=1}^{k} \frac{\T(s_i, t_\ast)}{(s_i - s_\ast)\prod_{\substack{j=1\\ j\neq i}}^{k} (s_i - s_j)}.\nn
	\end{align}
	Thus, the price of fixing the behavior at infinity is having to know the value of the amplitude evaluated at $k$ such points.
	
	In the above manipulations we made a few leaps of faith. Clearly, if the Euclidean region between $s_-$ and $s_+$ does not exist, i.e., there $u$- and $s$-channel cuts are not separated, then the above derivation fails. This happens for example when $|t_\ast|$ is not small or when the external particles are massless. For $n=5$, there are similar obstructions to writing single-variable dispersion relations \cite{Landshoff1961,PhysRev.132.902}.
	In the context of low-energy effective field theory, one often writes dispersion relations that are valid only approximately at a given energy scale; see, e.g., \cite{Adams:2006sv,Arkani-Hamed:2020blm,Bellazzini:2020cot,Tolley:2020gtv,Caron-Huot:2020cmc,Caron-Huot:2021rmr,Caron-Huot:2022ugt}. Their $(s,t,u)$-permutation-invariant version appeared in \cite{Auberson:1972prg,Sinha:2020win}.
	
	The goal of this section is to get a glimpse on the form of dispersion relations without any restrictions on momenta and masses.

	\subsection{Schwinger-parametric derivation}
	
	As was emphasized in Sec.~\ref{sec:im-disc}, in general the discontinuity of an amplitude does not have to coincide with its imaginary part.
	To make this statement more concrete, let us pick any linear combination of Mandelstam invariants $z$ and freeze all the remaining variables at some real values. For instance, in the simplest case $n=4$, we can take $z = s$ at fixed momentum transfer
	\be
	t = t_\ast <0,
	\ee
	or at fixed scattering angle
	\be
	\cos \theta_\ast = 1 + \frac{2t}{s-4M^2} \in [-1,1].
	\ee
	Similar slices through the kinematic space can be taken at higher multiplicity.
	Recall from Sec.~\ref{sec:im-disc} that in general the discontinuity between upper- and lower-half $z$-planes is given by
	\be
	\Disc_z \T = \lim_{\eps \to 0^+} \tfrac{1}{2i}\Big( \T(z+i\eps) - \T(z-i\eps)\Big),
	\ee
	while the imaginary part is simply
	\be
	\Im\, \T = \tfrac{1}{2i} \Big( \T(z) - \overline{\T(z)} \Big).
	\ee
	We therefore find ourselves with two possibilities for what a generalization of dispersion relations~\eqref{eq:dispersion} to arbitrary momenta and masses could mean: it could either relate $\T(z_\ast)$ to integrals over the discontinuity or the imaginary part. We will consider both options, working at the level of individual Feynman integrals $\I(z)$ in perturbation theory.

    \begin{figure}
    \centering
    \includegraphics[scale=1]{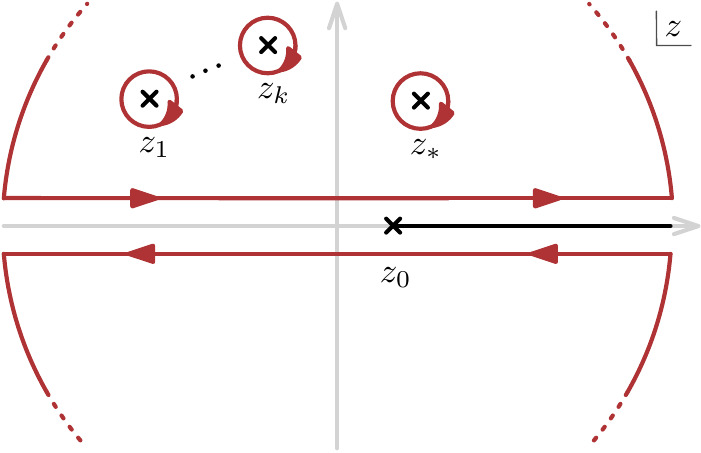}
    \caption{\label{fig:id1}Residue theorem for the identity~\eqref{eq:id1} in the $z$-plane needed to derive the generalized form of dispersion relations~\eqref{eq:generalized-dispersion}. In the illustrated case $\partial_z \V >0$, the possible branch cut starting at $z = z_0$ extends to the right, while for $\partial_z \V < 0$ it would stretch to the left.}
    \end{figure}
    
    \subsubsection{Discontinuity version}
    The idea is to perform all the manipulations under the integrand of the Schwinger-parametrized integral. Let us consider the following function for any real value of Schwinger parameters:
	\be\label{eq:disp-function}
	\frac{1}{2\pi i} \frac{1}{\prod_{i=1}^{k+1} (z - z_i) [-\V(\alpha_e; z)]^d},
	\ee
    where all the $z_i$'s are located away from the real axis and to shorten the notation we take $z_{k+1} = z_\ast$. More precisely, the implicit assumption is that none of the $z_i$'s lie on branch cuts or thresholds. This function has $k{+}1$ simple poles at $z=z_i$, and since the action $\V(\alpha_e;z)$ is linear in $z$, this function also has one additional pole/branch point at some $z = z_0(\alpha_e)$ determined by
    \be
    \V(\alpha_e; z_0) = z_0\, \partial_z \V(\alpha_e) + \V(\alpha_e; 0) = 0.
    \ee
    In the worst case, when $d$ is non-integer (for example when using dimensional regularization), there is a branch cut extending from $z = z_0$. Whether it goes to the left or right of the real $z$-axis depends on the sign of $\partial_z \V$ and it will not affect our derivation.
    
    Assuming that the number of subtractions in large enough such that~\eqref{eq:disp-function} vanishes at infinity, the residue theorem gives us
    \be\label{eq:id1}
    \sum_{i=1}^{k+1} \frac{1}{\prod_{\substack{j=1\\ j\neq i}}^{k+1} (z_i - z_j)} \frac{1}{[-\V(\alpha_e; z_i)]^d} = \frac{1}{\pi} \int_{-\infty}^{\infty} \frac{\d z}{\prod_{i=1}^{k+1}(z - z_i)} \Disc_z \frac{1}{[-\V(\alpha_e; z)]^d}. 
    \ee
    This identity holds for any value of Schwinger parameters and the remaining Mandelstam invariants. Note that here we do not want to move the $z$-contour entirely to the real axis, because we specifically want a formula that works regardless of the values of the Schwinger parameters.
    
    This is a good place to point out a subtlety with the asymptotics as $|z| \to \infty$. Let us first ask how to estimate the behavior of a given Feynman integral in this limit. Naively, one could conclude that the amplitude goes like $z^{-d}$ since $\V(\alpha_e;z)$ is linear in the variables $z$. This is not quite correct because in approximating $\V(\alpha_e; z) \approx z \partial_z \V(\alpha_e)$ we implicitly assumed that $\partial_z \V(\alpha_e)$ is of the same order as $\V(\alpha_e;0)$. But the aim is to integrate this quantity over \emph{all} positive values of the Schwinger parameters $\alpha_e$, and it might easily happen that $\partial_z\V(\alpha_e) \ll \tfrac{1}{z}$ or its complement $\V(\alpha_e,0) \gg z$ (as a toy model, consider approximating $\frac{1}{z} = \int_{0}^{\infty}\!\frac{\d \alpha}{(z\alpha + 1)^2}$) along the integration contour. Enhanced asymptotic behavior of the Feynman integral will therefore come from the neighborhood of such regions in the integration space. Hence in general $k>d$ subtractions are needed. There is presently no reliable way of predicting what $k$ should be for an arbitrary Feynman integral \cite{doi:10.1063/1.1703983,cmp/1103904351,doi:10.1063/1.1703919,doi:10.1063/1.1703918,doi:10.1063/1.523857,PhysRev.131.480,PhysRev.131.2373,Lam:1969xs}. At any rate, this discussion indicates that one needs to be extremely careful in checking that the limit $|z| \to \infty$ is handled correctly.
    
    We can check that this subtlety does not affect~\eqref{eq:id1}, which holds for any $\alpha_e$'s and even when $\partial_z \V(\alpha_e) \to 0$. To verify this, let us consider the right-hand side of~\eqref{eq:id1} in this limit. It becomes the distribution
    \be\label{eq:id11}
    \frac{\sgn[\partial_z \V(\alpha_e)]}{\Gamma(d)} \delta^{(d-1)}[\V(\alpha_e;0)] \int_{-\infty}^{\infty} \frac{\d z}{\prod_{i=1}^{k+1} (z - z_i)},
    \ee
    which converges if $k \geq 1$ and we applied~\eqref{eq:Im-identity2}. The above integral simply picks up the residues in the upper-half (or equivalently lower-half) plane. Similarly, the left-hand side of~\eqref{eq:id1} limits to the distribution
    \be\label{eq:id12}
    \frac{\sgn[\partial_z \V(\alpha_e)]}{\Gamma(d)} \delta^{(d-1)}[\V(\alpha_e; 0)] \sum_{i=1}^{k+1} \frac{1}{\prod_{\substack{j=1\\ j\neq i}}^{k+1} (z_i - z_j)} \frac{\sgn(\Im z_i )}{2},
    \ee
    where each term is weighted with a factor $\pm \tfrac{1}{2}$ depending on whether $z_i$ is located in the upper- or lower-half plane. One can check that, indeed,~\eqref{eq:id11} equals~\eqref{eq:id12}, meaning that---in a rather non-trivial way---the identity~\eqref{eq:id1} survives in the limit as $|z| \to \infty$. Note that we are still making an assumption that the $\alpha$- and $z$-integrals commute. This, in general, requires us to work in dimensional regularization.
    
    In the final stage we can simply integrate both sides against
    \be
    \Gamma(d) \int \frac{\d^\E \alpha}{\GL(1)} \frac{\widetilde{\N}}{\U^{\D/2}}\, \eqref{eq:id1},
    \ee
    which gives us the generalized dispersion relation:
    \be\label{eq:generalized-dispersion}
    \boxed{
    \sum_{i=1}^{k+1} \frac{\I(z_i)}{\prod_{\substack{j=1\\ j\neq i}}^{k+1} (z_i - z_j)} = \frac{1}{\pi} \int_{-\infty}^{\infty} \frac{\d z\, \Disc_z \I(z)}{\prod_{i=1}^{k+1}(z - z_i)},}
    \ee
    relating the Feynman integral evaluated at $k{+}1$ points to an integrated discontinuity in $z$. From this perspective, $\Disc_z \I(z)$ defined in~\eqref{eq:DiscI} gives a Schwinger-parametric representation of the dispersive function.
    We checked numerically that the generalized dispersion relations~\eqref{eq:generalized-dispersion} hold for the Feynman integrals studied in Sec.~\ref{sec:branch-cut-deformations} with $z=s$.
    
    Note that the above result seems extremely strong: we appear to have concluded that $\I(z)$ does not have any singularities away from the real $z$-axis (in fact, without even making use of the details of the function $\V$ other than its linearity in $z$). But notice that this was an \emph{assumption} rather than the result. The reason is that whenever $\V(\alpha_e; z) = 0$ we might be forced to deform the $\alpha$-contour, even on the real $z$-axis, in addition to giving $z$ a small imaginary part, if $\I(z)$ has singularities extending away from the real axis. This deformation can easily break down the above arguments. However, this perspective gives us a new tool for studying analyticity on the physical sheet. Currently, the most stringent results on the appearance of non-analyticities come from the work of Bros, Epstein, and Glaser for $2\to 2$ scattering in gapped theories \cite{Bros:1964iho,Bros:1965kbd}, which assert that there are no singularities at asymptotically-large $|s|$, though this result was never made precise in terms of the mass gap or the momentum transfer $t$; see also \cite{Bros:1972jh,Bros:1985gy} for extensions to $2\to 3$ scattering. For $st$-planar scattering amplitudes one can show that all the singularities are confined to the real $s$-axis in $2\to 2$ scattering without restrictions on the mass gap at every order in perturbation theory, and similar results hold at higher multiplicity \cite{Mizera:2021fap}.

    \subsubsection{Imaginary-part version}
    As emphasized in Sec.~\ref{sec:im-disc}, the discontinuity might no longer be equal to the imaginary part $\Im\, \I(z)$.
    Alternatively, we might aim to express the right-hand side of dispersion relations in terms of the imaginary part. To this end, let us work at the level of the Schwinger-parametric integrand and prove another identity:
    \be\label{eq:id2}
    \sum_{i=1}^{k+1} \frac{1}{\prod_{\substack{j=1\\ j\neq i}}^{k+1} (z_i - z_j)} \frac{\sgn [\partial_{z} \V(\alpha_e)]}{[-\V(\alpha_e; z_i)]^d} = \frac{1}{\pi}\lim_{\eps \to 0^+} \int_{-\infty}^{\infty} \frac{\d z}{\prod_{i=1}^{k+1}(z - z_i)} \Im \frac{1}{[-\V(\alpha_e; z) - i\eps]^d},
    \ee
    where we recall that $\sgn [\partial_{z} \V(\alpha_e)]$ is $z$-independent.
    The difference to~\eqref{eq:id1} is the $\Im$ in place of $\Disc_z$, as well as the sign function on the left-hand side.
    In order to prove this equation, let us consider the integrand before taking the imaginary part:
    \be\label{eq:im-int}
    \frac{1}{2\pi i}\frac{\d z}{\prod_{i=1}^{k+1}(z - z_i)} \frac{1}{[-\V(\alpha_e; z) - i\eps]^d}.
    \ee
    Apart from the simple poles at $z= z_i$, it has a pole/branch point at $z = z_\eps(\alpha_e)$ with
    \be
    z_{\eps} = -\frac{\V(\alpha_e;0) + i\eps}{\partial_z\V(\alpha_e)},
    \ee
    which is located just below the real axis if $\partial_z \V >0$ and just above it if $\partial_z \V < 0$. Likewise, the branch cut---if it exists---runs to the right and left, respectively.

\begin{figure}
        \centering
        \includegraphics[scale=1]{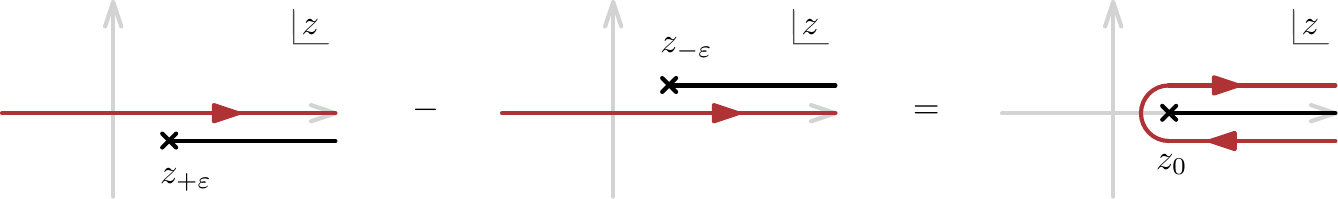}
        \caption{\label{fig:id2}Contour manipulations in the $z$-plane needed to derived the identity~\eqref{eq:id2}. As $\eps\to0^+$, the two branch cuts starting at $z = z_{\pm \eps}$ start coinciding at $z = z_0$ on the real axis. The difference between the two displaced contours is the discontinuity across the branch cut.}
    \end{figure}

    Let us consider the case $\partial_z \V > 0$ first.
    In order to compute the imaginary part, we take a difference between~\eqref{eq:im-int} and its complex conjugate, see Fig.~\ref{fig:id2}. The latter has a branch cut starting at $z_{-\eps}$ above the real axis. In the limit as $\eps \to 0^+$, we end up computing the discontinuity across the branch cut. At this stage we can deform the integration contour to enclose the remaining poles. The contribution at infinity vanishes by assumption. This proves the identity~\eqref{eq:id2} for $\partial_z \V > 0$. An analogous derivation can be repeated for $\partial_z \V <0$, where the only difference is that we end up with minus discontinuity and, as a result, minus the sum of residues. This fact is accounted for by the factor of $\sgn [\partial_z \V(\alpha_e)]$ on the left-hand side of~\eqref{eq:id2}.
    
    We are left with $\partial_z \V \to 0$, which might seem like an edge case, but is actually quite important to understand, given our earlier discussion. Indeed, one can use almost-identical logic to show that~\eqref{eq:id2} still holds in this limit, provided that $k \geq 1$.

    At this point, there are two ways we can apply
    the identity~\eqref{eq:id2}.
    The first one is to integrate it against the following Schwinger parametric measure weighted with the additional sign:
    \be
    \Gamma(d) \int \frac{\d^\E \alpha}{\GL(1)} \frac{\widetilde{\N}}{\U^{\D/2}}
    \sgn [\partial_z \V(\alpha_e)].\label{eq:dmu2}
    \ee
    In this way, the left-hand side simply becomes a sum over Feynman integrals, but the integrand on the right-hand side equals the discontinuity
    \be
    \sum_{i=1}^{k+1} \frac{\I(z_i)}{\prod_{\substack{j=1\\ j\neq i}}^{k+1} (z_i - z_j)} = \frac{1}{\pi} \int_{-\infty}^{\infty} \frac{\d z\, \Disc_z \I}{\prod_{i=1}^{k+1}(z - z_i)},
    \ee
    where we recognized the Schwinger-parametric formula for the discontinuity given in~\eqref{eq:DiscI}. We thus recover the same generalized dispersion relation as in~\eqref{eq:generalized-dispersion}.
    
    The second option is to integrate against the measure without the sign factor in~\eqref{eq:dmu2}. At this stage the integrand on the right-hand side is proportional to $\Im\, \I(z)$, but the quantities on the left-hand side are not Feynman integrals, but rather the combinations introduced in~\eqref{eq:I-UHP-LHP} when $d$ is a positive integer. Concretely, we have
    \be
    \sum_{i=1}^{k+1} \frac{\I^{+}(z_i) - \I^{-}(z_i)}{\prod_{\substack{j=1\\ j\neq i}}^{k+1} (z_i - z_j)} = \frac{1}{\pi} \int_{-\infty}^{\infty} \frac{\d z\, \Im\, \I(z)}{\prod_{i=1}^{k+1}(z - z_i)} \,.
    \ee
    Similar formulae can be derived in cases when $d$ is half-integer or non-positive.
    Since the left-hand side is no longer expressible in terms of Feynman integrals, this relation does not seem immediately useful. We confirmed its validity numerically on the examples studied in Sec.~\ref{sec:branch-cut-deformations} with $z=s$.

	\newpage
	\section{\label{sec:fluctuations}Fluctuations around classical saddle points}
	
	In this section, we study the expansion of the amplitude around its classical saddle points, for leading, subleading, and second-type Landau singularities. Combining this expansion with an assumption on the analyticity of the scattering amplitude, which implies the absence of isolated singularities of codimension-2, we show that isolated and non-degenerate saddles give rise to at most simple-pole singularities. We conjecture that higher-codimension branch points of scattering amplitudes always lie on codimension-1 curves, and demonstrate how this happens in a few examples. Finally, we explicitly compute expansions around all singularities of one-loop Feynman integrals and normal-threshold singularities at arbitrary loop order.
	
	\subsection{\label{sec:threshold-expansion}Threshold expansion}
	
	As explained in Sec.~\ref{sec:general}, singularities of the perturbative S-matrix can be interpreted as classical contributions to the path integral localizing on the worldline saddle points. They determine the singularity locus in the kinematic space, which can be written as
	\be\label{eq:Delta2}
	\Delta = 0,
	\ee
	if we assume it is of codimension-1. Here, $\Delta$ is an irreducible polynomial in the external kinematics; see Sec.~\ref{sec:general} and \cite{Mizera:2021icv} for examples.
	
	As with any other quantum-mechanical problem, we can study fluctuations around classical saddle points. Since saddles in the Schwinger-parameter space are already linked to the external kinematics via~\eqref{eq:Delta2}, studying fluctuations around them translates to probing the scattering amplitude in a neighborhood of this singularity. That is, we consider small perturbations
	\be\label{eq:deltaDelta}
	\Delta = 0 + \delta \Delta + \mc{O}\! \left[ \left( \delta \Delta\right)^2 \right]
	\ee
	and are interested in the leading behavior of Feynman integrals in $\delta \Delta = \Delta \to 0$. The result is given in~\eqref{eq:subleading-expansion} and is further illustrated on examples in Sec.~\ref{sec:sec7-examples}.
	As we will see in Sec.~\ref{sec:bound}, assuming analyticity of the amplitude will lead to a constraint on the types of singularities that can appear at any loop level: only simple poles, square-root singularities and logarithmic branch cuts are allowed in local quantum field theories.

	Let us emphasize up front that we are studying only those singularities that come from isolated and non-degenerate saddle points (as opposed to manifolds of saddles). It is well-known that there exist singularities not captured by these assumptions (see, e.g., \cite{doi:10.1063/1.1703752,Mizera:2021icv}), and in fact, they are rather ubiquitous when massless particles are involved. For simplicity we also do not give a more complicated discussion of dimensional regularization and/or renormalization, which means our discussion applies only to thresholds unaffected by UV and IR divergences (the Feynman integral itself does not necessarily need to be finite).
	
	Recall from Sec.~\ref{sec:general} that Feynman integrals can be written as
	\be
	\I = (-i\hbar)^{-d} \lim_{\eps \to 0^+} \int_0^\infty \frac{\d^{\E} \alpha}{\U^{\D/2}}\, \N \exp \left[ \frac{i}{\hbar} \left(\V + i\eps\, {\textstyle\sum_{e=1}^{\E}}\alpha_e\right)\right]
	\label{eq:FeynmanI}
	\ee
	with the overall degree of divergence $d = \E - \L\D/2 + d_\N$ and a numerator factor $\N$, which without loss of generality can be taken to be homogeneous of degree $d_\N \leq 0$, as shown in App.~\ref{app:parametric}. The absence of an overall UV divergence means $d > 0$. For the purpose of this section, it will be sufficient to work with the Feynman $i\eps$ prescription, since we will take the limit $\eps \to 0^+$ at the end anyway.
	
	Without loss of generality we also assume the Feynman integral is one-vertex irreducible (1VI), i.e., it cannot be disconnected by a removal of a single vertex. Since Feynman integrals of non-1VI diagrams simply factorize as a product of Feynman integrals of its 1VI components, one can apply the analysis of this section to each component separately and multiply the results.
	
	As a warm-up, we first consider the simplest case of bulk saddles, followed by a parallel, but more involved, discussion of boundary saddles.
	
	\subsubsection{Bulk saddles}
	
	Let us consider isolated bulk saddles, i.e., points $\alpha_e^\ast \in \C^\ast$ at which the first derivatives of the action vanish,
	\be
	\partial_{\alpha_{e}} \V^\ast = 0
	\ee
	for all internal edges $e$. We consider local fluctuations $\delta\alpha_e$ around such saddles,
	\be
	\alpha_e = \alpha_e^\ast + \delta\alpha_e + \mc{O}\! \left[ \left(\delta\alpha_e \right)^2 \right].
	\ee
	Since they are tied to a kinematic singularity at $\Delta=0$, we also need to expand the external kinematics according to~\eqref{eq:deltaDelta} in order to stay consistent.
	Recalling that the action evaluated at the saddle, $\V^\ast$, vanishes due to homogeneity, it can be expanded to leading orders as
	\be\label{eq:V-bulk}
	\V = 0 + \delta \Delta\, \partial_\Delta \V^\ast - \frac{1}{2} \sum_{e,e'=1}^{\E} \delta\alpha_e\, \delta\alpha_{e'}\, \mathbf{H}_{e e'}^\ast + \ldots,
	\ee
	where the entries of the Hessian matrix are
	\be\label{eq:Hessian}
	\mathbf{H}_{e e'}^\ast = -\partial_{\alpha_e} \partial_{\alpha_{e'}} \V^\ast.
	\ee
	The minus sign is included for later convenience. Due to the overall $\mathrm{GL}(1)$ redundancy, one of the $\delta\alpha_e$'s is not independent. For instance, fixing $\alpha_{\E}=1$ means we take $\delta \alpha_{\E} = 0$ and likewise $\partial_{\alpha_\E}\V^\ast = 0$. The assumption of non-degeneracy of the saddle means that $\mathbf{H}^\ast$ has rank $\E-1$. One can in principle also include terms proportional to $\delta \Delta\, \delta\alpha_e$, but we do not write them out here since they only contribute at subleading orders to the final answer.
	Using~\eqref{eq:deltaDelta}, we have $\delta \Delta = \Delta$ to leading orders.

	The difference to the traditional saddle-point analysis \cite{doi:10.1098/rspa.1997.0122,10.1215/S0012-9074-02-11221-6} is that the integral in~\eqref{eq:FeynmanI} actually has a $\GL(1)$ orbit of saddles, labeled by a continuous parameter $\lambda > 0$. There are two ways of proceeding: either we first integrate out $\lambda$ followed by localization on $\alpha_e^\ast$, or the other way around. While the first method
	in principle preserves the $\GL(1)$ orbits, it results in a more complicated analysis, so instead, we localize the $\alpha_e$ variables first and return back to an associated subtlety shortly.
	Accordingly, we separate the projective part of the integral from that over the overall-scale $\lambda$:
	\be\label{eq:I-lambda}
	\I = (-i\hbar)^{-d} \lim_{\eps\to0^+}  \int_0^\infty \frac{\d\lambda}{\lambda^{1-d}} \int \frac{\d^{\E} \alpha}{\GL(1)} \frac{\N}{\U^{\D/2}} \exp \left[ \frac{i\lambda}{\hbar}\left(\V + i\eps\, {\textstyle\sum_{e=1}^{\E}}\alpha_e\right)\right],
	\ee
	and
	treat $\lambda$ as a constant under the integral over $\alpha$'s.
	
	Since the action $\V$ is holomorphic in the Schwinger parameters, all its isolated critical points are saddles, meaning that there is a real $(\E{-}1)$-dimensional manifold $\cal J$ along which $\V$ decreases away from the saddle, and likewise a manifold $\cal K$ of the same dimension along which $\V$ increases. Therefore, the integral over the Lefschetz thimble $\cal J$ always converges and can be used to approximate the projective part of the integral close to the saddle:
	\begin{multline}
	\label{eq:thimble}
	    \int_{\cal J} \frac{\d^{\E} \alpha}{\GL(1)} \frac{\N}{\U^{\D/2}} \exp \left[ \frac{i\lambda}{\hbar} \left(\delta \Delta\, \partial_\Delta \V^\ast -\frac{1}{2} \sum_{e,e'=1}^{\E} \delta\alpha_e\, \delta\alpha_{e'}\, \mathbf{H}_{e e'}^\ast + i\eps\, {\textstyle\sum_{e=1}^{\E}}\alpha_e + \ldots \right) \right] \\ \approx 
	    \frac{(2\pi \hbar/\lambda)^{(\E-1)/2}\, \N^\ast}{(\U^\ast)^{\D/2} (i^{\E-1}\det' \mathbf{H}^\ast)^{1/2}} \exp \left[ \frac{i\lambda}{\hbar} \left(\delta \Delta\, \partial_\Delta \V^\ast + i\eps\, {\textstyle\sum_{e=1}^{\E}}\alpha_e^\ast\right) \right]\,.
	\end{multline}
	Every starred quantity means it is evaluated at the saddle.
	On the right-hand side, the determinant of the Hessian needs to be reduced to account for the action of $\GL(1)$ and is given by
	\be
	{\det}' \mathbf{H}^\ast = \frac{1}{(\alpha_e^\ast)^2} \det \mathbf{H}^\ast {}^{[e]}_{[e]},
	\label{eq:reducedH}
	\ee
	where the notation means that prior to computing the determinant we remove the $e$-th column and row from $\mathbf{H}^\ast$ for arbitrary $e$. In~\eqref{eq:thimble}, the square root of the reduced Hessian is evaluated on its principal branch.
	
	The resulting expression~\eqref{eq:thimble} is simple to understand physically: we have $\D$ copies of $\U^{-1/2}$ coming from localizing the path integral for the worldline fields, and the Hessian factor from localizing the leftover Schwinger parameters. It would be interesting to extend this analysis to subleading orders, but
	a consistent expansion around the saddles would, in that case, require including new effects such as boundary contributions, Stokes phenomena, and resurgence; see, e.g., \cite{Boyd1999}.
	
	At this stage we need to relate the integral over the thimble $\cal J$ to the original Feynman integral, which was over the $\alpha$-positive contour $\R_{+}^{\E-1}$. In general, this is an incredibly difficult procedure which lies at the heart of analyzing the analyticity of Feynman integrals. It amounts to asking whether a given saddle contributes to the scattering amplitude on a given sheet in the kinematic space (note that thimbles themselves depend on the external kinematics, as do the positions of saddles). A formal answer to this question is obtained by computing the intersection number $c$ between the original contour $\R_{+}^{\E-1}$ and the anti-thimble $\cal K$, as exemplified in Sec.~\ref{sec:thimbles}, though similar manipulations become prohibitively cumbersome for more complicated Feynman integrals.
	At any rate, we will denote by $c$ the integer counting if, and how many times, the saddle contributes to the original integral. An $\alpha$-positive saddle always gives $c=1$, while $c=0$ corresponds to a saddle that does not contribute on a given sheet.
	
	The subtlety we need to discuss now is that the above approximations break down if $\lambda$ is too small, since then the exponential is not actually sharply peaked. This means that we have introduced a spurious UV divergence, and before putting everything together, we need to cut off the $\lambda$ integral from the bottom at some fixed $\lambda_0 > 0$. This leaves us with
	\be
	\I \approx \lim_{\eps \to 0^+} \frac{c\, (2\pi i)^{(\E-1)/2}\, \N^\ast}{(\U^\ast)^{\D/2} (i^{\E-1} \det' \mathbf{H}^\ast)^{1/2}} (-i\hbar)^{\upgamma}  \int_{\lambda_0}^\infty \frac{\d\lambda}{\lambda^{1+\upgamma}} \exp \bigg[ \frac{i\lambda}{\hbar} \bigg(\Delta \, \partial_\Delta \V^\ast + i\eps\, \underbrace{{\textstyle\sum_{e=1}^{\E}}\alpha_e^\ast}_{\eta}\bigg)\bigg],
	\ee
	where we introduced
	\be
	\upgamma = \tfrac{\E-1}{2} - d = \frac{\L\D - \E - 2d_\N - 1}{2},
	\ee
	which is always $\upgamma \in \frac{\Z}{2}$.
	The phase $\partial_\Delta \V^\ast$ is crucial for determining the direction of the branch cut beginning at $\Delta=0$.
	
	At this stage, we can simply perform the final integration to give at the leading order as $\Delta \to 0$:
		\begin{align}
		(-i\hbar)^{\upgamma} \!\!\int_{\lambda_0}^\infty& \frac{\d\lambda}{\lambda^{1+\upgamma}} \exp \left[ \frac{i\lambda}{\hbar} \left(\Delta \, \partial_\Delta \V^\ast + i\eps\eta\right) \right]\nn\\ &\approx \begin{dcases}
			\Gamma(-\upgamma) \left[-\Delta \, \partial_\Delta \V^\ast - i\eps\eta\right]^\upgamma
			&\text{for } \upgamma < 0,\\
			- \log \left[-\Delta \sgn(\partial_\Delta \V^\ast) - i\eps \sgn(\eta)\right]
			\quad &\text{for } \upgamma = 0.
		\end{dcases}
		\end{align}
		In both cases, the leading behavior in $\Delta$ turns out to be independent of the cutoff $\lambda_0$. The sign of $\partial_\Delta \V^\ast$ dictates the direction of the branch cut (if present), while the sign of $\eta$ selects the causal branch. In the logarithmic case, $\rho=0$, only the signs of $\partial_\Delta \V^\ast$ and $\eta$ matter for the leading behavior as $\Delta \to 0$. For $\upgamma > 0$, the answer is non-singular, which of course does not exclude the possibility of a (numerically suppressed) branch cut extending from $\Delta=0$. In conclusion, singularities coming from the saddle points considered in this section can be (multiple) poles, as well as square-root and logarithmic singularities. This situation will be further improved in Sec.~\ref{sec:bound}, where we will put a lower bound on $\upgamma$.
		
		Note that for $\alpha$-positive singularities with real $\Delta$, the arguments simply acquire small negative imaginary parts, since then $\eta = \sum_{e=1}^{\E}\alpha_e^\ast > 0$. Additionally, the sign of $\partial_{\Delta}\V^\ast$ determines whether the branch cut runs in the positive or negative direction. It therefore selects a branch of the logarithm and square root:
		\be
		\frac{1}{\sqrt{-\Delta \partial_{\Delta}\V^\ast - i\eps}}, \qquad \log\left[ - \Delta \sgn(\partial_{\Delta}\V^\ast) - i\eps\right].
		\ee
		If $\Delta \partial_\Delta \V^\ast < 0$, the branch is the principal one, and hence the $i\eps$ can be omitted.
		
		To summarize, we found that fluctuation around a bulk saddle give the following approximation to the integral close to $\Delta=0$:
		\be\label{eq:expansion}
		\I \approx \frac{c\,(2\pi i)^{(\E-1)/2}\, \N^\ast}{(\U^\ast)^{\D/2} (i^{\E-1} \det' \mathbf{H}^\ast)^{1/2}} \begin{dcases}
			\Gamma(-\upgamma) \left[-\Delta\partial_\Delta \V^\ast - i\eps \eta \right]^\upgamma
			&\text{for } \upgamma < 0,\\
			- \log \left[-\Delta \sgn(\partial_\Delta \V^\ast) - i\eps \sgn(\eta) \right]
			\quad &\text{for } \upgamma = 0.
		\end{dcases}
		\ee
		For $\alpha$-positive singularities we set $c=1$ and $\eta=1$. 
		If multiple saddles are present, one needs to sum over their contributions.
		
	We can do an analogous expansion around leading second-type singularities, with $\U=0$ and $\F \propto \U$ on the saddle. Since the action $\V=\frac{\F}{\U}$ goes to a constant for second-type branch points, we must make sure to factor $\U$ out of the exponential before expanding. To this end, we start with a slightly modified version of the Schwinger-parametrized integral, where we have introduced an auxiliary parameter $\alpha_0$ to incorporate $\U$ as in Sec.~\ref{sec:general-thresholds};
	\begin{equation} 
	    \I = \frac{(-i\hbar)^{\D/2}}{
	    \Gamma\left(\D/2-d \right)}
	    \lim_{\eps\to0^+} \int \d^{\E+1} \alpha \, \alpha_0^{\D/2-d-1} \frac{\N}{\U^{\D/2}} \exp \left[ \frac{i}{\hbar}\left(\frac{\F+\alpha_0 \, \U}{\U} + i\eps\, {\textstyle\sum_{e=0}^{\E}}\alpha_e\right)\right].
	\end{equation}
	Next, we change variables by taking $\alpha_e \to \lambda \alpha_e$ for $e = 0,1,\ldots,\E$ followed by rescaling $\lambda \to \U \lambda$ to get
	\begin{equation}
	    \I = \frac{(-i\hbar)^{\D/2}}{
	    \Gamma\left(\D/2-d \right)}
	    \lim_{\eps\to0^+} \int_0^\infty \frac{\d\lambda}{\lambda^{1-\D/2}} \int \frac{\d^{\E+1} \alpha}{\GL(1)}\, \N \exp \left[ \frac{i\lambda}{\hbar}\left(\F + \alpha_0 \, \U + i\eps\, \U \, {\textstyle\sum_{e=0}^{\E}}\alpha_e\right)\right].
	\end{equation}
	Using this form of the integrand, we can expand around $\F+\alpha_0 \, \U$ in the same way as for first-type singularities above, with $\F+\alpha_0 \, \U \approx \Delta + \ldots$ (for simplicity we assume $\N$ remains finite). The analysis is analogous to before, except we have a different power of $\lambda$, and the integration in the $\alpha$'s is over $\E+1$ parameters instead of $\E$. As a result, in the vicinity of the second-type singularity, the integral behaves as $\Delta^\nu$ when $\nu < 0$ and as $\log \Delta$ when $\nu = 0$ with
	\be 
	    \nu = \frac{\E-\D}{2} \,.
	\ee 
	Recall that since second-type singularities are at the boundaries of physical regions and have vanishing $\U$, they are not singularities of the original integral with the integration contour over positive $\alpha$. The coefficient in front of this expansion therefore vanishes on the physical sheet, but the above behavior is the one expected on the sheets at which the integral becomes singular.

	\subsubsection{\label{sec:boundary-saddles}Boundary saddles}
				
	The above derivation can easily be extended to subleading Landau singularities with more bookkeeping. In this case, a subset of Schwinger parameters of the original Feynman diagram $G$ vanishes. Let us call the subdiagram formed by these edges $\gamma \subset G$. It will in general be edge-disconnected, i.e.,
                \be
                \gamma = \gamma_1 \sqcup \gamma_2 \sqcup \cdots \sqcup \gamma_{\CC_\gamma},
                \ee
				where $\CC_\gamma$ is the number of 1VI components. Following standard notation, we will refer to the reduced diagram, obtain by starting with $G$ and shrinking every $\gamma_i$ to a point, with $G/\gamma$. An example was given in Fig.~\ref{fig:reduced}. Since now we have plenty of different diagrams, we will add a subscript to distinguish between distinct quantities, e.g., $\V_G$ is the original action and $\V_{\gamma_i}$ is that of the subdiagram $\gamma_i$ on its own.

        	    Before starting a detailed analysis, let us explain the physical picture to guide our calculations. Directly at the threshold, the on-shell momenta of the reduced diagram $G/\gamma$ correspond to a classically-allowed scattering process \cite{Coleman:1965xm}. This forces the momenta of all the propagators within $\gamma$ to be rerouted in a specific way. It means that we expect to find a similar formula to the one for the leading singularity~\eqref{eq:expansion}, except for renormalization factors $\RR_{\gamma_i}$ attached to each effective vertex in the reduced diagram $G/\gamma$. Each $\RR_{\gamma_i}$ is the value of the Feynman integral associated to $\gamma_i$ with the flow of momenta determined by the reduced diagram. This logic predicts that the strength of singularity will be determined by
        	    \be\label{eq:nu-subleading}
        	    \boxed{
        	    \upgamma = \frac{\L_{G/\gamma}\D - \E_{G/\gamma} - 2d_{\N_{G/\gamma}} -1}{2},}
        	    \ee
        	    giving a $\Delta^{\upgamma}$ divergence when $\upgamma < 0$ and a $\log \Delta$ divergence when $\upgamma=0$. If $\upgamma>0$, the subleading singularity remains finite.
        	    
        	    Let us now turn this intuition into equations. In the above notation, subleading Landau equations determining boundary saddle points are
                \be
                \partial_{\alpha_e} \V^\ast_G  = 0 \qquad \text{for }e\in G/\gamma,
                \ee
                together with
                \be\label{eq:alpha-gamma}
                \alpha_e^\ast = 0 \qquad \text{for } e\in\gamma.
                \ee
                As before, we assume that all critical points are isolated and non-degenerate and the Feynman integral of the reduced diagram is UV/IR finite, which includes the conditions $d_{G/\gamma} > 0$ and that it does not have soft-collinear divergences or its generalizations \cite{Arkani-Hamed:2022cqe}. It is not required that the diagrams in $\gamma$ are finite. The results might apply even in the presence of UV/IR divergences in $G/\gamma$, but one has to be careful with commutation of limits as the dimension-regularization parameter $\epsilon$ is taken to zero.
                
                Near such a point, the action can be expanded as
                \be
                \V_G = 0 + \delta\Delta\, \partial_\Delta \V^\ast_G + \sum_{e \in \gamma} \delta \alpha_e\, \partial_{\alpha_e} \V^\ast_G - \frac{1}{2} \sum_{e,e' \in G/\gamma} \delta\alpha_e\, \delta\alpha_{e'}\, \mathbf{H}_{e e'}^\ast + \ldots,
                \ee
				where the Hessian $\mathbf{H}_{ee'}^\ast$ is defined as in~\eqref{eq:Hessian}. Compared to~\eqref{eq:V-bulk}, we have expanded the fluctuations $\delta\alpha_{e}$ to leading orders: linear for $e\in \gamma$ and quadratic for $e \in G/\gamma$. Given~\eqref{eq:alpha-gamma}, we have $\delta \alpha_e = \alpha_e$ for $e\in \gamma$. As before, the terms of order $\delta \Delta\, \delta \alpha_e$ will only contribute to a  subleading correction and hence are omitted here. 
				
				Let us now look at the behavior of the integrand as $\alpha_e \to 0$ for all $e \in \gamma$. In this limit we have
                \be\label{eq:U-expansion}
                \U_G = \U_{G/\gamma} \prod_{i=1}^{\CC_\gamma} \U_{\gamma_i} + \ldots.
                \ee
				This can be clearly seen from the definition of the Symanzik polynomial~\eqref{eq:U}. Out of the spanning trees $T$ contributing to $\U_G$, only those survive at the leading order that have most edges within each $\gamma_i$. This means we are left with only the $T$'s that also spanning trees of each $\gamma_i$, giving~\eqref{eq:U-expansion}. Using entirely analogous derivation, one can show that
				\be\label{eq:VG}
				\V_G = \V_{G/\gamma} + \ldots,
				\ee
				see, e.g., \cite[Sec.~4.2]{Mizera:2021icv} for details.
				We will also assume that the numerator factors similarly, i.e.,
                \be
                \N_G = \N_{G/\gamma} \prod_{i=1}^{\CC_\gamma} \N_{\gamma_i} + \ldots.
                \ee
                To summarize, the Feynman integral near the threshold can be approximated by
	            \begin{align}\label{eq:IG-subleading}
            	\I_G \approx &\; (-i\hbar)^{-d_G} \!\!\lim_{\eps\to0^+} \int_0^\infty \frac{\d\lambda}{\lambda^{1-d_G}} \prod_{i=1}^{\CC_\gamma} \int \frac{\d^{\E_{\gamma_i}}\alpha}{\U_{\gamma_i}^{\D/2}} \N_{\gamma_i} \exp \left[{\frac{i\lambda}{\hbar} \sum_{e \in \gamma_i} \alpha_e(\partial_{\alpha_e} \V_{G}^\ast {+} i\eps)}\right]\\
            	\times & \int \frac{\d^{\E_{G/\gamma}} \alpha}{\GL(1)} \frac{\N_{G/\gamma}}{\U_{G/\gamma}^{\D/2}} \exp \left[ \frac{i\lambda}{\hbar}\left( \Delta \partial_\Delta \V_{G/\gamma}^\ast - \frac{1}{2} \sum_{e,e' \in G/\gamma} \delta\alpha_e \delta\alpha_{e'} \mathbf{H}_{ee'}^\ast + i\eps\, \sum_{e \in G/\gamma} \alpha_e^\ast\right)\right].\nn
            	\end{align}
            	Note that the $\GL(1)$ mods out by only one overall scale, which we use to fix one of the Schwinger parameters belonging to $G/\gamma$. The $G/\gamma$ integrals can be localized using the exact same derivation as in~\eqref{eq:thimble}.
                
                The contribution from each $\gamma_i$ is captured entirely by the integrals
				\be\label{eq:Rgamma}
				\lim_{\eps\to 0^+} \int \frac{\d^{\E_{\gamma_i}} \alpha}{\U_{\gamma_i}^{\D/2}}\,\N_{\gamma_i}\, \exp \left[{\frac{i\lambda}{\hbar} \sum_{e \in \gamma_i} \alpha_e (\partial_{\alpha_e} \V_{G}^\ast + i\eps)}\right] = \left(\frac{-i\hbar}{\lambda}\right)^{\!d_{\gamma_i}} \RR_{\gamma_i}\,,
				\ee
				where $\RR_{\gamma_i}$ is defined by this equation. Note that since each $\partial_{\alpha_e} \V_{G}^\ast$ is well-defined whenever the masses $m_e$ are non-zero, the factors $\RR_{\gamma_i}$ are independent of $\lambda$.
 				The exponent features the combination
				\be
				\partial_{\alpha_e} \V_G^\ast = (q_e^\ast)^2 - m_e^2,
				\ee
				which is just the Feynman integral for $\gamma_i$ with the integral momenta $(q_e^\ast)^\mu$ pinned to the values determined by ths subleading Landau equations. This is precisely the sense in which the momenta on the anomalous threshold of $G/\gamma$ impose the kinematics within each $\gamma_i$ to be rerouted in a specific way. A quick trick to turn this into a calculable formula is to treat $\RR_{\gamma_i}$ as the original Feynman integral $\I_{\gamma_i}$, except for turning off the external momenta and absorbing the kinematic dependence into shifts of the internal masses: 
				\be
				\RR_{\gamma_i} = \I_{\gamma_i} \big|^{p_i^\mu = 0}_{m_e^2 = -\partial_{\alpha_e}\V_{G}^\ast},
				\ee
				where it is understood that the numerator remains untouched.
				For example, if $\gamma_i$ was a single edge $e$, we would have $\RR_{e} = -1/(\partial_{\alpha_e}\V_{G}^\ast + i\eps)$. To summarize, $\RR_{\gamma_i}$ simply renormalize the vertices to which they are contracted within $G/\gamma$.
				
				Finally, we are left with the overall-scale integral. Collecting powers of $\lambda$, from~\eqref{eq:IG-subleading} we have $\lambda^{d_G-1}$, from localizing the Schwinger parameters in $G/\gamma$ we get $\lambda^{(1-\E_{G/\gamma})/2}$, and $\lambda^{-d_{\gamma_i}}$ from every $\gamma_i$ contribution in~\eqref{eq:Rgamma}. Since the total number of loops and edges is conserved, we have $\sum_{i=1}^{\CC_\gamma} d_{\gamma_i} = d_{G} - d_{G/\gamma}$. The leftover $\lambda_0$-regulated integral is therefore
            	\begin{align}
            	\I_G \approx \lim_{\eps \to 0^+} &\frac{c\, (2\pi i)^{(\E_{G/\gamma}-1)/2}\, \N_{G/\gamma}^\ast \prod_{i=1}^{\CC_\gamma} \RR_{\gamma_i}}{(\U_{G/\gamma}^\ast)^{\D/2} (i^{\E_{G/\gamma}-1} \det' \mathbf{H}^\ast)^{1/2}}\\
            	&\times (-i\hbar)^{\upgamma}  \int_{\lambda_0}^\infty \frac{\d\lambda}{\lambda^{1+\upgamma}} \exp \left[ \frac{i\lambda}{\hbar} \bigg(\Delta \partial_\Delta \V_{G/\gamma}^\ast + i\eps \eta\bigg) + \ldots \right],\nn
            	\end{align}
				where $\eta = \sum_{e \in G/\gamma} \alpha_e^\ast$ and the integer $c$ measures how much the saddle contributes for a given kinematic point. The exponent $\upgamma = (\E_{G/\gamma} - 1)/2 - d_{G/\gamma}$ works out to be precisely~\eqref{eq:nu-subleading}. At this stage we can simply apply the identity~\eqref{eq:expansion}. The answer can be summarized as
				\be\label{eq:subleading-expansion}
				\boxed{
        		\I_G \approx \frac{c\,(2\pi i)^{(\E_{G/\gamma}-1)/2}\, \N_{G/\gamma}^\ast \prod_{i=1}^{\CC_\gamma} \RR_{\gamma_i}}{(\U_{G/\gamma}^\ast)^{\D/2} (i^{\E_{G/\gamma}-1} \det' \mathbf{H}^\ast)^{1/2}} \!\begin{dcases}
        			\Gamma(-\upgamma) \left[-\Delta \partial_\Delta \V_{G/\gamma}^\ast - i\eps \eta \right]^\upgamma  &\text{for } \upgamma < 0,\\
        			- \log \left[-\Delta \sgn(\partial_\Delta \V_{G/\gamma}^\ast) - i\eps \sgn(\eta) \right]  &\text{for } \upgamma = 0,
        		\end{dcases}}
        		\ee
        		where $\upgamma$ was given in~\eqref{eq:nu-subleading}.
				It is the same as the result for a leading singularity~\eqref{eq:expansion} of $G/\gamma$  multiplied by the renormalization factors of the shrunken subdiagrams. Recall that for a non-1VI diagram, the divergence is multiplicative across all its 1VI components.

				A couple of comments are in order. First, let us confirm $\GL(1)$-invariance of the formula~\eqref{eq:subleading-expansion}. Under rescaling all $\alpha_e \to \lambda \alpha_e$ with $\lambda > 0$, we have
				\begin{gather}
				\N_{G/\gamma}^\ast \to \lambda^{d_{\N_{G/\gamma}}} \N_{G/\gamma}^\ast, \qquad \U_{G/\gamma}^\ast \to \lambda^{\L_{G/\gamma}} \U_{G/\gamma}^\ast,\\
				{\det}' \mathbf{H}^\ast \to  \lambda^{-\E_{G/\gamma}-1} {\det}' \mathbf{H}^\ast, \qquad
				\partial_{\Delta} \V_{G/\gamma}^\ast \to \lambda\, \partial_{\Delta} \V_{G/\gamma}^\ast, \qquad \eta \to \lambda \eta,
				\end{gather}
				with all the remaining quantities unaffected. This transformation leaves~\eqref{eq:subleading-expansion} invariant. Note that in the logarithmic case the factors $\sgn(\partial_\Delta \V_{G/\gamma}^\ast)$ and $\sgn(\eta)$ are needed only to determine the direction of the branch cut and its causal branch.
				
				Notice that the exponent $\upgamma$ scales as minus the degree of divergence $d_{G/\gamma}$: this means the more UV divergent is, the lower $\rho$ and hence the singularity $\Delta^\rho$ becomes more severe. To be more precise, the most divergent contributions come from subleading singularities that minimize the number of loops and maximize the number of propagators. Presence of numerators cannot make the divergence worse. Naively, it appears that Feynman integrals can have arbitrarily-divergent singularities. 
				However, in Sec.~\ref{sec:bound} we will see that, assuming analyticity, one can in fact put a lower bound on $\rho$.
				
				As the simplest example, let us illustrate the formula~\eqref{eq:subleading-expansion} applied to one-particle exchanges. For example, a scalar particle $G/\gamma = e$ propagating between two subdiagrams $\gamma_L$ and $\gamma_R$ has
				\be
				\V_{G/\gamma} = \alpha_e (P_e^2 - m_e^2),
				\ee
				where $P_e^\mu$ is the total momentum flowing through the edge $e$ and $m_e$ its mass.
				The exponent counting~\eqref{eq:nu-subleading} gives 
				$\upgamma = -1$, which leads to a simple pole. All the remaining quantities equal to $1$. Hence, we find near the threshold $\Delta = P_e^2 - m_e^2$:
				\be\label{eq:exchange}
				\I \approx - \frac{\RR_{\gamma_L} \RR_{\gamma_R}}{P_e^2 - m_e^2 + i\eps}.
				\ee
				In Sec.~\ref{sec:threshold-examples} we will give a generalization of this result to a normal threshold with $\E$ exchanged particles. In Sec.~\ref{sec:one-loop} we study another family of examples: the one-loop anomalous thresholds.
				
				In the case of scalar Feynman integrals, Landau conjectured that an extension of~\eqref{eq:subleading-expansion} approximates the leading \emph{discontinuous} behavior near anomalous thresholds even when $\upgamma > 0$ with the behavior $\Delta^{\upgamma}\log\Delta$ if $\upgamma \in \Z_{\geq 0}$ and $\Delta^\upgamma$ otherwise \cite{Landau:1959fi,pham2011singularities}, see also \cite{Hannesdottir:2021kpd}.
				A similar result was argued for in \cite{Polkinghorne1960,Eden:1966dnq}, however the formula quoted there was not compatible with the effective field theory picture explained above (it used $\L_{G}$ instead of $\L_{G/\gamma}$) and hence incorrect unless $\gamma$ does not have loops.
				One can also derive similar expressions for second- and mixed-type singularities, whose divergence turns out to be characterized by a different exponent to $\upgamma$ \cite{nakanishi1971graph,Greenman1969}. Progress in generalizations to massless cases includes \cite{Kinoshita:1962ur,Kinoshita:1975ie}.
				In the more modern literature, threshold expansion is a part of the expansion-by-regions strategy for evaluating Feynman integrals
				\cite{Beneke:1997zp,Smirnov:2002pj,Ferroglia:2002mz,Semenova:2018cwy}, which has been implemented in numerical software such as \texttt{asy2.m} \cite{Jantzen:2012mw}.

				\subsection{\label{sec:bound}Bound on the type of singularities from analyticity}
				
				At this stage we are well-equipped to explain a surprising connection between analyticity of the S-matrix and the type of singularities that can potentially appear. We focus only on singularities of the type considered above, i.e., we assume all saddle points are isolated and non-degenerate.
				
				Let us first ask about the codimension of the singularity locus in the kinematic space. The saddle point equations give $\E_G$ constraints on $\E_G {-} 1$ independent Schwinger parameters (recall the $\GL(1)$ redundancy), leaving \emph{at least} one extra constraint on the external kinematics, meaning it translates to a codimension-$1$ or higher singularity.
				
				In order to quantify the above statement better, it is useful to reformulate the saddle-point conditions as Landau equations \cite{Bjorken:1959fd,Landau:1959fi,10.1143/PTP.22.128} (since the loop-integration is Gaussian, once can always go back and forth between the two formulations, cf.~\eqref{eq:qe}). Recall from Sec.~\ref{sec:general-thresholds} that together, they read
				\be\label{eq:on-shell}
				q_e^2 - m_e^2 = 0,
				\ee
				which are on-shell conditions for every internal edge $e \in G/\gamma$ and
				\be
				\sum_{e \in L} \pm \alpha_e q_e^\mu = 0,
				\ee
				which are the constraints of local interactions at every vertex, imposed for every loop $L$. Notice that the conditions~\eqref{eq:on-shell}
				are a subsystem of equations involving only the internal and external kinematics, but not the Schwinger parameters $\alpha_e$. It means that they put $\E_{G/\gamma}$ constraints on the $\L_{G/\gamma} \D$ components of the loop momenta by themselves. Hence, the codimension of the associated singularity is lower-bounded by
				\be
				\E_{G/\gamma} - \L_{G/\gamma} \D,
				\ee
				if it exists. If the above quantity is $\geq 2$, the Schwinger parameters have solutions lying on continuous manifolds, i.e., saddle points are no longer isolated.
				
				Now, for the S-matrix to be analytic, it can have at most codimension-$1$ singularities, i.e., those that can be written in the form~\eqref{eq:Delta2}. We suspect that all codimension-$2$ and higher singularities are subvarieties of more subleading singularities of lower codimension, in such as way that the union of all singularities is codimension-$1$. We verify this hypothesis on a couple of explicit examples in Sec.~\ref{sec:codimension}.

				Regardless of the explanation, we can simply \emph{assume} that the S-matrix is complex analytic, for which the necessary condition is that
				\be
				\label{eq:analytic-bound}
				\E_{G/\gamma} - \L_{G/\gamma} \D \leq 1
				\ee
				for every choice of $\gamma$ that leads to a singularity.
				At this stage it is useful to allow for non-localities in the form of higher-derivative propagators. After Schwinger-parametrizing propagators of the form $-1/(q_e^2 - m_e^2 + i\eps)^{p_e}$, one finds that the numerator is rescaled by
				\be\label{eq:codim-bound}
				\N_G \to \N_G \prod_{e \in G} \alpha_e^{p_e - 1},
				\ee
				up to constants. Hence, its degree changes by $d_{\N_{G/\gamma}} \to d_{\N_{G/\gamma}} + P_{G/\gamma}$, where
				\be
				P_{G/\gamma} = \sum_{e\in G/\gamma} (p_e - 1) \geq 0.
				\ee
				The inequality is strict for non-local Feynman integrals and saturated in the local cases. Using the fact that $d_{\N_{G/\gamma}} \leq 0$, along with the assumptions in~\eqref{eq:analytic-bound} and~\eqref{eq:codim-bound}, we get a bound on the exponent $\upgamma$ from~\eqref{eq:nu-subleading}:
				\be
				\upgamma \geq P_{G/\gamma} - 1.
				\ee
				Hence, when assuming complex analyticity of the S-matrix in local theories with $P_{G/\gamma} = 0$ we have
				\be
				\boxed{\upgamma \geq -1,}
				\ee
				which means the types of singularities we can get are only simple poles, inverse square-roots, and logarithmic singularities. Non-local interactions can introduce higher-degree poles in the S-matrix.
				
				\subsection{Anomalous thresholds that mimic particle resonances}
				
				The classifications of singularities explained in the previous subsections leaves us with an intriguing possibility: some anomalous thresholds give rise to simple pole singularities, as opposed to branch cuts, whenever $\upgamma = -1$.
				If such a threshold exists in a physical region, an experimentalist measuring the behavior of the S-matrix near such a divergence (say, at fixed impact parameter) would detect a resonance which looks indistinguishable from one mimicking a new particle in the spectrum!\footnote{We thank Nima Arkani-Hamed for pointing out this possibility.}
				
				Before working out examples in which this behavior can happen, let us understand why normal thresholds cannot produce poles. By definition, a normal threshold always corresponds to $G/\gamma$ which looks like a string of banana diagrams (each of them consist of two vertices connected by multiple edges). It suffices to consider one such banana diagram, which we call just $G/\gamma$.
				Since in this case $\E_{G/\gamma} = \L_{G/\gamma}{+}1$, we have
				\be
				\upgamma = \frac{(\D-1)\L_{G/\gamma}}{2} - d_{\N_{G/\gamma}} -1.
				\ee
				Recalling that $d_{\N_{G/\gamma}} \leq 0$, we can only achieve $\upgamma = -1$ for normal thresholds in $\D=1$ dimension, for scalar diagrams with $d_{\N_{G/\gamma}}=0$ (independently of the number of loops).
				
				In order to find the simplest example of this phenomenon beyond tree-level, let us consider $\L_{G/\gamma}=1$. Solving for $\upgamma = -1$ gives
				\be
				\E_{G/\gamma} = \D + 1 - 2d_{G/\gamma}.
				\ee
				The most straightforward solution is a scalar $(\D{+}1)$-gon in $\D$ dimensions. Since an example for $\D=4$ would be too involved to spell out explicitly here, let us focus on the case $\D=2$ as a toy model. For example, let us take the triangle with the same kinematics as in Sec.~\ref{sec:ExampleII}. Recall that it has the leading singularity at
				\be
				u = M^2 (4m^2 - M^2)/m^2.
				\ee
				Using momentum conservation, $s+t+u = 4M^2$, in terms of the $s$ Mandelstam invariant we have
				\be
				s = \underbrace{M^4/m^2 - t}_{m_\ast^2}.
				\ee
				As before, we consider the case $M > 2m > 0$, so that the anomalous threshold lies in the $s$-channel physical region with $t,u<0$ and $s>4M^2$. We consider $t$ to be fixed.
				
				Applying the formula~\eqref{eq:expansion} with $\D=2$, close to the anomalous threshold $s = m_\ast^2$ we find
				\be
				\I_{\triangle}^{\D=2} \approx \frac{ 
				2\pi i M}{m^2\sqrt{M^2 - 4m^2}} \frac{1}{s - m_\ast^2}.
				\ee
				One can confirm validity of this approximation using the full expression,
				\begin{align}
					\I_{\triangle}^{\D=2} = \frac{1}{(s-m_\ast^2) m^2} \bigg[& \frac{2}{\sqrt{1 - \frac{4m^2}{M^2}}} \log\left( \frac{-1+\sqrt{1 - \frac{4m^2}{M^2}}}{1+\sqrt{1 - \frac{4m^2}{M^2}}}\right) \\
					& + \frac{2M^2-u}{u \sqrt{1-\frac{4 m^2}{u}}} \log\left( \frac{-1+\sqrt{1 - \frac{4m^2}{u}}}{1+\sqrt{1 - \frac{4m^2}{u}}} \right)
					\bigg] \,,\nn
				\end{align}
				which is obtained by integrating~\eqref{eq:triintegral} directly.
				Simple- and double-pole singularities exist in integrable two-dimensional S-matrices, see, e.g., \cite{Coleman:1978kk,Mattsson:2001ts}.
				
				\subsection{\label{sec:codimension}Absence of codimension-\texorpdfstring{$2$}{2} singularities}
				
\begin{figure}
		\centering
		\includegraphics[scale=1]{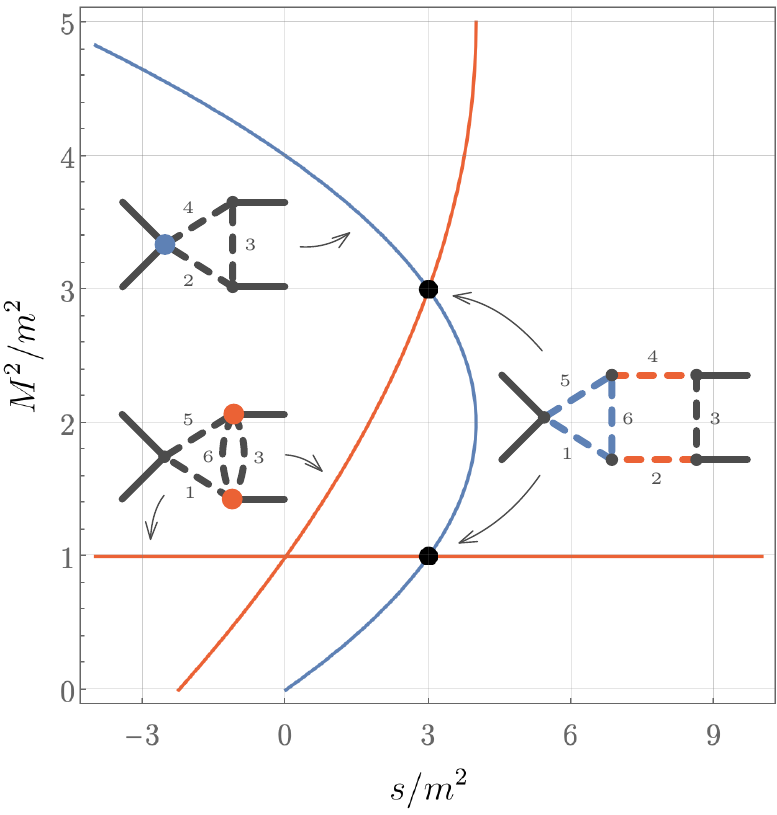}
		\caption{\label{fig:pltrb}Codimension-$2$ singularities of the triangle-box diagram (two black points) lie on the intersection of the codimension-$1$ singularities of the parachute diagram (red) and the triangle diagram (blue).}
	    \end{figure}

				Explicit computations of leading Landau equations reveal that solutions of codimension larger than one exist, i.e., they cannot be expressed as vanishing of a single polynomial $\Delta = 0$ \cite{Mizera:2021icv}. Arguably the simplest example is the triangle-box diagram illustrated in Fig.~\ref{fig:pltrb}, which has two codimension-$2$ solutions located at
				\be\label{eq:pltrb1}
				s = 3m^2 = M^2,
				\ee
				as well as
				\be\label{eq:pltrb2}
				s = 3m^2 = 3M^2,
				\ee
				where $m$ and $M$ are the masses of all the internal and external particles respectively.
				This seems in tension with analyticity, since one cannot write down a complex-analytic function with an isolated singularity like this (e.g., $\frac{\delta^+(s-3m^2)}{s-M^2}$ would not be analytic, while $\frac{1}{(s-3m^2)(s-M^2)}$ would not be codimension-$2$).
				As a resolution of this puzzle, let us demonstrate that these two solutions are really parts of codimension-$1$ subleading singularities of the same diagram.
				
				We first consider the point~\eqref{eq:pltrb1}. Solving the leading Landau equations at this kinematics does not localize the Schwinger parameters, but instead leaves us with a continuous family of solutions (on top of the usual $\GL(1)$ redundancy) that can be parametrized as
				\be
				(\alpha_1 : \alpha_2 : \alpha_3 : \alpha_4 : \alpha_5 : \alpha_6) = (x : 1{+}x : 1 : 1{+}x : x : -x).
				\ee
				Note that this singularity cannot be $\alpha$-positive. Here $x \in \C \setminus \{-1,0,\infty\}$ is not determined. In other words, the origin of the codimension-$2$ singularity is that one net constraint is put on the external kinematics instead of the Schwinger parameters. At this stage it is tempting to explore the limits $x \to -1,0,\infty$, which give candidate subleading singularities. It is actually guaranteed that solutions of the leading equations go to the subleading ones smoothly in this sense.\footnote{We thank Simon Telen for joint work on a rigorous proof of this statement.}
				
				To see this, let us consider a subset of edges $\gamma$ whose Schwinger parameters tend to zero as $\alpha_e \to \epsilon \alpha_e$. The action of the original diagram $G$ goes smoothly in this limit to that of $G/\gamma$:
				\be
				\V_G \to \V_{G/\gamma} + {\cal O}(\epsilon),
				\ee
				see, e.g., \cite[Sec.~4.2]{Mizera:2021icv}. Since the subleading Landau equations $\partial_{\alpha_{e'}}\V_{G/\gamma} = 0$ involve only the uncontracted edges $e' \in G/\gamma$, a solution of the subleading equations, if it exists, coincides with the $\epsilon\to0$ limit of the leading solution.
				
				We can exemplify this phenomenon on the above example. The case $x \to -1$ corresponds to the parachute diagram obtained by shrinking the edges $\alpha_2$ and $\alpha_4$, see Fig.~\ref{fig:pltrb}. The solution passing through this point turns out to be \cite{Mizera:2021icv}
				\be\label{eq:par}
				s = \frac{10M^2 m^2 - 9m^4 - M^4}{4m^2}.
				\ee
				On the other hand, the case $x \to 0$ means we shrink the edges $\alpha_1$, $\alpha_5$, and $\alpha_6$, resulting in a triangle diagram, see Fig.~\ref{fig:pltrb}. The corresponding solution is
				\be\label{eq:tri}
				s = - \frac{M^2 (M^2 - 4 m^2)}{m^2}.
				\ee
				Finally, there is the limit $x\to \infty$ which means the edge $\alpha_3$ shrinks to a point. The corresponding singularity does not have any solutions, because one cannot assign internal momenta to its internal edges such that without violating energy conservation, which can be diagnosed by checking that $\U_{G/\gamma}=0$ on this solution.
			    One can then verify that codimension-$2$ point $s = 3m^2 = M^2$ from~\eqref{eq:pltrb1} indeed lies on the intersection of~\eqref{eq:par} and~\eqref{eq:tri}.
				
				Repeating the same analysis for the point~\eqref{eq:pltrb2} gives us the continuous solution
				\be
				(\alpha_1 : \alpha_2 : \alpha_3 : \alpha_4 : \alpha_5 : \alpha_6) = (y : y{-}1 : 1 : y{-}1 : y : -y),
				\ee
				for $y \in \C \setminus \{0,1,\infty\}$. The limit $y \to 1$ gives the parachute diagram with the singularity at
				\be\label{eq:para2}
				m=M,
				\ee
				while $y \to 0$ corresponds to the triangle diagram with the same singularity as in~\eqref{eq:tri}. As before, the diagram at $y\to \infty$ obtained by shrinking $\alpha_3$ is non-singular. This means that the codimension-$2$ singularity $s = 3m^2 = 3M^2$ from~\eqref{eq:pltrb2} lies on the intersection of the parachute~\eqref{eq:para2} and triangle~\eqref{eq:tri} subleading singularities. This situation is illustrated in Fig.~\ref{fig:pltrb}.
				
				We have checked three other examples (see \cite[Fig.~1ghi]{Mizera:2021icv}), where codimension-$2$ leading singularities turn out to lie on intersections of subleading codimension-$1$ singularities. Based on this evidence, we conjecture that an arbitrary Feynman integral has only singularities of codimension-$1$, which is a prerequisite for its analyticity.
				
				\subsection{\label{sec:sec7-examples}Examples}
				
			    Before closing this section, we provide examples of the threshold expansion formula~\eqref{eq:subleading-expansion} for the families of banana and one-loop diagrams. 
				
				\subsubsection{\label{sec:threshold-examples}Normal thresholds}

				Let us consider leading $\alpha$-positive singularities of the banana diagrams with $\E$ edges from Fig.~\ref{fig:ngon-banana} (right), which model normal thresholds. We take all the internal masses $m_e$ to be strictly positive. The number of loops is $\L = \E-1$, which gives
				\be
				\upgamma = \frac{(\E-1)\D - \E -1}{2}.
				\ee
				Solving Landau equations tells us that the only $\alpha$-positive singularity is located at
				\be
				\Delta_{\E\text{-ban}} = s - ({\textstyle\sum}_{e=1}^{\E} m_e)^2 = 0, \qquad (\alpha_1^\ast : \alpha_2^\ast : \dots : \alpha_\E^\ast) = (\tfrac{1}{m_1} : \tfrac{1}{m_2} : \dots : \tfrac{1}{m_\E}),
				\ee
				as one may easily verify, see, e.g., \cite[Sec.~2.6]{Mizera:2021icv}. The $\U$ and $\V$ functions are given by
				\be
				\U_{\E\text{-ban}}  = \prod_{e=1}^{\E}\alpha_e \sum_{e'=1}^{\E}\frac{1}{\alpha_{e'}},\qquad
				\V_{\E\text{-ban}}  = \frac{s}{ {\textstyle\sum}_{e=1}^{\E}\tfrac{1}{\alpha_e}} - \sum_{e=1}^{\E} m_e^2 \alpha_e.
				\ee
			    Setting $\alpha_{\E}^\ast = \tfrac{1}{m_\E}$ and evaluating all the relevant quantities, we have
				\be\label{eq:UVH}
				\U_{\E\text{-ban}}^\ast = \frac{\sum_{e=1}^{\E} m_e}{\prod_{e=1}^{\E}m_e} ,\qquad
				\partial_{\Delta} \V^\ast_{\E\text{-ban}}  = \frac{1}{\sum_{e=1}^{\E} m_e},\qquad
				{\det}' \mathbf{H}^\ast = \frac{2^{\E-1} \prod_{e=1}^{\E}m_e^3 }{\sum_{e=1}^{\E} m_e}.
				\ee
				Here the saddle is non-degenerate only if all $m_e>0$, which we assume from now on. In such a case, all the quantities in~\eqref{eq:UVH} are strictly positive, and likewise the singularity is $\alpha$-positive so $c=1$. Plugging everything back into~\eqref{eq:expansion} we obtain the local expansion around the normal threshold:
				\be\label{eq:banana}
				\I_{\E\text{-ban}} \approx \frac{\pi^{(\E-1)/2} \left( \prod_{e=1}^{\E}m_e \right)^{\frac{\D-3}{2}} \RR_{\gamma_L} \RR_{\gamma_R} }{\left( \sum_{e=1}^{\E}m_e \right)^{\frac{\E(\D-1)}{2}-1}} \begin{dcases}
				\Gamma(-\upgamma) \left[ \left( {\textstyle\sum}_{e=1}^{\E} m_e \right)^{\!2} - s \right]^{\upgamma} \quad&\mathrm{for}\; \D < \tfrac{\E+1}{\E-1},\\
				-\log\left[ \left( {\textstyle\sum}_{e=1}^{\E} m_e \right)^{\!2} - s \right] \quad&\mathrm{for}\; \D = \tfrac{\E+1}{\E-1},
				\end{dcases}
				\ee
				where $\RR_{\gamma_L}$ and $\RR_{\gamma_R}$ account for possible contributions from the subdiagrams shrunken into two effective vertices of the banana diagram.
				
				Note that the leading contribution to the expansion around the normal threshold of any banana diagrams with $\E \geq 2$ is finite in $\D=4$ dimensions, so~\eqref{eq:banana} cannot  be used to determine its leading behavior.
				The case $\E=1$ reproduces the one-particle exchange behavior from~\eqref{eq:exchange}.
				Furthermore, the case $\E=2$ agrees with the previous analysis in Sec.~\ref{sec:thimbles}. Recall that our derivation was valid only for finite integrals, i.e., $\D < \frac{2\E}{\E-1}$. 
				
				Let us comment on a few more checks in the case of the sunrise diagram with $\E=3$.
				In $\D=2$, it is known to have a logarithmic singularity near the normal threshold \cite[Sec.~10]{Laporta:2004rb}. This illustrates a general principle, in which locally Feynman integrals posses a simple description, but combining this information into a global function over the whole kinematic space makes it complicated: in this case, the full Feynman integral is expressible in terms of elliptic functions. In the equal-mass case, i.e., $m_e = m$, the expansion of the sunrise integral around the threshold is given by (using \cite{Laporta:2004rb} adapted to our normalization and after Wick rotation to Lorentzian signature):
				\be
				\I_{3\text{-ban}}^{\D=2}\approx -\frac{\pi}{\sqrt{3} m^2}  \log \left(9m^2-s\right),
				\ee
				in agreement with~\eqref{eq:banana}.
				The threshold expansion of this integral with unequal masses can be alternatively verified directly from its differential equations \cite{Laporta:2004rb,Muller-Stach:2011qkg}. This, however, does not fix the constant multiplying the logarithm, which instead provides a boundary condition.
				Indeed, the above formulae for expansion around $\Delta=0$ can be thought of as providing boundary conditions for solving differential equations generally.

				\subsubsection{\label{sec:one-loop}One-loop anomalous thresholds}
		
		        Let us now apply~\eqref{eq:subleading-expansion} to the one-loop $n$-gon diagram $G=n\text{-gon}$ with $n$ external legs and $\E_G$ propagators in any dimension, see Fig.~\ref{fig:ngon-banana} (left). Its singularities are labelled by subdiagrams $\gamma$, which are collection of individual edges $\gamma_e = e$ shrunken to a point. Let us call the number of surviving (cut) propagators $\E_{G /\gamma}$, so that $|\gamma| = \E_{\gamma}$. For such a diagram, we have
		        \be
		        \U_{n\text{-gon}} = \sum_{i=1}^{n} \alpha_i, \qquad
		        \V_{n\text{-gon}} = \frac{1}{2 \, \U_{n\text{-gon}}} \sum_{i,j=1}^n \Y_{ij} \alpha_i \alpha_j,
		        \label{eq:U-V-one-loop}
		        \ee
		        where
		        \be
		        \Y_{ij} = p_{i,i+1,\ldots,j-1}^2 - m_i^2 - m_j^2.
		        \ee
		        For completeness, let us first solve the Landau equations explicitly. They impose
		        \be\label{eq:partial-V-star}
		        \partial_{\alpha_i}\V_{n\text{-gon}}^\ast =  \frac{1}{\U_{G/\gamma}^\ast} \left(  \sum_{j \in G/\gamma} \Y_{ij}\alpha_j^\ast
		        -\V^\ast_{n\text{-gon}} \right) = 0, \qquad i\in G/\gamma
		        \ee
		        for uncontracted edges and $\alpha_j^\ast = 0$ for every $j \in \gamma$. Let us assume that $\U^\ast_{G/\gamma} \neq 0$, since the converse will lead to second-type singularities only. Homogeneity of the action means $\V^\ast = 0$, so the above equation only imposes
		        \be\label{eq:Y-alpha}
		        \sum_{j \in G/\gamma} \Y_{ij} \alpha_{j}^\ast = 0.
		        \ee
		        This is a constraint on the minor $\Y_{[\gamma]}^{[\gamma]}$ of the matrix $\Y$ obtained by removing the columns and rows of the edges in $\gamma$. It has a solution only when
		        \be\label{eq:Delta-Ggamma}
		        \Delta_{G/\gamma} = \det \Y^{[\gamma]}_{[\gamma]} = 0,
		        \ee
		        which is a constraint on the external kinematics only and hence defines the Landau discriminant for this singularity.

                The solution for the Schwinger parameters is the null vector of the above minor, which is given by
		        \be\label{eq:alpha-ast}
		        \alpha_i^\ast = \eta_i \sqrt{ \lambda \det \Y_{[\gamma,i]}^{[\gamma,i]^\ast}} \,,
		        \ee
		      where $\Y_{[\gamma,j]}^{[\gamma,i]}$ denotes the minor of $\Y_{[\gamma]}^{[\gamma]}$, where row $j$ and column $i$ have been removed, $\lambda = \sgn \left( \det \Y_{[\gamma,i]}^{[\gamma,i]^\ast} \right)$, and the sign of each $\alpha_i$ is
		      \be
		      \eta_i = (-1)^{i+r} \sgn \left( \det \Y_{[\gamma,i]}^{[\gamma,r]^\ast} \right) \,,
		      \ee 
		      where $r$ is any reference vector. Using Sylvester's determinant identity,
		        \be
		        \label{eq:sylvester}
		            \det \Y_{[\gamma]}^{[\gamma]} \, \det \Y_{[\gamma,i,r_2]}^{[\gamma,i,r_1]} = \det \Y_{[\gamma,i]}^{[\gamma,i]} \, \det \Y_{[\gamma,r_2]}^{[\gamma,r_1]} - \det \Y_{[\gamma,i]}^{[\gamma,r_1]} \det \Y_{[\gamma,r_2]}^{[\gamma,i]} ,
		        \ee
		        for any references $r_1,r_2 \in (G/\gamma){\setminus} i$, noting that the left-hand side vanishes on the solution~\eqref{eq:Delta-Ggamma}, we can show by putting $r_1=r_2=j$ that $\lambda$ is independent of $i$.  
		        The condition for the $\alpha$-positive singularity is therefore that $\eta_i$ has the same sign for all $i \in G/\gamma$. 
		        To verify this is indeed a solution, we first use~\eqref{eq:sylvester}, and choose any reference vector $r\in (G/\gamma){\setminus} i$ to rewrite $\alpha_i^\ast$ as
		        \be
		        \label{eq:alpha-ast-ref}
		            \alpha_i^\ast
		            =
		            (-1)^{i+r}
		            \frac{\det \Y_{[\gamma,r]}^{[\gamma,i]^\ast}}{\sqrt{ \lambda \det \Y_{[\gamma,r]}^{[\gamma,r]^\ast}}} \,.
		        \ee
		       Naively, this expression appears to depend on $r$, but we can verify the projective invariance of $\alpha_e^\ast$ under choice of $r$: We use~\eqref{eq:sylvester} once again to evaluate the product of two sign factors:
		      \be 
		        \eta_i \eta_j = \lambda (-1)^{i+j} \sgn \left( \det \Y_{[\gamma,i]}^{[\gamma,j]^\ast} \right) \,.
		        \label{eq:etaprod}
		      \ee 
		      Comparing with~\eqref{eq:alpha-ast} we see that
		      the values of $\alpha_i^\ast$ are projectively invariant under the choice of $r$. So, we can use the Laplace expansion of the determinant and get:
		        \be
		        \sum_{j \in G/\gamma} \Y_{ij} \alpha_j^\ast =
		        \sum_{j \in G/\gamma} (-1)^{i+j}\, \Y_{ij} \frac{\det \Y^{[\gamma,j]^\ast}_{[\gamma,i]}}{\sqrt{\lambda \det \Y_{[\gamma,i]}^{[\gamma,i]^\ast}}}
		        =
		        0 \,,
		        \ee
		        where the last equation holds since
		        the right hand side is would be proportional to  $\det \Y^{[\gamma]}_{[\gamma]}$ if the condition $\Delta_{G/\gamma} = 0$ were not enforced in~\eqref{eq:alpha-ast-ref}.
		        Hence,~\eqref{eq:alpha-ast} is a solution of \eqref{eq:Y-alpha}.

		        In order to evaluate \eqref{eq:subleading-expansion}, we will need to compute a few quantities. First, we choose a variable $p_{ij}^2$ in which we compute the expansion around the singularity, i.e., we set $\Delta=p_{ij}^2-(p_{ij}^\ast)^2$. Then,
		        \be 
		            \partial_\Delta \V_{n\text{-gon}}^\ast = \frac{\alpha_i^\ast \alpha_j^\ast}{\U_{G/\gamma}^\ast}
		            =
		            \lambda (-1)^{i+j} \frac{\det \Y_{[\gamma,i]}^{[\gamma,j]^\ast}}{\U_{G/\gamma}^\ast},
		        \ee
		        where we have used Sylvester's determinant identity from~\eqref{eq:sylvester} and the values of the signs $\eta_i$ from~\eqref{eq:etaprod}.
		        Applying the derivative $\partial_{\alpha_i} \partial_{\alpha_j}$ to \eqref{eq:U-V-one-loop}, we find that the entries of the Hessian matrix read
		        \be
		        \mathbf{H}_{ij}^\ast = -\partial_{\alpha_i} \partial_{\alpha_j} \V^\ast_{n\text{-gon}} = -\frac{\Y_{ij}^\ast}{\U_{G/\gamma}^\ast},
		        \ee
		        for every $i,j \in G/\gamma$.
		        For the purpose of computing its reduced determinant defined in Eq.~\eqref{eq:reducedH}, we can pretend that $\alpha_r^\ast$ is fixed by $\GL(1)$ and we remove the $r$-th row and column: 
		        \be
		        {\det}' \mathbf{H}^\ast = \frac{1}{(\alpha_r^\ast)^2 (-\U^\ast_{G/\gamma})^{\E_{G /\gamma}-1}} \det \Y^{[\gamma,r]^\ast}_{[\gamma,r]} = \frac{\lambda}{ (-\U^\ast_{G / \gamma})^{\E_{G /\gamma}-1}}.
		        \ee
		        Finally, for the contracted edges $i \in \gamma$, we evaluate the renormalization factors
		        $\RR_i = 1/\partial_{\alpha_i} \V_{n\text{-gon}}^\ast$. We first note that by \eqref{eq:U-V-one-loop} and \eqref{eq:alpha-ast},
		        \be
		        \partial_{\alpha_i} \V_{n\text{-gon}}^\ast  = \frac{1}{\U_{G / \gamma}^\ast} \sum_{j \in G/\gamma} \Y_{ij}^\ast \eta_{j} \sqrt{ \lambda \det \Y_{[\gamma,j]}^{[\gamma,j]^\ast}} \,,
		        \ee
				and we can use the
				Sylvester determinant identity to write
				\be
		            \det \Y_{[\gamma,j]}^{[\gamma,j]} \, \det \Y_{[\gamma\setminus  i  ]}^{[\gamma \setminus i]} = \det \Y_{[\gamma]}^{[\gamma]} \, \det \Y_{[\gamma \setminus i,j]}^{[\gamma \setminus i ,j]} - \left(\det \Y_{[\gamma \setminus i,j]}^{[\gamma]}\right)^2 \,.
		        \ee
				On support of $\det \Y_{[\gamma]}^{[\gamma]}=0$, we get
				\be
				    \partial_{\alpha_i} \V_{n\text{-gon}}^\ast  = \frac{1}{\U_{G / \gamma}^\ast} \sum_{j \in G/\gamma} \Y_{ij}^\ast \left(-1\right)^{i+j} \frac{\det \Y_{[\gamma \setminus i,j]}^{[\gamma]^\ast} }{\sqrt{- \lambda \det \Y_{[\gamma \setminus i]}^{[\gamma \setminus i]^\ast}}} \,,
				\ee
				where we used that the sign of $\det \Y_{[\gamma\setminus  i  ]}^{[\gamma \setminus i]^\ast}$ is opposite to that of $\det \Y_{[\gamma,j  ]}^{[\gamma,j]^\ast}$ for all $i \in \gamma$ and $j \in G /\gamma$.
				We can trivially extend the sum to include $j=i$, by setting $\det \Y_{[\gamma \setminus i,i]}^{[\gamma]^\ast}=\det \Y_{[\gamma]}^{[\gamma]^\ast}=0$, and recognize the Laplace expansion of the determinant to write
				\be
				   \RR_{i} = \frac{1}{\partial_{\alpha_i} \V_{n\text{-gon}}^\ast}  = \frac{- \lambda \, \U_{G / \gamma}^\ast}{ \sqrt{- \lambda \det \Y_{[\gamma \setminus i]}^{[\gamma \setminus i]^\ast}}} \,.
				\ee
				Plugging everything back into \eqref{eq:subleading-expansion}, we find for $\alpha$-positive
				singularities of the form $\Delta=(p_{i_0 j_0}^2-(p_{i_0 j_0}^\ast)^2)$:
    \newpage
				\begin{align}
				\label{eq:one-loop}
        		\I_{n\text{-gon}} \approx& \frac{ (2\pi i)^{(\E_{G/\gamma}-1)/2}}{\left((-i)^{\E_{G/\gamma}-1}\lambda \right)^{1/2}\left[\sum_{j \in G/\gamma } ( \lambda \det \Y^{[\gamma,j]^\ast}_{[\gamma,j]})^{1/2}\right]^{\D-\E_{G/\gamma}-\E_\gamma} \!\prod_{i\in \gamma} \!\left[- (-\lambda \det \Y_{[\gamma \setminus i]}^{[\gamma \setminus i]^\ast}) ^{1/2} \right]} \nn \\
        		&\times
        		\begin{dcases}
        		\Gamma(-\upgamma) \left[-\lambda (-1)^{i_0+j_0} \det \Y_{[\gamma,i_0]}^{[\gamma,j_0]^\ast} 
        			(p_{i_0 j_0}^2-(p_{i_0 j_0}^\ast)^2)
        			\right]^\upgamma &\text{for } \upgamma < 0,\\
        			- \log \left[
        			(p_{i_0 j_0}^2-(p_{i_0 j_0}^\ast)^2) \right] \quad &\text{for } \upgamma = 0,
        		\end{dcases}
        		\end{align}
        		with $\upgamma = (\D-\E_{G/\gamma}-1)/2$ and $\lambda = \sgn \det \Y_{[\gamma,i]}^{[\gamma,i]}$.

				\paragraph{Bubble diagram.}
				
				As a cross-check, let us verify that \eqref{eq:one-loop} agrees with \eqref{eq:banana} in the special case of the leading singularity for the bubble diagram with $\E_{G} = 2$ and $\gamma = \varnothing$. The $\Y$ matrix is simply
				\be
				\Y = \begin{pmatrix}
                -2m_1^2 & s - m_1^2 - m_2^2\\
                s - m_1^2 - m_2^2 & -2m_2^2
                \end{pmatrix},
				\ee
				whose determinant gives the Landau discriminant
				\be
				\Delta = \det \Y = - \left[s - (m_1 + m_2)^2\right] \left[s - (m_1 - m_2)^2\right].
				\ee
				As we are interested in the first branch, near the normal threshold we can simplify it to
				\be
				\Delta = 4 m_1 m_2 [(m_1 + m_2)^2 - s ].
				\ee
				According to \eqref{eq:alpha-ast}, the $\alpha$-positive solution for the Schwinger parameters is
				\be
				(\alpha_1^\ast : \alpha_2^\ast ) = 
				(\sqrt{2} m_2: \sqrt{2} m_1) = (\tfrac{1}{m_1} : \tfrac{1}{m_2})\,.
				\ee
				The determinant of the submatrices obtained by removing rows and columns in $\Y$ is simply
				\be 
				    \sum_{j=1}^{2} \left(- \det \Y_{[j]}^{[j]} \right)^{1/2}
				    =
				    \sqrt{2}(m_1+m_2)\,,
				\ee 
				since $\lambda = \sgn \det \Y_{[j]}^{[j]}=-1$.
				The formula \eqref{eq:one-loop} therefore gives in $\D$ dimensions,
				\be
				    \I_{\text{bub}} \approx 
				    \frac{\sqrt{\pi} \, (m_1 m_2)^\rho }{ (m_1 + m_2)^{\D-2}} \Gamma(-\rho)  [(m_1 + m_2)^2 - s]^{\rho},
				\ee
				with $\rho=\frac{\D-3}{2}$, which is in agreement with \eqref{eq:banana}.
				
				\paragraph*{\bf Triangle diagram.}
				We can also check the asymptotic expansion around Landau loci on the triangle examples from Sec.~\ref{sec:ExampleII}. Setting the external masses to $M$ and the internal ones to $m$, we obtain
				\be 
				    \Y = \begin{pmatrix}
                    -2 m^2 & u -2  m^2 & M^2- 2 m^2\\
                    u - 2 m^2 & -2m^2 & M^2- 2 m^2\\
                    M^2- 2 m^2 & M^2- 2 m^2 & - 2 m^2\\
                    \end{pmatrix},
				\ee 
				and the Landau discriminant for the leading singularity is therefore
				\be
				    \Delta_{\triangle} = \det \Y =  2 u \left( m^2 u - 4 m^2 M^2 + M^4 \right) \,.
				\ee
				The solution at $u^\ast = -\frac{M^2(M^2-4m^2)}{m^2}$, or $s=s_\triangle = \frac{M^4}{m^2}-t$ is the one corresponding to $\alpha$-positive singularities, which are
				\be 
				    \left( \alpha_1^\ast : \alpha_2^\ast : \alpha_3^\ast \right) = \frac{M\sqrt{M^2-4 m^2}}{m^2}  \left( m^2 : m^2 : M^2{-}2m^2 \right).
				\ee 
				Since $\upgamma=0$ for this diagram, we can use~\eqref{eq:one-loop} to write
				\be \label{eq:triangle-thr}
				    \I_{\triangle} \approx -\frac{2 \pi i \,  m^2}{M^3 \sqrt{M^2-4 m^2}} \log \left[s_\triangle-s \right] \,.
				\ee 
				We can compare this expression to the one obtained by expanding the full form in~\eqref{eq:triLHP} around the branch point at $s - s_\triangle$, and find agreement with \eqref{eq:triangle-thr}  to leading order.
				
				\paragraph{Box diagram.}
				The box diagram studied in Sec.~\ref{sec:ExampleI}, with all equal internal masses $m$ and external masses set to zero corresponds to
				\be 
				     \Y = \begin{pmatrix}
                    -2 m^2 & -2  m^2 & s- 2 m^2 & -2 m^2\\
                    -2 m^2 & -2  m^2 & - 2 m^2 & u-2 m^2\\
                    s-2 m^2 & -2  m^2 & - 2 m^2 & -2 m^2\\
                    -2 m^2 & u-2  m^2 & - 2 m^2 & -2 m^2\\
                    \end{pmatrix} \,.
				\ee 
				The leading singularity at $\det \Y=0$ corresponds to $su(   s u + 4 m^2 t ) = 0$,
				and as discussed in Sec.~\ref{sec:ExampleI}, it has positive $\alpha$ whenever $s>4 m^2$ and $t = -s-u < -16 m^2$. For the singularity at $s-s_\boxx$ with $s_\boxx = \frac{4 m^2 u}{u-4 m^2}$, the expansion in~\eqref{eq:one-loop} gives in $\D=4$
				\be 
				    \I_{\boxx} \approx 
				    \frac{2 \pi^2 i}{u} \frac{1}{\sqrt{s_\boxx-s}}\,.\vspace{-0.4em}
				\ee
				This expression agrees with the limit of \eqref{eq:box-final} and direct numerical integration.

\newpage
\section{Conclusion}

In this work we asked whether S-matrix elements can be complexified in a way consistent with causality. While causality leaves many fingerprints on the analytic structure of the S-matrix, the most fundamental one is how to define the original matrix element $\T$ as a boundary value of an analytic function $\T_\C$. A practical handle on this question is provided by perturbation theory, where one can simply ask what are the consequences of decorating the propagators with the Feynman $i\eps$ factors. While in the loop-momentum space this problem looks daunting and riddled with subtleties, it becomes rather clear-cut in the worldline formalism. We formulated it in terms of algebraic criteria for the worldline action $\V$ and employed them to consistently ask which deformation of the external kinematic variables has the same effect as the Feynman $i\eps$. We showed that for $2\to 2$ scattering involving external unstable particles, or higher-multiplicity processes, one generically cannot expect to be able to define $\T_\C$ with branch cuts on the real axis. Moreover, there might exist an overall discontinuity across the real axis that grows with the decay widths of the scattered particles. Instead of insisting on branch cuts running along the real axis, we give a prescription for deforming them, in a way consistent with causality, without spoiling any other physical property. In the process, we closed some gaps in the literature that previously relied on the existence of the Euclidean region, which is not present for large momentum transfers or when massless particles are involved. These difficulties are rather expected in the context of broader discussion of anomalous thresholds, which are inevitable consequences of unitary in any quantum field theory. We would like to highlight a few specific directions where immediate progress can be made (but might require another \pageref*{LastPage} pages).

\paragraph{The case of unstable particles.}
Studying analytic properties of S-matrix elements with unstable particles is long overdue, especially given their importance in collider experiments. In this work, we outlined some aspects of the usual intuition that can break down, though we have by no means explored them exhaustively. Alternatively, we can think of such matrix elements as being embedded in a higher-point process, in which case we are probing analyticity of $5$-point amplitudes and higher.

In addition to applying these techniques to more realistic processes, such as $\mathrm{ZZ}\to \mathrm{ZZ}$, there are some concrete theoretical questions that can be asked.
For instance, the external-mass discontinuity due to the unstable particles has to become weaker and weaker as the decay widths decrease. Can we put bounds on this discontinuity to quantify the errors we would be making in always approaching the $s$-channel from the upper-half plane?

\paragraph{Analyticity in multiple Mandelstam invariants.}
One of the key proposals made in this work is that branch cut deformations allow one to reveal the physical sheet, where a causal amplitude is defined, in the neighborhood of the real $s$-axis, as in Fig.~\ref{fig:generic}. This would naively suggest that one can connect every non-singular point to any other by analytic continuation.
It is interesting to ask what obstruction could cause this to fail.
The most obvious one is that there could exist two singularities with mutually-exclusive $i\eps$ prescriptions. In fact, examples of this phenomenon are known \cite{Chandler:1969nd}. What saves the day is that one can also vary the Mandelstam invariant $t$ to resolve such singularities. It would be interesting to find examples where this cannot be done, or prove they do not exist.

Likewise, we have studied analyticity in the single variable $s$ for simplicity, even though, for the full understanding one needs to explore the two-complex-dimensional plane $(s,t)$. Along this vein, it would be especially interesting to find the analogue of the parametric formula for $\Disc_s \I$ given in \eqref{eq:DiscI} for the double-spectral function $\Disc_t \Disc_s \I$.
For $5$-point processes---where Mandelstam invariants are still not constrained by Gram conditions---one should explore whether a concrete statement can be made about the kinematic $i\eps$ prescription with only stable (or even, only massless) external particles, as an extension of Sec.~\ref{sec:analyticity-stable}. In these cases, one already has to take anomalous thresholds into account.

\paragraph{Generalized dispersion relations.}
As we have seen in Sec.~\ref{sec:dispersion}, a first-principles derivation of dispersion relations can be achieved with a trick: deriving the contour deformations in the $s$-plane at the level of the Schwinger-parametric \emph{integrand}, as opposed to integral.
It allows us to write down a generalization to any linear combination of Mandelstam invariants $z$.
This procedure gives a new handle on studying analyticity in the upper-half $z$-plane.

\paragraph{Threshold expansion.}
In Sec.~\ref{sec:fluctuations}, we demonstrated that studying fluctuations around classical saddle-point contributions to a scattering process is closely tied to the local behavior of Feynman integrals near their thresholds. Further, we showed that the assumption of analyticity leads to a bound on the type of singularities that can occur at thresholds. This analysis relied on some technical assumptions, and in particular was confined to isolated saddle points. However---especially when massless particles are involved---one often encounters continuous families of saddles. It would be fascinating to extend the analysis of Sec.~\ref{sec:fluctuations} to these cases. Here, one can use saddle-point localization in the ``orthogonal'' directions, but is left with genuine integrals in the directions ``along'' the saddle manifold. Our expectation is that the latter can at worst contribute logarithmic or power-law divergences in $\Delta$. Hence, we predict that the divergence close to the Landau discriminant $\Delta =0$ is always proportional to
\be
\Delta^\rho\, \log^{\sigma} \!\Delta.
\ee
In a fully-massless scattering process, one needs to take additional care with projectivity of the kinematic space, i.e., treat $(s : t)$ as projective variables.

Likewise, it is not unlikely that one can employ the strategy from Sec.~\ref{sec:fluctuations} to obtain a series expansion around $\Delta = 0$ beyond the leading order. In the same spirit, it would be interesting to study the Regge-limit asymptotics, $s \to \infty$, using saddle-point methods. This limit localizes on the critical points determined by
\be
\partial_{\alpha_e} \partial_s \V = 0
\ee
for internal edges $e$. In other words, the above conditions give a degenerate limit of Landau equations as $s \to \infty$.

\acknowledgments
We thank Nima Arkani-Hamed, Lance Dixon, Lorenz Eberhardt, Alfredo Guevara, Enrico Herrmann, Aaron Hillman, Andrew McLeod, Julio Parra-Martinez, Lecheng Ren, Matthew Schwartz, Marcus Spradlin, Simon Telen, Cristian Vergu, Anastasia Volovich, and Akshay Yelleshpur for useful discussions.
H.S.H. gratefully acknowledges support from the Simons Bridge Fellowship (816048).
S.M. gratefully acknowledges the funding provided by Frank and Peggy Taplin, as well as the grant DE-SC0009988 from the U.S. Department of Energy.

\appendix

\newpage
\section{\label{app:parametric}Review of Schwinger parametrization}

In this appendix we derive the Schwinger-parametric formula for an arbitrary Feynman integral, starting from its loop-momentum representation. We explain two separate routes that lead to the different expressions for the Symanzik polynomials $\mathcal{U}$ and $\mathcal{F}$, as well as the numerators $\mathcal{N}$ and $\widetilde{\mathcal{N}}$. They make use of two types of Schwinger tricks: bosonic and fermionic ones. Either representation might be more convenient depending on the specific application.

\subsection{Notation and review}

Recall from \eqref{eq:general-I} that a Feynman integral $\I$ is given by
\be\label{eq:general-I2}
\I = (-i\hbar)^{-d} \lim_{\eps \to 0^+} \int_0^\infty \frac{\d^{\E} \alpha}{\U^{\D/2}}\, \N\, \exp \left[ \frac{i}{\hbar} \left(\V + i\eps \textstyle\sum_{e=1}^{\E}\alpha_e \right)\right],
\ee
where $\E$ is the number of internal edges, $\L$ is the number of loops, and $\D$ is the space-time dimension. The action $\V = \F/\U$ is expressed in terms of the two Symanzik polynomials:
\be\label{eq:U2}
\U = \sum_{\substack{\mathrm{spanning}\\ \mathrm{trees }\,T}} \prod_{e \notin T} \alpha_e,
\ee
determined as a sum over spanning trees, as well as
\be\label{eq:F2}
\F = \sum_{\substack{\mathrm{spanning}\\ \text{two-trees}\\ T_L \sqcup T_R}} p_L^2 \prod_{e \notin T_L \sqcup T_R} \alpha_e - \U \sum_{e=1}^{\E} m_e^2 \alpha_e,
\ee
written as a sum over spanning two-tress. They are homogeneous polynomials of degree $\L$ and $\L{+}1$ respectively. See Sec.~\ref{sec:parametric} for more details and examples.

Finally, for Feynman integrals with non-trivial vertex interactions we have the numerator function $\N$. The main goal of this appendix is to explain how to obtain it algorithmically from the loop-momentum representation, which in principle is a textbook procedure, see, e.g., \cite[Ch.~2]{smirnov1991renormalization} or \cite[Ch.~2]{zavialov2012renormalized}. We base our discussion on the  streamlined derivation from \cite{AHHM}. This is particularly important for the conclusions of Sec.~\ref{sec:fluctuations}, which depend on the counting of the degree of $\N$. We will show that the most general numerator $\N$ can be written as a sum of rational functions with degrees $d_\N, d_\N{+}1, \ldots, 0$, where $d_\N \leq 0$.
The overall degree of divergence is therefore
\be
d = \E - \L\D/2 + d_{\N}.
\ee
In other words, adding a numerator can only make the diagram behave worse in the UV by decreasing $d$ compared to the scalar case.
After integrating out the overall scale, the Feynman integral \eqref{eq:general-I2} can be expressed as
\be\label{eq:I2}
\I = \Gamma(d) \int \frac{\d^{\E}\alpha}{\GL(1)} \frac{\widetilde{\N}}{\U^{\D/2} (-\V-i\eps)^{d}}, 
\ee
where $\widetilde{\N}$ is homogeneous with degree $d_{\N}$.

\subsection{Bosonic Schwinger tricks}

Our starting point will be a Feynman integral in the loop momentum space with $n$ external legs, $\E$ internal edges (propagators), $\L$ loops, and $\mathrm{V}$ vertices. In $\D$ space-time dimensions with Lorentzian signature it is written as
\be\label{eq:I}
\I =  \frac{(-1)^\E}{(i\pi^{\D/2})^{\L}} \lim_{\eps \to 0^+} \int \d^{\D\E} q\; \NN(q_e) \frac{\prod_{v=1}^{\mathrm{V}} \delta^\D(p_v + \sum_{e=1}^{\E} \eta_{ve} q_e) }{\prod_{e=1}^{\E} (q_e^2 - m_e^2 + i\eps)}.
\ee
Here $q_e$ and $m_e$ are the Lorentzian momentum and mass flowing through every edge $e= 1, 2, \ldots, \E$. For each vertex $v = 1,2,\ldots,\mathrm{V}$ we have a momentum-conserving delta function involving $p_v$, which denotes the total external momentum flowing into the vertex. If a given vertex $v$ does not have an external leg attached, we simply set $p_v = 0$. Here $\eta_{ve} = 1$ if the edge $e$ is directed towards $v$, $-1$ if it is directed away, and $0$ otherwise. The Feynman $i\eps$ enters the propagators and is sent to zero at the end of the computation. We also allow for a possibility of a numerator $\NN(q_e)$ which is taken to be a function of the internal momenta $q_e$ and other (external) parameters such as polarization vectors, gamma matrices and so on. We call its degree in the momenta $d_\NN$.

To each propagator we associate the Schwinger parameter $\alpha_e$ introduced using the identity
\be\label{eq:Schwinger-trick}
\frac{-1}{q_e^2 - m_e^2 + i\eps} = \frac{i}{\hbar} \int_{0}^{\infty} \d \alpha_e\, \exp\left[ \frac{i}{\hbar} (q_e^2 - m_e^2 + i\eps) \alpha_e \right].
\ee
From this perspective, the $i\eps$ ensures convergence as $\alpha_e \to \infty$.
Similarly, each delta function can be represented with
\be
\delta^\D(p_v + {\textstyle\sum}_{e=1}^\E \eta_{ve} q_e) = \frac{1}{(2\pi \hbar)^\D} \int_{\R^{1,\D-1}}\!\!\! \d^{\D} x_v\, \exp \bigg[\frac{i}{\hbar} (p_v + \sum_{e=1}^{\E} \eta_{ve} q_e)\cdot x_v \bigg],
\ee
where $x_v$ is the space-time position of the vertex $v$. They do, in principle, also need an $i\eps$ factor, but we suppress it from the notation for simplicity. The domains of integration will similarly be left implicit.

Similarly, we can represent the numerator factor with the help of auxiliary parameters $\beta_e^\mu$ as follows
\be
\NN(q_e) =  \NN (\tfrac{\partial}{\partial \beta_{e}}) \prod_{e=1}^{\E} \, \exp \left[ q_e \cdot \beta_e \right] \Big|_{\beta_e = 0},
\ee
where $\NN(\tfrac{\partial}{\partial \beta_{e}})$ on the right-hand side represents the same function but evaluated at $q_e^\mu \to \tfrac{\partial}{\partial \beta_{e,\mu}}$, which makes it a differential operator.
The parameters $\beta_e$ are always set to zero at the end of the computation and we will often omit this fact from the notation for readability.

Plugging in the above identities into $\I$ yields
\begin{align}\label{eq:A10}
\I = \;&c \lim_{\eps \to 0^+} \int \d^{\D\E} q\, \d^{\D\mathrm{V}} x \, \d^{\E}\alpha\, \NN(\tfrac{\partial}{\partial \beta_{e}}) \\
&\exp \left\{ \frac{i}{\hbar}\left[ \sum_{e=1}^{\E} \Big(q_e^2 \alpha_e + q_e {\cdot} ( \textstyle{\sum}_{v=1}^{\mathrm{V}} \eta_{ve} x_v - i\hbar\beta_e) - (m_e^2 - i\eps) \alpha_e\Big) + \displaystyle\sum_{v=1}^{\mathrm{V}} p_v {\cdot} x_v \right] \right\},\nn
\end{align}
where $c = 2^{-\D\VV}i^{\E-\L} \pi^{-(\L+2\VV)\D/2} \hbar^{-\E-\D\VV}$.
The goal is to integrate out the momenta $q_e$ and the positions of vertices $x_v$ and be left with only integrations over the Schwinger parameters $\alpha_e$. 

\begin{figure}
    \centering
    \includegraphics[scale=1.1]{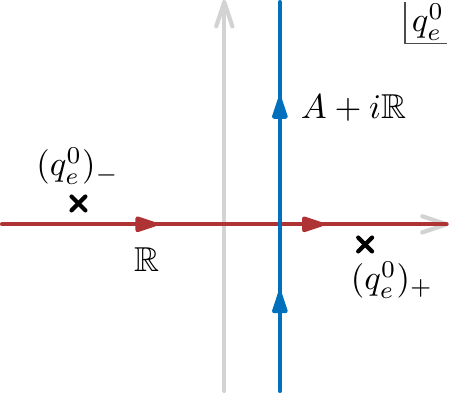}
    \caption{Wick rotation of the energy $q_e^0$ of the $e$-th internal particle. The original Lorentzian contour $\R$ (red) can be deformed into the Euclidean $A+i\R$ (blue) without crossing singularities, which are contained within the second and fourth quadrants for any values of Schwinger parameters.}
    \label{fig:Wick}
\end{figure}

We start with the integrals over the $q_e$'s, which look Gaussian. However, while the $i\eps$ factors guaranteed convergence for large $\alpha_e$'s, 
it is not clear that every integral over $q_e$ and $x_v$ converges. It is because in our conventions $q_e^2 = (q_e^0)^2 - \vec{q}_e^{\,2}$ and $x_v^2 = (x_v^0)^2 - \vec{x}_v^{\,2}$, so the timelike component naively looks like it might give a divergence. More precisely, this can happen in the integration over $q_e$ when each individual term in the sum over $e$ in \eqref{eq:A10} vanishes (since numerators are polynomial, we can set all $\beta_e = 0$ for these purposes). We thus have to look at the roots of $q_e^0$, which take the form
\be\label{eq:q0-roots}
(q_e^0)_\pm = A \pm \sqrt{B - i\eps}.
\ee
Here $A$ and $B$ are some functions of $\alpha_e$'s, $x_v$'s, and other variables that can be easily computed.
There are two options: either $B>0$ and both roots lie near the real axis on either side of $q_e^0 = A$, or $B<0$ and they lie near the imaginary line $A + i\R$ with positive and negative imaginary parts, see Fig.~\ref{fig:Wick}. These two directions, intersecting at $q_e^0 = A$, divide the complex $q_e^0$-plane into four quadrants. It is straightforward to see that the $i\eps$ deformation in \eqref{eq:q0-roots} moves the roots to the second and fourth quadrant. For $B>0$ we have $\Im\, (q_e^0)_\pm = \mp \tfrac{\eps}{2\sqrt{B}} + \ldots$, while for $B<0$ we get $\Re\, (q_e^0)_\pm = A \mp \tfrac{\eps}{2\sqrt{-B}} + \ldots$. If there are no UV divergences as $q_e^0$ tends to infinity, we can thus perform the Wick rotation: smoothly deform the $q_e^0$ contour from $\R$ counterclockwise to $A + i \R$ for any value of $\alpha_e$'s, $x_v$'s, and the remaining parameters. We might thus treat the internal energies $q_e^0 = A + i q_{e,E}^0$ as Euclidean $q_{e,E}^0 \in \R$, at the cost of picking up a Jacobian factor of $i$ for every $e = 1,2,\ldots,\E$.

The integrals over the Euclidean $q_e^\mu$ are Gaussian and diagonal, and straightforwardly give\footnote{
Recall that the normalization of Gaussian integrals works out to be:
		\be
		\int_{\R^d} \!\!\!\d^d \mathbf{x}\, e^{\frac{i}{\hbar} \left(\mathbf{x}^\intercal \mathbf{A}\, \mathbf{x} + \mathbf{x}^\intercal \mathbf{B} \right)} = \sqrt{\frac{(i\pi\hbar)^{d}}{\det \mathbf{A}}} e^{-\frac{i}{4\hbar} \mathbf{B}^\intercal \mathbf{A}^{-1} \mathbf{B}}
		\ee
		for a positive-definite
		matrix
		$\mathbf{A}$.
	}
\begin{align}\label{eq:int2}
\I = \,& c' \lim_{\eps \to 0^+} \int \!\!\frac{\d^{\D\mathrm{V}} x \, \d^{\E}\alpha}{(\prod_{e=1}^{\E} \!\alpha_e)^{\D/2}}\, \NN(\tfrac{\partial}{\partial \beta_{e}})\\
& \times \exp \left\{ \frac{i}{\hbar}\left[\sum_{e=1}^{\E} \left(- \frac{(\sum_{v=1}^{\mathrm{V}} \eta_{ve} x_v - i\hbar\beta_e)^2}{4\alpha_e} - (m_e^2 - i\eps) \alpha_e \right) + \sum_{v=1}^{\mathrm{V}} p_v {\cdot} x_v \right] \right\} ,\nn
\end{align}
where $c' = 2^{-\D\VV}(-1)^\E i^{-\E \D/2-\L} \pi^{-(\L-\E+2\VV)\D/2} \hbar^{\D( \E/2 -\VV)-\E}$.
At this stage it is useful to introduce matrix notation in which
\begin{align}
\I = \,&c'\lim_{\eps \to 0^+}\int \frac{\d^{\D\mathrm{V}} x \, \d^{\E}\alpha}{(\prod_{e=1}^{\E}\! \alpha_e)^{\D/2}}\, \NN(\tfrac{\partial}{\partial \beta_{e}}) \\
& \times \exp \left\{ \frac{i}{\hbar}\left[ - \tfrac{1}{4}\mathbf{x}^\intercal\, \mathbf{L}\, \mathbf{x} + \mathbf{x}^\intercal (\mathbf{p} + \tfrac{i\hbar}{2}\mathbf{M}\, \bm{\beta}) + \tfrac{\hbar^2}{4}\bm{\beta}^\intercal\, \mathbf{N}\, \bm{\beta} - \sum_{e=1}^{\E} (m_e^2 - i\eps) \alpha_e \right] \right\},\nn
\end{align}
where
\be
\mathbf{x} = (x_1, x_2, \ldots, x_\VV), \qquad \mathbf{p} = (p_1, p_2, \ldots, p_\VV), \qquad \bm{\beta} = (\beta_1, \beta_2, \ldots, \beta_{\E})
\ee
are vectors of the relevant variables and the matrices are given by (Lorentz index contractions are unambiguous and hence suppressed from the notation)
\be\label{eq:LMN}
\mathbf{L}_{v v'} = \sum_{e=1}^{\E} \frac{\eta_{ve} \eta_{v' e}}{\alpha_e}, \qquad \mathbf{M}_{ve} = \frac{\eta_{ve}}{\alpha_e}, \qquad \mathbf{N}_{e e'} = \frac{\delta_{ee'}}{\alpha_e}.
\ee
Here $\mathbf{L}$ is the Laplacian matrix for the Feynman diagram. Because of momentum conservation there is a zero mode corresponding to the length-$\mathrm{V}$ null vector $(1,1,\ldots,1)$ in $\mathbf{x}$ for $\mathbf{L}$ and $\mathbf{M}$ (from the left). Doing the integration over the linear terms in $\mathbf{x}$ on the support of this null vector gives the overall momentum conservation delta function $(2\pi \hbar)^{\D} \delta^\D({\textstyle\sum}_{v=1}^{\mathrm{V}} p_v)$. After integrating out the remaining
$x_v^\mu$'s, accounting for the Jacobian $(-i)^{\mathrm{V}-1}$ for the Wick rotation of the zero component, we obtain:
\begin{align}
\I = \,& (-i\hbar)^{\L\D/2-\E} \delta^\D({\textstyle\sum}_{v=1}^{\mathrm{V}} p_v) \lim_{\eps \to 0^+} \int \frac{\d^{\E}\alpha}{(\prod_{e=1}^{\E}\! \alpha_e)^{\D/2}({\det}^\prime \mathbf{L})^{\D/2}}\,\NN(\tfrac{\partial}{\partial \beta_{e}})\,\\
	& \exp \left\{ \frac{i}{\hbar}\left[(\mathbf{p}+ \tfrac{i\hbar}{2}\mathbf{M}\,\bm{\beta} )^\intercal\, \mathbf{L}^{-1} (\mathbf{p}+ \tfrac{i\hbar}{2}\mathbf{M}\,\bm{\beta}) + \tfrac{\hbar^2}{4} \bm{\beta}^\intercal\,\mathbf{N}\,\bm{\beta} - \sum_{e=1}^{\E}  (m_e^2 - i\eps) \alpha_e  \right] \right\} ,\nn
\end{align}
where we used $\VV - \E + \L = 1$ to simplify the overall normalization.
From now on $\mathbf{L}$ denotes the \emph{reduced} Laplacian, obtained by removing the last column and row from $\mathbf{L}$, making its inverse and determinant (denoted by ${\det}^\prime \LL$) well-defined. Similarly, contraction with $\mathbf{p}$ and $\mathbf{M}$ is only performed over the indices
$v,v'=1,2,\ldots,\mathrm{V}{-}1$. In the following we will omit the explicit momentum-conservation delta function from the notation.
In order to identify the individual ingredients in the integrand, it is useful to rewrite it as
\be
\I = (-i\hbar)^{\L\D/2-\E} \lim_{\eps \to 0^+} \int \frac{\d^{\E}\alpha}{\U^{\D/2}}\, \NN(\tfrac{\partial}{\partial \beta_{e}})\, \exp \left\{ \frac{i}{\hbar} \left[ \frac{\F_0 + \F_1 + \F_2}{\U} - \sum_{e=1}^{\E} (m_e^2 - i\eps) \alpha_e \right] \right\}.
\ee
In the exponent, we have separated terms $\F_i$, depending on their homogeneity degrees in the $\beta_e$'s. Explicitly,
\begin{gather}\label{eq:F01}
\F_0 = \U\!\! \sum_{1 \leq i, j < \VV} \!\!\! p_i {\cdot} p_j\, \LL^{-1}_{ij},\qquad \F_1 = i\hbar\,\U\!\! \sum_{\substack{1 \leq i < \VV\\ 1\leq e \leq \E}} p_i{\cdot}\beta_e (\mathbf{L}^{-1} \mathbf{M})_{ie},\\
\F_2 = -\tfrac{\hbar^2}{4}\,\U \!\!\sum_{1\leq e, f \leq \E} \!\!\!\beta_e {\cdot} \beta_f\, (\mathbf{M}^\intercal \mathbf{L}^{-1} \mathbf{M} - \mathbf{N})_{ef}.\label{eq:F22}
\end{gather}
In particular, the Symanzik polynomials can be identified as
\be\label{eq:U1}
\U = ({\textstyle\prod}_{e=1}^{\E} \alpha_e)\, {\det}^\prime \mathbf{L}, \qquad \F = \F_0 - \U \sum_{e=1}^{\E} m_e^2 \alpha_e.
\ee
The equivalence with the definitions \eqref{eq:U2} and \eqref{eq:F2} follows from a simple application of the matrix-tree theorem; see, e.g., \cite{AHHM}. We will return back to the Symanzik polynomials in App.~\ref{app:practical}, where we will provide practical formulae that can be quickly evaluated on a computer.

This shows \eqref{eq:general-I2} for scalar integrals, i.e., $\mathrm{N}=1$ and all $\beta_e = 0$. 
It now remains to derive an expression for the numerator $\N$ in \eqref{eq:general-I}. Since the dependence on $\beta_e$'s comes solely from $\F_1$ and $\F_2$, we can read off
\be\label{eq:tildeN}
\boxed{\N = (-i\hbar)^{d_\N}\, \NN(\tfrac{\partial}{\partial \beta_{e}})\, \exp \left\{ \frac{i}{\hbar} \left[ \frac{\F_1 + \F_2 }{\U} \right] \right\} \Bigg|_{\beta_e = 0}.}
\ee
From their definitions, $\F_1$ and $\F_2$ have degrees $\L$ and $\L{-}1$ in $\alpha_e$'s respectively. The numerator $\NN(\tfrac{\partial}{\partial \beta_{e}})$ with degree $d_\NN$ can be written as
\be
\NN(\tfrac{\partial}{\partial \beta_{e}}) = \sum_{k=0}^{d_\NN} \NN_k(\tfrac{\partial}{\partial \beta_{e}}),
\ee
where each term $\NN_k$ is a homogeneous polynomial in $\tfrac{\partial}{\partial \beta_{e}}$ with degree $k$.
Therefore \eqref{eq:tildeN} has to take the form
\be
\N = (-i\hbar)^{d_{\N}} \sum_{k=0}^{d_\NN} \sum_{\ell=0}^{\lfloor k/2 \rfloor} (-i\hbar)^\ell \N_{k,\ell}\, \U^{\ell - k},
\ee
where each $\N_{k,\ell}$ is a term containing exactly $\ell$ powers of $\F_2$ and $k {-} 2\ell$ powers of $\F_1$, meaning that it is a homogeneous polynomial with degree $k \L - \ell(\L{+}1)$. Therefore, each term in the sums has degree $-\ell$.
Furthermore, the normalization $(-i\hbar)^\ell$ ensures that each $\N_{k,\ell}$ is real and independent of $\hbar$. To conclude, $\N$ given in \eqref{eq:tildeN} can be written as a sum of homogeneous functions with degrees
\be
d_{\N} \leq - \ell \leq 0,
\ee
where the lowest degree is given by $d_\N = - \lfloor d_\NN / 2 \rfloor$.

Finally, we use the fact that all the factors above have simple covariance properties under the rescaling of all Schwinger parameters, $\alpha_e \to \lambda \alpha_e$ with $\lambda > 0$, namely
\be\label{eq:lambda-scaling}
\U \to \lambda^\L\, \U, \qquad \F \to \lambda^{\L +1} \F, \qquad \d^{\E} \alpha = \lambda^{\E-1} \frac{\d^{\E}\alpha}{\GL(1)} \, \d\lambda,
\ee
where, as before, the $\GL(1)$ redundancy is fixed by setting, say, $\alpha_{\E}=1$. Hence, the integral $\I$ can be written as
\begin{align}
\I = (-i\hbar)^{-d+d_\N} \lim_{\eps \to 0^+} &\int \frac{\d^{\E}\alpha/\GL(1)}{\U^{\D/2}}\, \sum_{k=0}^{d_\NN} \sum_{\ell=0}^{\lfloor k/2\rfloor} (-i\hbar)^\ell \N_{k,\ell}\, \U^{\ell-k} \\
\times &\int_0^{\infty} \!\!\d\lambda \lambda^{d - d_{\N} -\ell -1} e^{ \frac{i}{\hbar} \lambda \left(\V + i\eps \sum_{e=1}^{\E} \alpha_e \right)},\nn
\end{align}
where $d = \E - \L\D/2 + d_\N$. The integral in the second line evaluates to
\be
(-i\hbar)^{d - d_{\N} -\ell} \frac{\Gamma(d - d_{\N} -\ell)}{(-\V - i\eps)^{d - d_{\N} -\ell}} \,.
\ee
Reorganizing this expression yields
\be
\I = \Gamma(d) \lim_{\eps \to 0^+} \int \frac{\d^{\E}\alpha}{\GL(1)} \frac{\widetilde{\N}}{\U^{\D/2} (-\V-i\eps)^{d}}, 
\ee
with
\be
\widetilde{\N} =  \sum_{k=0}^{d_\NN} \sum_{\ell=0}^{\lfloor k/2\rfloor} \frac{\Gamma(d - d_\N - \ell)}{\Gamma(d)}\, \N_{k,\ell}\,\U^{\ell - k} (-\V - i\eps)^{d_\N + \ell}.
\ee
Note that the Gamma functions in the sum might introduce negative powers of $\epsilon$ in dimensional regularization $\D = 4-2\epsilon$. The resulting $\widetilde{\N}$ is a homogeneous function, but, in this convention, not necessarily a polynomial.
Its degree is
\be
d_{\widetilde\N} = d_\N,
\ee
making the integrand projective. This is precisely the parametric representation from \eqref{eq:I2}.

\subsection{Fermionic Schwinger tricks}

Here we provide another way of deriving the parametric form using fermionic Schwinger tricks \cite{AHHM}. We start with the representation \eqref{eq:I} after resolving the delta functions, which results in an integral purely in terms of the loop momenta of the form
\be
\I = \lim_{\eps \to 0^+} \int \prod_{a=1}^{\L}\frac{\d^\D \ell_a}{i\pi^{\D/2}}\, \frac{ (-1)^\E \NN(\ell_a)}{\prod_{e=1}^{\E} (q_e^2 - m_e^2 + i\eps)}.
\ee
We will express the numerator factor as a product of terms
\be
\NN(\ell_a) =  \prod_{b=1}^{\mathrm{B}} \mathsf{N}_{b}(\ell_a),
\ee
where each $\mathsf{N}_{b}$ is at most quadratic in the loop momenta and therefore can be parametrized as follows:
\begin{align}
\mathsf{N}_b(\ell_a) &= \Bell^{\intercal} \mathbf{Q}_b \Bell + 2\Bell^\intercal \mathbf{L}_b + c_b\\
&= \sum_{a,c=1}^{\L} (\mathbf{Q}_b)_{ac}\, \ell_a {\cdot} \ell_c + 2\sum_{a=1}^{\L} (\mathbf{L}_b)_a {\cdot} \ell_a + c_b\nn.
\end{align}
Here $\Bell = (\ell_1, \ell_2, \ldots, \ell_{\L})$ together with an $\L \times \L$ matrix $\mathbf{Q}_b$ and a length-$\L$ vector $\LL_b^\mu$ of Lorentz vectors. Lorentz index contractions are spelled out in the second line, but will, from now on, be suppressed from the notation for readability. For example, a numerator term $\mathsf{N}_b = \ell_1^\mu$ might be represented as $\mathbf{Q}_b = c_b = 0$ and $\mathbf{L}_b = (\eta^{\mu\nu}, 0, 0, \ldots, 0)$, where the index $\nu$ is contracted with $\Bell$.

At this stage we can apply the identity \eqref{eq:Schwinger-trick} together with its fermionic analogue:
\be
\mathsf{N}_b(\ell_a) = \int \d\beta_b \d\bar{\beta}_b\, e^{ \beta_b \bar{\beta}_b \mathsf{N}_b(\ell_a)},
\ee
where $\beta_b$, $\bar{\beta}_b$ are pairs of Grassmann variables. This gives us
\be
\label{eq:I-Schwinger-2}
\I = \frac{ i^{-\L}}{\pi^{\L\D/2} (-i\hbar)^{\E}} \lim_{\eps \to 0^+} \int \!\!\d^{\E}\alpha\, \d^{2\mathrm{B}}\beta\, \d^{\D\L} \ell\, \exp \left\{ \! \frac{i}{\hbar} \left[ \sum_{e=1}^{\E} \alpha_e (q_e^2 {-} m_e^2 {+} i\eps) - i\hbar\sum_{b=1}^{\mathrm{B}} \beta_b \bar{\beta}_b \mathsf{N}_b \right] \! \right\} \!.
\ee
Since the argument of the exponential is quadratic, the term in the square brackets can be written as
\be\label{eq:I-Schwinger}
\left[ \cdots \right] = \Bell^{\intercal} \mathbf{Q} \Bell + 2\Bell^\intercal \mathbf{L} + c + i\eps \sum_{e=1}^{\E} \alpha_e,
\ee
where we can separate the terms depending on the numerators as
\be
\mathbf{Q} = \mathbf{Q}_0 - i\hbar\sum_{b=1}^{\mathrm{B}} \beta_b \bar{\beta}_b \mathbf{Q}_b, \quad \mathbf{L} = \mathbf{L}_0 -  i\hbar\sum_{b=1}^{\mathrm{B}} \beta_b \bar{\beta}_b \mathbf{L}_b, \quad c = c_0 -  i\hbar\sum_{b=1}^{\mathrm{B}} \beta_b \bar{\beta}_b c_b.
\ee
The Gaussian integral over the loop momenta can be easily performed after Wick rotation similar to the one performed around Fig.~\ref{fig:Wick}, giving an additional factor of $i^\L$. This results in
\be
\I =  (-i\hbar)^{\L\D/2-\E} \lim_{\eps \to 0^+} \int  \frac{\d^{\E}\alpha\, \d^{2\mathrm{B}}\beta}{(\det \mathbf{Q})^{\D/2}}\, \exp \left\{ \frac{i}{\hbar} \left[ - \mathbf{L}^\intercal \mathbf{Q}^{-1} \mathbf{L} + c + i\eps {\textstyle\sum}_{e=1}^{\E} \alpha_e\right] \right\}.
\ee
By comparison with \eqref{eq:I2}, we can identify the $\beta$-deformed Symanzik polynomials
\be
\det \mathbf{Q} = \U_{\beta}, \qquad - \mathbf{L}^\intercal \mathbf{Q}^{-1} \mathbf{L} + c = \frac{ \F_\beta}{\U_\beta} = \V_\beta,
\ee
such that $\U_0 = \U$ and $\F_0 = \F$. Hence, the scalar case with $\B=0$ reduces to \eqref{eq:I2}. Once again, we can read-off the numerator
\be
\boxed{
\N =  (-i\hbar)^{d_\N} \int \d^{2\B} \beta \left(\frac{\U}{\U_\beta}\right)^{\D/2} \exp \left\{ \frac{i}{\hbar} \big[\V_\beta - \V\big] \right\}.}
\ee

At this stage we can integrate out the overall scale $\lambda$ by performing the scaling
\be
\alpha_e \to \lambda \alpha_e, \qquad \beta_b\bar{\beta}_b \to \lambda \beta_b\bar{\beta}_b.
\ee
The deformed Symanzik polynomials retain the same scaling as in \eqref{eq:lambda-scaling} and we only need to be careful with the additional Berezinian coming from the fermionic measure,
\be
\d^{2\mathrm{B}} \beta \to \lambda^{-\mathrm{B}}\, \d^{2\mathrm{B}} \beta.
\ee
This leaves us with
\be
\I = (-i\hbar)^{-d+d_\N} \lim_{\eps \to 0^+} \int  \frac{\d^{\E}\alpha\, \d^{2\mathrm{B}}\beta / \GL(1)}{\U_\beta^{\D/2}} \int_0^\infty \d\lambda\, \lambda^{d - d_{\N} - \mathrm{B}-1} e^{\frac{i}{\hbar} \lambda [\V_\beta + i\eps \sum_{e=1}^{\E} \alpha_e ] },
\ee
where the final integral evaluates to
\be
(-i\hbar)^{d - d_\N - \mathrm{B}} \frac{\Gamma(d - d_\N - \mathrm{B})}{\left(-\V_\beta - i\eps\right)^{d - d_\N - \mathrm{B}}}.
\ee
In practice, performing the fermionic integration is equivalent to taking first derivatives in all $\beta_b\bar{\beta}_b$ (treated as a single bosonic variable) and setting them to zero at the end. Therefore, we can rewrite the above expression in the familiar form
\be
\I = \Gamma(d) \lim_{\eps \to 0^+} \int \frac{\d^{\E}\alpha}{\GL(1)} \frac{\widetilde{\N}}{\U^{\D/2} (-\V - i\eps)^{d}}, 
\ee
where $\U = \U_\beta|_{\beta_b\bar{\beta}_b = 0}$, $\F = \F_\beta|_{\beta_b\bar{\beta}_b = 0}$, and the numerator $\N$ can be expressed as
\be
\widetilde{\N} =  (-i\hbar)^{-\B}\frac{\Gamma(d - d_\N - \B)}{\Gamma(d)}  \left(\prod_{b=1}^{\B}  \frac{\partial}{\partial \beta_b \bar{\beta}_b} \right) \frac{\U^{\D/2} (-\V {-} i\eps)^{d}}{ \U_\beta^{\D/2} (-\V_\beta - i\eps)^{d-d_\N - \B}} \bigg|_{\beta_b \bar{\beta}_b = 0},
\ee
which is an alternative representation for the numerator in \eqref{eq:I2}.

\subsection{\label{app:practical}Practical formulae}

The Laplacian matrix $\LL$ defined in \eqref{eq:LMN} can be alternatively constructed by starting with an empty matrix and adding a factor $1/\alpha_e$ to $\LL_{e_1 e_1}$ and $\LL_{e_2 e_2}$, as well as subtracting $1/\alpha_e$ from $\LL_{e_1 e_2}$ and $\LL_{e_2 e_1}$ for every edge $e = (e_1, e_2)$. Recall that we will be using only the reduced Laplacian obtained by removing, say, the $\VV$-th column and row. Note that $\LL$ is independent of the orientations of the edges.

The first Symanzik polynomial is already expressed in its simplest form
\be
\U = ({\textstyle\prod}_{e=1}^{\E} \alpha_e)\, {\det}^\prime \mathbf{L}.
\ee
The remaining manipulations will rely on using the identity
\be
({\det}^\prime \LL) \LL_{ij}^{-1} = (-1)^{i+j}\, {\det}^\prime \LL_{[i]}^{[j]},
\ee
where the right-hand side equals to the minor of $\LL$ obtained by removing columns $i$ and $\VV$, as well as rows $j$ and $\VV$, normalized by the sign $(-1)^{i+j}$. This straightforwardly leads to
\be
\F_0 = ({\textstyle\prod}_{e=1}^{\E} \alpha_e)\!\! \sum_{1 \leq i, j < \VV} \!\!\! (-1)^{i+j}\, p_i {\cdot} p_j\, {\det}^\prime \LL_{[i]}^{[j]}.
\ee
Similarly, we have
\be
\F_1 = i\hbar\, ({\textstyle\prod}_{e=1}^{\E} \alpha_e)\!\! \sum_{\substack{1 \leq i < \VV\\ 1\leq e \leq \E}} \frac{p_i{\cdot}\beta_e}{\alpha_e} \sum_{k=1}^{2} (-1)^{i+e_k+k}\, {\det}^\prime \LL_{[i]}^{[e_k]},
\ee
as well as
\be
\F_2 = -\tfrac{\hbar^2}{4}\,\U \!\!\sum_{1\leq e, f \leq \E} \!\frac{\beta_e {\cdot} \beta_f}{\alpha_e \alpha_f} \bigg[ \bigg( \sum_{k,\ell=1}^{2} (-1)^{e_k + f_\ell + k + \ell}\, {\det}^\prime \LL_{[e_k]}^{[f_\ell]} \bigg) - \delta_{ef} \alpha_e\, {\det}^\prime \LL\bigg],
\ee
where in both expressions we used the notation $e = (e_1, e_2)$, $f = (f_1, f_2)$. It is understood that terms involving vertex $\VV$ do not appear, i.e., ${\det}^\prime \LL_{[i]}^{[\VV]} = {\det}^\prime \LL_{[\VV]}^{[i]}= 0$. At this stage all the manipulations are entirely algorithmic and can be put on a computer.

\newpage
\addcontentsline{toc}{section}{References}
\bibliographystyle{JHEP}
\bibliography{references}
				
\end{document}